\documentclass[final,3p,times,fleqn]{elsarticle}
\usepackage{amssymb}
\usepackage{hyperref}
\usepackage{amsmath}
\usepackage{algorithm}
\usepackage{algorithmic}
\usepackage{multirow}
\usepackage{makecell}
\usepackage{booktabs} 
\usepackage{xcolor}
\usepackage{subcaption}

\newtheorem{theorem}{Thm}[section] 
\usepackage{amsmath,amssymb}
\usepackage{times}
\usepackage{setspace}
\usepackage{tabularx} 
\usepackage{contour}

\newcolumntype{L}[1]{>{\raggedright\let\newline\\\arraybackslash\hspace{0pt}}m{#1}}
\newcolumntype{C}[1]{>{\centering\let\newline\\\arraybackslash\hspace{0pt}}m{#1}}
\newcolumntype{R}[1]{>{\raggedleft\let\newline\\\arraybackslash\hspace{0pt}}m{#1}}

\usepackage{xcolor}
\usepackage[normalem]{ulem} 
\usepackage{xcolor}     
\definecolor{PRblue}{RGB}{0, 127, 172}  
\usepackage{url}        
\usepackage{bm}  
\usepackage{longtable} 
\renewcommand{\arraystretch}{1.15} 
\small 

\newcommand{\chapterdivider}[1]{%
  \hline
  \multicolumn{2}{l}{\textbf{#1}} \\
}

\newif\ifshowrev
\showrevfalse 

\ifshowrev
  \newcommand{\revdel}[2]{\textcolor{red}{\sout{#1}【C#2】}}
  \newcommand{\revadd}[2]{\textbf{\textit{#1【C#2】}}}

\else
  \newcommand{\revdel}[2]{}     
  \newcommand{\revadd}[2]{#1}    
\fi

\begin{document}
\begin{frontmatter}
\title{Predictability of Complex Systems}
\author[label1]{En Xu\fnref{fn1}}
\author[label2]{Yilin Bi\fnref{fn1}}
\author[label3]{Hongwei Hu}
\author[label3]{Xin Chen}
\author[label4,label5]{Zhiwen Yu} 
\author[label1]{Yong Li}
\author[label3]{Yanqing Hu}
\author[label2]{Tao Zhou}
\affiliation[label1]{organization={Department of Electronic Engineering, Tsinghua University},city={Beijing},postcode={100084},country={China}}

\affiliation[label2]{organization={CompleX Lab, University of Electronic Science and Technology of China},city={Chengdu},postcode={611731},country={China}}

\affiliation[label3]{organization={Department of Statistics and Data Science, Southern University of Science and Technology},city={Shenzhen},postcode={518055},country={China}}

\affiliation[label4]{organization={School of Computer Science, Northwestern Polytechnical University},city={Xi'an},postcode={710072},country={China}}

\affiliation[label5]{organization={Harbin Engineering University},city={Harbin},postcode={150001},country={China}}

\fntext[fn1]{These authors contributed equally to this work.}

\begin{abstract}

The study of complex systems has attracted widespread attention from researchers in the fields of natural sciences, social sciences, and engineering. Prediction is one of the central issues in this field. Although most related studies have focused on prediction methods, research on the predictability of complex systems has received increasing attention across disciplines—aiming to provide theories and tools to address a key question: What are the limits of prediction accuracy? Predictability itself can serve as an important feature for characterizing complex systems, and accurate estimation of predictability can provide a benchmark for the study of prediction algorithms. This allows researchers to clearly identify the gap between current prediction accuracy and theoretical limits, thereby helping them determine whether there is still significant room to improve existing algorithms. More importantly, investigating predictability often requires the development of new theories and methods, which can further inspire the design of more effective algorithms. Over the past few decades, this field has undergone significant evolution. In particular, the rapid development of data science has introduced a wealth of data-driven approaches for understanding and quantifying predictability. This review summarizes representative achievements, integrating both data-driven and mechanistic perspectives. After a brief introduction to the significance of the topic in focus, we will explore three core aspects: the predictability of time series, the predictability of network structures, and the predictability of dynamical processes. Finally, we will provide extensive application examples across various fields and outline open challenges for future research.

\end{abstract}


\end{frontmatter}

\tableofcontents 

\section{Introduction}
To date, there is no standard definition of a "complex system". However, the academic community widely recognizes that a complex system generally refers to a system composed of a large number of interacting and interrelated individuals (or components), where the overall behavior far exceeds the simple superposition of individual behaviors \cite{newman2011complex, ladyman2013complex, bar2019dynamics}. A complex system is neither a "simple system" -- like a mechanical clock -- where the interactions between components are simple and can be accurately predicted, nor a "random system" -- like rolling dice -- where outcomes are completely unpredictable; instead, it lies between these two types of systems. Complex systems exist extensively in fields such as economics \cite{mantegna1999introduction, hidalgo2021economic}, society \cite{gao2019computational, jusup2022social}, biology \cite{adami2000evolution, barabasi2004network}, and so on.

Complex systems possess core characteristics such as self-organization, emergence, nonlinearity, adaptability, and openness. (i) \textit{Self-organization} -- A system does not require a "central controller"; instead, it can spontaneously form order structures or show global statistical regularities solely through local and simple interaction rules between components \cite{bak2013nature}. For instance, in flocks of birds and schools of fish, each bird or fish only needs to follow a few simple local rules (e.g., cohesion, alignment and separation), yet the entire group can exhibit ordered collective movement \cite{vicsek2012collective}. (2) \textit{Emergence} -- New properties and behaviors that do not exist at the individual level emerge at the system level, which cannot be deduced by analyzing individual components \cite{holland2000emergence}. For example, a single ant only performs simple behaviors like foraging for food and secreting pheromones; however, when tens of thousands of ants gather, intelligent behaviors that individual ants do not possess -- such as collective nest-building and division of labor -- emerge \cite{johnson2002emergence}. Another example is that a single neuron can only receive and transmit electrical signals, but when nearly $100$ billion neurons interact through synaptic connections, advanced cognitive functions that cannot be explained by analyzing a single neuron emerge \cite{wang2019}. (3) \textit{Nonlinearity} -- The complex positive and negative feedback interactions between components result in a nonlinear response of the system to external stimuli. A small difference in input may lead to a huge difference in the final outcome, while a large difference in input may result in nearly identical outcomes \cite{bertuglia2005nonlinearity}. The well-known "butterfly effect" -- a butterfly flapping its wings in South America may eventually trigger a hurricane in North America through nonlinear interactions in atmospheric circulation -- is a typical nonlinear effect \cite{lorenz1972predictability}. (4) \textit{Adaptability} -- The system as a whole or its components can perceive environmental changes through "learning and feedback", and adjust their own dynamical rules or the interaction patterns between components accordingly to adapt to the environment and maximize fitness \cite{holland2006studying}. In some scenarios, the system can also alter the environment itself through interaction with it; the changed environment then exerts a different influence on the system, leading to more complex and unpredictable behaviors \cite{lam1995active, schweitzer2003brownian}. A classic example is ecosystem, where species modify their behaviors (such as predation, avoidance, and reproduction) in response to environmental changes (e.g., climate change or biological invasion) and changes in the quantity and behavior of other species, to maximize their ability to survive and reproduce \cite{traveset2014mutualistic, morone2019k}. (5) \textit{Openness} -- A system is not closed and isolated, but continuously exchanges matter, energy, and information with the outside world, and maintains an ordered (low-entropy) structure based on this exchange. Once the interaction between the system and the outside world is cut off, the system will gradually collapse \cite{prigogine1978time}. For example, living organisms need to take in food from the environment and excrete waste; if these exchanges are interrupted, death will soon follow \cite{kaneko2006life}. Another example is city: as a whole, it needs to import energy, water, food, and other resources from the outside; once these supplies are cut off, the city will quickly become paralyzed \cite{zhang2025metacity}. The reason why complex systems exhibit the above characteristics -- or in other words, the reason for their complexity -- may stem from the complexity of the dynamics \cite{ott2002chaos, barabasi2005origin}, or from the interactions between components \cite{kauffman1992origins, ball2006critical}; more often than not, both factors contribute \cite{barabasi2007architecture}.

Why do we study complex systems? Or rather, what can we gain from the study of complex systems? First, some phenomena that once puzzled us can now be explained convincingly through the study. For instance, in the field of transportation, there were two well-known phenomena that remained unexplained for a long time: one is "phantom traffic jams", referring to the phenomenon where traffic jams come and dissipate without obvious causes such as accidents or road construction \cite{treiterer1974hysteresis}; the other is the "Braess’s paradox", which describes a scenario where adding a new road or expanding the capacity of an existing one to increase traffic capacity may instead lead to reduced travel efficiency and more frequent congestion across the entire network \cite{braess1968paradoxon}. Both phenomena have now been well explained within the framework of complex systems research: the former is a typical self-organized critical behavior driven by stop-and-go waves \cite{chowdhury2000statistical, helbing2001traffic}, while the latter comes from the dual complexity of human decision-making and traffic networks \cite{rapoport2009choice, gounaris2024braess}. Another example is the volatility clustering \cite{cont2007volatility} and market crashes \cite{johansen2010shocks} in financial markets, which are difficult to be explained under the framework of the efficient market hypothesis but are well understood by introducing complex financial networks and the spreading dynamics of risks \cite{haldane2011systemic, elliott2014financial, acemoglu2015systemic, bardoscia2017pathways}.

Second, the findings from complex systems research can be used to achieve better system design and intervention, thereby yielding practical benefits. For example, once we fully understand the mechanism behind the Braess’s paradox, we can accurately identify the roads that cause this paradox and close them under specific conditions \cite{ma2018link}, or convert a small number of roads into tidal flow roads \cite{ampountolas2019motorway, striewski2022adaptive}. This allows us to increase the throughput of the entire traffic network and reduce congestion rates without building new roads. Similarly, when we recognize that interactions within financial networks may lead to the rapid amplification of risks or even market crashes, we need to monitor these interactions (such as overnight interbank lending) for anomalies. Furthermore, we should promote a shift in financial risk control from focusing on individual risks to systemic risks, incorporating the degree of connectivity between financial institutions into monitoring and supervision, rather than solely focusing on the capital adequacy ratio of individual institutions \cite{battiston2012debtrank, squartini2013early}. There are many similar examples: for instance, complex systems research has driven the design of power grids to shift from "component redundancy" to "network resilience" -- including adding backup power sources for hub substations and implementing real-time load distribution monitoring and early warning systems -- thereby significantly reducing the frequency of global blackouts \cite{pagani2013power, amani2021power, liu2022network}. In ecological conservation, the focus has expanded from endangered species to hub species (such as bees and bats in tropical rainforests, and shrimp in freshwater ecosystems), with special attention paid to ecosystem connectivity and the construction of ecological corridors linking fragmented habitats \cite{montoya2006ecological, landi2018complexity}. For example, by building ecological corridors, China's Giant Panda National Park has connected previously isolated giant panda habitats, promoted genetic exchange between different populations, and thus significantly increased the number of wild giant pandas and their survival resilience \cite{he2024suitable}.

Third, some concepts, methods, and theories developed in the study of complex systems have played important roles in many other fields. The most typical example is \textit{complex networks}. The early development of complex networks research is mainly contributed by complexity science community \cite{mitchell2006complex}, while complex networks have now become highly frequently used concept and tools in fields such as life sciences, ecology, sociology, management science, computer science, and physics \cite{barabasi2016, newman2018}. In fact, since complex systems usually involve systems with many-body nonlinear interactions, and networks are the optimal tool for representing the relationships of many-body interactions, most complex systems also fall within the scope of complex network research. Other theories in the center of complex systems research, such as chaos theory, fractal theory, and self-organization theory, have also found applications in a very wide range of fields \cite{siegenfeld2020introduction}.

The study of complex systems can generally be divided into three phases: "explanation, prediction, and intervention", with their respective tasks as follows: (1) proposing theoretical models to explain observed phenomena; (2) predicting unobserved data or phenomena (which may be missing data or events that will occur in the future); (3) implementing specific interventions on real systems to achieve preset goals \cite{batty2005modelling, san2012challenges, kwapien2012physical}.

An explanation that seems correct for a phenomenon does not prove that the corresponding theory, model, or mechanism is correct; it merely asserts a possibility of being correct. In fact, the credibility of such an explanation is insufficient. Much scientific research is typically hindsight bias: it is always possible to propose some theoretical models that provide qualitatively correct or even quantitatively accurate explanations for the event that has already happened. There is a saying in the field of economics, roughly meaning "There is not a single theory that can predict an economic crisis, but theories that can explain an economic crisis are everywhere". In this sense, prediction is generally more difficult than explanation, and a correct prediction also provides stronger endorsement for the validity of a theory than an explanation. It can be said that prediction is the bridge connecting explanation and intervention, and a rehearsal before conducting intervention. If a theory cannot make relatively accurate predictions about the future, the probability of failure in interventions based on this theory will certainly be high.

Prediction is one of the central issues in the study of complex systems. However, at the same time, complex systems possess inherent unpredictability, otherwise, they would not be called complex systems \cite{grieves2016digital}. This unpredictability may originate from complex nonlinear dynamics, particularly the characteristics of chaos, where tiny errors accumulate and amplify over time, making short-term prediction feasible while long-term prediction impossible \cite{krishnamurthy2019predictability, chang2023hybrid}. It may also arise from complex interactions: interactions can cause stimuli to either amplify or diminish, meaning that although the overall statistical laws of "stimulus-response" can be derived, the exact intensity of the response triggered by a single stimulus cannot be accurately predicted \cite{bak1987self, jalili2017information}.

Precisely because complex systems possess inherent unpredictability, quantifying the predictability of complex systems (i.e., the upper limit of prediction accuracy) has become a problem of great significance. Theoretically, predictability characterizes the position of a system between complete regularity and complete randomness, and can itself serve as a feature for describing system complexity \cite{boffetta2002predictability}. Predictability also provides a unique perspective for observing complex systems: significant changes in a system’s predictability often correspond to major changes in the system’s environment or in the system’s own organizational and evolutionary mechanisms. Examples include the impact of natural disasters on the predictability of human mobility \cite{lu2012predictability}, and the impact of environmental changes on the predictable time horizon of the climate system \cite{li2020pacific}, among others.

Practically, estimating the upper limit of prediction accuracy helps us evaluate the performance of current algorithms and decide whether it is necessary to invest more time and resources in optimizing them: if there is a large gap between the upper limit of prediction accuracy and the highest accuracy of existing algorithms, a significant further improvement in algorithm accuracy is promising; conversely, if the upper limit of prediction accuracy is already very close to the highest accuracy of existing algorithms, then even if the current accuracy is not high, the room for algorithm improvement is limited, and at this point, it is not cost-effective to devote efforts to algorithm optimization. Similarly, by assessing the predictable time horizon (the average length of time a prediction can be made while ensuring accuracy does not fall below a certain threshold), we can also decide whether to invest resources in studying long-term prediction algorithms. For instance, if weather forecasts require an 80\% accuracy rate to guide agricultural production, and the corresponding predictable time horizon is 10 days, then investing substantial resources in monthly or quarterly weather forecasts would be of no value for agricultural production.

Recently, study on the predictability of complex systems has received increasing attention. The purpose of this review is to compile and introduce representative achievements in this field. From Sections 2 to 4, this review will present research progresses on the predictability for time series, complex networks, and dynamical systems, respectively. Next, Section 5 will select applications of the study of predictability in various fields, including human behavior, financial markets, the Earth system, as well as politics, culture, art, and so on. Finally, we will discuss the current opportunities and challenges in the study on the predictability of complex systems, and attempt to propose some interesting questions that can serve as directions for future research.

\section{Predictability of Time Series}
A time series is an ordered sequence of data points that records the evolution of a system's state over time \cite{hamilton2020time}. As a fundamental form of data representation, time series are ubiquitous across diverse domains, including meteorological observations \cite{karevan2020transductive}, financial markets \cite{tsay2005analysis}, healthcare \cite{caballero2015dynamically}, as well as infrastructure monitoring and the operational states of Internet of Things (IoT) devices \cite{macchiarulo2022monitoring,esling2012time}. By capturing system dynamics along the temporal dimension, time series exhibit distinctive characteristics compared to other data modalities, providing a crucial entry point for understanding the regularities of complex systems. The core value of time series lies in their dual capability to both describe and predict system behaviors \cite{hlavavckova2007causality}. On the one hand, the ordered records of time series reveal trends, periodicities, and fluctuations in system operations \cite{li2015trend}, thereby enabling researchers to uncover the underlying mechanisms of the system \cite{hilliard2014strategy}. On the other hand, time series offer the most direct empirical basis for forecasting, making it possible to infer future dynamics from historical states \cite{fan2019multi}.

In the study of complex systems, time series are not only an essential form for recording system dynamics but also a core vehicle driving the transition from observation to modeling and inference. First, complex systems are typically composed of a large number of interacting and nonlinearly coupled microscopic units, whose collective behavior exhibits pronounced dynamical and emergent properties \cite{BATTISTON20201}. By recording behavioral evolution along the temporal dimension, time series provide a fundamental data basis for understanding the dynamic mechanisms of such systems. Second, the inherent temporal dependencies of time series offer direct cues for analyzing causal relations and feedback mechanisms, enabling empirical dynamic modeling based solely on observational data \cite{hlavavckova2007causality}. Classical theoretical tools, such as Takens' embedding theorem and empirical dynamic modeling (EDM), rely on the continuity and delayed embedding properties of time series, allowing the reconstruction of state spaces and effective prediction without the need for explicitly modeling microscopic mechanisms \cite{takens2006detecting,sugihara1990nonlinear,ye2015equation}.

With the widespread application of time series data across diverse domains \cite{box2015time}, the methodological landscape of time series modeling and forecasting has undergone continuous evolution and multiple paradigm shifts. Early efforts were dominated by statistical models built on prior assumptions, followed by the rise of data-driven machine learning \cite{masini2023machine} and deep learning techniques \cite{benidis2022deep}, and more recently, the rapid emergence of the foundation model paradigm \cite{liang2024foundation}. This trajectory reflects a transition from modeling paradigms centered on prior assumptions and explicit structures toward data-driven paradigms emphasizing large-scale data and joint model optimization.

Initially, the statistical community proposed a series of classical parametric models to characterize the dynamic properties of time series, including the Autoregression (AR) model \cite{hannan1979determination}, the Autoregressive Integrated Moving Average (ARIMA) model \cite{contreras2003arima}, the Vector Autoregression (VAR) model \cite{stock2001vector}, as well as the Autoregressive Conditional Heteroskedasticity (ARCH) model \cite{engle1982autoregressive} and its generalized extension, the Generalized Autoregressive Conditional Heteroskedasticity (GARCH) \cite{bollerslev1986generalized}. These models operate on both the mean and variance levels, thereby providing a basic description of expected trends and volatility patterns in time series. Their strengths lie in the clarity of model structure and strong interpretability, making them suitable for short-term, linear, and relatively stationary sequences. Nevertheless, when faced with the nonlinearity, multi-scale dependencies, and high-dimensional interactions that are ubiquitous in real-world systems, the expressive capacity and applicability of traditional statistical models are severely constrained~\cite{WANG20161}.

With advances in computational capacity, machine learning methods have demonstrated unique advantages in modeling such complexities, particularly in handling nonlinearities, multi-scale dependencies, and heterogeneous data sources, thereby becoming indispensable tools for time series modeling \cite{masini2023machine}. Non-parametric methods such as Support Vector Machines (SVM) \cite{GestelSBLLVMV01}, Random Forests (RF) \cite{lin2017random}, and Gradient Boosting Decision Trees (GBDT) \cite{lee2018interpretable} break away from the reliance on stationarity and linearity assumptions, offering stronger nonlinear fitting capacity and generalization performance. This stage is characterized by a feature-engineering-driven paradigm, in which models rely on the design and extraction of handcrafted sequence features and leverage machine learning to capture nonlinear patterns for modeling complex temporal regularities \cite{tavenard2020tslearn}. Nevertheless, these methods lack end-to-end modeling capabilities when dealing with long-range dependencies and dynamic feedback structures, often resorting to sliding windows or manually crafted features as indirect representations of temporal information.

In the deep learning era, time series modeling has progressively advanced toward end-to-end architectures centered on neural networks. Recurrent Neural Networks (RNNs) \cite{khaldi2023best}, Long Short-Term Memory networks (LSTMs) \cite{siami2019performance}, and Gated Recurrent Units (GRUs) \cite{zhang2024novel} leverage hidden-state mechanisms to effectively capture long-range dependencies in time series. Convolutional Neural Networks (CNNs) and Causal Convolutions \cite{livieris2020cnn} provide complementary strengths in local pattern extraction and computational efficiency. More recently, Transformer architectures, owing to their global dependency modeling capacity and efficient parallel computation \cite{wen2023transformers}, have achieved outstanding performance in forecasting, classification, and anomaly detection tasks, becoming the dominant architecture within the deep learning paradigm \cite{zamanzadeh2024deep}. In parallel, hybrid modeling strategies, sequence decomposition, and attention mechanism refinements have emerged to address the challenges of non-stationarity, multi-scale dynamics, and sparsity in time series \cite{wang2023wavelet}.

Large-scale pretrained models, which have demonstrated strong cross-task transferability and unified modeling capabilities in natural language processing and computer vision, have also been introduced into time series analysis. This paradigm shift is propelling the field toward foundation models (FMs) that emphasize large-scale pretraining and unified representation learning \cite{liang2024foundation}. Inspired by the success of large language models (LLMs) and vision foundation models \cite{wang2024sam}, Time Series Foundation Models (TSFMs) \cite{li2025tsfm} employ self-supervised pretraining on massive sequence data, endowing them with powerful cross-task transferability and few-shot adaptation abilities. These models not only exhibit superior performance in conventional time series tasks but also extend to spatio-temporal sequences, trajectory data, and event streams, thereby fostering a paradigm shift from “task-specific models” to a “unified modeling framework” in time series analysis \cite{liang2025foundation}.

As time series analysis methods have continued to evolve—from early statistical approaches grounded in prior assumptions and structural models \cite{box2015time,contreras2003arima}, to machine learning and deep learning techniques centered on data-driven feature learning \cite{benidis2022deep,masini2023machine}, and more recently, to the paradigm of foundation models characterized by large-scale pretraining and unified representation learning \cite{liang2024foundation,li2025tsfm}—the capacity of time series modeling has steadily advanced, substantially enhancing the ability to fit and forecast complex dynamical systems. Nevertheless, improvements in algorithms and model architectures do not imply that predictive accuracy can grow without bound \cite{song2010limits}. In reality, time series not only encapsulate the structural regularities of systems but are also inevitably subject to stochastic fluctuations, environmental variations, and observational noise—all of which jointly determine the intrinsic upper limit of predictability \cite{song2010limits}.

In the study of complex systems, time series simultaneously reflect predictable evolutionary patterns and irreducible randomness; the coexistence of regularity and stochasticity is an inherent characteristic of system evolution \cite{barabasi2005origin}. Thus, although innovations in model architectures and learning paradigms have improved the capacity to capture system regularities, the ultimate performance of predictive models remains constrained by the intrinsic uncertainties of system dynamics \cite{song2010limits}. As model scales and computational resources continue to increase, the marginal gains from structural stacking and parameter expansion have shown clear diminishing returns.
For this reason, researchers have increasingly shifted their focus from merely improving predictive models to investigating the fundamental boundaries of time series predictability itself \cite{song2010limits}. Clarifying the theoretical upper limits of predictability, and further uncovering its intrinsic origins, has become one of the central challenges in contemporary time series analysis and complex system modeling.

This section addresses the central question of "to what extent time series can be predicted" by constructing an analytical framework comprising three classes of approaches: information-theoretic bounds, complexity-based metrics, and the equivalence between predictability $\Pi$ and the Bayes error rate (BER). Information-theoretic methods rigorously define the theoretical limits of prediction, with Fano scaling as a representative technique, which establishes a quantitative mapping between entropy and optimal accuracy. These methods have been further refined and extended through approaches such as reachability-based candidate set estimation, incorporation of Top-$\ell$ probability structures, and the use of cross-entropy and context entropy, thereby tightening and interpreting the upper bound. Metric-based methods complement this by addressing scenarios involving numerical sequences, short series, and noisy environments. Typical examples include permutation entropy and the $\kappa$-index, which directly capture predictability through measures of structural complexity and provide lightweight tools adaptable to diverse data conditions. Finally, we show a strict equivalence between time series predictability and the BER, demonstrating that the predictability limit can be understood as the optimal achievable performance of a classification task, while offering alternative pathways to approximate this limit via non-parametric estimation. Collectively, these three classes of methods constitute the theoretical and methodological foundations of time series predictability research.

\subsection{Information-Theoretic Methods}
As an important data form for describing the dynamic evolution of systems, time series are generally categorized into two types based on the representation of their state space: symbolic and numerical \cite{lin2003symbolic}. Symbolic time series consist of a finite set of discrete symbols or state labels, highlighting sequential dependencies and structural patterns in state transitions; numerical time series, in contrast, record system states as continuous real-valued measurements, emphasizing variations in magnitude and amplitude \cite{zeng2023transformers}. Despite their discrete–continuous differences in data form, both types can usually be handled within a unified modeling framework through appropriate input–output representation mappings. At the input stage, symbolic sequences can be encoded or embedded into continuous numerical vectors, which are then processed alongside raw numerical series within unified architectures such as RNNs, Transformers, or Foundation Models \cite{siami2019performance,wen2023transformers,liang2024foundation}. At the prediction stage, continuous outputs generated by the model can be discretized back into symbolic states of the original state space \cite{agarap2018neural}. Hence, from input representation to predictive output, existing algorithmic frameworks do not require explicit distinctions between symbolic and numerical time series, as unified modeling and forecasting can be achieved through representational transformations alone.

However, when it comes to measuring time series predictability, existing methodological systems cannot be uniformly applied across symbolic and numerical sequences. For symbolic sequences, the state space is finite and discrete, and their predictability can be directly characterized by the size of the candidate set for the next state, the transition structure, and the entropy measures of the symbolic process. In particular, entropy, through the Fano inequality, rigorously defines the theoretical upper limit of predictability \cite{song2010limits}. For numerical sequences, by contrast, the state space is continuous and, in principle, admits infinitely many candidate states, making it impossible to quantify predictability by simple state enumeration. This fundamental difference in state space definitions and information quantification precludes a straightforward methodological unification of predictability metrics across symbolic and numerical time series. Consequently, dedicated measurement frameworks and theoretical systems have been established separately for symbolic and numerical time series.

A core characteristic of symbolic time series lies in the finiteness and enumerability of their state space. The evolutionary process of such sequences can be viewed as transition paths and combinatorial patterns among symbols. This discreteness and structural nature not only makes them more analytically tractable in theory \cite{lin2003symbolic}, but also allows their predictability to be directly characterized through state transition structures and information-theoretic measures such as entropy. Leveraging these properties, symbolic time series are widely observed across diverse complex systems: in social systems, human mobility trajectories can be discretized into state sequences of finite locations, whose transition patterns reflect individual preferences and mobility regularities \cite{song2010limits}; in biological systems, DNA, RNA, and protein sequences consist of finite alphabets of nucleotides or amino acids, with strong couplings between sequence structures and biological functions \cite{bar2012studying}; in digital and information systems, user clickstreams, shopping paths, and operation logs can be represented as sequences of discrete action symbols, supporting preference modeling and behavior prediction \cite{zhao2012empirical,zhang2015daily,zha2016unfolding}; and in engineering and technological systems, symbolic sequences of equipment alarms, system log events, and cyber-attack chains play a critical role in monitoring and anomaly detection \cite{karim2020adversarial}. For these reasons, symbolic time series occupy a central position in the study of predictability in complex systems.

This section provides a systematic review of information-theoretic methods for measuring the predictability of time series, with a particular focus on the theoretical modeling, metric estimation, and methodological extensions of symbolic sequences. First, we introduce the formal definition of predictability and its quantitative relationship with sequence entropy, elaborating the scaling method based on Fano's inequality and its seminal application to human mobility research. Next, we discuss refined approaches that incorporate geographic and topological constraints, as well as scaling strategies integrating Top-$\ell$ probability information, in order to tighten and enhance the applicability of predictability bounds. Finally, we summarize methods such as cross-entropy, context entropy, and conditional entropy, highlighting the development of the information-theoretic framework toward multi-source data integration, context-aware adjustments, and interpretability analysis.

\subsubsection{Foundations}
This subsection introduces the theoretical foundations of time series predictability, which center on characterizing the theoretical limit of correctly forecasting the next state based on historical observations under an optimal strategy. Within this framework, Fano's inequality establishes the relationship between entropy and predictability, and the Fano scaling method proposed by Song \textit{et al.} \cite{song2010limits} provides a computable estimate of predictability. Empirical studies on large-scale datasets have validated the effectiveness of this approach and systematically revealed the connections between predictability and multiple factors such as spatial range of activity, location preference, data completeness, spatio-temporal resolution, and exploration behavior. These findings laid the groundwork for subsequent optimizations and extensions of the method.

To formally characterize the predictability of a time series, we begin with the basic formalization of the sequence prediction task. Consider a symbolic time series with a finite and enumerable state space. Let the time series be denoted as $\mathcal{Z} = \{Z_1, Z_2, \dots, Z_L\}$, where $Z_i \in \Omega$ represents the state at time step $i$, and $L$ is the sequence length. The finite state space is denoted as $\Omega = \{z_1, z_2, \dots, z_C\}$, where $z_i$ represents a specific candidate state and $C$ is the total number of states. For any time step $t \in \{1,2,\dots,L\}$, the history $h_{t-1} = \{Z_{t-1}, Z_{t-2}, \dots, Z_1\}$ denotes the sequence of observations from the initial moment up to step $t-1$. The core problem of interest is: given the history $h_{t-1}$, what is the maximum achievable probability of correctly predicting the next state $Z_t$?
Formally, we define:
\begin{align}
	\pi(h_{t-1}) = \sup_{z \in \Omega} Pr[Z_t = z \mid h_{t-1}], \label{eqn_pi}
\end{align}
where $\pi(h_{t-1})$ represents the probability of the most likely state among the $C$ candidates, given the historical record $h_{t-1}$. This quantity thus characterizes the theoretical upper bound of predictability for the sequence under the condition of history $h_{t-1}$.

In practical prediction, any forecasting algorithm $\alpha$ generates, given a history $h_{t-1}$, a probability distribution $P_{\alpha}(z \mid h_{t-1})$ over the next state $Z_t$, while the true distribution of the sequence is denoted by $P(z \mid h_{t-1})$. Thus, the probability that algorithm $\alpha$ correctly predicts the next state at time $t$ can be expressed as:
\begin{align}
	Pr_{\alpha}[Z_t = \hat{Z}_t \mid h_{t-1}] = \sum_{t} P(z \mid h_{t-1}) P_{\alpha}(z \mid h_{t-1}),
\end{align}
where $\hat{Z}_t$ denotes the true next state at time $t$. According to Eq.~(\ref{eqn_pi}), for any state $z$ we have $P(z \mid h_{t-1}) \leq \pi(h_{t-1})$, which leads to:
\begin{align}
	Pr_{\alpha}[Z_t = \hat{Z}_t \mid h_{t-1}] 
	\leq \sum_{z} \pi(h_{t-1}) P_{\alpha}(z \mid h_{t-1}) 
	= \pi(h_{t-1}).
\end{align}
This inequality indicates that regardless of the prediction algorithm employed, the accuracy of forecasting the next state, given the historical record $h_{t-1}$, can never exceed $\pi(h_{t-1})$.

Furthermore, to verify the attainability of the upper bound $\pi(h_{t-1})$, consider an ideal prediction algorithm $\alpha^\star$, which always selects the most probable state as its prediction, namely:
\begin{align}
	P_{\alpha^{\star}}(z \mid h_{t-1}) = 
	\begin{cases}
		1, & \text{if } z = z_{\textup{MS}}, \\
		0, & \text{otherwise}, \\
	\end{cases}
\end{align}
where $z_{\textup{MS}}$ denotes the most probable state corresponding to $\pi(h_{t-1})$. In this case, the prediction accuracy of the ideal algorithm $\alpha^\star$ is:
\begin{align}
	Pr_{\alpha^\star}[Z_t = \hat{Z}_t \mid h_{t-1}] = \pi(h_{t-1}),
\end{align}
which demonstrates that $\pi(h_{t-1})$ is not only the theoretical upper bound of predictability but can also be achieved under the optimal prediction strategy.

To characterize the predictability of a sequence as a whole, the upper bounds of predictability under different historical records should be aggregated through a weighted average. Specifically, when the history length is $t-1$, the average predictability across all possible histories is defined as:
\begin{align}
	\Pi(h_{t-1}) = \sum_{h_{t-1}} P(h_{t-1}) \pi(h_{t-1}),
\end{align}
where $P(h_{t-1})$ denotes the probability of observing the history $h_{t-1}$.

Finally, as the sequence length tends to infinity, the overall predictability of the sequence can be defined as:
\begin{align}
	\Pi = \lim_{t \to \infty} \frac{1}{t} \sum_{i=1}^{t} \Pi(h_{i-1}).
\end{align}
The metric $\Pi$ represents the theoretical average accuracy achievable under an optimal prediction strategy on an infinitely long sequence. It reflects the intrinsic limit of predictability determined by the interplay between regularity and uncertainty within the sequence \cite{song2010limits}.

The above discussion provides a formal definition of time series predictability. However, in practical research, a core question arises: how can one evaluate the theoretical upper bound of predictability on real-world data? In the study of symbolic time series, a close relationship exists between the information entropy of the sequence and its predictability. Intuitively, lower entropy implies stronger structural regularity within the sequence, and thus higher predictability. Nevertheless, entropy essentially quantifies the uncertainty of a sequence and does not directly specify the theoretical limit of prediction accuracy in an actual forecasting task. 

To address this gap, Song \textit{et al.} \cite{song2010limits}, introduced a scaling method based on Fano's inequality, which established a quantitative mapping between entropy and predictability \cite{feder2002relations}. The core idea of this method lies in structurally simplifying the true state distribution of the sequence and relaxing the entropy bound to construct an analytically tractable Fano function. This enables the estimation of the upper bound of prediction accuracy directly from the entropy of the sequence. Using human mobility trajectories as an example, Song \textit{et al.} empirically demonstrated that the predictability of human mobility can reach as high as 93\% \cite{song2010limits}.

The core problem addressed by Song \textit{et al.} is: given a user's historical trajectory $h_{t-1}$, what is the optimal accuracy for predicting the next location? To this end, they first defined the true distribution of the next state as $P(Z_t \mid h_{t-1})$, where $Z_t$ denotes the location at the next time step.  
For the sake of theoretical tractability, Song \textit{et al.} introduced a simplified approximation strategy for the probability distribution: among all $C$ candidate locations, they preserved the probability $p$ of the most likely location $z_{\mathrm{MS}}$ under the true distribution, while uniformly distributing the remaining probability mass among the other $C-1$ locations as $\tfrac{1-p}{C-1}$. The simplified distribution is then expressed as $P^{'}(Z \mid h) = \left(p, \tfrac{1-p}{C-1}, \dots \right)$.
It should be noted that the loss introduced by the uniform approximation essentially depends on the distribution of the states other than the one with the maximum probability. If the probabilities of these states vary considerably, i.e., the distribution is highly skewed, then the bias caused by the uniform approximation becomes larger. Conversely, when the probability ratios among these states remain consistent, the degree of skewness is the same, and different values of $C$ merely correspond to differences in sampling granularity of the distribution; in such cases, the effect is limited to the discrepancy between the true and uniform distributions at different granularities. Moreover, it is evident that the entropy of the flattened distribution is greater than that of the original distribution, thus the sequence entropy satisfies:
\begin{align}
	S(Z_t \mid h_{t-1}) \leq S(Z^{'} \mid h_{t-1}),
\end{align}
where $ S(Z^{'} \mid h_{t-1})$ is the entropy of the flattened distribution, which can be expanded as:
\begin{align}
	S(Z^{'} \mid h_{t-1}) 
	&= -p \log_2 p - (C-1) \cdot \frac{1-p}{C-1} \log_2 \frac{1-p}{C-1} \notag \\
	&= -p \log_2 p - (1-p) \log_2 (1-p) + (1-p) \log_2 (C-1).
\end{align}

\begin{figure}[!t]
	\centering
	\includegraphics[width=0.8\textwidth]{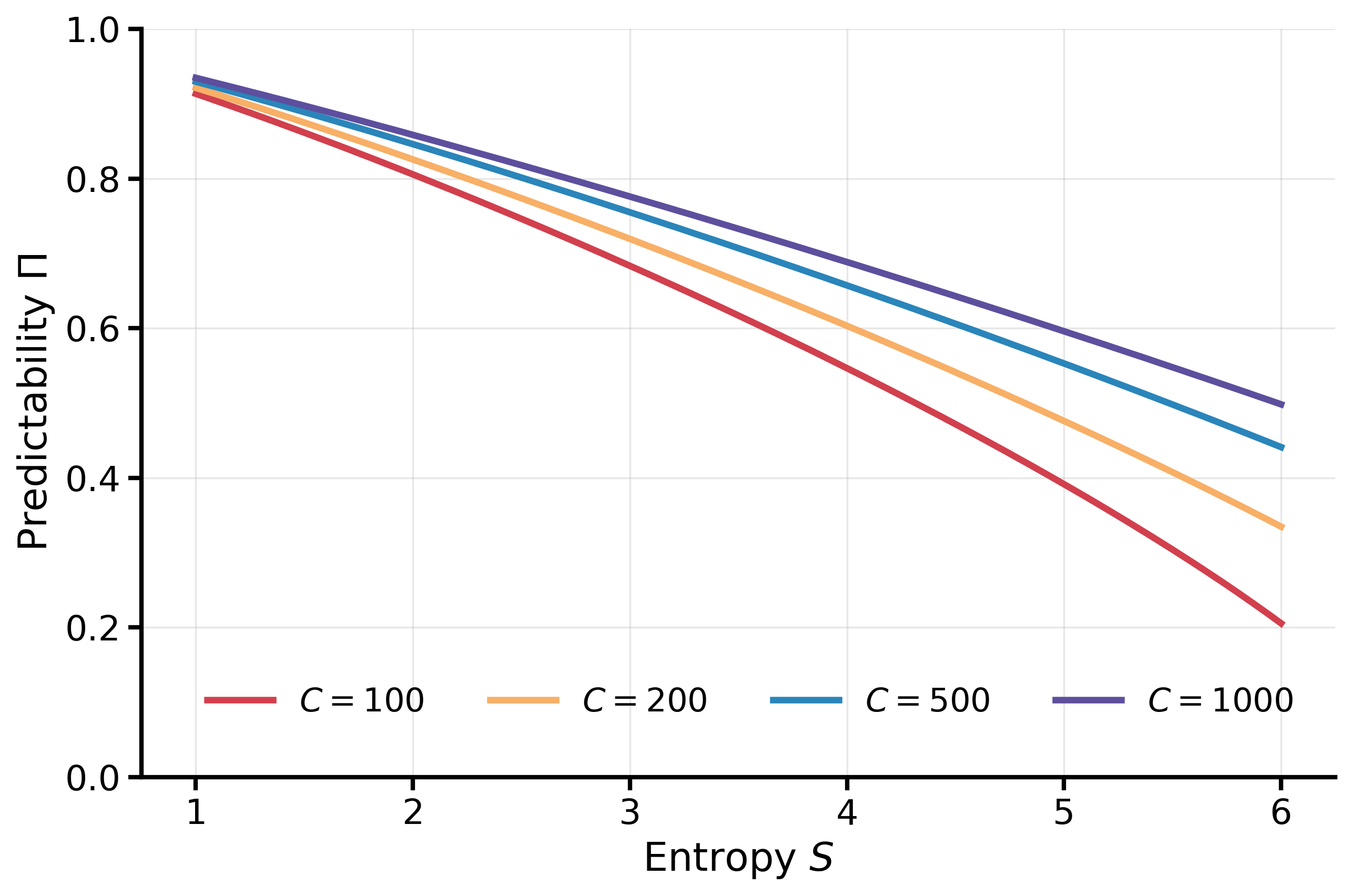}
	\caption{Illustration of the relationship between entropy $S$ and predictability $\Pi$ as mapped by the Fano function $S_F(p)$ under different candidate set sizes $C$. The curves demonstrate the monotonic decreasing and concave properties of $S_F(p)$. obviously, the relationship between entropy and predictability is not linear.}
	\label{fig:fano_function}
\end{figure}

The above function is the Fano function, denoted by $S_F(p)$. As $p$ is the probability of the most likely location in the next step, which is exactly same to $\pi(h_{t-1})$, we have $S_F(p) = S_F(\pi(h_{t-1}))$. $S_F(p)$ is a concave function of $p$, monotonically decreasing over the interval $p \in \left[\tfrac{1}{C}, 1 \right)$. Fig.~\ref{fig:fano_function} illustrates examples of the Fano function $S_F(\cdot)$ under different candidate set sizes $C$. As shown, $S_F(\cdot)$ is monotonically decreasing and concave within its domain, visually demonstrating the nonlinear relationship between entropy and predictability. This illustration helps clarify how Fano scaling transforms sequence entropy into an upper bound on predictability.

Taking the expectation of the sequence entropy over all historical records $h_{t-1}$ yields the entropy of the sequence at history length $t-1$:
\begin{align}
	S(t) = \sum_{h_{t-1}} P(h_{t-1}) S(Z_t \mid h_{t-1}) 
	\leq \sum_{h_{t-1}} P(h_{t-1}) S_F(\pi(h_{t-1})).
\end{align}
Exploiting the concavity of the Fano function, Jensen's inequality gives:
\begin{align}
	S(t) \leq S_F\left( \sum_{h_{t-1}} P(h_{t-1}) \pi(h_{t-1}) \right) = S_F(\Pi(t)),
\end{align}
where $\Pi(t)$ denotes the average predictability over all histories of length $t-1$. Furthermore, by extending entropy and predictability from single-step instances to the entire time series as $t \rightarrow \infty$, we obtain:
\begin{align}
	S = \lim\limits_{t\rightarrow \infty}\frac{1}{t} \sum_{i=1}^{t} S(i) 
	\leq S_F\left( \lim\limits_{t\rightarrow \infty}\frac{1}{t} \sum_{i=1}^{t} \Pi(i) \right) 
	= S_F(\Pi).
\end{align}

Denote $\Pi^{\max}$ as the unique solution to the following equation:
\begin{align}
	S &= S_F(\Pi^{\mathrm{max}}) \label{eq:fano_max} \notag \\
	&= -\Pi^{\mathrm{max}} \log_2 \Pi^{\mathrm{max}} 
	- (1-\Pi^{\mathrm{max}}) \log_2 (1-\Pi^{\mathrm{max}}) 
	+ (1-\Pi^{\mathrm{max}}) \log_2 (C-1). 
\end{align}
Since $S_F(\Pi)$ is monotonically decreasing for $\Pi \in \left(\tfrac{1}{C},\,1\right]$, and $S \leq S_F(\Pi)$ holds, it follows that $\Pi^{\max} \geq \Pi$. Therefore, $\Pi^{\max}$ represents the theoretical upper bound of the sequence predictability.

Within this theoretical framework, Song \textit{et al.} \cite{song2010limits} further proposed the concept of real entropy $S^{\mathrm{real}}$ to evaluate the entropy of mobility trajectories, defined as:
\begin{align}
	S^{\mathrm{real}} = - \sum_{Z' \subset Z} P(Z') \log_2 P(Z'),
\end{align}
where $Z$ denotes a user's trajectory sequence, $Z'$ is an ordered subsequence of $Z$, and $P(Z')$ represents the probability of observing $Z'$ in the trajectory. Compared with frequency-based measures, $S^{\mathrm{real}}$ captures not only the frequency of location visits but also explicitly encodes the order and temporal dependencies, thereby accounting for spatio-temporal patterns in the trajectory. Since direct computation of $S^{\mathrm{real}}$ requires knowledge of the full distribution, Song \textit{et al.} introduced a non-parametric estimation method based on Lempel–Ziv compression \cite{ziv1977universal}, as:
\begin{align}
	S^{\mathrm{est}} = \left( \frac{1}{t} \sum_i \Lambda_i \right)^{-1} \ln t, \label{eq:entropy_est}
\end{align}
where $t$ is the length of the sequence, the summation runs over $i = 1, \dots, t$, and $\Lambda_i$ denotes the length of the shortest new substring starting at position $i$. If all substrings starting from position $i$ have already appeared in the history, $\Lambda_i$ is conventionally defined as the maximum match length plus one (i.e., the remaining sequence length $+1$), ensuring that the estimator is always well-defined \cite{kontoyiannis2002nonparametric}. It has been theoretically proved that as $t \to \infty$, $S^{\mathrm{est}}$ converges to $S^{\mathrm{real}}$. Therefore, in practice, $S^{\mathrm{est}}$ is regarded as an approximation of $S^{\mathrm{real}}$, and is substituted into the relation $S = S_F(\Pi^{\max})$ to compute the theoretical upper bound of predictability $\Pi^{\max}$.

\begin{figure*}[!t]
	\centering
	\setlength{\abovecaptionskip}{0pt}
	\setlength{\belowcaptionskip}{0pt}
	\includegraphics[width=\textwidth]{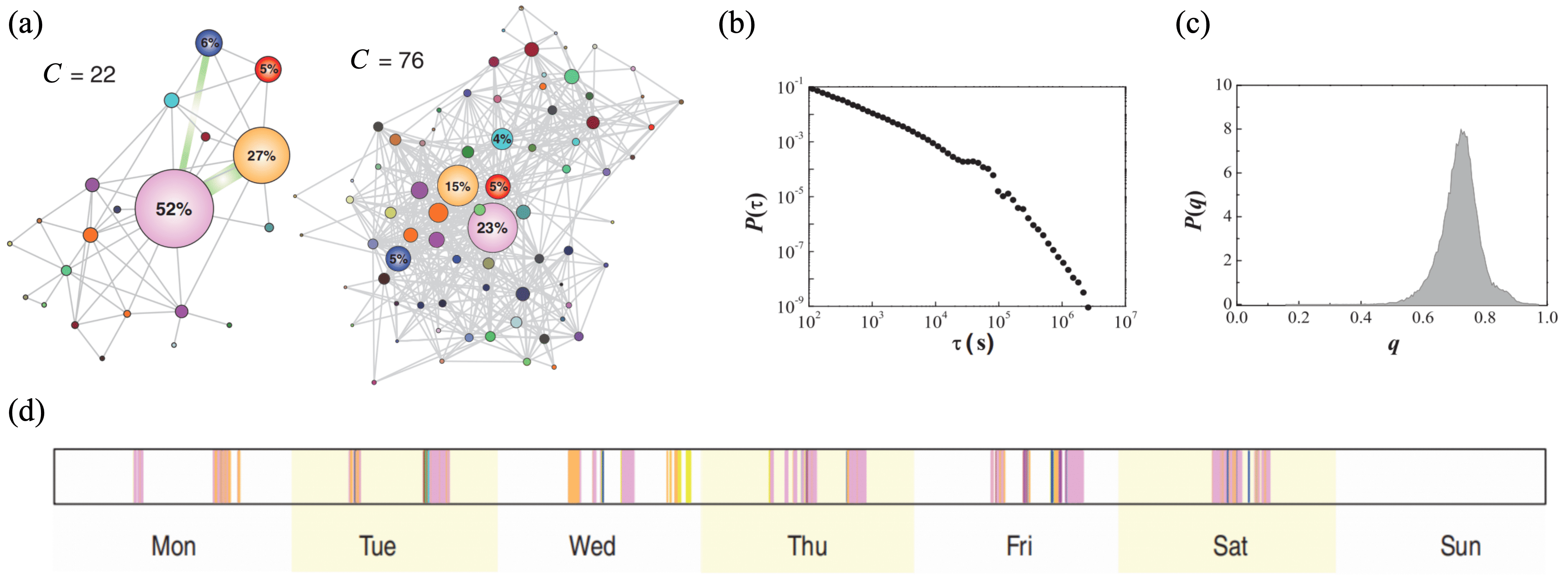} 
	\caption{Illustrative examples of the network structure and temporal patterns of human mobility. 
		(a) Example of a user mobility network, where nodes represent cell tower locations, node size is proportional to visit frequency, and edge width indicates the observed frequency of direct transitions between towers. 
		(b) Distribution of inter-call times $P(\tau)$ for all users, where $\tau$ denotes the time interval between two consecutive calls, showing that call activities exhibit bursty temporal patterns. 
		(c) Distribution $P(q)$ of the hourly proportion of unknown locations, where $q$ represents the fraction of time within an hour during which a user makes no calls and thus location is unobserved. 
		(d) Example of a weekly call sequence, where each vertical bar represents a call and colors correspond to the base station at which the call occurred, reflecting the temporal dependencies of location visits. \\
		\textit{Source}: The figure is reproduced from Ref. \cite{song2010limits}.}
	\label{fig:mobility_network_patterns}
	\vspace{0pt}
\end{figure*}

Song \textit{et al.} conducted an empirical analysis based on large-scale human mobility data. Fig.~\ref{fig:mobility_network_patterns}(a) shows the structure of the mobility network, revealing the spatial distribution of cell towers (in that study, the locations are cell towers) and the frequency of edges between them, which intuitively reflects the intensity of user transitions across locations. Fig.~\ref{fig:mobility_network_patterns}(b) demonstrates the distribution of inter-call times $P(\tau)$, which follows a heavy-tailed pattern, indicating that call activities exhibit significant burstiness in the temporal dimension, in line with other known empirical studies on human communications \cite{hong2009heavy,wu2010evidence,zhao2011empirical}. Fig.~\ref{fig:mobility_network_patterns}(c) introduces the data completeness parameter $q$, which measures the proportion of unknown location information in the observed sequence. The results show that for most users, the value of $q$ is below 0.8, meaning that at least 20\% of the time intervals contain observable location information, thereby ensuring the validity of subsequent entropy and predictability analyses. Finally, Fig.~\ref{fig:mobility_network_patterns}(d) presents a weekly call sequence, further highlighting the temporal dependencies of location visits. Different colors correspond to the base stations where calls occurred, emphasizing the periodic regularities of daily behavior.

\begin{figure*}[!t]
	\centering
	\setlength{\abovecaptionskip}{0pt}
	\setlength{\belowcaptionskip}{0pt}
	\includegraphics[width=0.70\textwidth]{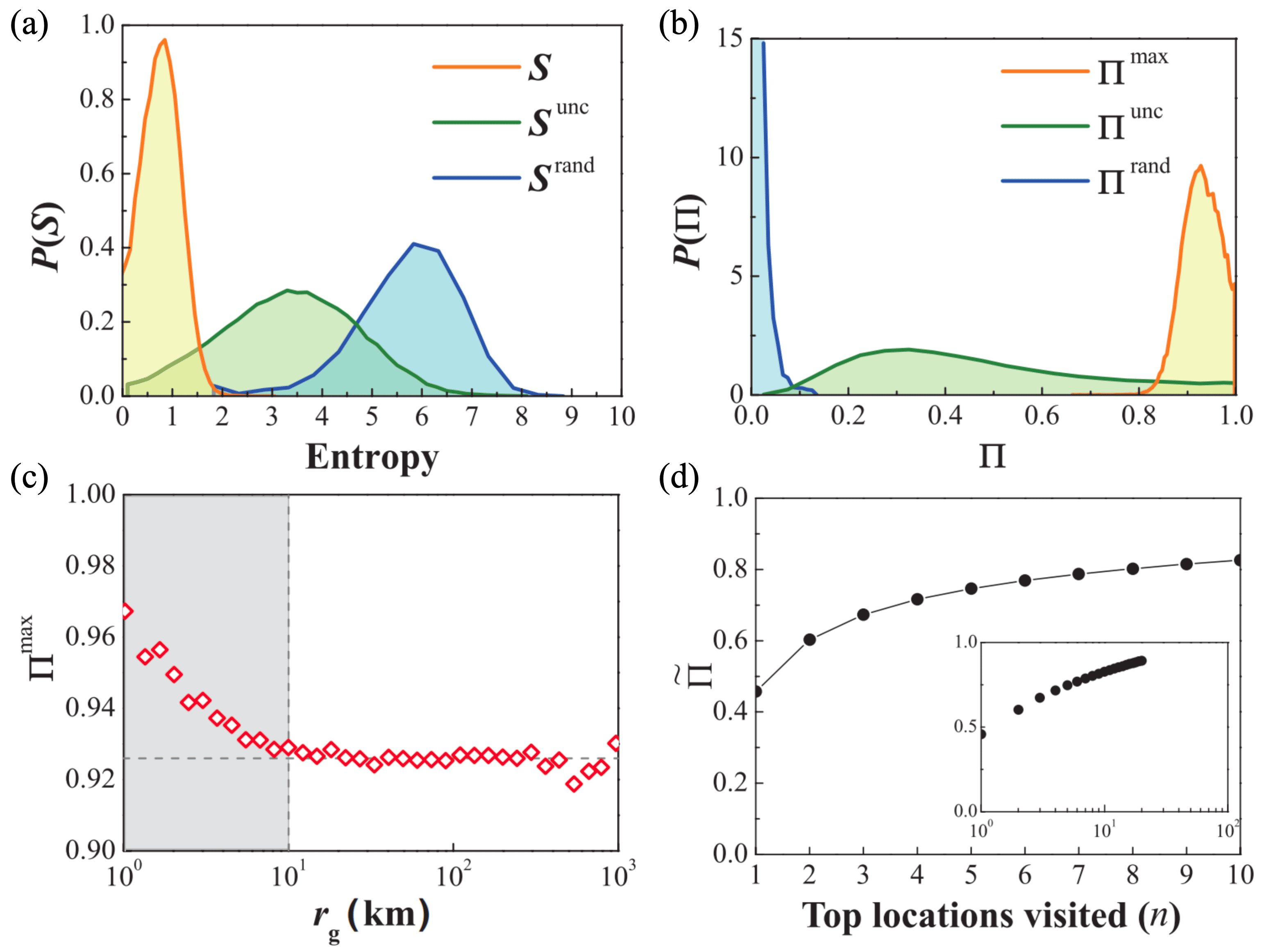} 
	\caption{Entropy distributions, predictability limits, and influencing factors of human mobility. 
		(a) Distributions of the real entropy $S$ (denoted as $S^{\mathrm{real}}$ in the main text), the uncorrelated entropy $S^{\mathrm{unc}}$, and the random entropy $S^{\mathrm{rand}}$ across 45,000 users. 
		(b) Corresponding distributions of the maximum predictability $\Pi^{\mathrm{max}}$, the uncorrelated predictability $\Pi^{\mathrm{unc}}$, and the random predictability $\Pi^{\mathrm{rand}}$. 
		(c) Relationship between the maximum predictability $\Pi^{\mathrm{max}}$ and the radius of gyration $r_g$, which characterizes the typical spatial scale of user activity. When $r_g > 10\ \mathrm{km}$, $\Pi^{\mathrm{max}}$ becomes nearly independent of $r_g$ and saturates at approximately 0.93. 
		(d) Fraction of time $\tilde{\Pi}$ spent by users in their Top-$c$ most frequently visited locations, used as an approximation to the upper bound $\Pi^{\mathrm{max}}$. When $c=2$, $\tilde{\Pi}$ is about 0.6; as $c$ increases, $\tilde{\Pi}$ grows approximately logarithmically. \\
		\textit{Source}: The figure is reproduced from Ref. \cite{song2010limits}.}
	\label{fig:entropy_predictability}
	\vspace{0pt}
\end{figure*}

In terms of entropy measures, Song \textit{et al.} defined three indicators to quantify the uncertainty of user trajectories: the random entropy $S^{\mathrm{rand}} = \log_2 C$ (assuming $C$ locations are visited with equal probability), the uncorrelated entropy $S^{\mathrm{unc}} = -\sum p_i \log_2 p_i$ (accounting for the frequency distribution but not the visiting order), and the real entropy $S^{\mathrm{real}}$, which incorporates both spatial frequency and temporal dependencies. As shown in Fig.~\ref{fig:entropy_predictability}(a), $S^{\mathrm{real}}$ (in the figure, the real entropy is simply denoted by $S$) is significantly lower than the other two, indicating that user behavior exhibits strong spatio-temporal regularities, which are the fundamental source of high predictability. Correspondingly, the distributions of the maximum predictability $\Pi^{\mathrm{max}}$ and the two baseline cases are shown in Fig.~\ref{fig:entropy_predictability}(b). Furthermore, the authors introduced the radius of gyration $r_g$ (the average deviation between the trajectory center and the actual positions) to capture the spatial scale of individual mobility ranges. Although $r_g$ follows a power-law heavy-tailed distribution across the population, reflecting substantial heterogeneity in individual mobility ranges, the upper bound of predictability $\Pi^{\mathrm{max}}$ converges to a stable value (approximately 0.93) for users with $r_g > 10\ \mathrm{km}$ (see Fig.~\ref{fig:entropy_predictability}(c)). This suggests that regardless of the size of the activity range, individual mobility patterns remain highly predictable, revealing intrinsic regularities of human behavior.

In addition, they analyzed the contribution of hotspot location preferences to predictability (Fig.~\ref{fig:entropy_predictability}(d)). The results indicated that when considering only the two most frequently visited locations ($c=2$, e.g., “home” and “office”), the upper bound of predictability is about 0.6. As $c$ increases, the predictability $\tilde{\Pi}$ exhibits an approximately logarithmic growth, implying that high predictability arises not only from a few dominant hotspots but also from other regular activity patterns.

\begin{figure*}[!t]
	\centering
	\setlength{\abovecaptionskip}{0pt}
	\setlength{\belowcaptionskip}{0pt}
	\includegraphics[width=1.0\textwidth]{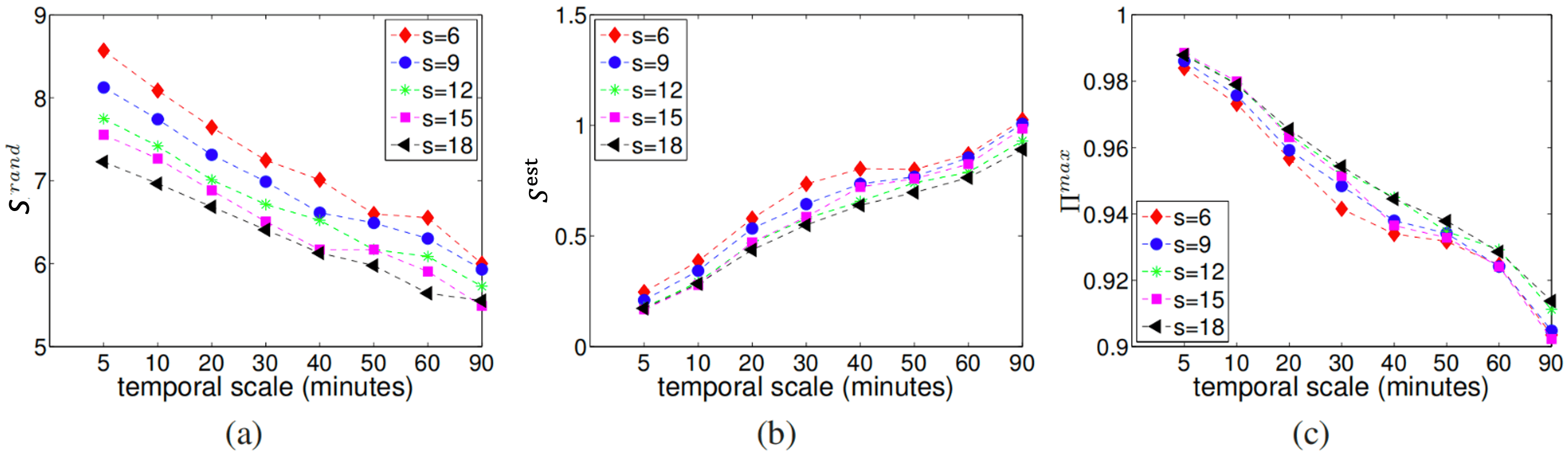}
	\caption{Entropy measures and predictability under different spatio-temporal resolutions for an individual with a radius of gyration of about 15 kilometers. 
		(a) Random entropy $S^{\mathrm{rand}}$, 
		(b) estimated entropy $S^{\mathrm{est}}$, and 
		(c) maximum predictability $\Pi^{\mathrm{max}}$ across varying spatio-temporal scales. 
		Temporal resolution ranges from 5 minutes to 90 minutes, and spatial resolution $s$ ranges from 6 to 18. \\
		\textit{Source}: The figure is reproduced from Ref.  \cite{lin2012predictability}.}
	\label{fig:entropy_predictability_15km}
	\vspace{0pt}
\end{figure*}

Lin \textit{et al.}~\cite{lin2012predictability} systematically evaluated the impact of spatial and temporal resolution on individual mobility predictability using high-resolution GPS trajectory data. They mapped trajectories into grid cells controlled by a scale parameter $s$: when $s$ is small (e.g., $s=4$, approximately $0.15\ \mathrm{km}^2$), the spatial resolution is high; when $s$ is large (e.g., $s=18$, approximately $3\ \mathrm{km}^2$), the spatial resolution is coarse.  
As shown in Fig.~\ref{fig:entropy_predictability_15km}(a–c), as the spatial scale increases from $s=4$ to $s=18$, the number of distinct locations $C$ decreases substantially, leading to a reduction in random entropy $S^{\mathrm{rand}} = \log C$. Meanwhile, because locations are merged into coarser grid cells, trajectory sequences become more repetitive, which reduces the estimated entropy $S^{\mathrm{est}}$. Consequently, the maximum predictability $\Pi^{\mathrm{max}}$ derived from the Fano framework increases overall, rising from about 0.90 to 0.93. This indicates that individual trajectories appear more regular and thus easier to predict at coarser spatial granularity. Nevertheless, even at building-level high resolution, predictability remains around 90\%, underscoring the intrinsically high predictability of human mobility.  

Moreover, the authors observed an approximate invariance relationship between $\Pi^{\mathrm{max}}$ and spatial uncertainty $\mu = \tfrac{1}{s}$, expressed as $\mu^\beta \Pi^{\mathrm{max}} \approx \kappa$. This finding reveals an inherent trade-off between spatial granularity and prediction accuracy: finer resolution inevitably reduces the predictability limit, whereas coarser resolution improves it. This insight provides important theoretical guidance for choosing scales and optimizing performance in predictive algorithms under varying precision requirements.  

Regarding temporal resolution, when the sampling interval increases from 5 minutes to 90 minutes, short stops are merged, reducing the number of distinct locations $C$ and thereby decreasing the random entropy $S^{\mathrm{rand}}$. At the same time, as temporal granularity becomes coarser, each sampling point covers a longer duration, weakening or even breaking up long dwell periods (consecutive identical symbols). This significantly reduces long-term correlations in the sequence, causing the estimated entropy $S^{\mathrm{est}}$ to increase slightly. As a result, the maximum predictability $\Pi^{\mathrm{max}}$ decreases with increasing sampling intervals. Contrary to the spatial trend, this indicates that coarser temporal resolution masks fine-grained behavioral patterns and reduces trajectory predictability.

Although prior studies have demonstrated the high predictability of human mobility trajectories, most of them mainly focused on repetitive visiting patterns within a finite set of locations, while overlooking the exploration of new places in behavioral sequences. To address this gap, Cuttone \textit{et al.} \cite{cuttone2018understanding} proposed the metric of exploration rate that quantifies the frequency of visiting new locations in a sequence, and examined its relationship with trajectory predictability. Experimental results revealed a significant negative correlation between predictability and exploration rate: the stronger the exploratory behavior, the lower the predictability of the sequence. Despite individual differences in exploration behavior, the exploration–exploitation balance was identified as a core factor shaping individual predictability. This study emphasized that predictability is not only associated with the size of activity ranges or hotspot preferences, but also inherently influenced by exploratory behavior, which introduces additional uncertainty and thus depress the upper bound of prediction accuracy.  

Teixeira \textit{et al.} investigated the impact of multi-source social media check-in data on user location predictability \cite{teixeira2018predictability}, aiming to explore whether cross-platform data integration could enhance predicting ability. Their study leveraged combined check-in data from Instagram and Foursquare to evaluate the predictability of next-location. The results showed that simply aggregating data across platforms did not significantly improve the predictability. The primary reason lies in the pronounced differences in user behavior patterns between platforms: Foursquare check-ins tend to capture routine mobility trajectories, whereas Instagram check-ins often reflect selective stays at socially or aesthetically valued locations. This heterogeneity in usage context and behavioral patterns introduces additional uncertainty when the datasets are combined, thereby reducing the overall predictability of the sequence. This study highlighted that increasing the volume of data does not necessarily improve predictability. Instead, behavioral consistency and contextual compatibility between data sources are critical factors determining the effectiveness of data fusion.

\begin{figure*}[!t]
	\centering
	\setlength{\abovecaptionskip}{0pt}
	\setlength{\belowcaptionskip}{0pt}
	\includegraphics[width=\textwidth]{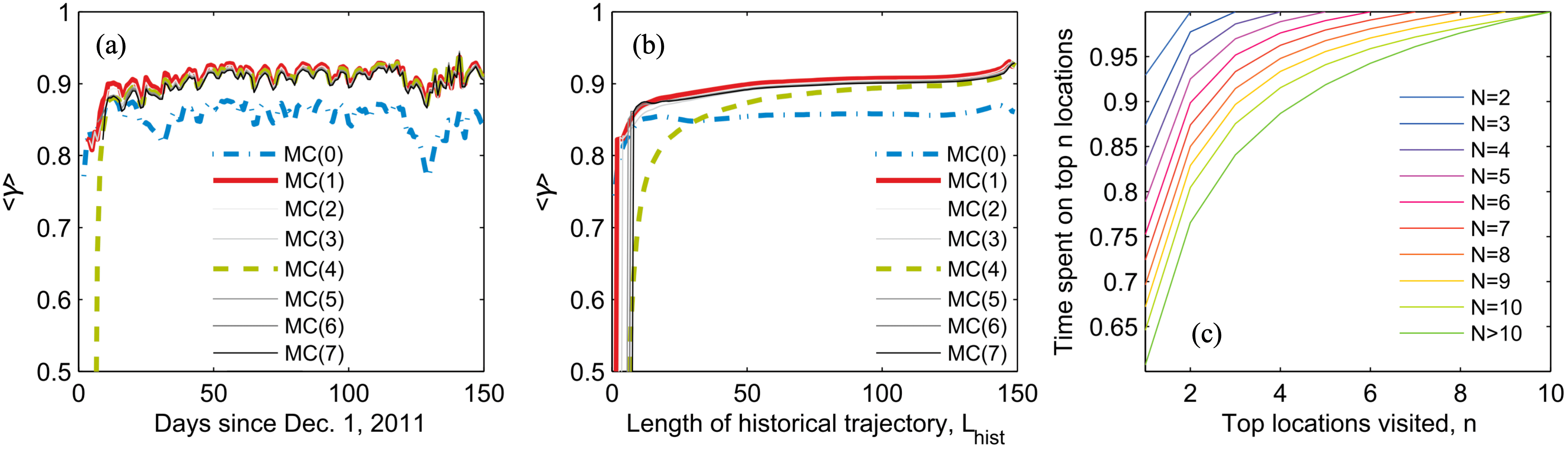} 
	\caption{Relationship between visiting behavior and prediction accuracy. 
		(a) Day-to-day prediction accuracy based on historical trajectory data (excluding users with no activity on a given day), comparing Markov chain models of different orders (MC(0)–MC(7)). 
		(b) Trends of prediction accuracy as a function of historical trajectory length $L_{\mathrm{hist}}$, showing that all Markov chain models improve in performance as the history length increases. 
		(c) Distribution of the proportion of time spent at the top-$n$ most frequently visited locations. Users are grouped into 10 categories according to their total number of distinct visited locations $N$. \\
		\textit{Source}: The figure is reproduced from Ref. \cite{lu2013approachingTest}.}
	\label{fig:visiting_behavior_prediction_accuracy}
	\vspace{0pt}
\end{figure*}

Lu \textit{et al.}~\cite{lu2013approachingTest} argued that the maximum predictability is not only an information-theoretic limit but can also be empirically approached through concrete modeling methods. Using mobile phone data from 500,000 users in Côte d'Ivoire, they employed higher-order Markov chain models to systematically evaluate the predictive performance of practical algorithms under different user behavioral patterns, thereby verifying the attainability of the theoretical bound. As shown in Fig.~\ref{fig:visiting_behavior_prediction_accuracy}(a), Markov chains of different orders (MC(0) as a frequency-based baseline model, and MC(1)–MC(7) as first- and higher-order Markov chains) exhibit a general trend of increasing accuracy with longer historical trajectories in day-to-day prediction. The improvement from MC(0) to MC(1) is particularly pronounced—accuracy rises sharply from about 80\% to above 90\%—indicating that even a simple first-order Markov chain can capture the core sequential regularities in user mobility trajectories.  

Furthermore, Fig.~\ref{fig:visiting_behavior_prediction_accuracy}(b) shows the overall trend of prediction accuracy as a function of the history length $L_{\mathrm{hist}}$. As $L_{\mathrm{hist}}$ increases, the performance of Markov models gradually saturates and eventually approaches the theoretical upper bound of predictability. The additional improvements from MC(2) to higher-order models are limited, suggesting that first-order dependencies already account for the majority of the predictable structure.  

Finally, Fig.~\ref{fig:visiting_behavior_prediction_accuracy}(c) shows the distribution of dwell times at the most frequently visited locations from the perspective of user behavioral preferences. The results reveal that although the total number of locations $N$ visited by users varies considerably, more than 80\% of activity time is concentrated in the top-5 most frequently visited locations. This strong preference for core hotspot locations provides an essential structural basis for achieving high-accuracy predictions with Markov models and explains why such models can empirically approach the theoretical upper bound of prediction accuracy.

Wang \textit{et al.} \cite{wang2020predictability}, building on an understanding of the mechanisms underlying sequence predictability, proposed a mobility prediction method designed to address behavioral pattern switching and the sparsity of trajectory data. They argued that the error in estimating the predictability of individual mobility sequences primarily stems from two factors: (i) strong sequential dependencies in behavior, especially reflected in stage-based daily activity patterns such as commuting, working, and resting; and (ii) the sparsity of trajectory data, which weakens the expression of regular structural features, leading to an overestimation of entropy and underestimation of predictability. To address these issues, they divided a day into multiple behavioral stages and constructed multi-chain Markov models to characterize state transitions within each stage, thereby capturing users' temporal switching patterns. In addition, they proposed a joint inference mechanism based on Gibbs sampling \cite{geman1984stochastic} to simultaneously impute missing locations and optimize state transition probabilities, thereby maximally restoring structural information obscured by data sparsity.  
\begin{figure*}[!t]
	\centering
	\setlength{\abovecaptionskip}{0pt}
	\setlength{\belowcaptionskip}{0pt}
	\includegraphics[width=\textwidth]{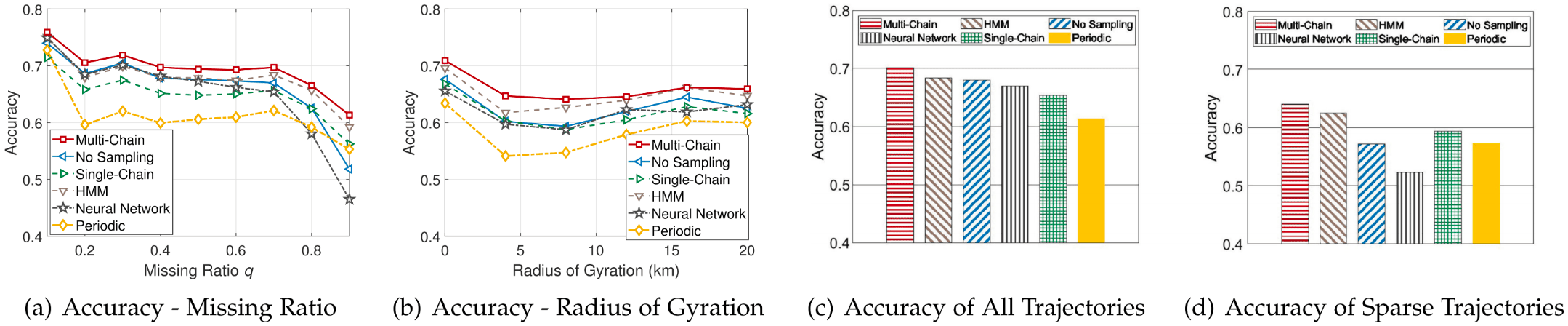} 
	\caption{Performance comparison of different prediction algorithms under various conditions. 
		(a) Prediction accuracy under different levels of missing data proportion $q$, comparing the robustness of Multi-Chain, No Sampling, single-chain Markov model, HMM, Neural Network, and Periodic models; 
		(b) trend of prediction accuracy with respect to the radius of gyration, reflecting the influence of user activity range on the performance of different algorithms; 
		(c) comparison of average prediction accuracy across complete trajectories; 
		(d) comparison of average prediction accuracy across sparse trajectories. 
		For details of these methods, please refer to \cite{wang2020predictability}. \\
		\textit{Source}: The figure is reproduced from Ref. \cite{wang2020predictability}.}
	\label{fig:comparison_prediction_algorithms}
	\vspace{0pt}
\end{figure*}

As shown in Fig.~\ref{fig:comparison_prediction_algorithms}, the proposed method outperforms baseline approaches—including single-chain Markov models, hidden Markov models (HMMs), neural networks, and periodic models—in terms of prediction accuracy across different levels of missing data proportion $q$, demonstrating superior performance. The method also maintains its advantage under varying activity ranges (radius of gyration). In both complete and sparse trajectory datasets, it achieves the highest average prediction accuracy, making it the empirical approach closest to the theoretical bound of the prediction of mobility sequences.

\begin{figure*}[!t]
	\centering
	\setlength{\abovecaptionskip}{0pt}
	\setlength{\belowcaptionskip}{0pt}
	\includegraphics[width=1.0\textwidth]{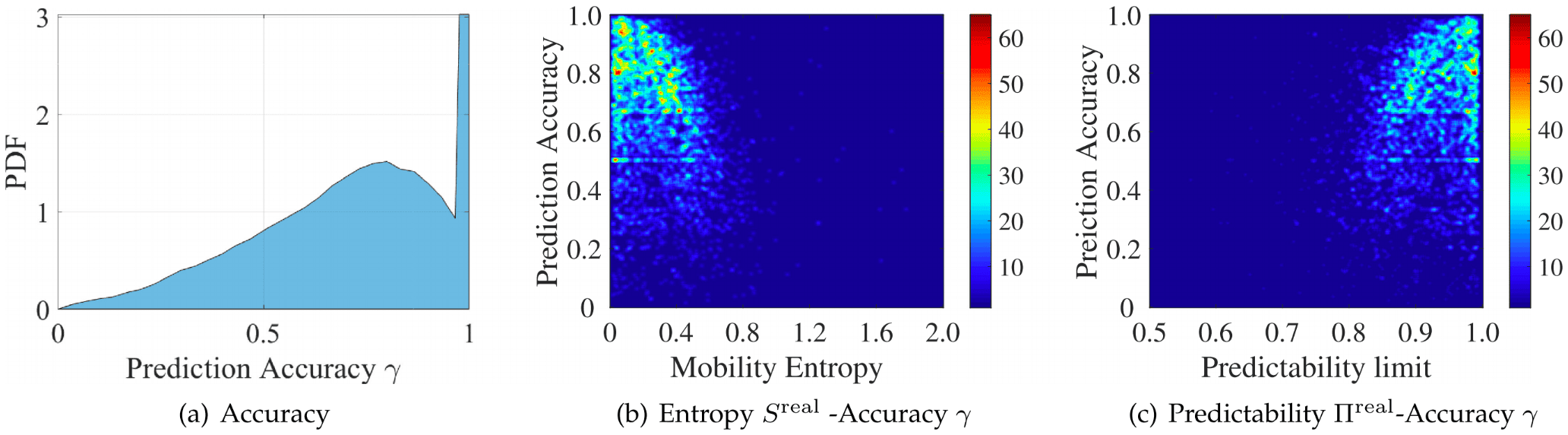} 
	\caption{Relationships among prediction accuracy, entropy, and predictability. 
		(a) Probability density distribution (PDF) of user prediction accuracy $\gamma$; 
		(b) relationship between the actual mobility entropy $S^{\text{real}}$ and prediction accuracy $\gamma$, where color indicates the density of users in each region; 
		(c) relationship between the actual predictability $\Pi^{\text{real}}$ and prediction accuracy $\gamma$, with color representing user density. \\
		\textit{Source}: The figure is reproduced from Ref. \cite{wang2020predictability}.}
	\label{fig:correlation_accuracy_entropy_predictability}
	\vspace{0pt}
\end{figure*}

In addition, Wang \textit{et al.} analyzed the relationship between prediction accuracy, the real entropy $S^{\mathrm{real}}$, and the theoretical upper bound of predictability $\Pi^{\mathrm{real}}$ (see Fig.~\ref{fig:correlation_accuracy_entropy_predictability}). Specifically, (a) shows the probability density distribution of user prediction accuracy $\gamma$; (b) illustrates the two-dimensional density relationship between $S^{\mathrm{real}}$ and $\gamma$; and (c) presents the correspondence between $\Pi^{\mathrm{real}}$ and $\gamma$. The results demonstrate a strong overall positive correlation between $\Pi^{\mathrm{real}}$ and $\gamma$, with prediction accuracy for some highly predictable sequences already approaching the theoretical limit. Nevertheless, for a large number of sequences with medium-to-high predictability, there remains considerable room for improvement in prediction accuracy.

\subsubsection{Optimizations}
After the establishment of the framework based on Fano's inequality, improving estimation accuracy has become an important research direction. Prior studies have introduced two main lines of optimization: (i) refining the estimation of candidate set size by incorporating geographic and topological constraints to tighten the state space; and (ii) integrating top-$\ell$ candidate probability information into the Fano scaling formula. These approaches reduce uncertainty through topological constraints and probability information fusion, respectively, thereby yielding more accurate predictability estimates. Within the Fano-based predictability framework (see Eq.~\eqref{eq:fano_max}), the key parameters determining predictability $\Pi$ include the sequence entropy $S$ and the size of the candidate state space $C$. Here, $S$ reflects the uncertainty of the sequence, while $C$ denotes the number of possible next states given historical information. The accuracy of estimating $C$ directly affects the precision of the final predictability estimate.  

In early studies, Song \textit{et al.}~\cite{song2010limits} directly used the number of distinct locations visited in a user's historical trajectory as an estimate of the candidate set size, denoted as $C_s$. This estimation method implicitly assumes that the user may visit any previously visited location at any time, without considering the geographic and topological constraints imposed by the user's current position.

To address the above issue, Smith \textit{et al.}~\cite{smith2014refined} proposed a candidate set estimation method with geographic and topological constraints, based on the concept of reachable locations. They argued that in reality, users cannot “instantaneously jump” from their current position to all previously visited locations, since many locations require passing through intermediate locations to be reached. Accordingly, the authors introduced the concept of a “one-step reachable location set” as a more realistic definition of the next-state candidate set, and used its maximum value $C_r$ as an approximate upper bound of the candidate set size, formally defined as:
\begin{align}
	C_{r} = \max_{v \in \Omega} \lvert \{z_{i+1} : z_{i} = v\} \rvert, \label{formula_7}
\end{align}
where $\Omega$ denotes the set of all locations that appear in the user's trajectory. Based on this definition, the number of distinct successor states for each state $z_i$ is computed, and $C_r$ is the maximum value. Compared with $C_s$, $C_r$ is theoretically closer to the real behavioral boundaries of users, and it provides a direction for incorporating spatial structural constraints into predictability evaluation in subsequent studies.

\begin{figure*}[!t]
	\centering
	\setlength{\abovecaptionskip}{0pt}
	\setlength{\belowcaptionskip}{0pt}
	\includegraphics[width=0.65\textwidth]{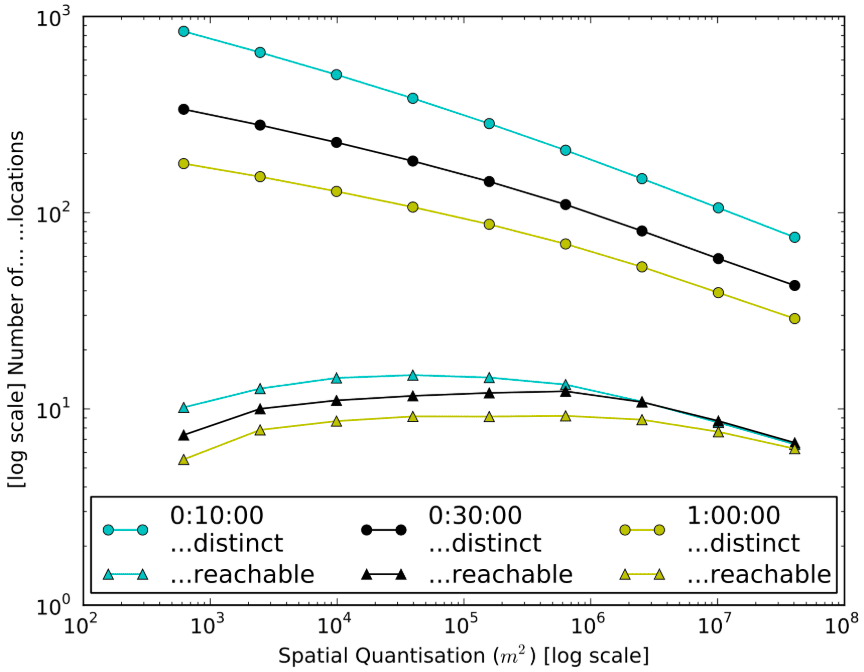}
	\caption{Comparison between the total number of distinct locations and the number of reachable locations across different spatial quantization scales under varying temporal sampling intervals, illustrating the gap caused by topological constraints. \\
		\textit{Source}: The figure is reproduced from Ref. \cite{smith2014refined}.}
	\label{fig:HF_N_relation}
\end{figure*}

Furthermore, Fig.~\ref{fig:HF_N_relation} compares the total number of distinct locations and the number of reachable locations under different spatial quantization scales across varying temporal sampling intervals. The results indicate that due to geographic and topological constraints, the number of actually reachable locations is far smaller than the number of distinct locations visited in the history. This correction significantly tightens the candidate set size $C$, thereby yielding a stricter upper bound on predictability for the same entropy. Based on the Fano scaling relation (Eq.~\eqref{eq:fano_max}), let the predictability upper bound be denoted as $\Pi^{\max}_C$ to explicitly indicate the dependence on candidate set size $C$. Since $S_F(\Pi, C)$ is monotonically decreasing with respect to $\Pi$ over the interval $\Pi \in \left(\tfrac{1}{C},1\right]$ and monotonically increasing with respect to $C$ (see, for example, Fig.~\ref{fig:fano_function}), when reachability constraints are introduced such that $C_r \leq C_s$, it follows that:
\begin{align}
	\Pi^{\max}_{C_r} \;\le\; \Pi^{\max}_{C_s}.
\end{align}
This refinement demonstrates that by imposing reasonable constraints on the state space, the theoretical upper bound of predictability can be tightened, making it closer to the actual predictability level of behavioral sequences.  

\begin{figure*}[htbp]
	\centering
	\setlength{\abovecaptionskip}{0pt}
	\setlength{\belowcaptionskip}{0pt}
	\includegraphics[width=\textwidth]{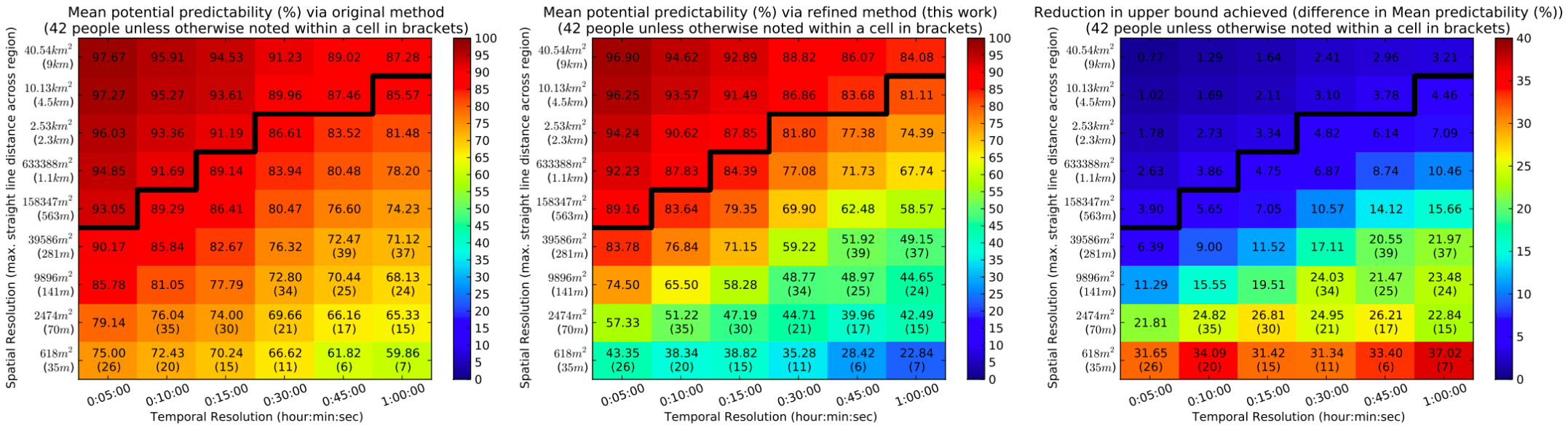} 
	\caption{Upper bounds of human mobility predictability under different spatio-temporal quantization conditions. The thick black line indicates the spatial resolution threshold that can be crossed within a single time step at a preferred walking speed of 5 km/h. 
		(a) Average potential predictability upper bound computed using the method proposed by Song \textit{et al.}~\cite{song2010limits}; 
		(b) average potential predictability upper bound computed using the refined method proposed by Smith \textit{et al.}~\cite{smith2014refined}; 
		(c) difference between the two methods (i.e., (a) minus (b)), reflecting the reduction in the upper bound after incorporating geographic reachability constraints. \\
		\textit{Source}: The figure is reproduced from Ref. \cite{smith2014refined}.}
	\label{fig:predictability_upper_bounds}
	\vspace{0pt}
\end{figure*}

Furthermore, Smith \textit{et al.} compared the differences in estimated upper bounds between the traditional method and the reachable-locations method under various spatio-temporal resolutions. As shown in Fig.~\ref{fig:predictability_upper_bounds}, the original approach \cite{song2010limits}, which is based on the candidate set size $C_s$, produces relatively high values of average upper bounds across different spatial and temporal resolutions, while the reachable-locations method suggests lower upper bounds. Fig.~\ref{fig:predictability_upper_bounds}(c) illustrates the distribution of differences between the two methods (i.e., (a) minus (b)), where in all spatio-temporal resolution settings, the reachable-locations method significantly reduces the estimated predictability, with the maximum decrease exceeding 30\%. This effect is particularly pronounced under high spatial resolution and short temporal sampling intervals, highlighting the critical role of the reachable-locations method in avoiding overestimation of predictability in fine-grained mobility prediction scenarios.  

The Fano scaling formula has also been widely used to compute the theoretical upper bound of predictability in recommender systems. In its standard form (i.e., Eq.~\eqref{eq:fano_max}), $\Pi^{\max}$ represents the hit rate of the most probable item in the candidate set, and $C$ is the size of the candidate set. This formulation indeed characterizes the predictability upper bound in the top-1 case. However, in practical recommendation scenarios, performance evaluation often focuses on top-$\ell$ results. A straightforward modification is to treat the top-$\ell$ outcomes as a whole and replace $C-1$ with $C-\ell$ in the formula, yielding:
\begin{equation}
	S = -\Pi^{\max} \log_2 \Pi^{\max} + \Pi^{\max} \log_2 \ell 
	- (1 - \Pi^{\max}) \log_2 (1 - \Pi^{\max}) 
	+ (1 - \Pi^{\max}) \log_2 (C-\ell). \label{eq:fano_topk}
\end{equation}

\begin{figure*}[htbp]
	\centering
	\setlength{\abovecaptionskip}{0pt}
	\setlength{\belowcaptionskip}{0pt}
	\includegraphics[width=1.0\textwidth]{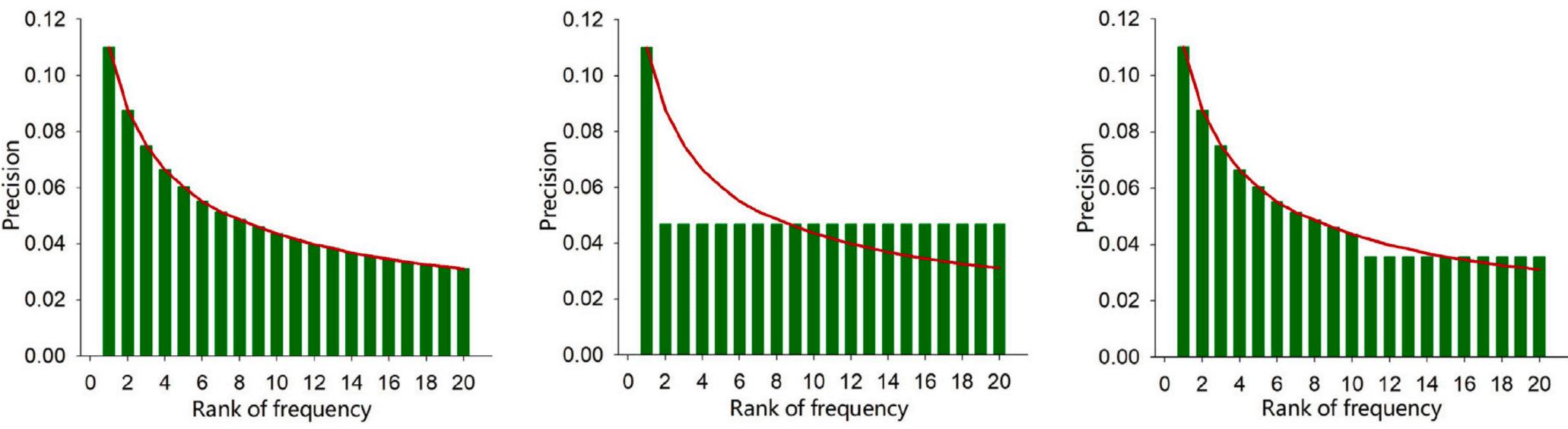} 
	\caption{Impact of different scaling forms on the predictability in recommender systems. 
		(a) True probability distribution of user-preferred items; 
		(b) Top-1 predictability computed by the classical method; 
		(c) The scaling form proposed in~\cite{xu2024limits}. \\
		\textit{Source}: The figure is reproduced from Ref. \cite{xu2024limits}.}
	\label{fig:scaling_forms_predictability}
	\vspace{0pt}
\end{figure*}

However, this approach has an inherent limitation: under the same entropy $S$ and candidate set size $C$, the resulting predictability is nearly identical because $\ell \ll C$. To overcome such limitation, Xu \textit{et al.} proposed a refined scaling strategy that explicitly incorporates the probability associations among top-$\ell$ candidates during computation \cite{xu2024limits}. As shown in Fig.~\ref{fig:scaling_forms_predictability}(a), the true probability distribution of user-preferred items often exhibits a pronounced long-tail pattern, and the probability differences among the top-$\ell$ items contain important information. In contrast, when computing top-1 predictability, the classical method retains only the maximum probability item (see Fig.~\ref{fig:scaling_forms_predictability}(b)), leading to an overly coarse approximation of the true distribution.
To more accurately capture the probability distribution,Xu \textit{et al.} introduced a set of scaling coefficients $\{r_1, r_2, \dots, r_\ell\}$ into the generalized scaling formula \eqref{eq:fano_topk}, where $r_i$ denotes the ratio of the probability of the $i$-th candidate to that of the most probable item. Based on this, they derived the formula of the top-$\ell$ predictability as:
\begin{equation}
	\begin{aligned}
		S_{F_C}(\Pi_1^{\max}) = & -\sum_{i=1}^\ell r_i \Pi_1^{\max} \log_2(r_i \Pi_1^{\max}) 
		- \left( 1 - \sum_{i=1}^\ell r_i \Pi_1^{\max} \right) 
		\log_2 \left(1 - \sum_{i=1}^\ell r_i \Pi_1^{\max} \right) \\
		& + \left( 1 - \sum_{i=1}^\ell r_i \Pi_1^{\max} \right) \log_2 (C - \ell).
	\end{aligned}
	\label{eq:top-l}
\end{equation}
By retaining richer information from the original distribution, this refined upper bound becomes tighter and more realistic. When $\ell=1$, it naturally reduces to the traditional top-1 case.

\begin{figure*}[!t]
	\centering
	\setlength{\abovecaptionskip}{0pt}
	\setlength{\belowcaptionskip}{0pt}
	\includegraphics[width=1.0\textwidth]{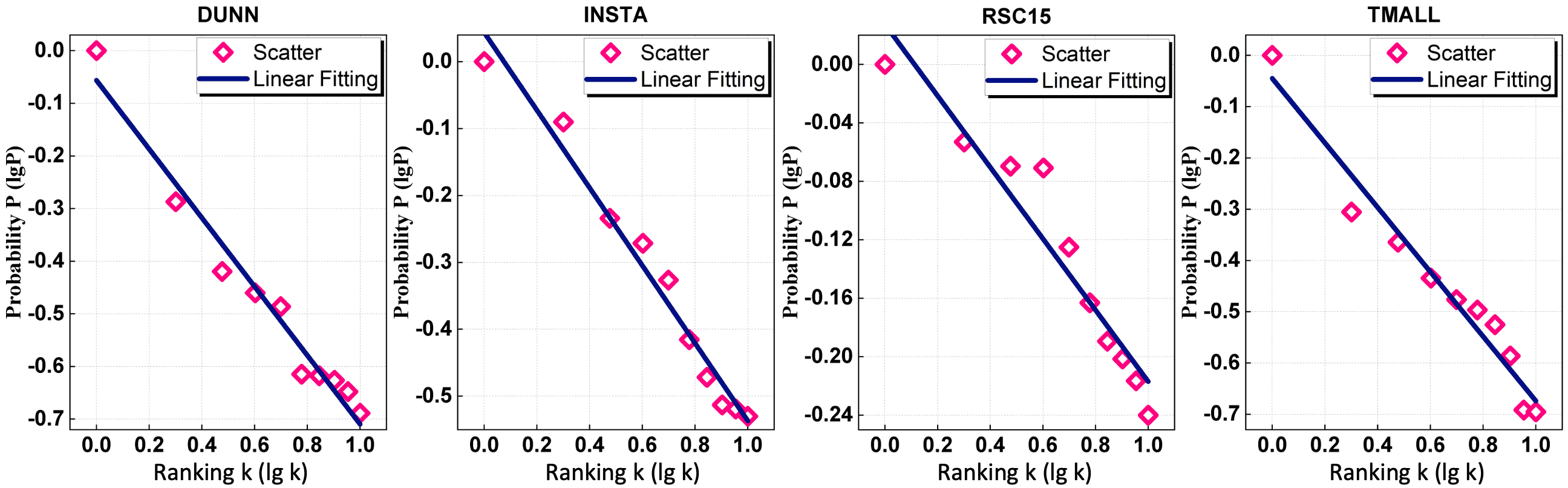} 
	\caption{Probability ratio distributions of user-preferred items across different datasets. The horizontal axis denotes the item rank in the frequency table ($\log k$), and the vertical axis denotes the item occurrence probability ($\log P(k)$). Red diamonds represent the observed probability distribution, while the blue solid line shows the linear fitting result. The results indicate that item frequencies decay with rank following a power-law relationship. \\
		\textit{Source}: The figure is reproduced from Ref. \cite{xu2024limits}.}
	\label{fig:prob_rank_dist}
	\vspace{0pt}
\end{figure*}

In terms of probability relationships, Xu \textit{et al.} conducted statistical analyses on multiple real-world datasets and found that the probability distribution of top-$\ell$ items follows Zipf's law \cite{zipf2016human,lu2010zipf}:
\begin{equation}
	P(k) \propto k^{-\xi},
\end{equation}
where $k$ ($1 \leq k \leq \ell$) denotes the rank of item in the top-$\ell$ list, and the fitting values of the Zipf's exponent $\xi$ are consistently close to 0.6 across different datasets (see Fig.~\ref{fig:prob_rank_dist}). 

Based on the fitted Zipf's distribution, one can calculate the predictability according to Eq.~\ref{eq:top-l}. Xu \textit{et al.} further analyzed $S_{F_\ell}\!\big(\Pi^{\max}_{1,\ell}\big)$ and $S_{F_{\ell-1}}\!\big(\Pi^{\max}_{1,\ell-1}\big)$, where $\Pi^{\max}_{1,\ell}$ and $\Pi^{\max}_{1,\ell-1}$ denote the predictability upper bounds obtained by incorporating the probability distribution of the top-$\ell$ and top-$(\ell-1)$ candidate sets, and $S_{F_\ell}\!\big(\Pi^{\max}_{1,\ell}\big)=S_{F_{\ell-1}}\!\big(\Pi^{\max}_{1,\ell-1}\big)=S^{\mathrm{est}}$. To facilitate the comparison between $\Pi^{\max}_{1,\ell}$ and $\Pi^{\max}_{1,\ell-1}$, Xu \textit{et al.} introduced an intermediate quantity $S_{F_{\ell-1}}\!\big(\Pi^{\max}_{1,\ell}\big)$, which adopts the top-$(\ell-1)$ aware scaling form while maintaining the same predictability value as in $S_{F_\ell}\!\big(\Pi^{\max}_{1,\ell}\big)$:
\begin{align}
	S_{F_{\ell}}\!\big(\Pi^{\max}_{1,\ell}\big)
	&= S\!\left(
	\Pi^{\max}_{1,\ell},\, r_{2}\Pi^{\max}_{1,\ell},\, \dots,\, r_{\ell}\Pi^{\max}_{1,\ell},\,
	\underbrace{\tfrac{1-\sum_{i=1}^{\ell} r_i \Pi^{\max}_{1,\ell}}{C-\ell}, \dots}_{C-\ell \text{tail terms evenly distributed}}
	\right) \label{eq:sfk} \\
	&\le S\!\left(
	\Pi^{\max}_{1,\ell},\, r_{2}\Pi^{\max}_{1,\ell},\, \dots,\, r_{\ell-1}\Pi^{\max}_{1,\ell},\,
	\underbrace{\tfrac{1-\sum_{i=1}^{\ell-1} r_i \Pi^{\max}_{1,\ell}}{C-(\ell-1)}, \dots}_{C-(\ell-1) \text{tail terms evenly distributed}}
	\right) \label{eq:sfk-1} \\
	&= S_{F_{\ell-1}}\!\big(\Pi^{\max}_{1,\ell}\big). \label{eq:sfk-1mid}
\end{align}
Equations~\eqref{eq:sfk} and \eqref{eq:sfk-1} share the same first $\ell-1$ terms, with the only difference lying in the tail allocation: in the top-$(\ell-1)$ aware case, the $\ell$-th candidate is merged into the tail distribution, making the overall distribution more uniform and thus yielding a larger entropy value. Consequently, we have
$S_{F_{\ell}}\!\big(\Pi^{\max}_{1,\ell}\big) \leq S_{F_{\ell-1}}\!\big(\Pi^{\max}_{1,\ell}\big)$.

Combining the known condition 
$S^{\mathrm{est}} = S_{F_\ell}(\Pi^{\max}_{1,\ell}) = S_{F_{\ell-1}}(\Pi^{\max}_{1,\ell-1}) \leq S_{F_{\ell-1}}\!\big(\Pi^{\max}_{1,\ell}\big)$ 
with the property that $S_{F_{\ell-1}}(\cdot)$ is strictly monotonically decreasing over $\Pi \in (\frac{1}{C},1]$, we obtain
\begin{equation}
	\Pi^{\max}_{1,\ell-1} \geq \Pi^{\max}_{1,\ell},
\end{equation}
which obviously leads to the following chain relation:
\begin{equation}
	\Pi^{\max}_{1,1} \;\ge\; \Pi^{\max}_{1,2} \;\ge\; \cdots \;\ge\; \Pi^{\max}_{1,\ell} \;\ge\; \Pi_{1},
\end{equation}
where $\Pi_{1}$ denotes the theoretical top-1 predictability, and $\Pi^{\max}_{1,1}$ corresponds to the upper bound obtained under the traditional scaling method. This result indicates that as $\ell$ increases, the method proposed by Xu \textit{et al.} yields progressively tighter upper bounds. In practical applications, when $\ell$ becomes too large, the probability mass of the items in the tail of the top-$\ell$ list tends to be very small, so that those items contribute marginally to the overall predictability but introduce additional noise. Therefore, to balance the coverage of the major probability mass and avoid amplification of estimation errors, Xu \textit{et al.}~\cite{xu2024limits} suggest using $\ell \leq 10$.

\subsubsection{Extensions}
The aforementioned studies have primarily focused on establishing the theoretical framework of predictability and improving the accuracy of estimation. Another line of research emphasizes applying this framework to broader contexts and analyzing the influence of additional features on predictability. Representative approaches include: employing cross entropy to evaluate the informational contribution across different subjects or sequences; using context entropy to quantify the constraining effect of contextual factors on the predictability bound; and adopting conditional entropy to examine how the predictability of the original sequence changes given other available information. These methods, while preserving theoretical rigor, extend predictability analysis to complex scenarios such as multi-source data integration, contextual modulation, and feature interaction mechanisms, thereby broadening the applicability of the theoretical framework.

\begin{figure*}[!t]
	\centering
	\setlength{\abovecaptionskip}{0pt}
	\setlength{\belowcaptionskip}{0pt}
	\includegraphics[width=\textwidth]{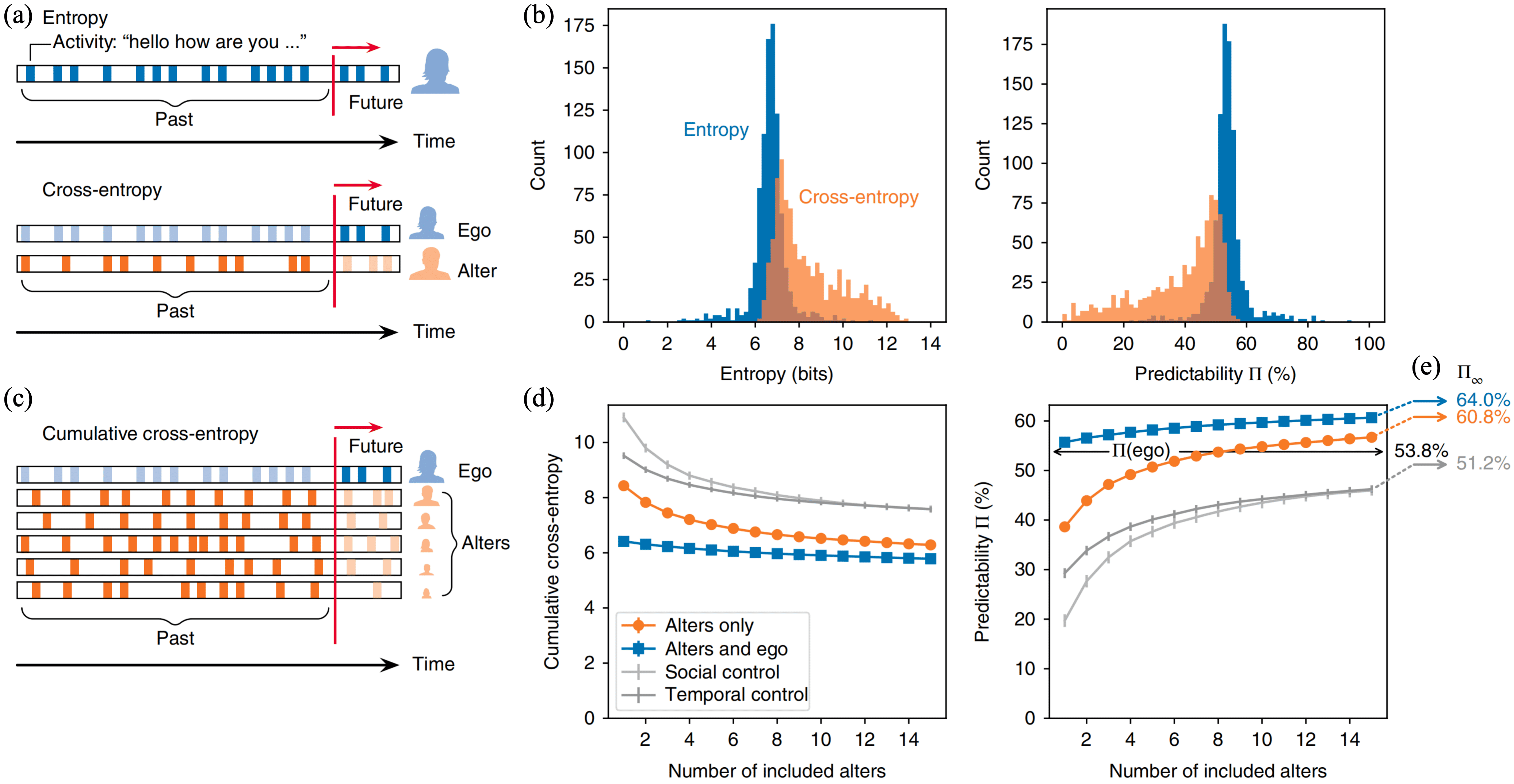} 
	\caption{Information and predictability analysis in online social activities. 
		(a) Illustration of entropy and cross-entropy computation for user posting time series: entropy measures the uncertainty of predicting future text given past text, while cross-entropy quantifies the amount of information about the ego's future text contained in the past text of another user (alter). 
		(b) Distributions of user entropy, cross-entropy, and the corresponding predictability, showing that most users fall within a limited range of predictability. 
		(c) Illustration of cumulative cross-entropy: incorporating the historical activities of multiple alters provides additional predictive information. 
		(d) As the number of included alters increases, cumulative cross-entropy decreases and predictability increases; results are also compared across settings using only alters, alters+ego, temporal control, and social control. 
		(e) Extrapolated estimates of the upper bound of online activity predictability $\Pi_{\infty}$, highlighting differences across information source conditions. 
		Error bars represent mean $\pm95\%$ confidence intervals. \\
		\textit{Source}: The figure is reproduced from Ref. \cite{bagrow2019information}.}
	\label{fig:info_predictability_social}
	\vspace{0pt}
\end{figure*}

\textbf{Cross Entropy}. Bagrow \textit{et al.}~\cite{bagrow2019information} proposed an information-theoretic framework to quantify the upper limit of behavioral predictability in social media. The core question is: in the absence of an individual's own behavioral data, is it possible to accurately predict that individual's future behavior solely based on the historical activities of her/his social contacts? To this end, the authors utilized Twitter's Gardenhose data stream (a random sample of approximately 10\% of all public tweets) and, after filtering and preprocessing, constructed a structured dataset consisting of 927 users (ego) and their 15 most frequent contacts (alter). This dataset contained over 30 million temporally ordered tweets spanning at least one year. During preprocessing, case distinctions, punctuation, and URLs were removed so that tweet sequences could be modeled as continuous linguistic text. The core methodology of the study is built upon estimating the entropy of language sequences. Given a sequence of words $\{w_1, w_2, \ldots, w_N\}$, its entropy $S^{\mathrm{est}}$ represents the average uncertainty per word (see Eq.~\eqref{eq:entropy_est}).

To quantify the predictive power of friends (alters) over a target user (ego), the authors extended the above entropy to the cross-entropy $S^{\mathrm{est}}(A|B)$, which estimates the average amount of information required to encode the future word sequence of $A$ given only the historical text of $B$. The cross-entropy is defined as:
\begin{equation}
	S^{\mathrm{est}}(A|B) = \frac{N_A \log_2 N_B}{\sum_{i=1}^{N_A} \Lambda_i(A|B)},
\end{equation}
where $\Lambda_i(A|B)$ denotes the length of the shortest substring starting at the $i$-th word in $A$ that has not previously appeared in $B$'s historical text; “historical” is strictly limited to the content of $B$ prior to the timestamp of the current word in $A$, thereby capturing the directionality of information flow.

Building on this, the authors further introduced the concept of cumulative cross-entropy to measure the joint predictive capability of multiple alters. Let the set of alters be $\mathcal{B}$, then the multi-source cross-entropy is defined as:
\begin{equation}
	S^{\mathrm{est}}(A|\mathcal{B}) = \frac{N_A \log_2 N_{A\mathcal{B}}}{\sum_{i=1}^{N_A} \Lambda_i(A|\mathcal{B})},
\end{equation}
where $\Lambda_i(A|\mathcal{B}) = \max_{B \in \mathcal{B}} \Lambda_i(A|B)$ represents the maximum matching length among all alters at position $i$, and
\begin{equation}
	N_{A\mathcal{B}} = \frac{\sum_{B \in \mathcal{B}} w_B N_B}{\sum_{B \in \mathcal{B}} w_B},
\end{equation}
where $N_B$ is the length of sequence $B$, and $w_B$ denotes the number of times alter $B$ provides the longest match for the ego sequence during prediction, thereby measuring $B$'s contribution to the overall information computation. This metric reflects the trend of increasing cumulative information as the size of the social circle expands. After estimating cross-entropy, the authors applied Fano's inequality and followed the approach of Song \textit{et al.}~\cite{song2010limits} to convert it into an upper bound of predictability.

Empirical results show that when relying only on the ego's own historical text, the average entropy of users is about 6.6 bits, corresponding to a theoretical predictability upper bound of 53.8\%. Figure~\ref{fig:info_predictability_social}(d) illustrates the trend of increasing prediction accuracy as the number of alters is expanded: relying on only the Top 8–9 most frequent alters is sufficient to achieve performance comparable to, or even exceeding, that based on the ego's own text. Furthermore, Figure~\ref{fig:info_predictability_social}(e) presents nonlinear fitting and extrapolation of predictability curves for alters-only (orange) and alters+ego (blue). The estimated saturation limits of social behavior predictability are $\Pi_{\infty} = 60.8\%$ and $64.0\%$, respectively. This result suggests that information embedded in social ties is sufficient to recover more than 95\% of an individual's predictive capacity, indicating that predictability arises not only from the temporal patterns of personal behavior but is also deeply embedded in the structure of social networks, providing important implications for modeling social behavior and understanding privacy leakage mechanisms.

\begin{figure*}[!t]
	\centering
	\setlength{\abovecaptionskip}{0pt}
	\setlength{\belowcaptionskip}{0pt}
	\includegraphics[width=0.95\textwidth]{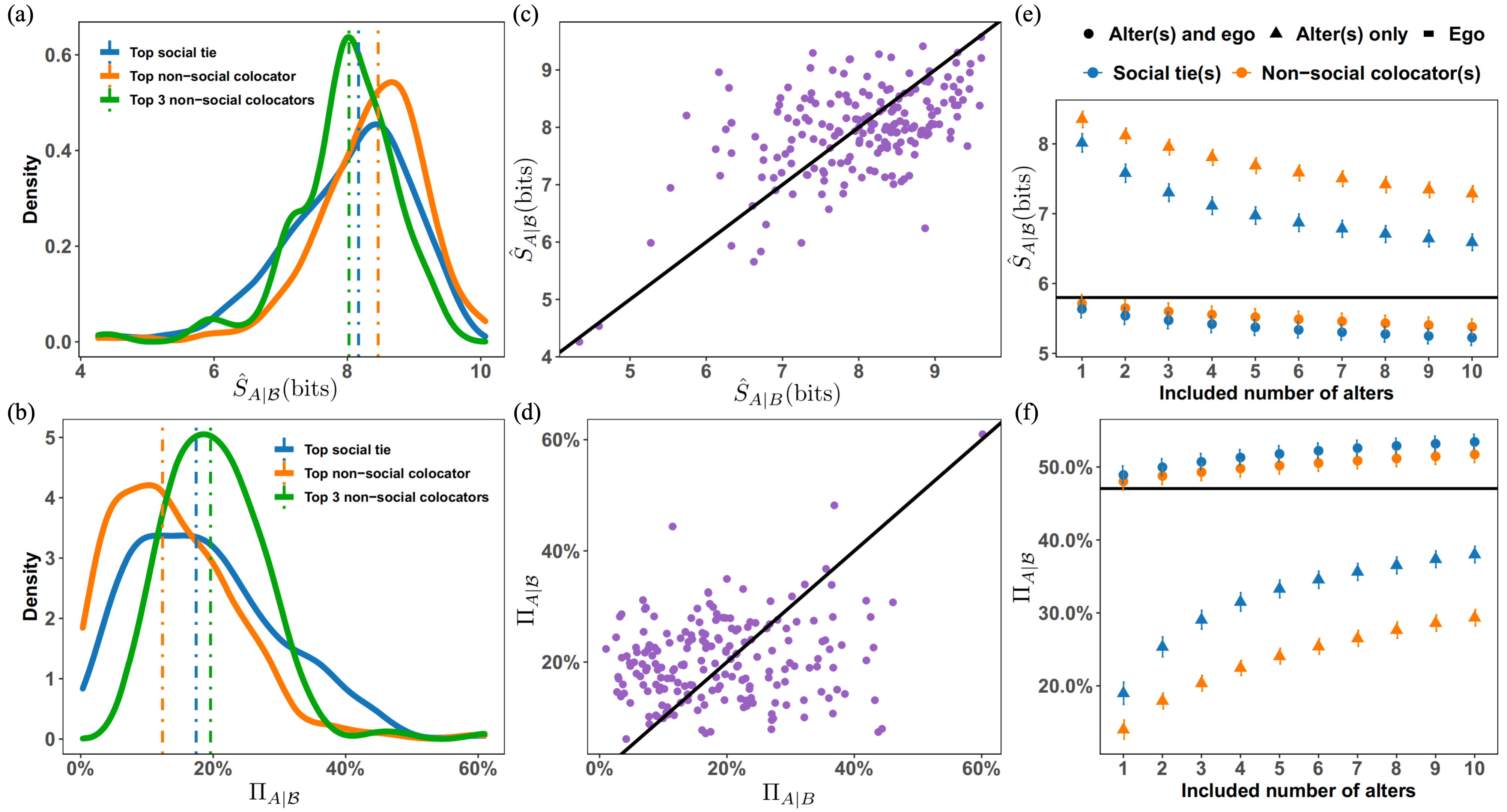} 
	\caption{Comparison of cross-entropy and predictability between social ties and non-social colocators. 
		(a) Distribution of conditional cross-entropy $\hat{S}_{A|B}$ in the Weeplaces dataset for the Top-1 social tie (median 8.17 bits), the Top-1 non-social colocator (median 8.46 bits), and the Top-3 non-social colocators (median 8.02 bits); 
		(b) Corresponding distribution of predictability $\Pi_{A|B}$: Top-1 social tie (median 17.43\%), Top-1 non-social colocator (median 12.35\%), and Top-3 non-social colocators (median 19.60\%); 
		(c) Each point corresponds to one ego, with the x-axis representing the cross-entropy $\hat{S}_{A|B}$ of the Top-1 friend and the y-axis representing the cumulative cross-entropy of the Top-3 colocators; the black solid line $y=x$ denotes equality; 
		(d) Similar to (c), but with axes showing $\Pi_{A|B}$; 
		(e) Variation of $\hat{S}_{A|B}$ with the number of alters after cumulatively including the Top 10 alters, comparing social ties and non-social colocators. Circles indicate inclusion of self-history, while triangles indicate exclusion; 
		(f) Corresponding variation of predictability $\Pi_{A|B}$, with horizontal lines denoting the mean entropy of the ego (5.80 bits) and the ego's predictability (47.05\%). 
		Error bars represent mean $\pm95\%$ confidence intervals. \\
		\textit{Source}: The figure is reproduced from Ref. \cite{chen2022contrasting}.}
	\label{fig:cross_entropy_social_vs_nonsocial}
	\vspace{0pt}
\end{figure*}

The above finding raises a critical question: does this phenomenon stem from genuine information carried by social ties, or is it merely the result of behavioral similarity among users under the same platform environment? To address this issue, Chen \textit{et al.}~\cite{chen2022contrasting} conducted a systematic comparison of the roles of social ties and non-social colocators in behavioral predictability, based on data from nearly 6,000 active Twitter users. In their study, “friends” were defined as accounts with explicit interactions with the ego (e.g., mentions, retweets). Three prediction scenarios were constructed accordingly: (1) self-only: using only the ego's own historical behavioral sequence; (2) friends-only: using only the behavioral sequences of the ego's social ties (friends); and (3) non-friends-only: using only the behavioral sequences of colocators—users who appeared at the same locations as the ego but had no direct interaction. In all three scenarios, Lempel–Ziv entropy \cite{kontoyiannis2002nonparametric} estimation combined with Fano's inequality was applied to convert cross-entropy $S^{\mathrm{est}}(A|B)$ into predictability $\Pi(A|B)$.

\begin{figure*}[!t]
	\centering
	\setlength{\abovecaptionskip}{0pt}
	\setlength{\belowcaptionskip}{0pt}
	\includegraphics[width=0.90\textwidth]{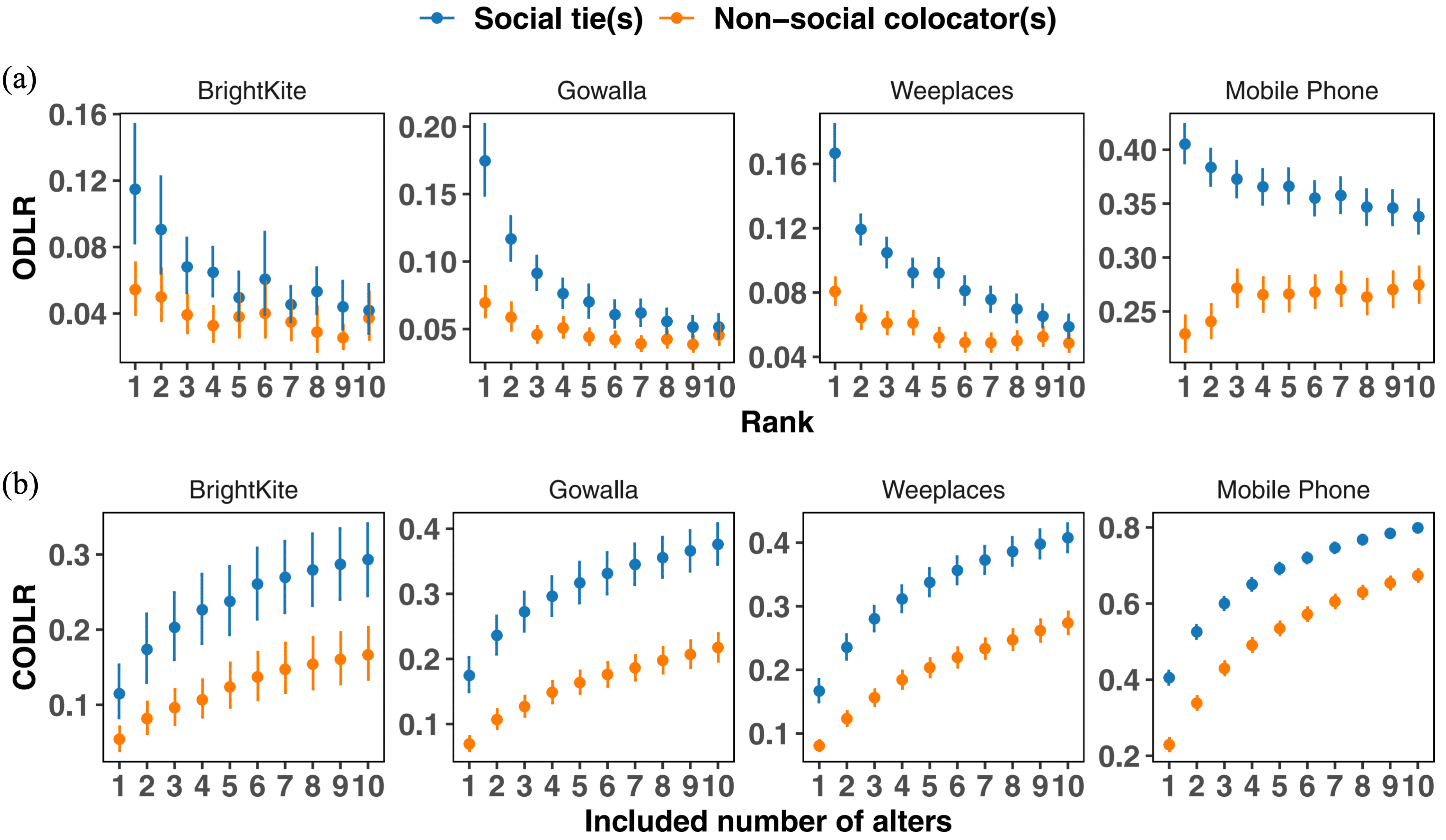} 
	\caption{Comparison of distinct location overlap between the ego and alters. 
		(a) Distribution of distinct location overlap ratios for alters of different ranks. Results show that higher-ranked alters share more unique locations with the ego, with the Top-ranked (rank-1) alter exhibiting the highest overlap. This trend is markedly stronger for social ties (blue) than for non-social colocators (orange). 
		(b) Cumulative proportion of distinct location overlap as a function of the number of alters added sequentially by rank. As alters are incorporated, the proportion of location overlap increases, though the growth rate gradually diminishes. \\
		\textit{Source}: The figure is reproduced from Ref. \cite{chen2022contrasting}.}
	\label{fig:ego_alter_location_overlap}
	\vspace{0pt}
\end{figure*}

Figures~\ref{fig:cross_entropy_social_vs_nonsocial}a and \ref{fig:cross_entropy_social_vs_nonsocial}b first compare the performance under single-source information. Results show that the cross-entropy (median 8.17 bits) and predictability (17.43\%) of a single friend are both superior to those of a single non-friend colocator (8.46 bits, 12.35\%). However, when the top three colocators are combined, their cross-entropy decreases to 8.02 bits and predictability increases to 19.60\%, indicating that non-social colocators can substantially supplement predictive information when aggregated. Furthermore, Figures~\ref{fig:cross_entropy_social_vs_nonsocial}c and \ref{fig:cross_entropy_social_vs_nonsocial}d pair the Top-1 friend with the top three colocators, with the x- and y-axes denoting cross-entropy and predictability under the two sources, respectively; the reference line $y=x$ marks equal performance. One can see that a considerable number of points lie above the $y=x$ line, showing that the joint predictive power of multiple non-friend colocators can surpass that of a single strongest social tie. Finally, Figures~\ref{fig:cross_entropy_social_vs_nonsocial}e and \ref{fig:cross_entropy_social_vs_nonsocial}f present the cumulative trends: as the number of alters increases, friends-only cross-entropy and predictability improve markedly and gradually converge toward the ego's own predictability upper bound (47.05\%); in contrast, improvements for non-friends-only rely more heavily on quantity accumulation. Although overall performance remains lower than that of friends-only, in some cases, non-friends-only prediction can exceed that of a small number of social ties. These findings reveal the general superiority of social ties in prediction, while also highlighting the complementary value of non-social colocators when aggregated at scale. Figure~\ref{fig:ego_alter_location_overlap} shows information contribution from the perspective of location overlap. One can observe that high-ranking friends share a greater number of unique locations with the ego, enabling a small set of friends to reconstruct individual trajectories with high fidelity. Overall, the study by Chen \textit{et al.}~\cite{chen2022contrasting} highlights the central role of micro-level social structures in behavioral prediction but also raises new challenges for data anonymization, privacy protection, and information governance on social platforms.

\begin{figure*}[!t]
	\centering
	\setlength{\abovecaptionskip}{0pt}
	\setlength{\belowcaptionskip}{0pt}
	\includegraphics[width=0.80\textwidth]{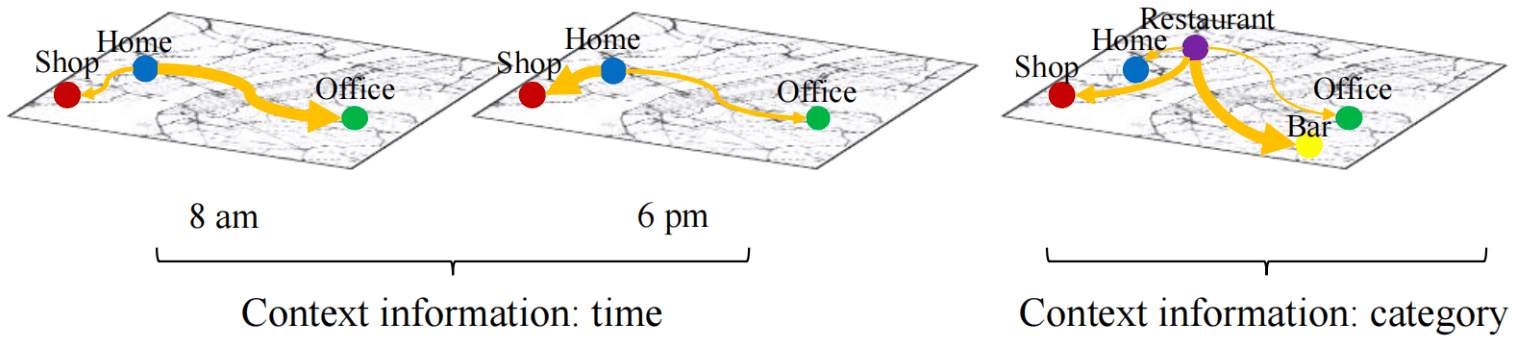} 
	\caption{Illustration of human mobility patterns under contextual information, highlighting the critical role of context factors such as time and location categories in characterizing and predicting human mobility behavior. \\
		\textit{Source}: The figure is reproduced from Ref. \cite{zhang2022beyond}.}
	\label{fig:human_mobility_context}
	\vspace{0pt}
\end{figure*}

\textbf{Context Entropy}. As shown in Figure~\ref{fig:human_mobility_context}, contextual information such as time and location categories plays a critical role in modulating human mobility behavior. However, existing methods generally fail to account for the impact of contextual factors on mobility predictability. To address this gap, Zhang \textit{et al.}~\cite{zhang2022beyond} proposed a framework based on context entropy to evaluate the contribution of contextual factors to the predictability of human mobility.

Specifically, let a user's trajectory sequence be $X = \{X_1, X_2, \dots, X_L\}$, representing the sequence of visited locations; the corresponding contextual information is denoted as $Y = \{Y_1, Y_2, \dots, Y_L\}$, such as the hour of the day or the day of the week of the visits. Traditional methods employ Shannon entropy $S(X)$ to quantify the overall uncertainty of the trajectory, whereas Zhang \textit{et al.} employed the conditional entropy $S(X|Y)$ to capture the uncertainty of the trajectory given contextual conditions. The conditional entropy is defined as the difference between trajectory entropy and the mutual information with the context:
\begin{align}
	S(X|Y) = S(X) - I(X;Y),
\end{align}
where the trajectory Shannon entropy is computed from the occurrence probabilities of all transition pairs $P(x_i \to x_j)$:
\begin{align}
	S(X) = -\sum_{i,j} P(x_i \to x_j)\log_2 P(x_i \to x_j),
\end{align}
and the mutual information term $I(X;Y)$ quantifies the informational constraints imposed by the context on the trajectory sequence, defined as:
\begin{align}
	I(X;Y) = \sum_{x \in X}\sum_{y \in Y} P(x,y)\log \frac{P(x,y)}{P(x)P(y)}.
\end{align}
By incorporating conditional entropy into Fano's inequality, the theoretical upper bound of predictability under contextual constraints, $\Pi^{\text{Context}}_{\text{Transition}}$, can be derived, which characterizes the structural strength of behavioral patterns and their predictability limits.

\begin{figure*}[!t]
	\centering
	\setlength{\abovecaptionskip}{0pt}
	\setlength{\belowcaptionskip}{0pt}
	\includegraphics[width=\textwidth]{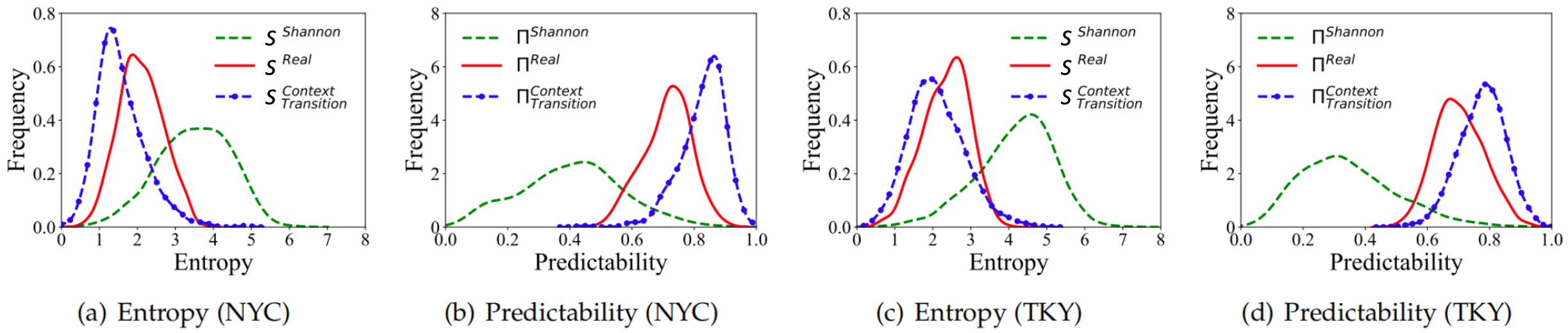} 
	\caption{Comparison of entropy and predictability distributions of users in New York City (NYC) and Tokyo (TKY). 
		(a)(c) Entropy distributions of users in the two cities under different computation methods, including $S^{\text{Shannon}}$ (green dashed line), $S^{\text{Real}}$ (red solid line), and $S^{\text{Context}}_{\text{Transition}}$ (blue dash-dotted line); 
		(b)(d) Corresponding distributions of predictability, including $\Pi^{\text{Shannon}}$, $\Pi^{\text{Real}}$, and $\Pi^{\text{Context}}_{\text{Transition}}$. 
		The blue, red, and green curves respectively denote the probability density functions under context-aware transitions, real transitions, and the Shannon entropy assumption. \\
		\textit{Source}: The figure is reproduced from Ref. \cite{zhang2022beyond}.}
	\label{fig:entropy_predictability_nyc_tky}
	\vspace{0pt}
\end{figure*}

To validate the effectiveness of the context entropy approach, the authors conducted an empirical analysis using mobility data from New York City (NYC) and Tokyo (TKY). Figures~\ref{fig:entropy_predictability_nyc_tky}(a) and \ref{fig:entropy_predictability_nyc_tky}(c) present the entropy distributions of users in the two cities under three computation methods: the green dashed line represents Shannon entropy $S^{\text{Shannon}}$, the red solid line represents real transition entropy $S^{\text{Real}}$, and the blue dash-dotted line represents context transition entropy $S^{\text{Context}}_{\text{Transition}}$. It can be observed that the introduction of contextual conditions reduces overall entropy values and yields more concentrated distributions. Correspondingly, Figures~\ref{fig:entropy_predictability_nyc_tky}(b) and \ref{fig:entropy_predictability_nyc_tky}(d) show the predictability distributions derived from the three entropy measures ($\Pi^{\text{Shannon}}$, $\Pi^{\text{Real}}$, and $\Pi^{\text{Context}}_{\text{Transition}}$). Predictability under contextual transitions is significantly higher than under the other two methods and also exhibits stronger concentration. This indicates that contextual factors can substantially reduce behavioral uncertainty and enhance the predictability. The trend remains consistent in both NYC and TKY, demonstrating that the method possesses generalizability and robustness across different urban environments.

\textbf{Conditional Entropy}. Traditional entropy estimation methods based on compression algorithms (e.g., Lempel–Ziv compression \cite{ziv1977universal,kontoyiannis2002nonparametric}), although effective in quantifying the theoretical upper bound of individual mobility predictability, suffer from a lack of interpretability. In particular, the method of Song \textit{et al.}~\cite{song2010limits} does not reveal the intrinsic connections between behavioral features and predictability. To address this gap, Teixeira \textit{et al.}~\cite{teixeira2021impact} investigated three influencing factors of predictability: stationarity, regularity, and context, and proposed an interpretable information-theoretic modeling framework. They represented the temporally ordered location sequence as $X = (X_1, X_2, \dots, X_L)$ and introduced two indicators. First, stationarity $st(X)$ denotes the proportion of consecutive timesteps where the location remains unchanged, capturing short-term consistency of dwelling behavior:
\begin{align}
	st(X) = \frac{1}{L-1} \sum_{i=1}^{L-1} \mathbf{1}\{X_i = X_{i+1} \}.
\end{align}
Second, regularity $reg(X)$ measures the concentration of location visits and is defined as:
\begin{align}
	reg(X) = 1 - \frac{|\mathcal{U}(X)|}{L},
\end{align}
where $\mathcal{U}(X)$ denotes the set of distinct locations appearing in sequence $X$. Together, these two measures characterize behavioral features from the perspectives of local temporal consistency and global location preference.

Teixeira \textit{et al.} further fitted trajectory entropy $S(X)$ using the following linear regression model to evaluate the explanatory power of $st(X)$ and $reg(X)$ for predictability, as
\begin{align}
	S(X) \approx \alpha \cdot reg(X) + \beta \cdot st(X) + \gamma \cdot reg(X) \cdot st(X) + \epsilon, \label{eqn:SR_model}
\end{align}
where $\alpha$, $\beta$, and $\gamma$ are regression coefficients, and $\epsilon$ is the residual term. Experimental results showed that joint modeling with both indicators significantly improved the fitting performance, revealing the complementary roles of stationarity and regularity in behavioral structures, and providing a more transparent explanatory framework beyond traditional compression-based entropy methods.  

To further analyze the moderating effect of contextual factors on predictability, Teixeira \textit{et al.} introduced a conditional entropy measure. Different from the mutual-information-based definition adopted in \textit{et al.}~\cite{zhang2022beyond}, they employed the joint-entropy difference method. Let $X = \{X_1, X_2, \dots, X_L\}$ denote the location sequence, and $Y = \{Y_1, Y_2, \dots, Y_L\}$ the corresponding contextual feature sequence (e.g., temporal information). The conditional entropy is defined as
\begin{align}
	S(X \mid Y) = S(X,Y) - S(Y),
\end{align}
where $S(\cdot)$ denotes Shannon entropy. Substituting $S(X \mid Y)$ into the Fano inequality then allows estimation of the predictability upper bound under contextual constraints, denoted as $\Pi(X \mid Y)$.

This section reviews three representative extensions of predictability measures, including cross entropy based on compression principles, context entropy derived from mutual information, and conditional entropy derived from joint entropy. Each of the three measures has distinct characteristics in terms of computational logic, data dependence, applicable scenarios, as well as advantages and limitations. Table~\ref{tab:entropy_methods} provides a comparative summary of these measures.

\begin{table}[!t]
	\centering
	\small  
	\caption{Comparison of three predictability measures: cross entropy, context entropy, and conditional entropy.}
	\label{tab:entropy_methods}
	\renewcommand{\arraystretch}{1.1}
	\scalebox{0.9}{
		\begin{tabular}{m{1.3cm} m{5.0cm} m{5cm} m{5cm}}
			\hline
			& \textbf{Cross Entropy} & \textbf{Context Entropy} & \textbf{Conditional Entropy} \\
			\hline
			Method
			& Modified from Lempel–Ziv entropy; compress sequence $X$ using sequence $Y$ to obtain the upper bound $S(X\mid Y)$ 
			& Derived from mutual information: $S(X\mid Y)=S(X)-I(X;Y)$, measuring the information gain of contextual features 
			& Derived from joint entropy: $S(X\mid Y)=S(X,Y)-S(Y)$, directly \revadd{estimating}{80} the joint distribution \\
			
			Input 
			& Two symbolic sequences (original $X$ and reference $Y$) 
			& Main sequence $X$ + contextual features $Y$ (e.g., time, location) 
			& Main sequence $X$ + conditional variables $Y$ (e.g., time, location) \\
			
			Scenarios 
			& Inter-sequence predictability influence (\revadd{information flow or interaction prediction}{67}) 
			& High-dimensional or continuous context; analyzing feature contributions to predictability 
			& Low-dimensional, discrete settings with sufficient samples \\
			
			Advantages 
			& Parameter-free; compression-based; naturally characterizes information flow 
			& Strong interpretability; suitable for high-dimensional continuous features; consistent with the mutual information framework 
			& Conceptually intuitive; robust when \revadd{categories are limited}{67} and sample size is sufficient \\
			
			Limitations 
			& Requires long sequences; unstable on short or sparse data; limited interpretability 
			& Biased estimation under small sample sizes or weak dependencies 
			& Severe sparsity and bias in high-dimensional settings or with large alphabets \\
			\hline
		\end{tabular}
	}
\end{table}

\subsection{Metric-Based Methods}
The aforementioned information-theoretic approaches for predictability estimation are primarily suited for symbolic time series with a finite and discrete state space, whereas in many complex systems, data points often appear in continuous numerical form, such as precipitation, temperature, and pressure in atmospheric science, carbon emissions in environmental systems, electricity load in energy systems, and EEG signals in physiological systems.
For such numerical sequences, the candidate state set is theoretically infinite, making it difficult to directly determine its size within the Fano scaling framework. Discretization looks fine, but it inevitably introduces unquantifiable information loss. To address this challenge, researchers have proposed metric-based methods for predictability assessment, which rely on specifically designed statistics that capture the structural complexity and regularity of sequences, thereby directly characterizing their predictive potential. Representative examples include permutation entropy, which evaluates sequence randomness and regularity by analyzing the ordering structure of local patterns, and the $\kappa$-index, which quantifies predictability by comparing prediction error differences between the original and randomized sequences. These methods do not require the construction of a complete entropy-scaling model and are well-suited for numerical sequences, short time series, and scenarios where the entropy cannot be precisely estimated.

\subsubsection{Permutation Entropy}
Different from previous approaches that establish indirect \revadd{relationships}{70} between entropy and predictability via Fano's inequality or empirical functions, Scarpino and Petri~\cite{scarpino2019predictability} proposed to directly measure the predictability of epidemic outbreak time series using permutation entropy $S^{\mathrm{perm}}$. The core idea is that if the value-ordering patterns in a sequence are highly repetitive, the sequence exhibits stronger regularity and lower randomness, thereby implying higher predictability. Conversely, if the occurrence probabilities of all ordering patterns are nearly uniform, the sequence approaches a random process and becomes more difficult to predict. Based on this intuition, they defined predictability as the complement of permutation entropy:
\begin{equation}
	\Pi = 1 - S^{\mathrm{perm}},
\end{equation}
where $S^{\mathrm{perm}}$ reflects the complexity of ordering patterns in the time series. After normalization, $S^{\mathrm{perm}}$ lies within $[0,1]$, with smaller value indicating stronger structural regularity and thus higher predictability.

Specifically, for a time series $\{x_t\}_{t=1,\dots,N}$ indexed by positive integers, given an embedding dimension $d$ and time delay $\tau$, consider a subsequence of length $d$:
\begin{equation}
	s = \{x_t, x_{t+\tau}, \dots, x_{t+(d-1)\tau}\}.
\end{equation}
This subsequence is reordered according to its values, yielding the corresponding permutation pattern:
\begin{equation}
	\widetilde{s} = \pi(s) = [x_{t_1}, \dots, x_{t_d}], \quad x_{t_i} \le x_{t_j} \ \text{if and only if} \ t_i \le t_j.
\end{equation}
For equal values, ties are broken either by preserving the original temporal order or by adding small noise, which has no substantive effect on the final result~\cite{zunino2017permutation}. On this basis, the permutation entropy of the time series is defined as the Shannon entropy over the probability distribution of permutation patterns:
\begin{equation}
	S^{\mathrm{perm}}_{d,\tau}(\{x_t\}) = - \sum_{\pi} p_{\pi} \log p_{\pi},
\end{equation}
where $p_{\pi}$ denotes the frequency of occurrence of pattern $\pi$ in the sequence. The value of $S^{\mathrm{perm}}$ depends on the embedding dimension $d$, the delay $\tau$, and the window size $W$, where $W$ represents the length of the sliding window used in the computation to capture local regularities in different time periods. Scarpino and Petri characterized the maximum predictability of a sequence by taking the minimum permutation entropy across all $(d,\tau)$ combinations:
\begin{equation}
	S^{\mathrm{perm}}(\{x_t\}) = \min_{d,\tau} S^{\mathrm{perm}}_{d,\tau}(\{x_t\}).
\end{equation}

However, due to the presence of unobserved permutation patterns in high-dimensional cases for finite-length sequences, $S^{\mathrm{perm}}$ theoretically decreases continuously as $d$ increases and thus does not yield a true minimum. To address this, Scarpino and Petri adopted the method in~\cite{brandmaier2015pdc}, which produces a stable minimum for $S^{\mathrm{perm}}$ under finite samples. In addition, $S^{\mathrm{perm}}$ is normalized by dividing it by $\log(d!)$, standardizing its range to $[0,1]$. It is important to note that $1-S^{\mathrm{perm}}$ here is not a strictly derived upper bound on predictability as in the case of Fano's inequality, but rather an empirical measure. Normalization provides a clear interval interpretation: $S^{\mathrm{perm}}=0$ indicates a fully regular and most predictable sequence, while $S^{\mathrm{perm}}=1$ corresponds to an approximately random and least predictable sequence. Thus, $1-S^{\mathrm{perm}}$ can be regarded as an empirical estimate of the maximum prediction accuracy, enabling a continuous characterization of prediction difficulty and serving as an index for predictability in empirical studies. That is to say, the absolute value of $1-S^{\mathrm{perm}}$ has no clear meaning, but the relative value of $1-S^{\mathrm{perm}}$ can be used to compare the predictability.

\begin{figure*}[!t]
	\centering
	\setlength{\abovecaptionskip}{0pt}
	\setlength{\belowcaptionskip}{0pt}
	\includegraphics[width=0.98\textwidth]{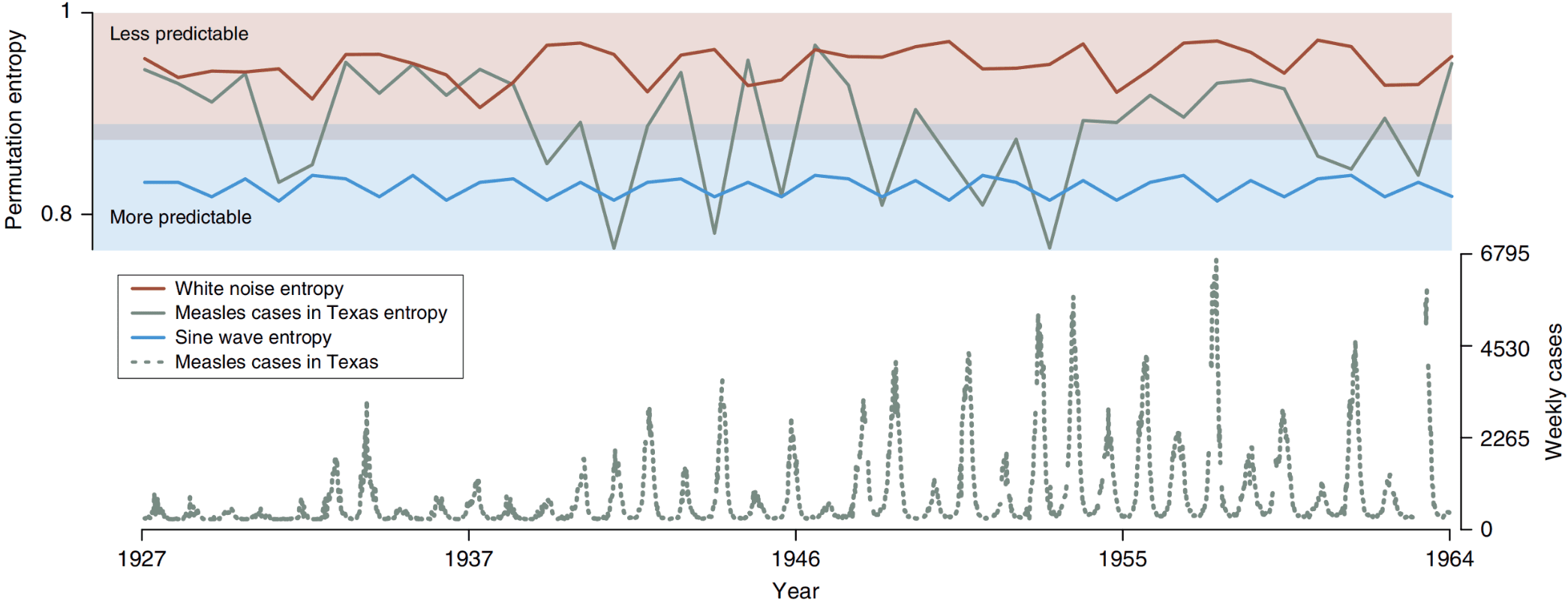}
	\caption{Permutation entropy of the Texas measles case time series. Permutation entropy is computed using a 52-week sliding window and compared against two control sequences: dark red denotes Gaussian white noise ($\mu=0, \sigma=1$); blue denotes a sine wave with added Gaussian noise ($\mu=0, \sigma=0.01$); and gray denotes weekly measles cases in Texas from 1927--1965 (gray dashed line, bottom panel). The blue region (closer to the sine wave) indicates that higher prediction accuracy could be achieved by proper model, while the red region (closer to white noise) suggests lower predictive performance when relying solely on time-series statistics. \\
		\textit{Source}: The figure is reproduced from Ref. \cite{scarpino2019predictability}.}
	\label{fig:permutation_entropy_measles}
	\vspace{0pt}
\end{figure*}

Scarpino and Petri further utilized real and simulated epidemic surveillance data to analyze the epidemic predictability and its influencing factors. Figure~\ref{fig:permutation_entropy_measles} illustrates the variation of permutation entropy for weekly measles cases in Texas from 1927 to 1965 (gray dashed line in the bottom panel), computed with a sliding window of 52 weeks (gray solid line), and compared against two reference sequences: Gaussian white noise (dark red) and a sine wave with weak Gaussian noise ($\mu=0, \sigma=0.01$) superimposed (blue). As shown in the figure, the permutation entropy of measles cases exhibits significant temporal fluctuations. When the entropy values fall into the blue shaded region, the corresponding predictability is higher; in contrast, when entropy values lie within the red shaded region, the sequence resembles a random process and prediction based solely on time-series statistics may be less effective. This result indicates that epidemic predictability varies dynamically with time, rather than being stable.

\begin{figure*}[!t]
	\centering
	\setlength{\abovecaptionskip}{0pt}
	\setlength{\belowcaptionskip}{0pt}
	\includegraphics[width=0.98\textwidth]{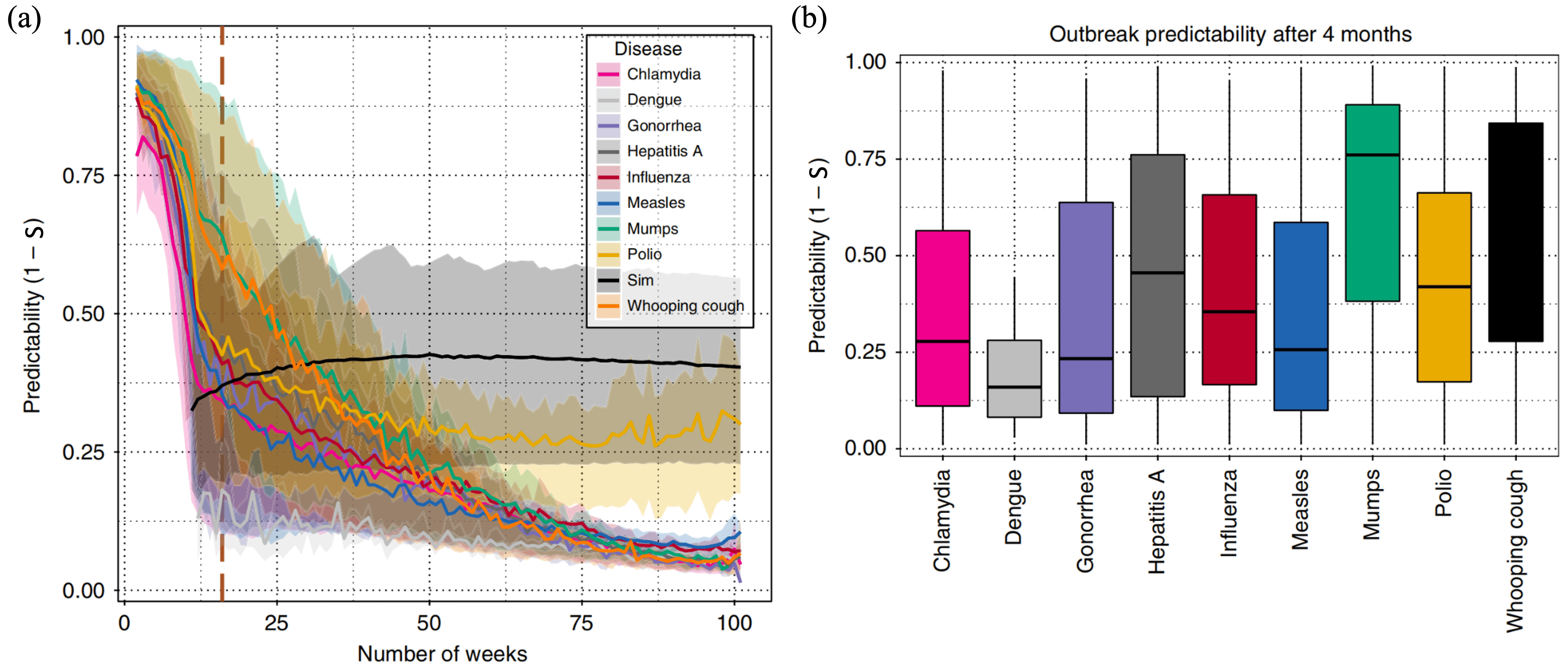} 
	\caption{Predictability analysis of nine infectious diseases across U.S. states. 
		(a) Average predictability $(1-S^{\mathrm{perm}})$ as a function of time series length, with curves denoting mean values and shaded regions representing interquartile ranges across states and starting times. The black solid line represents the median permutation entropy of the SIR model. The dark brown dashed line marks the time point (16 weeks) selected for the results shown in (b). 
		(b) Predictability distributions of different diseases at 4 months (16 weeks). Box plots represent the median, interquartile range, and full range. Results indicate that while the trend of predictability over time is consistent across diseases, substantial differences exist in predictability levels at the same time point. \\
		\textit{Source}: The figure is reproduced from Ref. \cite{scarpino2019predictability}.}
	\label{fig:disease_predictability}
	\vspace{0pt}
\end{figure*}

Figure~\ref{fig:disease_predictability} characterizes the patterns of epidemic predictability across different diseases. Panel (a) shows that nine infectious diseases generally exhibit high predictability $(1-S^{\mathrm{perm}})$ close to 1 in the early stages of outbreaks (shorter time-series lengths), which then rapidly declines and stabilizes as the number of weeks increases. Although the decay rates differ among diseases, the overall trend consistently follows a “high in the early stage, low in the later stage” pattern. The black dashed line indicates the median permutation entropy obtained from SIR model simulations, serving as a reference for the feasibility of dynamical model prediction. The dark brown dashed line marks the 16-week point corresponding to the cross-sectional analysis in panel (b). Panel (b) compares the predictability distributions of different diseases after four months of outbreak, where the boxplots clearly display differences in medians, interquartile ranges, and overall ranges.

\begin{figure*}[!t]
	\centering
	\setlength{\abovecaptionskip}{0pt}
	\setlength{\belowcaptionskip}{0pt}
	\includegraphics[width=0.98\textwidth]{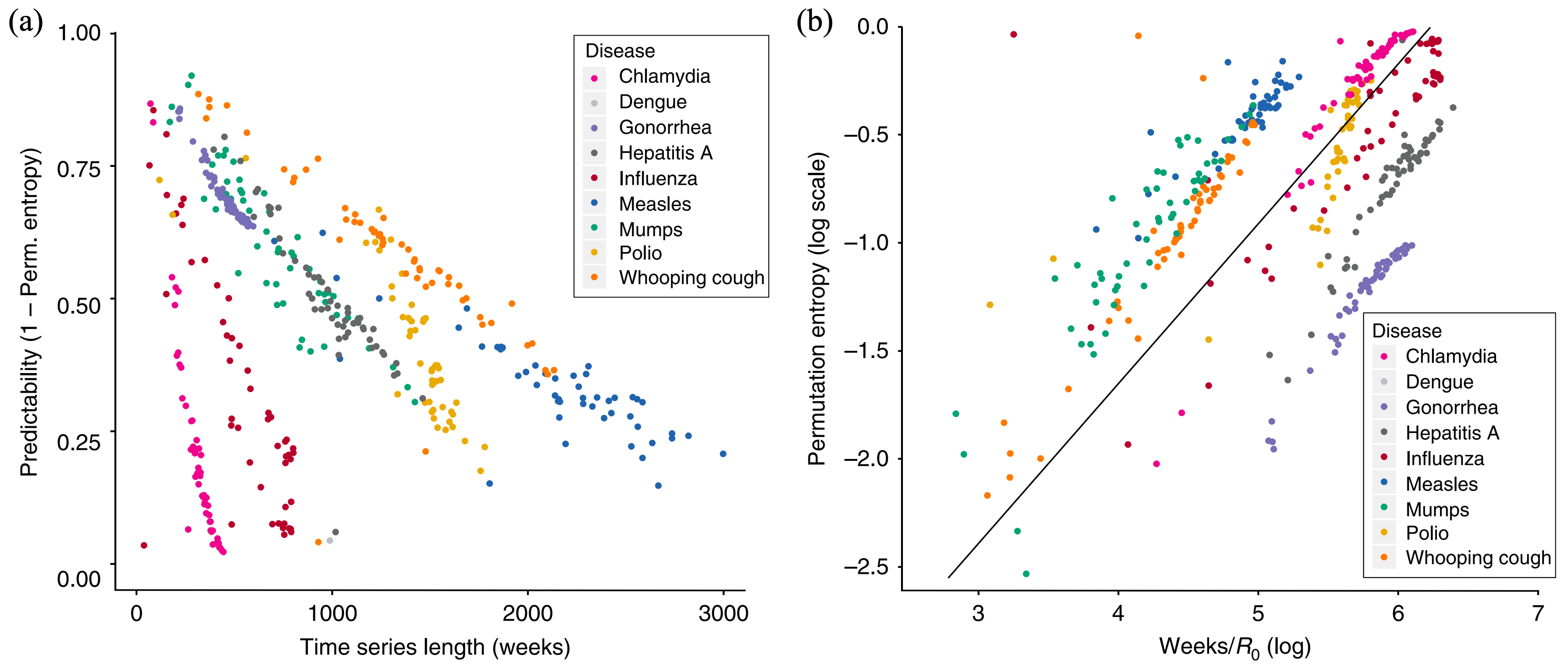} 
	\caption{Relationship between predictability and time-series characteristics across different diseases. 
		(a) Average predictability $(1-S^{\mathrm{perm}})$ of state-level time series as a function of sequence length (in weeks). Although slopes differ across diseases, all exhibit the general trend that longer sequences correspond to lower predictability, with clear clustering patterns among diseases in terms of slope. 
		(b) Time series length normalized by the reported average basic reproduction number $R_{0}$, plotted against permutation entropy on a logarithmic scale; the straight line represents a fitted reference. Each point corresponds to the predictability of one disease at the full state-level time span.  \\
		\textit{Source}: The figure is reproduced from Ref. \cite{scarpino2019predictability}.}
	\label{fig:disease_entropy_relationship}
	\vspace{0pt}
\end{figure*}

Figure~\ref{fig:disease_entropy_relationship} shows the relationship between predictability and epidemic transmission characteristics. Panel (a) shows the trends of average predictability at the state level for different diseases as a function of sequence length (in weeks). Although the slopes differ across diseases, a general pattern emerges in which longer time series correspond to lower predictability, and the slope differences form disease-specific clusters in the scatter distribution. Panel (b) normalizes sequence length by the average basic reproduction number $R_{0}$ reported in the literature and compares it with permutation entropy on a logarithmic scale. The results indicate that transmission speed, as characterized by $R_{0}$, is a key driver of the decaying pattern of predictability. This normalization analysis reveals common regularities in predictability variation across diseases, laying the groundwork for establishing a unified framework for predictive assessment.

\subsubsection{$\kappa$ Index}
In early studies of time series predictability, Kaboudan introduced the $\kappa$ index based on Genetic Programming (GP) to measure sequence predictability~\cite{kaboudan2000genetic}. The core idea is to compare the difference in prediction errors between the original sequence and a randomly shuffled sequence:
\begin{equation}
	\kappa = 1 - \frac{\mathrm{SSE}_Y}{\mathrm{SSE}_S},
\end{equation}
where $\mathrm{SSE}_Y$ denotes the sum of squared prediction errors for the original sequence, and $\mathrm{SSE}_S$ is that for the randomly shuffled sequence. If a sequence is highly predictable, the ratio $\frac{\mathrm{SSE}_Y}{\mathrm{SSE}_S}$ will be much smaller than 1; otherwise, it will approach 1. However, this method suffers from two major drawbacks: (1) \textit{Sensitivity to sequence length} — the $\kappa$ value for a long sequence is significantly higher than that of its subsequences, with the increase mainly attributed to non-stationarity rather than genuine improvements in predictability; (2) \textit{Insufficient resolution} — in the case of long sequences, even randomized sequences can yield $\kappa$ values above 0.9, causing predictability estimates to concentrate within the narrow interval $[0.9, 1.0]$, which makes it difficult to effectively distinguish between different sequences.

Duan and Povinelli~\cite{duan2001estimating} proposed an improved version. They divided a long sequence into multiple subsequences of length $Q$, $\{y_t, t = i, \dots, i+Q-1\}$, and repeated the computation of $\kappa$ by shifting $\tau$ time steps each time, thus obtaining a series of local predictability values. The overall predictability was then defined as their average. This approach eliminates the bias introduced by sequence length through the use of a unified sampling window, and also captures the temporal variation of predictability. Furthermore, to address the second limitation, the authors introduced a square-root correction:
\begin{equation}
	\kappa = 1 - \sqrt{\frac{\mathrm{SSE}_Y}{\mathrm{SSE}_S}},
\end{equation}
which effectively expands the dynamic range of $\kappa$, ensuring that the predictability of a completely random sequence is closer to 0.

In the experimental validation, Duan and Povinelli tested three types of time series with distinct characteristics. For the deterministic series generated by the Mackey–Glass equation, both methods yielded predictability values close to 1 (0.999 for Kaboudan's method and 0.997 for Duan-Povinelli method), confirming their high predictability. For random walk sequences, Kaboudan's method produced an unrealistically high value of 0.875, whereas Duan-Povinelli method significantly reduced it to 0.140, which better reflected its randomness. For real-world stock price series (CPQ and GE),  Duan-Povinelli method produced intermediate values of 0.818 and 0.415, respectively, effectively distinguishing them from both completely random and fully deterministic sequences. These results demonstrate that the improved method not only preserves the intuitive interpretation of the $\kappa$ index but also substantially enhances its discriminative power across different types of time series, while scaling numerical series predictability more reasonably within the $[0,1]$ range.

\subsection{Equivalence to Bayes Error Rate}

The aforementioned information-theoretic approaches estimate predictability upper bounds through entropy estimation combined with the Fano inequality~\cite{song2010limits}, and have been widely applied in the analysis of symbolic time series. However, such methods often rely on implicit Markov assumptions, making it difficult to capture long-range dependencies; moreover, they are highly sensitive to sequence length, with significant estimation bias in short sequences~\cite{xu2019predictability,xu2023equivalence}. Metric-based methods, while more suitable for numerical and short-sequence scenarios, lack a direct correspondence to prediction accuracy.

To overcome these limitations, Xu \textit{et al.}~\cite{xu2023equivalence} rigorously proved the exact mathematical equivalence between time series predictability $\Pi$ and the Bayes error rate (BER, denoted as $\mathcal{R}_B$), expressed as $\Pi = 1 - \mathcal{R}_B$. The BER is a measure of the “unavoidable error” in classification, supported by a well-developed theoretical foundation and mature estimation techniques. Establishing this equivalence not only provides a novel computational pathway for predictability, but also bridges predictability analysis with classification theory, enabling bidirectional transfer and integration between the two domains, thereby significantly extending their methodological and application frontiers.

\begin{figure*}[!t]
	\centering
	\setlength{\abovecaptionskip}{0pt}
	\setlength{\belowcaptionskip}{0pt}
	\includegraphics[width=1.0\textwidth]{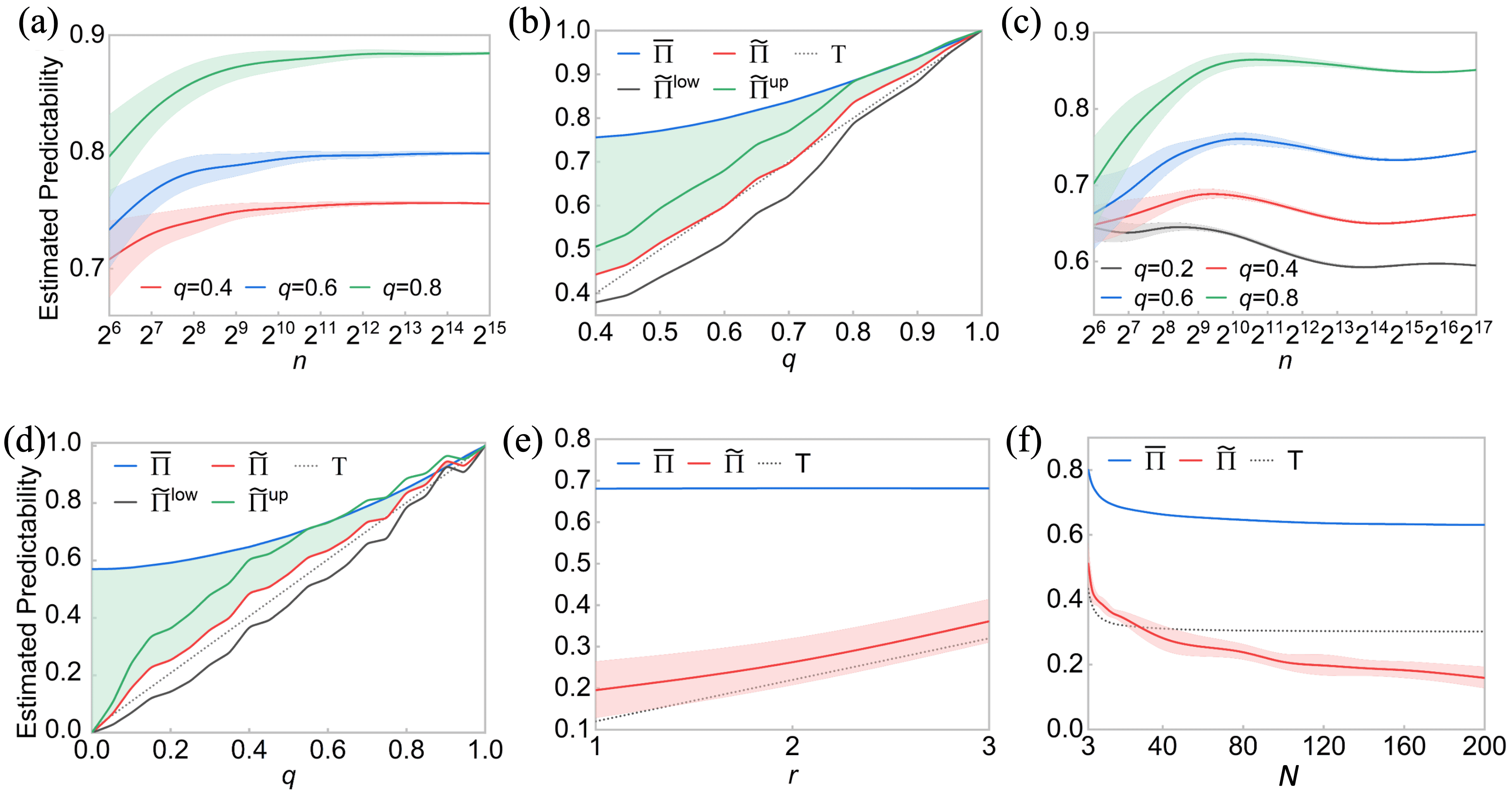} 
	\caption{Comparison of entropy-based methods and BER-heuristic methods in predictability estimation under three types of sequence generators. 
		(a, c) Results of entropy-based methods as a function of sequence length $n$ for the first and second generators, respectively. 
		(b, d) Under the same generators, performance comparison between entropy-based methods and the BER-heuristic method \revadd{\cite{wisler2016empirically}}{78}, where $\tilde{\Pi}^{\text{low}}$ and $\tilde{\Pi}^{\text{up}}$ denote the lower and upper bounds, $\tilde{\Pi} = \tfrac{1}{2}(\tilde{\Pi}^{\text{low}} + \tilde{\Pi}^{\text{up}})$ represents the estimate, and the shaded region indicates where the BER method outperforms the entropy-based method. 
		(e) Results for the third generator with $N=20$ fixed, varying parameter $r$. 
		(f) Results for the third generator with $r=3$ fixed, varying $C$. 
		All results are averaged over 10 independent runs at $n=2^{15}$, with shaded regions representing standard errors. \\
		\textit{Source}: The figure is reproduced from Ref. \cite{xu2023equivalence}.}
	\label{fig:entropy_vs_ber_methods}
	\vspace{0pt}
\end{figure*}

Xu \textit{et al.} formalized time series prediction as an equivalent multi-class classification problem. Let the sequence $\{\omega^1\omega^2\dots\omega^{n}\}$ represent the evolution of an $C$-state system. At each time step, the prediction problem can be expressed as: given the historical states $\{x_{n-1} = \omega^1 \omega^2 \dots \omega^{n-1}\}$, predict the current state $\omega^n$. Here, $x_{n-1}$ can be regarded as the input feature in a classification task, and $\omega^n$ as the class label. The optimal single-step prediction accuracy is defined as
\begin{equation}
	\pi(x_{n-1}) = \sup_{\omega} \left\{ \Pr[\omega^n = \omega \mid x_{n-1}] \right\},
\end{equation}
and predictability $\Pi$ is then defined as the expectation of this maximum prediction probability over the historical state space:
\begin{equation}
	\Pi = \lim_{n \to \infty} \frac{1}{n} \sum_{i=1}^{n} \sum_{x_{i-1}} P(x_{i-1}) \pi(x_{i-1}).
\end{equation}

On the other hand, the BER is defined as the minimum error rate achievable by the optimal classifier:
\begin{equation}
	\mathcal{R}_B = 1 - \sum_{j=1}^{M} \int_{\Gamma_j} p(\omega_j) p(x \mid \omega_j) \,\mathrm{d}x,
\end{equation}
where $\Gamma_j$ denotes the region of the observation space assigned to class $\omega_j$. Using Bayes' theorem, the integral can be reformulated as:
\begin{equation}
	\mathcal{R}_B = 1 - \sum_{j=1}^{M} \int_{\Gamma_j'} p(x) p(\omega_j \mid x) \,\mathrm{d}x,
\end{equation}
with the region $\Gamma_j'$ defined as:
\begin{equation}
	\Gamma_j' = \left\{ x \in X' \mid p(\omega_j \mid x) > \max_{k \neq j} p(\omega_k \mid x) \right\}.
\end{equation}

Xu \textit{et al.} further proved that for any $x$, the maximum of $p(\omega_j \mid x)$ is precisely $\pi(x)$, and by aligning the expectation over the integral with the averaging over historical states, they rigorously derived the equivalence between predictability and the BER:
\begin{equation}
	\Pi = 1 - \mathcal{R}_B.
\end{equation}

To validate the effectiveness of the theoretical equivalence, Xu \textit{et al.} designed three types of synthetic sequence generators with known true predictability, each used to examine the effects of sequence length, state-space size, and memory depth on estimation methods. The first generator is a one-step Markov chain, where the next state remains in the target state with probability $q$ and transitions randomly to other states with probability $1-q$. Its true predictability is $T=q$, mainly used to test estimation bias under different sequence lengths $n$. The second generator introduces two-step dependencies in a large state space, where the next state is determined by the previous two states with added noise disturbance. Its true predictability is approximately $T \approx q$, used to analyze the effect of candidate set size $C$ on estimation performance. The third generator combines multi-step memory with random perturbations: the next state copies one of the previous 1–3 states with probabilities $q_1, q_2, q_3$, and with the remaining probability randomly samples from $C$ states. Its true predictability is given by $T=\max\{q_1,q_2,q_3\}+(1-q_1-q_2-q_3)/C$, designed to investigate the combined effects of memory depth $r$ and state-space size $C$.

As shown in Figs.~\ref{fig:entropy_vs_ber_methods}(a) and \ref{fig:entropy_vs_ber_methods}(c), under the first and second generators, traditional entropy-based methods exhibit strong fluctuations with increasing sequence length $n$, and in some parameter settings deviate significantly from the true predictability $T$. In contrast, Figs.~\ref{fig:entropy_vs_ber_methods}(b) and \ref{fig:entropy_vs_ber_methods}(d) demonstrate that the BER method yields upper and lower bounds $\tilde{\Pi}^{\text{low}}$ and $\tilde{\Pi}^{\text{up}}$ that stably envelope the true value, with their mean $\tilde{\Pi}$ closely matching $T$, significantly outperforming the entropy-based estimate $\overline{\Pi}$. Furthermore, Figs.~\ref{fig:entropy_vs_ber_methods}(e) and \ref{fig:entropy_vs_ber_methods}(f), under the third generator with different parameter settings (fixing $C=20$ while varying $r$, or fixing $r=3$ while varying $C$), show that the BER method maintains higher accuracy and robustness even under complex dependency structures. These results verify that the utilization of the equivalence between predictability and BER substantially improves estimation performance in practice, providing a new tool for accurately characterizing the predictability of complex systems.

\subsection{Summary}
This chapter addresses the central question of “to what extent a time series can be predicted” by constructing an analytical framework consisting of three categories of methods: information-theoretic upper bounds, metric-based predictability measures, and the equivalence between predictability $\Pi$ and the BER. Information-theoretic methods rigorously define the predictive limit from a theoretical perspective, with Fano scaling at the core, establishing the relation between entropy and predictability~\cite{song2010limits}. Metric-based methods target numerical time series and propose corresponding predictability evaluation metrics based on complexity quantification. Xu \textit{et al.}~\cite{xu2023equivalence} further established a strict equivalence between predictability and the Bayes error rate, showing that the predictability of a time series can be reformulated as the minimal unavoidable error in a classification task.

Information-theoretic methods provide an upper-bound characterization for symbolic sequences. Centered on Fano scaling, they establish a quantitative relationship between entropy and predictability, and their foundational applications and assumptions are reviewed in representative scenarios such as human mobility. Subsequently, refinements are introduced by re-estimating the candidate space with geographic reachability and behavioral order constraints, and by aligning with top-$\ell$ evaluations, thereby tightening the upper-bound estimates. Finally, extensions such as cross entropy, context entropy, and conditional entropy are incorporated to support multi-source information fusion and contextual modulation, enhancing both interpretability and adaptability of the framework across scenarios.

Metric-based predictability measures are suitable for numerical time series, with two main lines of thought. The first is complexity-oriented, where nonparametric measures (such as normalized permutation entropy) quantify the randomness of a sequence, with their complements serving as approximations of predictability potential. The second is relative-performance-oriented, where indices such as the $\kappa$ score define predictability as the normalized error improvement relative to reference baselines (e.g., random, persistence, or seasonal decomposition baselines). The former answers “how much exploitable structure exists in the data,” while the latter addresses “what relative gain can be achieved under the chosen baseline and loss function.”

Traditional entropy-based and Fano-based frameworks, while effective for symbolic sequences, rely on implicit Markov assumptions and are highly sensitive to sequence length, making them less accurate and unstable for short sequences. Metric-based methods, though applicable to numerical series, lack a direct correspondence to predictability values. By leveraging the equivalence $\Pi = 1 - \mathcal{R}_B$, time series predictability is reformulated as the minimal unavoidable error in a classification task, which can be directly computed using well-established BER estimators. This equivalence not only provides a new computational pathway but also opens possibilities for methodological integration and boundary expansion between the two fields.

As a key carrier of complex system dynamics, the studies on time series predictability offer critical insights into understanding the intrinsic regularities and uncertainties of systems. Looking ahead, in the face of real-world challenges such as high-dimensionality, non-stationarity, and sparsity, integrating multi-source information, introducing context awareness, and leveraging the representational power of foundation models hold promise for alleviating applicability bottlenecks and advancing towards a more general framework for predictability assessment.

\section{Predictability of Complex Networks}

\subsection{Preliminaries} \label{sec1}

\subsubsection{Complex Networks}

In the contemporary era, marked by heightened interconnectedness, complex systems have become pervasive across all facets of human activity. It is evident that a multitude of phenomena in modern society exhibit characteristics of nonlinearity, dynamic evolution, and multi-agent interaction. These phenomena include, but are not limited to, daily human communication and interaction (e.g., making phone calls, sending WeChat messages, and engaging in face-to-face conversations) \cite{fernando2007, mailk2022}, the outbreak and global spread of infectious diseases (e.g., the rapid transmission of the COVID-19 pandemic) \cite{sun2020a, silva2022}, and international political and economic competition (e.g., information warfare in the Russia-Ukraine conflict \cite{maneejuk2024} and drastic fluctuations in the global trade landscape \cite{york2025}). Confronted with such complexity, traditional analytical methodologies often struggle to effectively capture the system’s overall structure and functional interrelationships.

It is important to note that in the early stages of research, the complexity exhibited by systems was widely believed to stem primarily from the individual dynamics of their components. For instance, chaos theory focused on complex behaviors within a single system \cite{may1976}. However, with advances in multi-agent system research, a growing body of studies has shown that many observed complex phenomena do not originate from the inherent complexity of individual components, but rather from interactions between these components \cite{mitchell2009}. In summary, even if the behavior of each individual is relatively simple, complex connection patterns between individuals can lead the entire system to exhibit highly complex behavioral patterns. Consequently, conventional approaches predicated on individual dynamics are often inadequate for analyzing such systems, making it imperative to shift our focus to the system’s internal connection structure. In this context, the "node-edge" abstract model from graph theory has emerged as a powerful tool for describing the structure of complex systems: the basic building blocks of a system are abstracted as nodes, and their interactive relationships are represented as edges. As early as the 18th century, Euler laid the foundation for graph theory while solving the "Seven Bridges of Königsberg" problem \cite{biggs1986}. In its subsequent development, graph theory primarily focused on structures and problems that could be precisely described using mathematical language—such as regular graphs, extremal graphs, embedded graphs, and infinite random graphs \cite{west2001}. Such research objects typically possess favorable mathematical properties, which can be analyzed precisely through concise definitions and reasoning.

However, real-world complex networks, such as social networks \cite{smith2008}, technological networks \cite{albert1999}, and biological networks \cite{jeong2001}, are far more complex than the ideal graph structures commonly studied in graph theory. These networks often exhibit universal and non-trivial topological properties, such as small-worldness \cite{watts1998}, local clustering \cite{watts1998}, scale-freeness \cite{barabasi1999}, modularity \cite{newman2004}, and hierarchical structure \cite{ravasz2003}. Unlike regular graphs (which are uniform) and random networks (which have statistical stability), real-world complex networks lack both properties.

Drawing on the aforementioned observations and modeling practices, researchers have sought theoretical foundations and inspiration from mathematics; analytical tools and methods from computer science and statistical physics; and real-world scenarios and metrics from the social sciences, life sciences, and management studies. This interdisciplinary integration has given rise to the field of network science \cite{chen2015, barabasi2016, newman2018}. As an interdisciplinary research field, network science is dedicated to analyzing the topological structures, evolutionary mechanisms, and functional behaviors of complex systems, thereby offering a new systemic research paradigm. From a macro perspective, network science primarily encompasses three core aspects as follows. (i) \textbf{Network evolution} \cite{mendes2002, barthelemy2011, holme2012, bianconi2021}: This involves studying how nodes and edges grow and adjust over time, and how these changes influence the network’s overall topological characteristics; (ii) \textbf{Network dynamics} \cite{nishikawa2003, nowak2006, yan2006, arenas2008, motter2002, pastor2015}: Under a given network structure, this focus examines how network topology regulates dynamic behaviors such as information propagation, disease transmission, and opinion formation; (iii) \textbf{Network mining} \cite{grivan2002, lv2011, lv2016}: This aims to extract key features, identify critical nodes, predict trend directions, and support real-world applications—including public opinion monitoring, risk prevention and control, and recommendation systems—through the analysis of large-scale real-world network data.

In the early stages of network science research, the field advanced from empirical observation to theoretical modeling through the systematic analysis of large-scale real-world networks \cite{mendes2002, albert2002, newman2003, boccaletti2006}. This gradual progression led to the establishment of a comprehensive conceptual framework and analytical toolkit \cite{barabasi2012, vespignani2012}. These advancements have not only deepened our understanding of complex systems but also provided a robust theoretical basis for addressing real-world problems from a network perspective.

In parallel with exhaustive analyses of the statistical regularities of real-world networks, scientists have constructed various artificial network models to reproduce observed structural properties. Among the numerous artificial network models, three classic models are of particular significance. The earliest one is the Erdos-Renyi (ER) Network, which primarily includes two forms: Firstly, the fixed-edge-number model $G(N,M)$, in which $M$ edges are randomly distributed in among the $N(N-1)/2$ node pairs \cite{erdos1959}. Secondly, the fixed-linkage-probability model $G(N,p)$ , in which any two nodes are connected with probability $p$ \cite{gilbert1959}. The ER network has a simple structure, with degree distributions that approximate a Poisson distribution. This implies that the majority of nodes possess similar degrees. Despite the absence of heterogeneity and clustering structures, which are hallmarks of real-world networks, this model offers invaluable insights that inform the development of subsequent network theory.

The Watts-Strogatz (WS) Network was proposed by Watts and Strogatz in 1998 \cite{watts1998} to describe the "small-world" phenomenon in real-world networks. The model commences by constructing a regular ring network, subsequently randomly reconnecting a proportion of edges with a specified probability $p$, thus facilitating a transition between regularity and randomness. The WS network successfully reproduces the two important properties of real-world networks: namely, high clustering coefficient and short average path length. It is particularly suitable for systems exhibiting the "six degrees of separation" effect, such as social networks \cite{stanley1967}.

The Barabasi-Albert (BA) Network was proposed by Barabasi and Albert in 1999 \cite{barabasi1999}, and is utilised to elucidate the pervasive power-law distribution characteristics that are observable in real-world networks. The model introduces the "preferential attachment" mechanism, whereby new nodes tend to connect to existing highly connected nodes, leading to an evolutionary pattern of "the rich getting richer". This mechanism gives rise to scale-free properties in the network, whereby a small number of nodes become highly connected "hubs", thus making it widely applicable in fields such as the internet and social networks \cite{albert1999, barabasi2003, caldarelli2007}.

Complex networks are not merely a high-level abstraction of the structures of various real-world systems, but also a crucial tool for exploring their evolutionary patterns and functional mechanisms. The integration of modeling and analysis with theoretical reasoning, grounded in the principles of artificial network models, has resulted in the emergence of a novel paradigm for research and thinking.

\subsubsection{Link Prediction}
Link prediction is a significant research branch within the broader field of complex network, boasting both considerable theoretical value and broad application prospects. The objective of this study is to estimate likelihoods of missing links, future links, and temporal links, based on known topology \cite{lv2011, wang2015, martinez2016, kumar2020, zhou2021, arrar2024}. This problem is also one that engages multiple disciplines, including graph theory \cite{bondy2008}, statistical physics \cite{sator2023}, and machine learning \cite{zhouzh2021}. The field of research concerned with link prediction originated in the domain of social network analysis. For example, Adamic and Adar proposed a similarity measurement method based on the number of common neighbors, known as the Adamic-Adar (AA) index \cite{adamic2003}. This method assesses the likelihood of a potential link between two nodes by measuring the rarity of their shared neighbors. Subsequently, Liben-Nowell and Kleinberg conducted a systematic validation of the Common Neighbor (CN) index based on local structure \cite{liben2007}, further advancing the development of prediction methods based on network topological structure. 

The early development of link prediction can be broadly divided into two main streams: similarity indices based on local structure and maximum likelihood methods via probabilistic models. In the first category, in addition to the AA and CN metrics mentioned earlier, Zhou \textit{et al.} proposed a more physically inspired similarity metric—Resource Allocation (RA) \cite{qu2007, zhou2009}. This method assumes that each node evenly distributes “resources” to its neighbors, which are then passed on to the target node. The similarity between two nodes is equal to the sum of the reciprocals of their common neighbors' degree values. The RA index has a clear physical interpretation, and is particularly suitable for identifying potential links in sparse networks. Furthermore, a number of enhanced local similarity indices have been proposed, including the local Path (LP) index \cite{zhou2009, lu2009similarity} and the Local Community Paradigm (LCP) index \cite{cannistraci2013}, thus further enhancing the prediction framework based on local information. The second category primarily relies on maximum likelihood, such as the Hierarchical Structure Model (HSM) proposed by Clauset \textit{et al.} \cite{clauset2008}, and the Stochastic Block Model (SBM) proposed by Guimera \textit{et al.} \cite{guimera2009}. These methods utilise the modular structure of networks to infer the probability distribution of potential links, thereby enhancing the robustness and generalisation capabilities of predictions. 

The advent of deep learning techniques, most notably the development of graph neural networks (GNNs) \cite{scarselli2008}, has precipitated a paradigm shift in link prediction, with the focus now shifting towards feature learning. Methods based on graph embedding \cite{wang2018} can automatically extract high-order structural features of networks, significantly improving prediction performance \cite{perozzi2014, grover2016, qiu2018, ahn2021}. In recent years, models such as GraphSAGE \cite{hamilton2017}, Graph Attention Networks (GAT) \cite{velickovic2017}, and Variational Graph Autoencoders (VGAE) \cite{kipf2016} have been proposed, gradually shifting link prediction from traditional heuristic methods to efficient, flexible data-driven paradigms \cite{bian2025}.

Link prediction has demonstrated significant value in various practical applications. For instance, in the context of social networking platforms, the leveraging of existing interaction relationships between users has been shown to enhance the user experience and platform activity by predicting individuals who may be connected but have not yet established connections \cite{freeman2004, shahmohammadi2016, martincic2017}. Similarly, in biological networks, including protein interaction networks and gene regulatory networks, link prediction has been demonstrated to uncover potential functional associations, thereby providing theoretical foundations for disease mechanism analysis and drug target discovery \cite{herrgard2008, lei2013, csermely2013, musawi2025}. Furthermore, in e-commerce or recommendation systems, modeling the latent connections between users and items has been shown to enable more precise product recommendations and service customization \cite{resnick1997, zhou2007bipartite, zhou2010solving, lv2012, lakshmi2024}. The success of these applications serves to validate the effectiveness of link prediction algorithms and to drive the continuous development and optimization of related algorithms.

For a given undirected network $G(V,E)$, where $V$ and $E$ denote the sets of nodes and edges, respectively, it can be theoretically demonstrated that the maximum possible number of edges in a network containing $N=|V|$ nodes is equal to $\frac{N(N-1)}{2}$. The set of all possible edges is denoted by $U$. The primary objective of the link prediction task is to identify existing but unobserved missing links or future links that may appear in the future from the potential edge set $U-E$. To evaluate the performance of different link prediction methods, the original edge set $E$ is typically divided into a training set $E^T$ and a test set $E^P$, such that $E=E^T \cup E^P$ and $E^T \cap E^P= \emptyset$. The ideal link prediction algorithm should demonstrate an ability to accurately identify the true links in the test set that belong to $E^P$ from the candidate set $U-E^T$. 

In order to perform a more scientific evaluation of the performance of link prediction algorithms, researchers utilise various evaluation metrics to standardize the assessment of prediction effectiveness. The most prevalent of these are as follows: Precision \cite{buckland1994}, Matthews Correlation Coefficient (MCC) \cite{matthews1975}, Normalized Discounted Cumulative Gain (NDCG) \cite{jarvelin2002}, Area Under the ROC Curve (AUC) \cite{hanley1982}, Area Under the Precision-Recall Curve (AUPR) \cite{davis2006}, and AUC-Precision \cite{muscoloni2022}. These metrics offer a multifaceted evaluation of a specific algorithm's predictive performance on a given network \cite{zhou2023, jiao2024, bi2024, wan2024}. 

\subsubsection{Network Predictability}
\begin{figure}[htbp]
	\centering
	\includegraphics[width=1\textwidth]{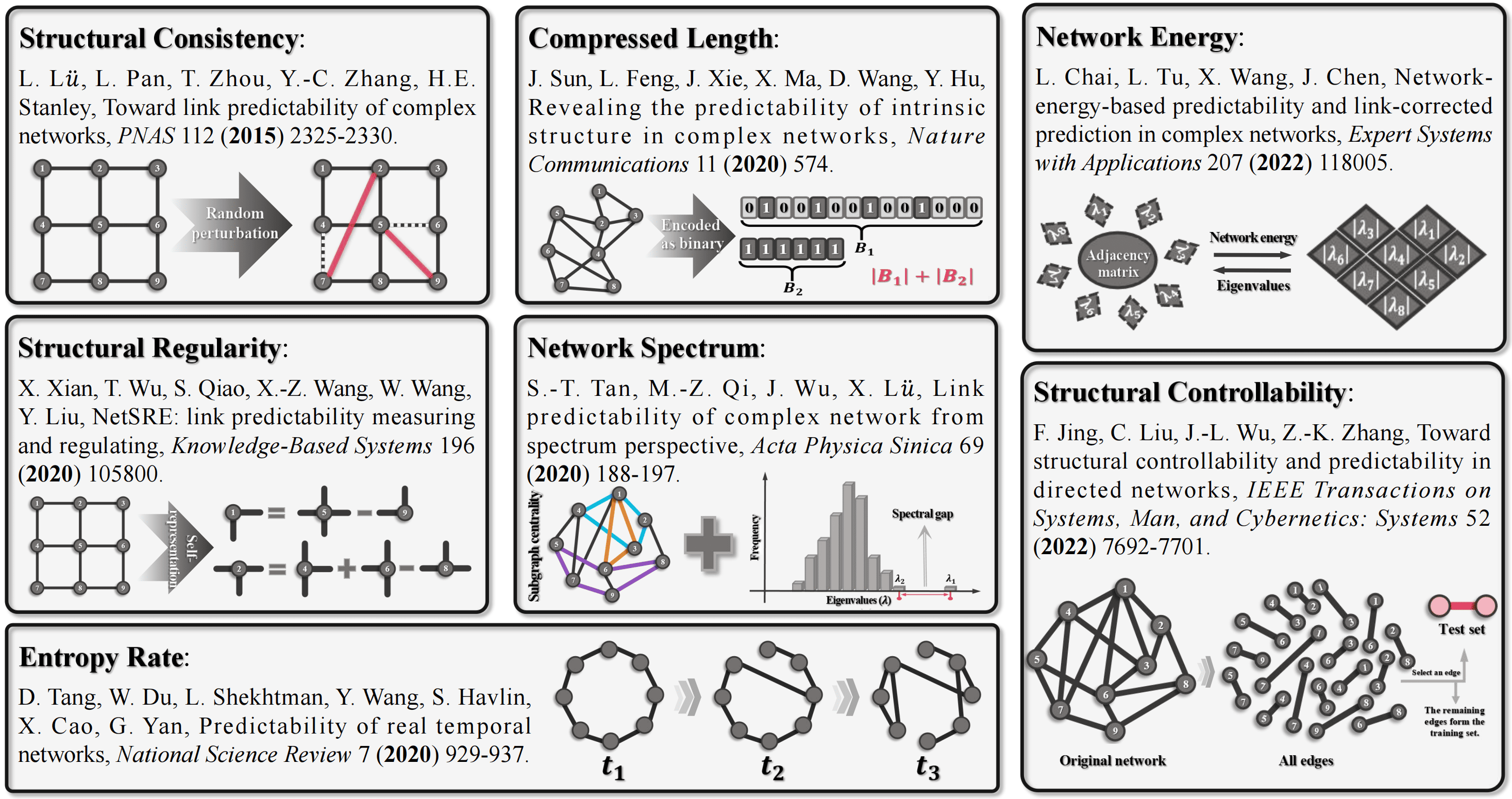} 
	\caption{A schematic overview of research related to network predictability. Each box represents a representative work.}
	\label{fig:conclusion} 
\end{figure}

In certain real-world networks, even when employing the most advanced link prediction algorithms currently available, performance may not meet expectations. This phenomenon gives rise to a critical question: What is the theoretical upper limit of link prediction accuracy in a given network? In other words, in some networks, no matter which algorithm is used, the prediction results may still be unsatisfactory. Is this due to the need for further improvements in existing algorithms, or does it reflect an inherent unpredictability in the network structure itself? In order to provide a response to this question, it is necessary to transcend the confines of particular algorithms and adopt a more fundamental approach, seeking to ascertain whether there exist underlying patterns in network evolution that can be predicted. This prompts the identification of a significant research direction: network predictability. The objective of this direction is twofold: firstly, to quantify the regularities in the  network structure without reliance on any specific prediction algorithm; and secondly, to estimate the highest accuracy that network structure prediction can achieve. This perspective not only helps reveal the theoretical limits of link prediction but also provides new ideas and directions for algorithm design. An overview of representative works in this direction is demonstrated in \autoref{fig:conclusion}.

Network predictability is naturally defined as the extent to which potential structural features (e.g., the existence of missing links) can be inferred based on existing network structures or node attributes \cite{bezbochina2023}. In this review, we focus on the predictability of the most basic structural element in a network, say links (so this problem is called link predictability). This concept is distinct from the performance evaluations of link prediction algorithms \cite{hofman2017, ren2018structure}, as it does not depend on any specific algorithm. Instead, it is derived from the network's inherent structure \cite{martin2016}, thereby characterizing the degree of determinism in its evolutionary process. Network predictability provides a theoretical foundation for understanding the internal connection mechanisms of complex systems and sets performance upper bounds for predictive model design \cite{lorenz1969}.

From a practical application perspective, the predictability of a network is not merely a theoretical issue but also an important reference indicator for guiding practical applications. Considering a scenario in which it is evident that the target network exhibits high predictability, yet the prevailing prediction accuracy remains low. So that we believe the prevailing algorithm still has a huge space of improvement and justifies the allocation of additional resources towards optimization. Conversely, if the network itself demonstrates low predictability, it would be imprudent to attribute poor prediction performance solely to inadequate algorithm design. Rather, the prediction task or data collection methods should be subjected to re-evaluation \cite{jing2024}. In reality, real-world networks frequently contain both regular and irregular structures. Regular structures (e.g. community structures in social networks, hierarchical routing structures in the Internet, or modular organisation in protein interaction networks) exhibit strong intrinsic regularity and can be modeled and explained. At the same time, irregular structures may arise from noise, disturbances, or system complexity, thereby increasing prediction difficulty. A recent work indicates that the larger the proportion of regular structures in a network, the higher its predictability \cite{ming2019}. One might consider an intuitive approach to assessment to be one which quantifies the proportion of regular structures in a network, with a view to measuring its predictability. Building on this, researchers have proposed various methods to assess network predictability from multiple perspectives. We will systematically review relevant research from the perspectives of spectral analysis, information theory, and structural controllability. The aim of this review is to provide theoretical support and methodological guidance for understanding and characterizing network predictability.

\subsection{Spectral Methods}
In the study of complex networks, spectral analysis provides a powerful mathematical tool, enabling the extraction of key information about the structural regularities and evolutionary mechanisms of networks from the eigenvalues and eigenvectors of the adjacency matrix or Laplacian matrix. Consequently, spectral analysis has become an indispensable tool for comprehending network functionality and dynamic behavior \cite{chung1997, guo1998, farkas2001, van2003,  chung2003, mieghem2023}. For instance, eigenvectors reveal node centrality \cite{estrada2012} and network bipartisity \cite{newman2006a}, and the eigenvalue distribution characterises network hierarchical structure and modularity \cite{xing2006, wang2019}. Furthermore, it has been shown that spectral information provides a theoretically bounded and computationally feasible modeling framework for link prediction tasks \cite{lee2021}.

This chapter will explore three core methods based on spectral information: (i) Structural Consistency, which quantifies the regularity of network structure by analyzing the response of eigenvectors and eigenvalues to perturbations in the network structure; (ii) Network Spectrum, which leverages the properties of the spectral gap and adopts the form of subgraph centrality to measure the predictability of a network; (iii) Network Energy, which integrates information from all eigenvalues to quantify the overall stability and predictability of a network. Together, these three approaches constitute a theoretical framework for understanding network predictability from a spectral analysis perspective, thereby providing a robust mathematical foundation for subsequent algorithm design.

\subsubsection{Structural Consistency}
Recent studies have demonstrated that the predictability of networks is contingent not only on their local connectivity patterns but is also closely related to their global structural properties. Among these, the Structural Consistency method \cite{lvzhou2015} presents a novel approach to quantifying the predictability of network links by means of graph perturbations. The fundamental principle underlying this approach is articulated as follows: Should a network demonstrate the capacity to sustain the stability of its eigenvectors in the face of minor structural perturbations, it is indicative of a high degree of regularity and modelability, consequently exhibiting a high degree of predictability. Conversely, if minor perturbations trigger significant changes, this suggests a disordered network structure and greater prediction difficulty. This concept bears a close relation to perturbation theory in physics \cite{kato2013} and has been effectively implemented mathematically in the field of graph analysis.

The adjacency matrix of the original network, denoted by $G = (V, E)$, is represented by the matrix $\bm{A}$. A random sample proportion, denoted by $p^H$, of edges is selected to form the perturbation set, denoted by $\Delta E$. The remaining edges constitute the residual edge set, denoted by $E^R = E - \Delta E$. The corresponding adjacency matrices are denoted by $\Delta \bm{A}$ and $\bm{A}^R$ respectively, and obviously $\bm{A} = \bm{A}^R + \Delta \bm{A}$. It is evident that, since the matrix $\bm{A}^R$ is symmetric, it possesses a complete orthogonal eigenvector basis $\bm{x}_i$ and corresponding eigenvalues $\lambda_i$ satisfying the following conditions:
\begin{equation}
	\bm{A}^R \bm{x}_i = \lambda_{i} \bm{x}_i.
	\label{eq:AR}
\end{equation}
At the same time, the matrix $\bm{A}^R$ can also be represented as a diagonal matrix:
\begin{equation}
	\bm{A}^{R}=\sum_{i=1}^{N}\lambda_{i}\bm{x}_{i}\bm{x}_{i}^{T}.
	\label{eq:eigenvector}
\end{equation}
When introducing the perturbation $\Delta \bm{A}$, we have $\bm{A} = \bm{A}^R + \Delta \bm{A}$. The eigenvalues and eigenvectors become $\lambda_{i} + \Delta \lambda_{i}$ and $\bm{x}_{i} + \Delta \bm{x}_{i}$ respectively, satisfying:
\begin{equation}
	(\bm{A}^R+\Delta \bm{A})(\bm{x}_{i}+\Delta \bm{x}_{i})=(\lambda_{i}+\Delta \lambda_{i})(\bm{x}_{i}+\Delta \bm{x}_{i}).
	\label{eq:AR_Delta_A}
\end{equation}
Utilizing first-order perturbation theory \cite{dirac1981}, expanding both sides of Equ.(\ref{eq:AR_Delta_A}) and neglecting the second-order terms (e.g., $\Delta \bm{A} \Delta \bm{x}_{i}$, $\Delta \lambda_{i} \Delta \bm{x}_{i}$), it simplifies to:
\begin{equation}
	\bm{A}^R \Delta \bm{x}_{i}+\Delta \bm{A} \bm{x}_{i} \approx \lambda_{i} \Delta \bm{x}_{i} + \Delta \lambda_{i} \bm{x}_{i},
\end{equation}
Multiplying both ends of the above equation by $\bm{x}^{T}_{i}$ on the left yields an approximate modified eigenvalue $\Delta\lambda_i$ as follows:
\begin{equation}
	\Delta\lambda_i \approx \bm{x}_i^T \Delta \bm{A} \bm{x}_i.
	\label{eq:delta}
\end{equation}
Using the perturbed eigenvalues $(\lambda_i + \Delta\lambda_i)$ and combining them with the preceding Eq. (\ref{eq:eigenvector}), we can reconstruct an approximate perturbed adjacency matrix $\bm{\tilde{A}}$:
\begin{equation}
	\bm{\tilde{A}} = \sum_{i=1}^{N} (\lambda_i + \Delta\lambda_i)\bm{x}_i \bm{x}_i^T.
	\label{eq:tildeA}
\end{equation}

If the network structure is stable, the change in eigenvectors before and after perturbation should be minimal. It can thus be concluded that the eigenvectors $\bm{x}_i$ of the observed matrix $\bm{A}^R$ and the eigenvectors $\bm{x}_i + \Delta \bm{x}_i$ of the perturbed matrix $\bm{A}^R + \Delta \bm{A}$ should be almost identical. It is evident from Eq. (\ref{eq:tildeA}) that the matrix $\bm{A}$ should be in close proximity to $\bm{A}^R + \Delta \bm{A}$. Conversely, for highly regular networks, the random removal of a set of edges, denoted by the symbol $\Delta E$, does not result in a substantial alteration of the structural characteristics. It can thus be deduced that the vectors $\bm{A}$ and $\bm{\tilde{A}}$ should be in close proximity to each other.

\begin{figure}[htbp]
	\centering
	\includegraphics[width=1\textwidth]{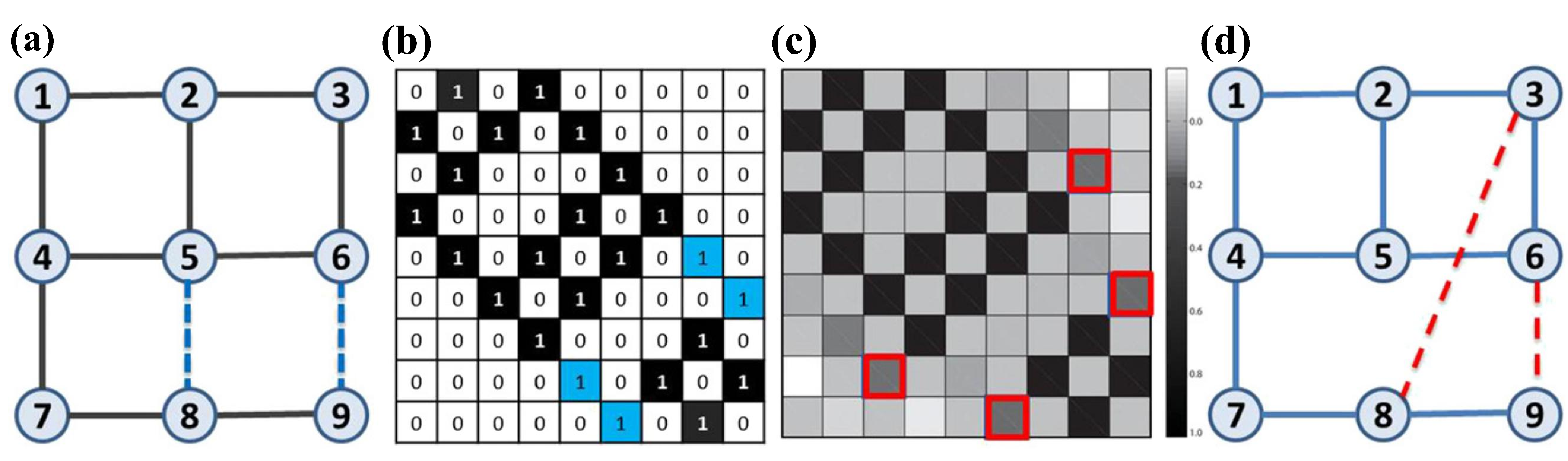} 
	\caption{An illustration of how to calculate the structural consistency. (a) is the given network, the blue dashed links constitute the perturbation set $\Delta E=\{(5,8), (6,9)\}$ (corresponding to $\Delta \bm{A}$), while the solid links constitute the set $E^R$ (corresponding to $\bm{A}^R$); (b) is the adjacency matrix $\bm{A}$ of the given network, where the number in each square is the corresponding value of the matrix element. The black and blue squares represent the links in $E^R$ and $\Delta E$, respectively. To calculate the consistency, we perturb $\bm{A}^R$ with $\Delta \bm{A}$. The perturbed matrix $\bm{\widetilde{A}}$ is shown in (c), from which we derive the perturbed network in (d), where the red dashed lines are outcome links selected by ranking all links in $U - E^R$ in descending order according to their corresponding values in $\bm{\widetilde{A}}$. Since there are two links in $\Delta E$, then $L = 2$, and the set $E^L=\{(3,8), (6,9)\}$. In this case, only one of the two blue links is recovered by perturbation; then we have $\sigma_c = 0.5$. \\
		\textit{Source}: The figure is reproduced from Ref. \cite{lvzhou2015}.}
	\label{fig:sc-illustration} 
\end{figure}

In particular, for all potential edges not in $E^R$ (i.e. the set $U - E^R$), the edges are sorted in descending order based on the corresponding element values in $\bm{\tilde{A}}$. The initial step in the process is the selection of the first $L = | \Delta E|$ edges to form the prediction set $E^L$. Structural Consistency is then defined as follows:
\begin{equation}
	\sigma_c = \frac{|E^L \cap \Delta E|}{|\Delta E|},
\end{equation}
The numerator of the metric signifies the number of edges concurrently present in $E^L$ and the perturbation edge set $\Delta E$, denoted by the denominator. This metric is indicative of the efficacy of perturbation edge identification during reconstruction. A larger $\sigma_c$ indicates greater structural consistency in the network, making its latent missing edges easier to predict. The conceptual flow of this method is illustrated in \autoref{fig:sc-illustration}.

\begin{figure}[htbp]
	\centering
	\includegraphics[width=1.0\textwidth]{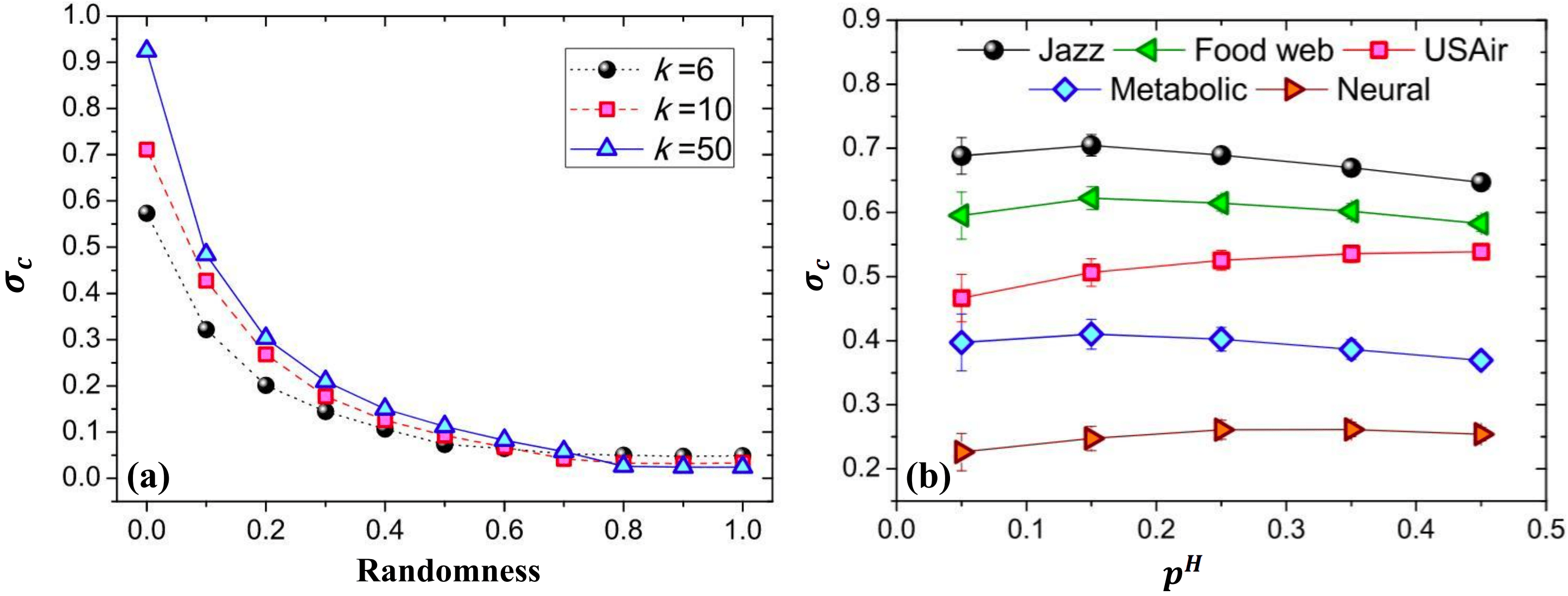}
	\caption{Structural consistency of modeled and real networks: 
		(a) WS networks with $N=1000$, varying $k$ and rewiring probability; (b) Dependence on perturbation set size. 
		Error bars is the standard deviation over 100 realizations. \\
		\textit{Source}: The figure is reproduced from Ref. \cite{lvzhou2015}.}
	\label{fig:sc_network}
\end{figure}

In order to validate the effectiveness of this method, Lü \textit{et al.} conducted systematic tests on multiple artificial networks. As shown in \autoref{fig:sc_network}(a), it can be observed that for the WS network, as the reconnection probability increases (i.e., the network becomes more random), the structural consistency significantly decreases. This finding lends further support to the assertion that "highly random networks are unpredictable". Furthermore, experimental findings on multiple real-world networks indicate that even when altering the sampling proportion $p^H$, the computed structural consistency remains largely unchanged (as demonstrated in \autoref{fig:sc_network}(b)). It is evident that the property under discussion confers upon the method two significant advantages: firstly, robust performance and, secondly, strong generalization capabilities in practical applications. In a word, the structural consistency not only provides an effective tool for quantitatively assessing the predictability of network links, but more importantly, it reveals the deep connection between graph stability and network regularity, and forms a foundation for spectral analysis in network predictability. 

\subsubsection{Network Spectrum}
As previously stated, the eigenvalues $\lambda_i$ and the eigenvectors are of pivotal significance in the characterization of the collective connection strength of a network. This is particularly evident in the demonstration of their efficacy in preliminary modeling for the assessment of link predictability. Beyond the eigenvalues themselves, the intervals between eigenvalues offer additional information, typically containing key insights into community structure, mode distribution, and topological stability \cite{newman2006a}. In recent years, researchers have attempted to apply spectral analysis to link prediction tasks, seeking to quantify a network's latent predictability through the statistical properties of its characteristic spectrum. Tan \textit{et al.} \cite{tan2020} propose a quantitative metric $\lambda_s$ based on the characteristic spectrum of the adjacency matrix. This metric is employed to assess the consistency of network structure and its predictive capability for missing edges.

For an undirected, unweighted graph $G$, let the eigenvalues of the adjacency matrix $\bm{A}$ be denoted by the sequence of real numbers, namely, $\lambda_1 \geq \lambda_2 \geq \cdots \geq \lambda_N$. The set of eigenvalues $\{\lambda_i\}$ is then called the eigenvalue spectrum of $G$. Early studies indicate that a significant proportion of networks (particularly scale-free networks) exhibit a pronounced spectrum gap \cite{chung2003, dorogovtsev2003, tan2016}. The largest eigenvalue $\lambda_1$ is considerably greater than the next largest one $\lambda_2$, say $\lambda_1 \gg \lambda_2$. This phenomenon is indicative of the presence of a dominant global structural component within the network, with the remaining eigenvalues reflecting more local heterogeneity \cite{lee2021}. In light of the foregoing observation, Tan \textit{et al.} \cite{tan2020} raise a hypothesis that the quantification of the disparity between the largest eigenvalue and the remaining eigenvalues in the eigenvalue spectrum has the potential to facilitate a quantitative depiction of network predictability. This approach bears resemblance to the concept of pulse diagnosis (qiemai) in traditional Chinese medicine. This study asserts that, guided by this concept, a novel metric for network predictability is introduced.

In the domain of network mining, a plethora of metrics exist for the evaluation of network structural properties. Among these, subgraph centrality (SC) is a notable metric positing that shorter path lengths within closed loops indicate more efficient information transmission within the loop, tighter connections between nodes, and thus a greater contribution to node centrality. The subgraph centrality of node $i$ is hereby defined as follows \cite{estrada2005}:
\begin{equation}
	\operatorname{SC}(i) = \sum_{j=1}^{N}(x_j^i)^2 e^{\lambda_j},
	\label{eq:SC}
\end{equation}
where $\lambda_j$ is the $j$th eigenvalue of the network adjacency matrix $\bm{A}$, and $x_j^i$ signifies the $i$th component of the corresponding eigenvector $\bm{x}_j$.

According to the relationship between exponential and hyperbolic functions: $e^{\lambda_j}=sinh(\lambda_j)+cosh(\lambda_j)$, where the hyperbolic cosine and hyperbolic sine are defined as: $cosh(\lambda_j)=\frac{e^{\lambda_j} + e^{-\lambda_j}}{2}$, $sinh(\lambda_j)=\frac{e^{\lambda_j} - e^{-\lambda_j}}{2}$, a Taylor series expansion to $e^{\lambda_j}$ yields:
\begin{equation*}
	e^{\lambda_j}=\sum_{k=0}^{\infty} \frac{(\lambda_j)^k}{k!}= \sum_{m=0}^{\infty} \frac{\lambda_j^{2m}}{(2m)!} + \sum_{m=0}^{\infty} \frac{\lambda_j^{2m+1}}{(2m+1)!},
\end{equation*}
Substituting this formula into Eq. (\ref{eq:SC}) yields:
\begin{equation}
	\operatorname{SC}(i) = \sum_{m=0}^{\infty} \frac{1}{(2m)!} \left[ \sum_{j=1}^{N} (\lambda_j)^{2m} (x_j^i)^2 \right] + \sum_{m=0}^{\infty} \frac{1}{(2m+1)!} \left[ \sum_{j=1}^{N} (\lambda_j)^{2m+1} (x_j^i)^2 \right].
	\label{eq:SC1}
\end{equation}

Since the adjacency matrix $\bm{A}$ can be decomposed spectrally: $\bm{A} = \sum_{j=1}^{N} \lambda_j \bm{x}_j \bm{x}_j^T$, its $k$th power can be expressed as: $(\bm{A}^k) = \sum_{j=1}^N (\lambda_j)^k \bm{x}_j \bm{x}_j^T$. Therefore, $(\bm{A}^k)_{ii}$ denotes the number of closed paths of length $k$ originating from node $i$ and returning to itself, i.e. $(\bm{A}^k)_{ii}=\sum_{j=1}^{N} (\lambda_j)^k (x_j^i)^2$. Substituting this into Eq. (\ref{eq:SC1}) yields:
\begin{equation}
	\begin{aligned}
		\operatorname{SC}(i) & =\sum_{m=0}^{\infty} \frac{1}{(2m)!} (\bm{A}^{2m})_{ii} + \sum_{m=0}^{\infty} \frac{1}{(2m+1)!} (\bm{A}^{2m+1})_{ii} \\
		& =\operatorname{SC}_{\mathrm{even}}(i) + \operatorname{SC}_{\mathrm{odd}}(i).
	\end{aligned}
\end{equation}

In the first term, $\operatorname{SC}_{\mathrm{even}}(i)$, the expression $(\bm{A}^{2m})_{ii}$ denotes the total number of closed paths of length $2m$ originating from node $i$. Thus, $\operatorname{SC}_{\mathrm{even}}(i)$ represents the weighted sum of all closed paths of even length. The second term, $(\bm{A}^{2m+1})_{ii}$ in $\operatorname{SC}_{\mathrm{odd}}(i)$, represents the total number of closed paths of length $(2m+1)$, corresponding to the weighted contribution of all odd-length closed paths. Odd-length paths are given greater emphasis since they are hypothesised to better reflect direct connections between a node and its neighbours:
\begin{equation}
	\operatorname{SC}_{\mathrm{odd}}(i)=(x_1^i)^2\sinh(\lambda_1)+\sum_{j=2}^{N}(x_j^i)^2\sinh(\lambda_j).
	\label{eq:sc_odd}
\end{equation}

By utilizing the spectral gap assumption $\lambda_1 \gg \lambda_2$, we have 
\begin{equation}
	(x_1^i)^2\sinh(\lambda_1) \gg \sum_{j=2}^{N}(x_j^i)^2\sinh(\lambda_j),
\end{equation}
At this point, it is reasonable to approximate:
\begin{equation}
	\operatorname{SC}_{\mathrm{odd}}(i) \approx (x_1^i)^2\sinh(\lambda_1).
	\label{eq:sc_ideal}
\end{equation}
Therefore, the following conclusion can be drawn:
\begin{equation}
	\widetilde{x}_{1}^{i} = \sqrt{ \frac{\operatorname{SC}_{\mathrm{odd}}(i)}{\sinh(\lambda_1)}}.
	\label{eq:minus}
\end{equation}

Should the observed value of $x_1^i$ deviate significantly from the standard value, denoted by $\widetilde{x}_{1}^{i}$, then this may be indicative of a network structure which is not consistent with the standard model. In such cases, prediction may be rendered more challenging. In order to address this issue, the authors constructed a deviation metric:
\begin{equation}
	\Delta\log x_{1}^{i}\!=\!\log ( \frac{x_{1}^{i}}{\widetilde{x}_{1}^{i}}) \!=\!\log\left\{\!\!\frac{\left(x_{1}^{i}\right)^{2}\sinh\left(\lambda_{1}\right)}{\mathrm{SC}_{\mathrm{odd}}(i)}\!\right\}^{0.5},
\end{equation}
Finally, an indicator $\sigma_s \in [0,1]$ is defined, which measures how close $\Delta\log x_{1}^{i}$ is to 0. The mathematical expression for this indicator is as follows:
\begin{equation}
	\sigma_s=\exp\left[-\sum\sqrt{\left(\frac{\mathrm{SC}_{\mathrm{odd}}(i)-\left(x_{1}^{i}\right)^{2}\sinh\left(\lambda_{1}\right)}{N \cdot \mathrm{SC}_{\mathrm{odd}}(i)}\right)^{2}}\right].
\end{equation}

\begin{figure*}[!t]
	\centering
	\includegraphics[width=0.65\textwidth]{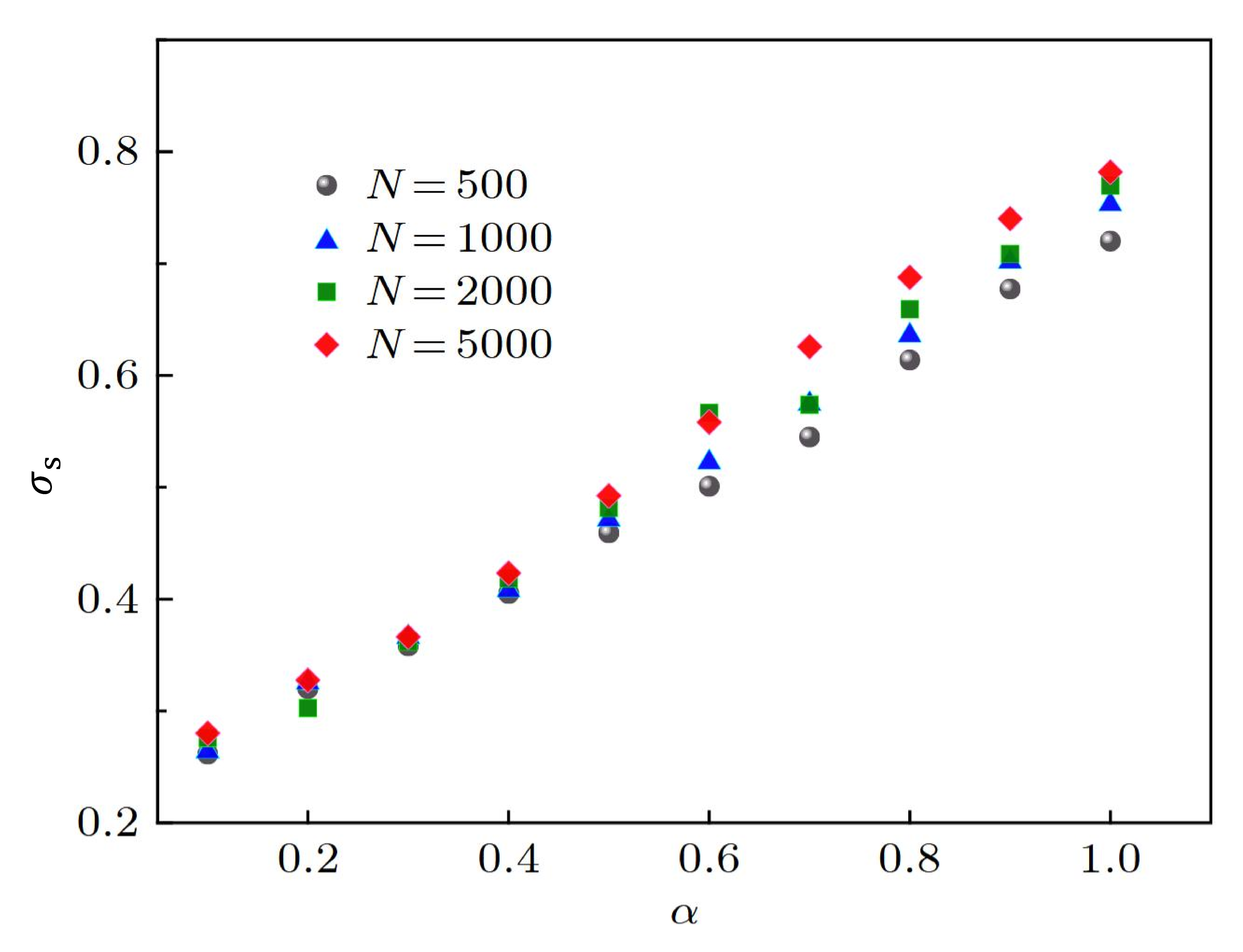} 
	\caption{\texorpdfstring{
			The Predictability $\sigma_s$ of model networks versus $\alpha$ with different network sizes $N$. \\
			\textit{Source}: The figure is reproduced from Ref.~\protect\cite{tan2020}.
		}{
			The Predictability sigma_s of model networks versus alpha with different N. 
			Source: Figure reproduced from Ref. [tan2020].
	}}
	\label{fig:ns_BA} 
\end{figure*}

\begin{figure}[H]
	\centering
	\includegraphics[width=0.6\textwidth]{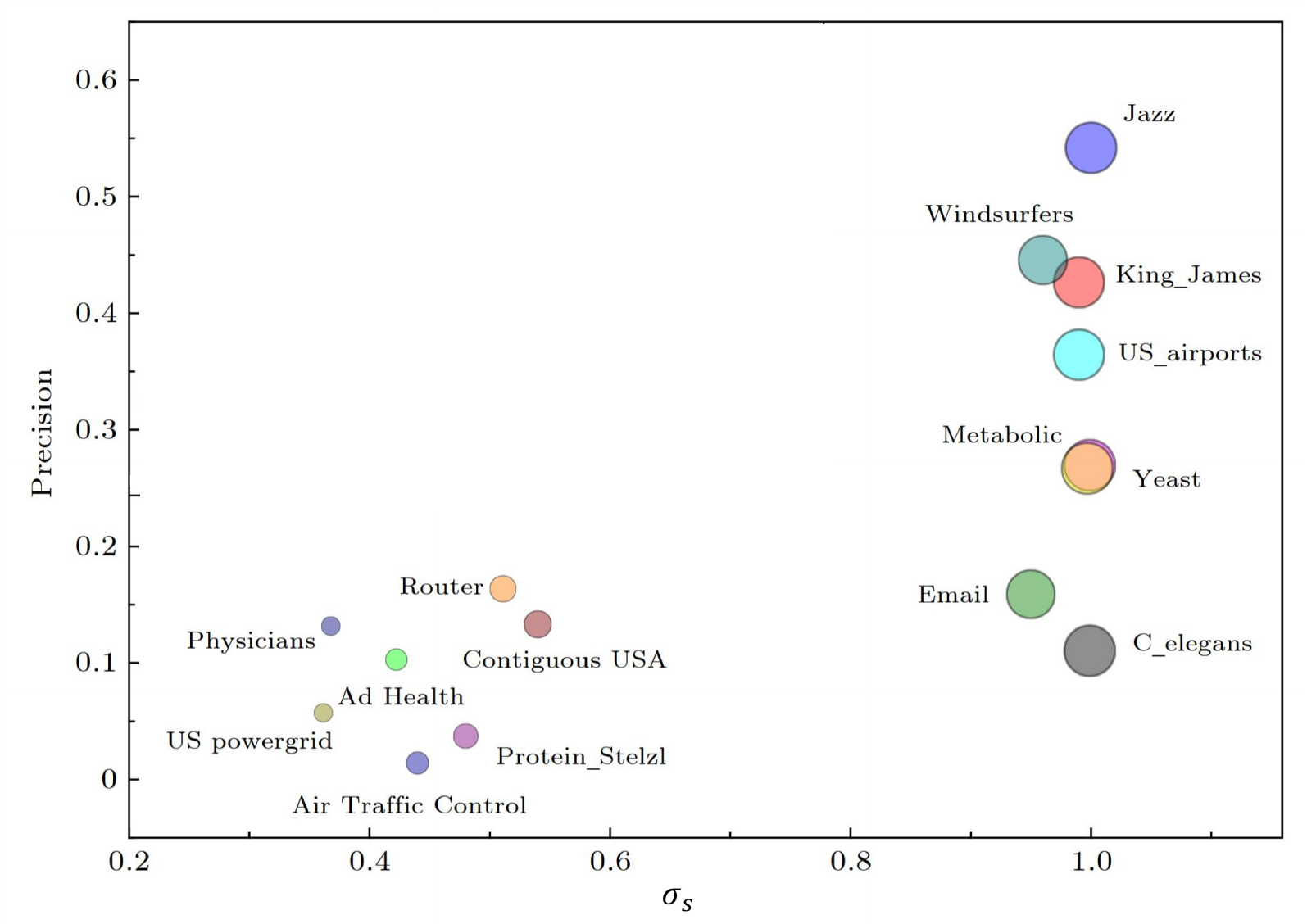} 
	\caption{Performance of precision and $\sigma_s$ in real networks. Different colored circles represent different real networks, and the size of each circle is proportional to the network predictability $\sigma_s$. The $x$-axis represents the value of $\sigma_s$, and the $y$-axis represents the Max Precision of link prediction algorithms—the higher the value, the better the algorithm's maximum accuracy. \\
		\textit{Source}: The figure is reproduced from Ref. \cite{tan2020}.}
	\label{fig:ns_Real_network} 
\end{figure}

The predictability of the network can be indirectly assessed through the magnitude of its value.  If the deviation from the ideal case approaches 0, then $\sigma_s$ approaches $1$, indicating excellent link predictability in the network. Conversely, if the network's standard deviation, denoted by $\sigma_s$, is found to be minimal, this is indicative of poor predictability. Applying this method to a generalized BA model, where the structural regularity of the network can be controlled by adjusting a preferential attachment parameter $\alpha$ (the larger $\alpha$ corresponds to the higher regularity). As demonstrated in \autoref{fig:ns_BA}, the enhancement of preferential attachment characteristics is observed with increasing values of $\alpha$, consequently resulting in an increase in $\sigma_s$ values. This finding indicates that the metric is capable of effectively capturing the strengthening of structural regularity during network evolution. A comparison of BA networks with ER networks reveals that ER networks exhibit significantly lower $\sigma_s$ (see  \autoref{tab:ns_BA_Random}). In addition, experimental findings on real-world networks (see \autoref{fig:ns_Real_network} and Table 3) demonstrate that networks with higher values of $\sigma_s$ generally have greater link prediction precisions, thereby partially validating the positive correlation between this metric and network predictability. However, it is noteworthy that $\sigma_s$ exhibits limited resolution: on certain real-world networks with distinct structures, its values are identical. This phenomenon exposes the limitations of this method and indicates that it requires further optimization to characterize network predictability, demonstrating significant potential for improvement.

This section elucidates the methodology for using the spectral gap between the largest and second-largest eigenvalues to construct a novel metric, denoted as $\sigma_s$, which reflects network structural consistency and predictability. As a key component of spectral analysis methods, this approach shows two primary advantages: firstly, it deepens our understanding of network structural properties; secondly, it provides a new theoretical framework for link prediction tasks. Nevertheless, this method is not without its limitations. For example, in networks with insufficient spectral gaps, this metric may fail to accurately reflect structural coherence. Additionally, its reliance on computing global eigenvectors introduces substantial computational complexity, which in turn restricts its direct applicability in the context of large-scale networks.

\begin{table}[htbp]
	\centering
	\caption{Performance of link prediction algorithms in model networks. The Precision of 12 similarity-based link prediction algorithms on these two synthetic networks are presented. In the experiments, random sampling was used to split the dataset into training and test sets with a ratio of $9:1$. When calculating Precision, the number of edges in the test set was used as the threshold. \\
		\textit{Source}: The table is reproduced from Ref. \cite{tan2020}.}
	\label{tab:ns_BA_Random}  
	\begin{tabularx}{\textwidth}{@{}l *{13}{>{\centering\arraybackslash}X}@{}}
		\toprule
		\textbf{Networks} & $\sigma_s$ & CN & AA & RA & LP & IA & CAR & PA & Katz & RWR & ACT & LRW & LHNII \\
		\midrule
		BA network & 0.975 & 0.423 & 0.324 & 0.272 & 0.415 & 0.271 & 0.283 & 0.594 & 0.412 & 0.136 & 0.507 & 0.085 & 0.003 \\
		ER network & 0.543 & 0.015 & 0.008 & 0.009 & 0.008 & 0.009 & 0.000 & 0.030 & 0.008 & 0.002 & 0.020 & 0.000 & 0.001 \\
		\bottomrule
	\end{tabularx}
\end{table}

\begin{table}[htbp]
	\centering
	\caption{Precisions of link prediction algorithms in real networks. \textit{Source}: The table is reproduced from Ref. \cite{tan2020}. }
	\label{tab:ns_Real_table}
	\begin{tabularx}{\textwidth}{@{}l *{13}{>{\centering\arraybackslash}X}@{}}
		\toprule
		\textbf{Networks} & $\sigma_s$ & CN & AA & RA & LP & IA & CAR & PA & Katz & RWR & ACT & LRW & SRW \\
		\midrule
		C\_elegans & 0.999 & 0.100 & 0.107 & 0.105 & 0.101 & 0.108 & 0.094 & 0.058 & 0.101 & 0.105 & 0.055 & 0.110 & 0.108 \\
		Windsurfers & 0.999 & 0.379 & 0.396 & 0.413 & 0.370 & 0.393 & 0.381 & 0.214 & 0.369 & 0.360 & 0.247 & 0.402 & 0.426 \\
		Adolescent health & 0.422 & 0.103 & 0.103 & 0.088 & 0.089 & 0.101 & 0.094 & 0.003 & 0.088 & 0.053 & 0.008 & 0.042 & 0.047 \\
		Jazz & 1.000 & 0.502 & 0.523 & 0.542 & 0.489 & 0.535 & 0.517 & 0.133 & 0.489 & 0.352 & 0.168 & 0.342 & 0.393 \\
		USAirport & 0.998 & 0.333 & 0.336 & 0.364 & 0.332 & 0.332 & 0.330 & 0.280 & 0.332 & 0.087 & 0.294 & 0.076 & 0.080 \\
		Metabolic & 0.999 & 0.137 & 0.195 & 0.269 & 0.141 & 0.168 & 0.132 & 0.104 & 0.141 & 0.196 & 0.092 & 0.214 & 0.215 \\
		Yeast & 0.998 & 0.154 & 0.177 & 0.267 & 0.158 & 0.161 & 0.148 & 0.094 & 0.174 & 0.073 & 0.211 & 0.045 & 0.059 \\
		US powergrid & 0.362 & 0.054 & 0.032 & 0.028 & 0.058 & 0.047 & 0.037 & 0.000 & 0.057 & 0.016 & 0.034 & 0.015 & 0.018 \\
		Physicians & 0.368 & 0.119 & 0.126 & 0.121 & 0.117 & 0.122 & 0.106 & 0.014 & 0.117 & 0.119 & 0.015 & 0.132 & 0.127 \\
		Air Traffic Control & 0.480 & 0.036 & 0.024 & 0.018 & 0.037 & 0.021 & 0.025 & 0.007 & 0.037 & 0.002 & 0.015 & 0.002 & 0.002 \\
		Contiguous USA & 0.540 & 0.096 & 0.130 & 0.132 & 0.005 & 0.000 & 0.000 & 0.012 & 0.004 & 0.067 & 0.053 & 0.133 & 0.121 \\
		Email & 0.950 & 0.144 & 0.158 & 0.143 & 0.142 & 0.159 & 0.145 & 0.018 & 0.141 & 0.065 & 0.024 & 0.052 & 0.051 \\
		King James Bible & 0.960 & 0.167 & 0.270 & 0.446 & 0.163 & 0.256 & 0.176 & 0.078 & 0.163 & 0.186 & 0.069 & 0.197 & 0.224 \\
		Protein Stelzl & 0.441 & 0.001 & 0.002 & 0.001 & 0.001 & 0.002 & 0.006 & 0.014 & 0.001 & 0.006 & 0.013 & 0.006 & 0.006 \\
		Router & 0.511 & 0.051 & 0.029 & 0.020 & 0.056 & 0.031 & 0.055 & 0.022 & 0.055 & 0.006 & 0.164 & 0.005 & 0.005 \\
		\bottomrule
	\end{tabularx}
\end{table}

\subsubsection{Network Energy}
The concept of network energy was first proposed by Gutman \textit{et al.} within the framework of Chemical Graph Theory \cite{gutman2001, gutman2009}, originating from research on the stability of molecular $\pi$-electron systems. It aims to describe the stability of molecular structures and their reactivity. Mathematically, the total energy of a graph is defined as the sum of the absolute values of all its eigenvalues in its adjacency matrix. This serves to quantify the overall activity or structural complexity of the network. In recent years, this method has found extensive application in complex network analysis, demonstrating particular strengths in revealing structural features during network evolution and in predictive potential.

\begin{figure}[htbp]
	\centering
	\includegraphics[width=0.75\textwidth]{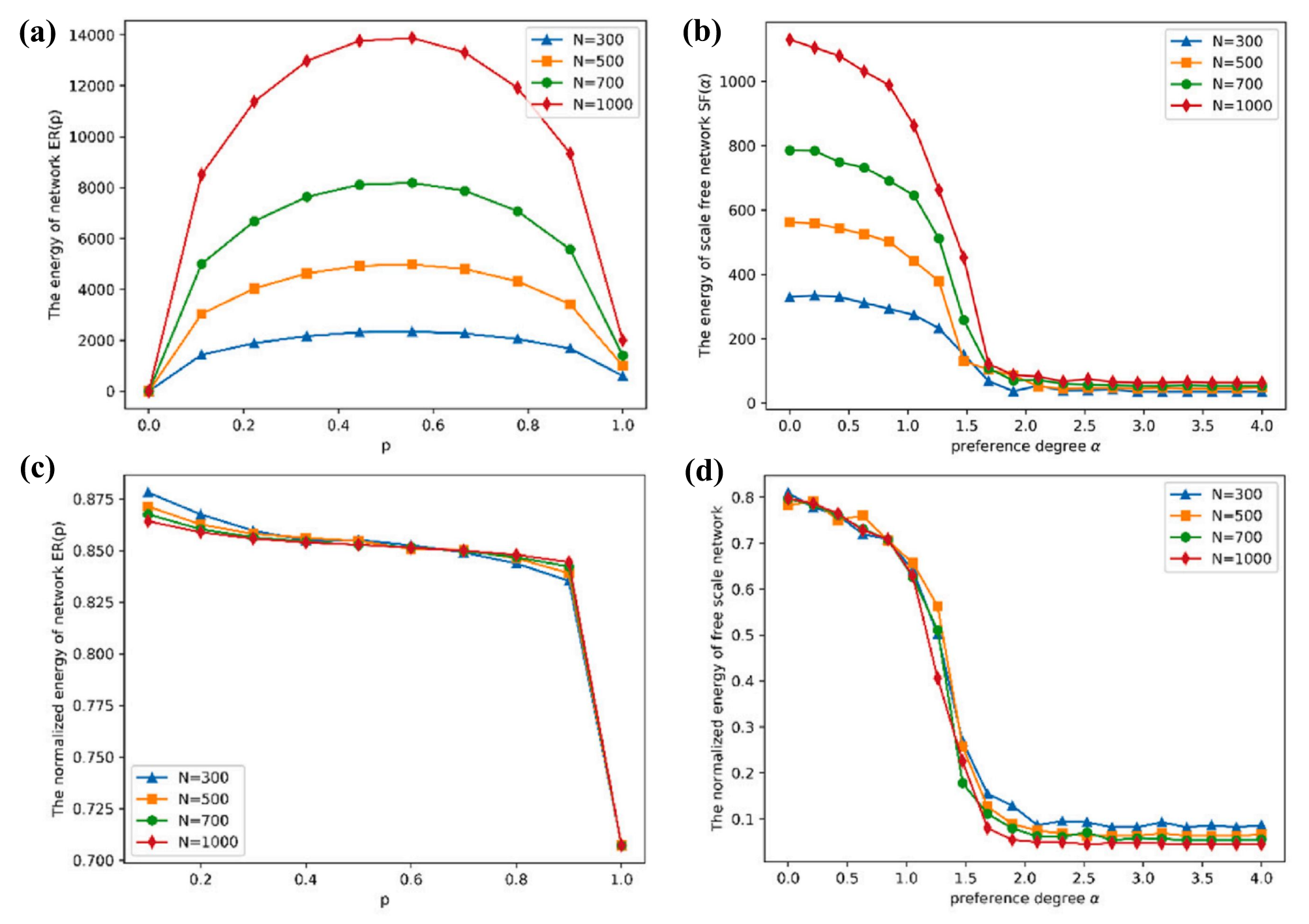} 
	\caption{Original network energy and normalized network energy at different sizes.  (a) The network energy $E(G)$ of ER networks at different sizes; (b) the network energy $E(G)$ of scale-free networks at different sizes;  (c) The network energy $\widehat{E}(G)$ of ER networks at different sizes; (d) the network energy $\widehat{E}(G)$ of scale-free networks at different sizes. \\
		\textit{Source}: The figure is reproduced from Ref. \cite{chai2022network}.}
	\label{fig:ne_energy} 
\end{figure}

For an undirected, unweighted graph $G$, its energy is \cite{gutman2009}
\begin{equation}
	E(G)=|\lambda_1|+|\lambda_2|+\dots+|\lambda_N|=\sum_{i=1}^{N}|\lambda_i|.
	\label{eq:energy}
\end{equation}
As shown in \autoref{fig:ne_energy}(a)(b), the network energy $E(G)$ is contingent not only on the topological structure, but also on the number of nodes $N$ and edges $M$. To eradicate these extrinsic factors and enhance the comparability among disparate networks, Chai \textit{et al.} \cite{chai2022network} proposed a normalized form:
\begin{equation}
	\widehat{E}(G)=\frac{E(G)}{E_{max}(G)},
\end{equation}
where $E_{max}(G)$ represents the maximum energy value achievable for a graph with the same number of nodes $N$ and edges $M$. Its expression is divided into two cases based on the relationship between $2M$ and $N$:
\begin{equation*}
	E_{max}(G)=
	\begin{cases}
		\sqrt{2MN} & \text{for} \quad 2M < N, \\ 
		\frac{2M}{N}+\sqrt{(N-1)\left[2M-\left(\frac{2M}{N}\right)^{2}\right]} & \text{for} \quad 2M \geq N.
	\end{cases}
\end{equation*}

The above definitions originate from the classical inequality derivations by McClelland \cite{mcclelland1971} and Koolen \textit{et al.} \cite{koolen2001}. For random networks with varying edge probability $p$ and scale-free networks with different preference degree exponents $\alpha$ (used to adjust the probability of new edge formation), their normalized energy values are shown in \autoref{fig:ne_energy}(c)(d).

Chai \textit{et al.} \cite{chai2022network} further show a close relationship between network energy and link predictability. Specifically, when networks manifest stronger regularity (such as BA networks with pronounced preferential attachment), their normalized energy $\widehat{E}(G)$ tends to decrease. Conversely, networks displaying higher randomness yield higher $\widehat{E}(G)$ values. This phenomenon suggests that network energy can serve as an effective metric for assessing a network's potential predictability. Based on this, Chai \textit{et al.} \cite{chai2022network} proposed a new predictability metric $\sigma_{e}$, defined as follows:
\begin{equation}
	\sigma_{e}=[\widehat{E}(G)]^{-1}=\frac{E_{max}(G)}{E(G)}.
\end{equation}
The value range of this metric is $(1, \infty)$. A higher value indicates a more regular network structure and greater predictability.

Moreover, the author puts forward a novel metric, denoted here as $\sigma_{e}^{'}$, which integrates network energy with structural consistency information, denoted here as $\sigma_c$:
\begin{equation}
	\sigma_{e}^{'}=\sigma_{e} \cdot \sigma_c = \frac{E_{max}(G)}{E(G)} \cdot \sigma_c.
\end{equation}
This fusion metric aims to combine global spectral information with local structural patterns to obtain richer predictive evidence.

\begin{figure*}[!t]
	\centering
	\includegraphics[width=0.95\textwidth]{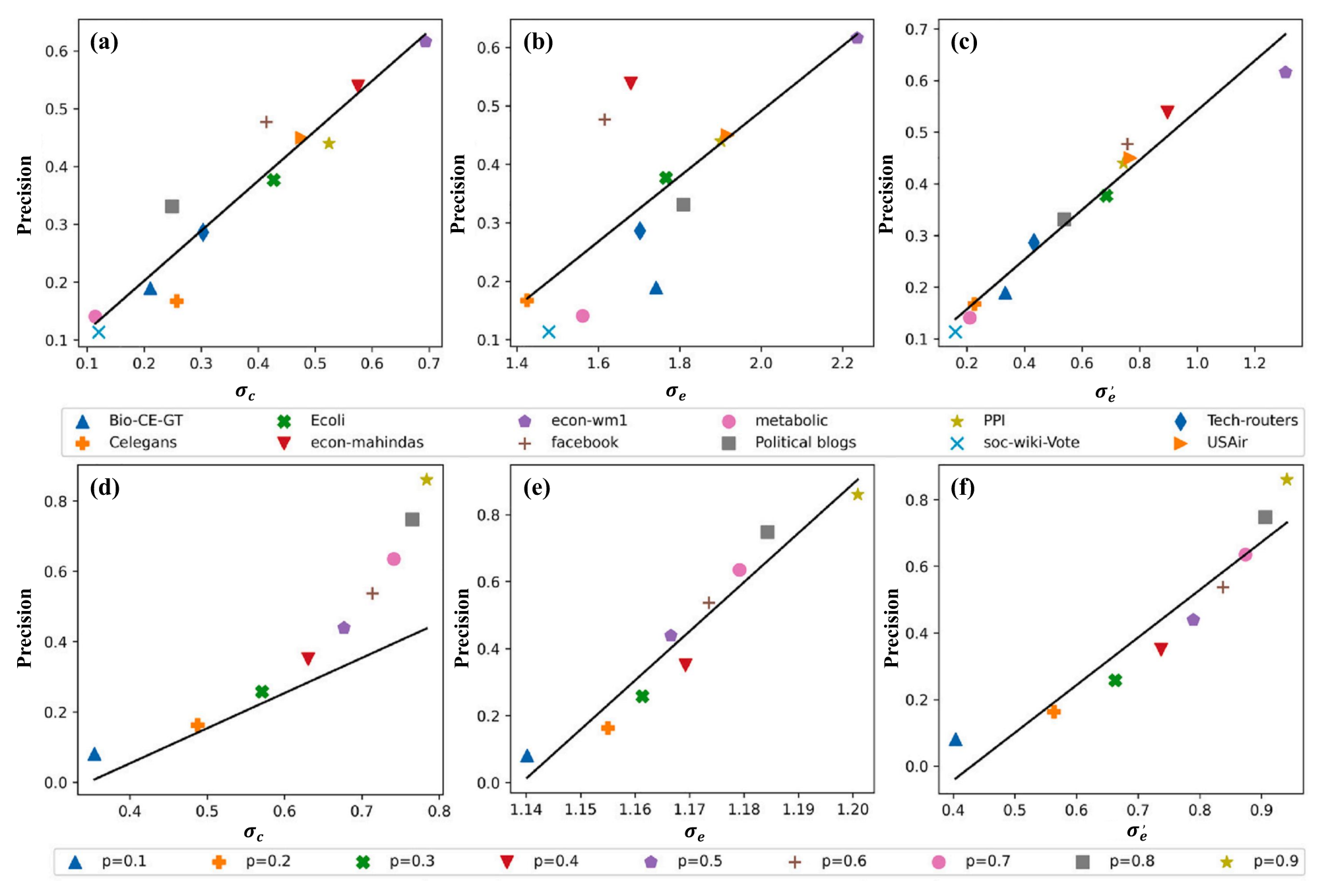} 
	\caption{The relationships between the prediction of real and model networks for $\sigma_c$, $\sigma_e$, $\sigma_e^{'}$. The $y$-axis represents the precision of link-corrected prediction algorithm (LCPA), a method based on random perturbation and structural consistency. The detailed algorithmic procedure is described in Reference \cite{chai2022network}. And the $x$-axis represents the network predictability metric. (a)-(c) show the performance of the $\sigma_c$,  $\sigma_e$, $\sigma_e^{'}$ metrics on real networks, respectively. (d)-(f) show the performance of the $\sigma_c$, $\sigma_e$, $\sigma_e^{'}$ metrics on ER networks, respectively. Each point is the average of 10 independent experimental results, and the solid black line represents the linear fitting curve of the scatter plot. \\
		\textit{Source}: The figure is reproduced from Ref. \cite{chai2022network}.}
	\label{fig:ne_real_model_network} 
\end{figure*}

What is the relationship between different network predictability metrics and the accuracy of link prediction algorithms? As shown in \autoref{fig:ne_real_model_network}, it illustrates the correlation between $\sigma_c$, $\sigma_e$, and $\sigma_e^{'}$ with the performance of link prediction algorithms on ER networks. The results reveal that all three metrics exhibit a linear trend with the prediction accuracy (Precision) based on the link-corrected prediction algorithm (LCPA) \cite{chai2022network}. This correlation is especially evident in artificial networks (e.g., ER networks). However, in some real-world networks, the correlation weakens due to factors such as structural complexity and noise interference. It is important to note that, despite LCPA being a relatively advanced link prediction algorithm, there are certain limitations to directly using it as the "gold standard" for network predictability. Consequently, reliance on the precision output of the model alone to validate the effectiveness of metrics is inadequate. It is recommended that future research incorporate a greater number of benchmark datasets and multi-source validation strategies.

As a comprehensive spectral information metric, network energy incorporates multiple eigenvalues while addressing the limitations of local spectral gap analysis. This metric thereby characterizes the structural features of networks at a higher level. It has been demonstrated that the proposed system enables the identification of highly regular network structures and further opens new avenues for developing a unified cross-network prediction and evaluation framework. Integrating insights from the preceding two subsections (Structural Consistency and Network Spectrum), we observe that spectral analysis methods are gradually shifting from a focus on individual eigenvalues to the examination of eigenvalue distributions, and are further advancing toward modeling global spectral energy. This progression provides a robust mathematical foundation for understanding the structure-function mapping in complex networks, while also laying a critical groundwork for future advancements in link prediction, network reconstruction, and evolutionary modeling.

\subsection{Information-Theoretic Methods}
Since Shannon introduced information theory in 1948 \cite{shannon1948}, it has become an indispensable tool for quantifying uncertainty and modeling complex systems. In recent years, information theory methods have been used to characterize the regularity, redundancy, and evolutionary uncertainty of network structures. This provides a formal, computable, and physically meaningful metric system for network predictability \cite{sole2004}. In contrast to the spectral analysis methods based on the distribution of eigenvalues, the information-theoretic approaches reveal hidden regularities and redundancies within network structures from a more fundamental perspective of probabilistic modeling and coding efficiency. For instance, a highly regular network typically exhibits low information entropy or short compressed representations, indicating that its structure is amenable to modeling and prediction. Conversely, a highly random or heterogeneous network displays higher entropy values or longer compressed lengths, signifying its inherent unpredictability. This section focuses on two representative methods: Compressed Length and Entropy Rate. The former focuses on the structural redundancy while the latter accounts for the temporal evolution uncertainty. These two methods not only enrich our understanding of network predictability, but also provide new theoretical foundations and practical tools for a unified evaluation system across disparate networks.

\subsubsection{Compressed Length}
In the field of information theory, the complexity of a system is often directly quantified as the minimum amount of information necessary for its description. From this perspective, a network can be regarded as a structured information object, the complexity of which can be characterized by the length of information required to efficiently encode its topological structure. If a network exhibits highly regular and repetitive patterns, its structure can be compressed into a shorter binary string. Conversely, if its structure is highly random and devoid of discernible patterns, its encoding length will be long, indicating higher intrinsic uncertainty and unpredictable link evolution. Sun \textit{et al.} \cite{sun2020revealing} proposed a network predictability quantification method based on information theory and data compression techniques. Drawing inspiration from lossless compression concepts in computer science, this approach converts network structures into binary strings and employs efficient compression algorithms to extract latent structural features. This process yields a quantitative metric named \textit{compressed length}, which is both highly interpretable and computationally efficient.

\begin{figure*}[!t]
	\centering
	\includegraphics[width=0.8\textwidth]{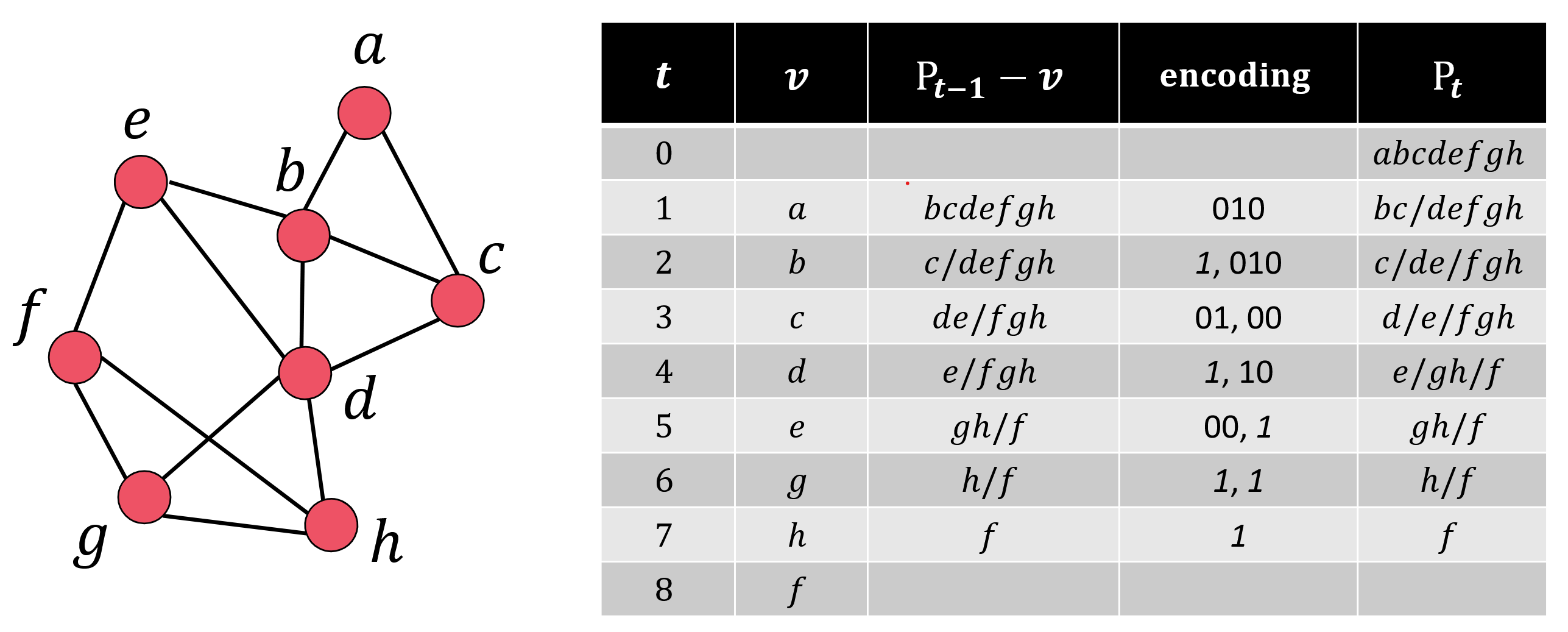} 
	\caption{An example of the topological encoding. Let $v$ be the chosen node to be encoded in the $t$ step and $\bm{P}_{t}$ be the partition of the set of remaining nodes after the $t$ step. For each of subsets of $\bm{U}$ in each line of the third column $\bm{P}_{t-1}-v$, we use $\lceil \log_2 (|\bm{U}|+1) \rceil$ bits to encode the number of neighbors of $v$ in $\bm{U}$, where $\lceil x \rceil$ means that $x$ is rounded up to the nearest integer. At the beginning, we first arbitrarily select a node. For example, as shown in the table on the right, we choose node $a$ as the starting point. We first encode the number of its neighbors in binary form — since $k_a = 2$, and since the element count of the set $\bm{P}_{0}-a$ to which node $a$ belong is $|\bm{U}|=7$, we have $\lceil \log_2 (|\bm{U}|+1) \rceil=3$. Therefore, the binary representation requires $3$ bits, resulting in the binary number "010". Then, based on whether the remaining nodes are neighbors of $a$, we divide them into two groups: $\{b, c\}$ and $\{d, e, f, h, g\}$. At this point, the encoding of node $a$ is completed. Next, we select a new node from the neighbors of $a$, for instance node $b$, and proceed to encode it. Similarly, we store two numbers in binary form: the number of neighbors of $b$ within each of the two previously defined groups. Since node $b$ has $1$ neighbor within the set $\{ b, c\}$ and $2$ neighbors within the set $\{d, e, f, h, g\}$, its corresponding encoded values are "1" and "010", respectively. Then, we further partition the remaining nodes into four groups (excluding empty sets) based on whether they are neighbors of both $a$ and $b$. This completes the encoding of node $b$. Following this procedure, we continue encoding each node in the network one by one until all nodes have been processed. During the encoding process, those of length more than one bit (\textit{i.e.}, $|\bm{U}|>1$) are appended to sequence $\bm{B}_1$, while those of length exactly one bit (\textit{i.e.}, $|\bm{U}|=1$) are appended to sequence $\bm{B}_2$. After $8$ steps, $\bm{B}_1$ and $\bm{B}_2$ are $01001001001000$ and $111111$, respectively. \\
		\textit{Source}: The figure is reproduced from Ref. ~\protect\cite{sun2020revealing}.}
	\label{fig:cl_topological_encoding} 
\end{figure*}

This study builds upon the Structural Zip (SZIP) compression algorithm proposed by Choi \textit{et al.} in 2012 \cite{choi2012}, which encodes network structures as binary strings with minimized information loss. Within this framework, the complexity of a network is naturally reflected in the length of its corresponding binary string: Shorter strings are indicative of simpler, more regular network structures, while longer strings signify greater complexity. \autoref{fig:cl_topological_encoding} illustrates an example of how to encode a network as binary string by SZIP, and one can find with more details and explanations in Ref. \cite{choi2012}.  

As is well known, if a bijective function $f : V_1 \rightarrow V_2$ (where $V_1$ and $V_2$ represent the node sets of the two networks $G_1(V_1,E_1)$ and $G_2(V_2,E_2)$), such that for any two nodes $u, v \in V_1$, the edge $(u,v) \in E_1$ holds if and only if the edge $(f(u), f(v)) \in E_2$ holds, then these two networks are isomorphic, which have exactly the same topology \cite{biggs1986}. SZIP can ensure that from the encoded binary string, a network isomorphic to the original network can be obtained \cite{choi2012}. To further eliminate redundant information and correlations within the structure, we leverage dependencies between symbols in binary characters \cite{skretting1999} to compress the original binary representations $\bm{B}_1$ and $\bm{B}_2$ (see \autoref{fig:cl_topological_encoding}), yielding the compressed representations $\widehat{\bm{B}}_1$ and $\widehat{\bm{B}}_2$. The compression length is then defined as:
\begin{equation}
	\sigma_l=|\widehat{\bm{B}}_1|+|\widehat{\bm{B}}_2|.
\end{equation}

\begin{figure*}[!t]
	\centering
	\includegraphics[width=1\textwidth]{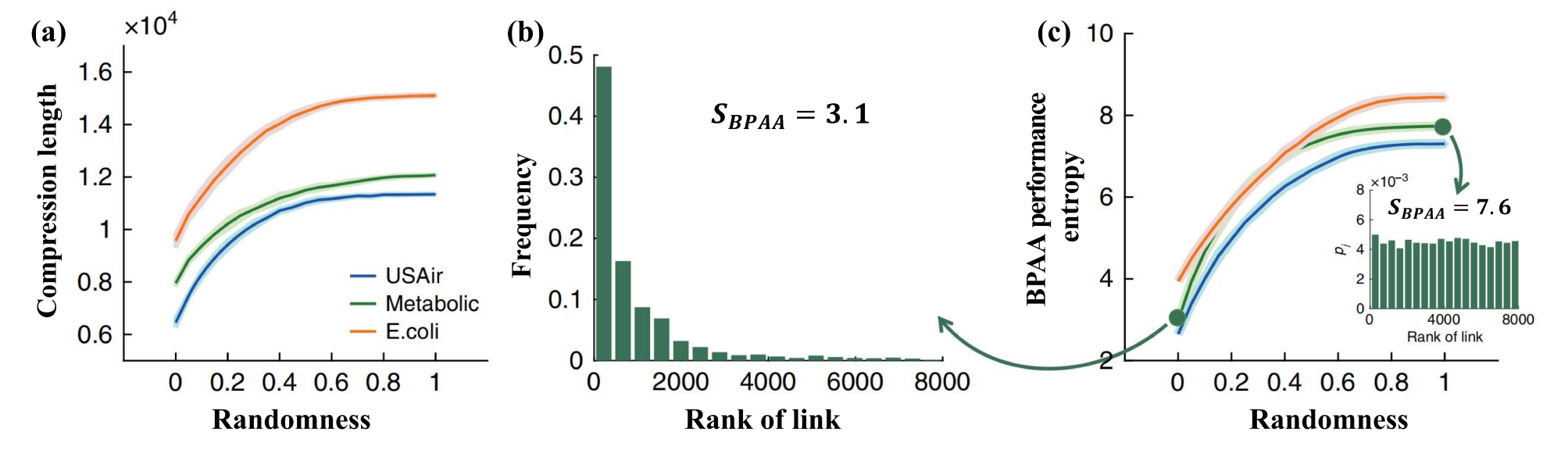} 
	\caption{The process of network compression. (a) The shortest compression length of the shuffled networks. Here the shuffling operation is that we randomly pick Randomness fraction of links from the original network and rewire them randomly. The compression length increases monotonically with Randomness until $Randomness = 1$, when the network has the same property to the corresponding ER network. (b) Distribution of $r_i$ of Metabolic network \cite{jeong2000}. Here we use RA index \cite{zhou2009} to do the prediction and each bar has the width equal to the network size $N$. (c) The BPAA performance entropy $S_{BPAA}$ vs. the link shuffling fraction Randomness. We can see a monotonic trend similar to (a) is present for all of the three networks. Each line is the average of $50$ simulations and the shaded region denotes the standard deviation. (Inset) Distribution of $r_i$ of ER network with the same number of nodes and links as Metabolic network. \\
		\textit{Source}: The figure is reproduced from Ref. \cite{sun2020revealing}.}
	\label{fig:cl_compression_length} 
\end{figure*}

As shown in \autoref{fig:cl_compression_length}(a), by randomly reconnecting some edges in real networks, it can be observed that the higher the probability of random reconnection, the greater the compression length. This demonstrates that the compression length is able to capture the change of randomness. However, the compression length is influenced not only by the inherent randomness of the network itself, but also by its size. It has been demonstrated that networks with more nodes and edges have statistically longer compression lengths, compared to small networks under the similar level of randomness. To eliminate the influence of network size, we normalized $\sigma_l$ by the maximum compression length $R$
\begin{equation}
	\sigma_{l^{*}}=\frac{\sigma_l}{R},
	\label{eq:l_normalized}
\end{equation}
where $R = \binom{N}{2}h(w) - N \log N$, $w = \frac{E}{\binom{N}{2}}$ denotes the prior probability of a link being formed between any pair of nodes, and $h(w) = -w \log w - (1-w) \log(1-w)$ represents the binary entropy. All instances of $\log$ in this section denote logarithms to base 2. According to Shannon's Source Coding Theorem \cite{cover2001}, the smaller the value of $\sigma_{l^{*}}$, the more regular the network structure, and the higher the predictability of the corresponding network. Conversely, a larger value of $\sigma_{l^{*}}$ is indicative of a more random network structure and lower predictability.

In the domain of network science, predictability is conventionally employed to denote the capacity to precisely anticipate future alterations in network configuration, based on historical data. Since the dynamics of real-world networks are frequently unknown and thus cannot be utilised to evaluate the performance of algorithms, Sun \textit{et al.} have devised a methodology for approximating network predictability, employing the "Best Prediction Algorithm Available (BPAA)" approach, that is to say, to direct estimate the upper bound of prediction accuracy by the performance of the best-performed algorithm among a huge number of state-of-the-art algorithms \cite{ghasemian2020stacking,muscoloni2023stealing}. Specifically, the Leave-One-Out Approach \cite{shirer2012, wong2015} is employed to test each edge: Remove an edge $e_i$, compute the connection probability of the remaining node pairs using a specific algorithm, rank them by score, and record the rank $r_i$ of $e_i$. The set of all edge rankings $\bm{D} = \left\{ r_1, r_2, \dots, r_M \right\}$ forms a distribution, and the entropy of distribution $\bm{D}$ serves as a good overall metric for algorithm performance. For example, in a highly predictable network, an ideal algorithm would yield $\bm{D} = \left\{ r_1 = r_2 = \dots = r_M = 1 \right\}$, resulting in the lowest distribution entropy. Conversely, for a low-predictability network, even an ideal algorithm will yield vastly different values in $\bm{D}$, resulting in higher distribution entropy. When calculating the performance entropy $S$ of this algorithm, the values of $r_i$ can vary within the range $ 1 \leq r_{i} \leq (\frac{N(N-1)}{2}-\frac{\langle k\rangle N}{2} + 1) \approx \frac{N^{2}}{2}$, where $\frac{N(N-1)}{2}-\frac{\langle k\rangle N}{2} + 1$ is the total number of unconnected node pairs and $\langle k \rangle$ is the network's average degree. Consequently, this range can be subdivided into $N$ equal-width intervals, thereby negating the impact of network size $N$ on the outcomes. The Shannon entropy $S$ can be calculated based on the probability distribution of $\frac{N}{2}$ intervals as follows: 
\begin{equation}
	S = - \sum_{j=1}^{\frac{N}{2}}p_{j} \log p_{j}, 
\end{equation}
where $p_j$ is the probability of $r_i$ occurring in interval $j$.

On this basis, the minimum entropy value achieved by the best-performing mainstream link prediction algorithm is defined as the optimal prediction algorithm entropy $S_{BPAA}$ for the network. To further eliminate the influence of network size $N$ and average degree $\langle k \rangle$, the following is obtained after additional normalization:
\begin{equation}
	S_{BPAA}^{*} = \frac{S_{BPAA}}{\log N -1}.
\end{equation}
This approach preserves the original network structure to the greatest extent possible while maintaining its internal predictability. However, when performing specific computations, the computational complexity tends to be high if the network is particularly large (e.g., when $N$ takes a large value) or extremely dense.

\begin{figure}[htbp]
	\centering
	\includegraphics[width=0.70\textwidth]{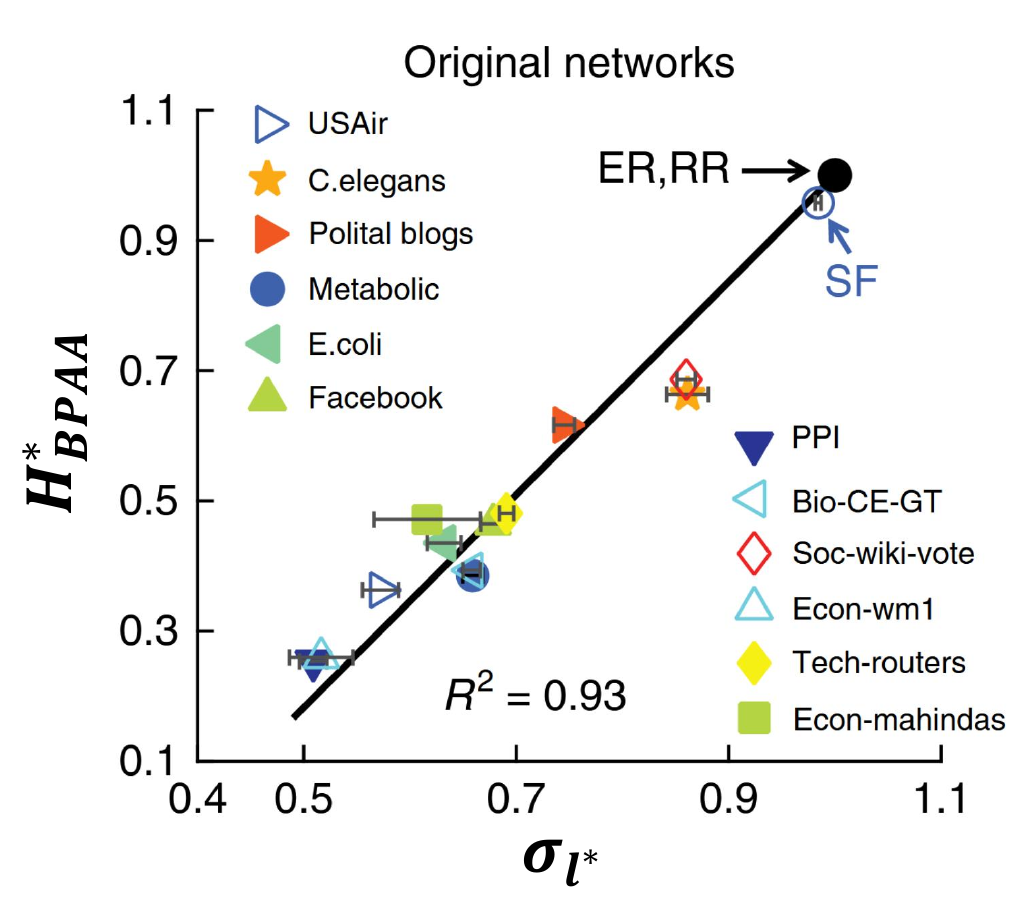} 
	\caption{For empirical networks drawn from different fields, the values $(\sigma_{l^{*}}, S_{BPAA}^{*})$ fall on the linear relationship in the black straight line given by Eq. (\ref{eq:sigma_l*}). The horizontal error bars indicate the standard deviation of the compression length across $50$ calculations. The standard deviation of the slope is $0.06$ obtained by the bootstrap method and the coefficient of determination $R^2$ is $0.93$. \\
		\textit{Source}: The figure is reproduced from Ref.  \cite{sun2020revealing}.}
	\label{fig:CL_Equement} 
\end{figure}

Sun \textit{et al.} \cite{sun2020revealing} discovered a significant linear relationship between the normalization $\sigma_{l^*}$ and the entropy $S_{BPAA}^*$ of the best prediction algorithm across multiple artificial and real-world networks:
\begin{equation}
	S_{BPAA}^* \approx 1.63 \sigma_{l^*}-0.63.
	\label{eq:sigma_l*}
\end{equation}
This relationship indicates that the compressed length predictability $\sigma_{l^*}$ exhibits a positive linear relationship with the approximated network predictability $S_{BPAA}^*$, and the high value $R^2=0.93$ demonstrates that this linear relationship is exceptionally strong (as shown in \autoref{fig:CL_Equement}). This implies that when $\sigma_{l^*}$ is small, the corresponding $S_{BPAA}^*$ is also small, indicating higher network predictability. Consequently, the upper bound of network predictability can be estimated through structural compression alone, obviating the necessity for extensive link prediction algorithms. Furthermore, Sun \textit{et al.} showed that the linear relationship remains consistent even when network edges are randomly perturbed, added, or removed (for further details, see \cite{sun2020revealing}). Nevertheless, this method still has several limitations. Firstly, in large-scale sparse or dense networks, the computational complexity of compression and entropy calculation is relatively high. Furthermore, the validation of $\sigma_l$ is checked by some selected link prediction algorithms, so that there is still no ground truth to support the main conclusion. Future research should concentrate on the development of more efficient compression strategies, the enhancement of the method's robustness, and the exploration of its extension to multi-layer and multi-relational networks.

\subsubsection{Entropy Rate}
In the real world, many complex systems not only possess non-trivial topological structures but also exhibit interaction patterns that evolve over time. Such systems are commonly referred to as temporal networks, where the connections between nodes are no longer static but exhibit significant temporal dependence \cite{holme2012, masuda2013, scholtes2014, li2017}. In comparison with the predictability analysis of static networks or single time series, the evolutionary mechanisms of temporal networks are more complex and challenging. Quantifying their dynamic uncertainty from an information-theoretic perspective and revealing their underlying prediction upper bounds have become key research directions in recent years.

\begin{figure}[h]
	\centering
	\includegraphics[width=0.95\textwidth]{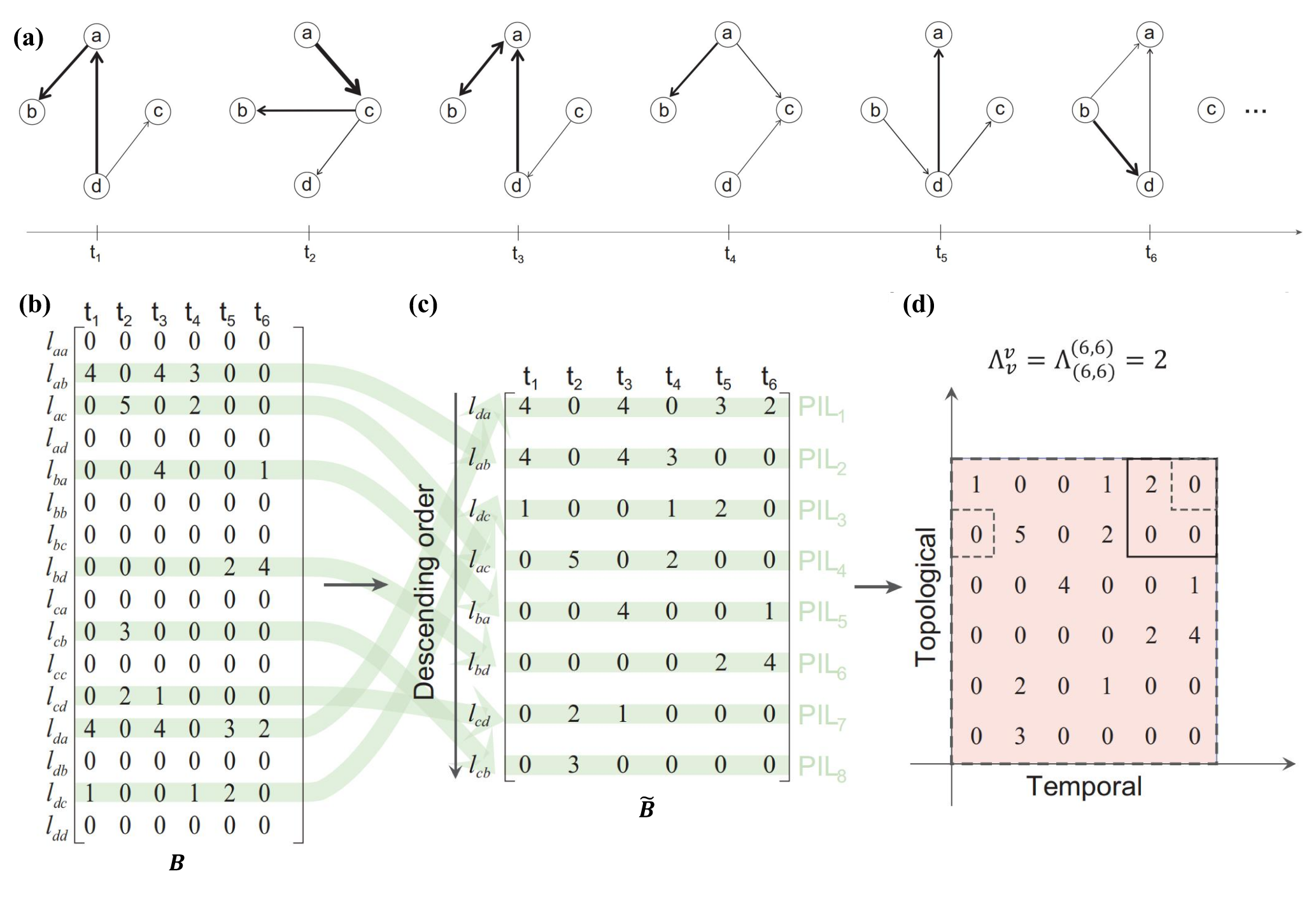} 
	\caption{Quantifying the predictability of a temporal network. (a) The time-unfolded representation of a temporal network with four nodes. Each snapshot is a weighted directed network where the thickness of links represents their weights. (b)-(c) Matrix $B$ encodes the time evolution of each potential link, where each column embodies the structure of a snapshot. Links that rarely appear within the whole duration are removed from the matrix, resulting in matrix $\widetilde{B}$, which captures the meaningful part of $B$. The rows of $\widetilde{B}$ are sorted into descending order according to the number of occurrences. (d) Calculation of $\Lambda_v^{v}$ for a part of $\widetilde{B}$. Note that $\widetilde{B}_{C(k)}$ is defined as a $2D$ square with side $k$, where $C(k)=\{v=(t, s)\in \bm{Z}^{2}: 0 \leq t \leq k, 0 \leq s \leq k\}$ denotes the set of coordinates for elements in $\widetilde{B}$. The quantity $\Lambda_{v}^{v}$ is given by $\inf \{k \geq 1 \mid \widetilde{B}_{u-C(k)} \neq \widetilde{B}_{v-C(k)},\ \forall u \in [0, v], u \neq v\}$, which means that finding the smallest integer $k$ such that the submatrix $\widetilde{B}_{v-C(k)}$ starting at $v$ in $\widetilde{B}$ is unique within the rectangle $[0, v]$. For example, to calculate $\Lambda_{(6,6)}^{(6,6)}$: When $k=1$, then $C(1)=\{(0,0), (0,1), (1,0), (1,1) \}$ and $\widetilde{B}_{(6,6)-C(1)} = \left[\begin{smallmatrix} 0 & 0 \\ 0 & 0 \end{smallmatrix}\right]$. Since there exists another submatrix within the rectangle $[0, (6,6)]$ of $\widetilde{B}$ that is identical to $\widetilde{B}_{(6,6)-C(1)}$, we need to consider larger values of $k$. When $k=2$, then  $\widetilde{B}_{(6,6)-C(2)} = \left[\begin{smallmatrix} 0 & 2 & 4 \\ 1 & 0 & 0 \\ 0 & 0 & 0 \end{smallmatrix}\right]$. This submatrix does not appear elsewhere within the rectangle $ [0, (6,6)]$, so $\Lambda_{(6,6)}^{(6,6)}$ is $2$. \\
		\textit{Source}: The figure is reproduced from Ref.  \cite{tang2020predictability}.}
	\label{fig:er_example} 
\end{figure}

Tang \textit{et al.} \cite{tang2020predictability} proposed a unified framework for the estimation of the upper bound of predictability for real-world temporal networks. The fundamental concept underpinning this framework is to treat the temporal network as a joint stochastic process, with the entropy rate of its future states serving as a metric for its unpredictability. A temporal network comprising $T$ time snapshots can be represented as a two-dimensional adjacency matrix $\bm{B} \in \{ 0,1 \}^{N^2 \times T}$, where each column corresponds to the adjacency matrix of a single time snapshot, and each row represents the state of a potential edge $e_{ij}$ across all time snapshots (whether it exists or not, or its weight). Since most real-world networks are sparse—meaning the vast majority of node pairs are unconnected—this matrix is typically highly sparse. In order to enhance the accuracy of subsequent entropy rate estimation, it is necessary to filter the original matrix. Specifically, it firstly calculates the frequency of occurrence (activation rate) of each edge across the entire time range, and then sorts all edges in descending order by activation rate and retain the most active portion. The activation rates (in descending order) of all edges are denoted by $a_1, a_2, \dots, a_{N^2}$. The filtered matrix $\bm{\widetilde{B}}$ is defined as:
\begin{equation}
	\bm{\widetilde{B}}=\begin{cases}\left\{\ell_{1},\,\ell_{2},\ldots,\,\ell_{b}\right\}|b=\inf\left\{b\in\left(1,2,\ldots,N^2\right)|\sum\limits_{i=1}^{b}a_{i}\geq 0.6 \sum\limits_{i=1}^{N^2}a_{i}\right\},\,\,b<b_{\theta}\\ \left\{\ell_{1},\,\ell_{2},\ldots,\,\ell_{b}\right\}|b=\inf \left\{ b \in \left(1,2,\ldots,N^2\right)|a_{b}\geq 0.1 \right\},b \geq b_{\theta}
	\end{cases}
\end{equation}
Here, $b_{\theta}$ denotes the set threshold (that is set to be 1000 in \cite{tang2020predictability}) to ensure sufficient active edges are retained while mitigating noise interference. The processed matrix thus captures the primary evolutionary characteristics of the original network while effectively reducing the impact of dimensionality and sparsity. As demonstrated in \autoref{fig:er_example}(a)-(d) provide a concrete illustration.

It is possible to characterize the dynamic uncertainty of the filtered matrix $\bm{\widetilde{B}}$ by means of the entropy rate $S_R(\bm{\widetilde{B}})$. The entropy rate is a rigorous metric for measuring the average uncertainty per unit time in a stochastic process, defined as follows:

\begin{equation}
	S_R(\bm{\widetilde{B}})=\frac{N^{2}\log N^{2}}{\sum_{v\in C(N)}(\Lambda_{v}^{v})^{2}},
	\label{eq:entropy_rate}
\end{equation}
where $C(N)$ denotes the set of all positions whose coordinates do not exceed $N$. The symbol $\Lambda^{v}_v$ denotes the maximum window size required for a new pattern to first appear at position $v$. This formula originates from the generalized Lempel-Ziv compression algorithm \cite{kontoyiannis2002nonparametric}, which approximates the true entropy rate by quantifying the frequency of local patterns in the statistical matrix. Once the entropy rate $S_R(\bm{\widetilde{B}})$ has been obtained, the upper bound on the network's predictability, denoted by $\Pi_{\bm{\widetilde{B}}}^{max}$, can be further determined. This is achieved by satisfying the following equation:
\begin{equation}
	S_{R}(\bm{\widetilde{B}}) = 
	-\left[ \Pi_{\bm{\widetilde{B}}}^{\max} \log \Pi_{\bm{\widetilde{B}}}^{\max} 
	+ (1 - \Pi_{\bm{\widetilde{B}}}^{\max}) \log (1 - \Pi_{\bm{\widetilde{B}}}^{\max}) \right] 
	+ (1 - \Pi_{\bm{\widetilde{B}}}^{\max}) \log(s - 1),
\end{equation}
where $s$ denotes the number of distinct values in matrix $\bm{\widetilde{B}}$. The upper bound $\Pi_{\bm{\widetilde{B}}}^{\max}$ is a theoretical upper limit on the achievable prediction accuracy for any algorithm. This is referred to as $\sigma_t$ and is used as the primary metric for evaluating the predictability of temporal networks. It is important to note that the entropy rate obtained using the generalized Lempel-Ziv algorithm serves as a progressive measure of randomness, and Eq. (\ref{eq:entropy_rate}) becomes more accurate as the number of time snapshots increases. Therefore, when finite data snapshots from a real-world time-series network are considered, the computed value of $\Pi_{\bm{\widetilde{B}}}^{\max}$ should be regarded as an asymptotic estimate of the upper bound on predictability.

\begin{figure}[htbp]
	\centering			
	\includegraphics[width=0.75\textwidth]{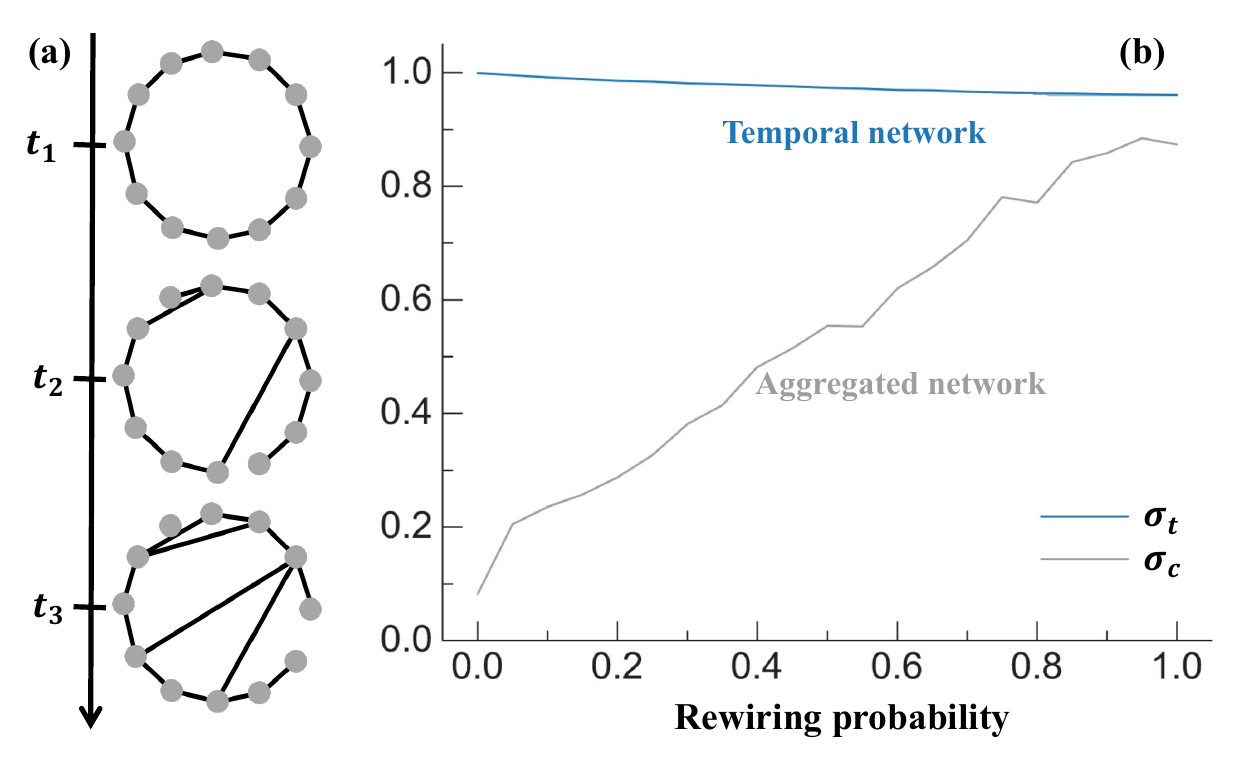} 
	\caption{(a) The synthetic temporal networks. To investigate the impact of network topology on predictability, an evolving small-world network model is employed. The first snapshot is a ring network; subsequent topologies of the network are generated by randomly rewiring a fraction $p$ of links in the previous snapshot. (b) Predictability of synthetic temporal networks. Predictabilities of evolving small-world networks against rewiring probability. The networks are generated by the model in \autoref{fig:er_model} with $50$ nodes and average degree $2$. Structural consistency is an existing predictability measure, which we primarily introduce in the "Structural Consistency" section \cite{lvzhou2015}. \\
		\textit{Source}: The figure is reproduced from Ref.  ~\protect\cite{tang2020predictability}.}
	\label{fig:er_model} 
\end{figure}

In order to validate the effectiveness of this method, Tang \textit{et al.} \cite{tang2020predictability} conducted systematic tests on both artificial temporal networks and multiple real temporal networks. As demonstrated in \autoref{fig:er_model}, increasing network randomness (for example, controlled by the random reconnection ratio $p$) leads to a significant decrease in the proposed predictability metric $\sigma_t$. This indicates that the network becomes more difficult to predict. When subjected to identical conditions, the traditional predictability metric for aggregated static networks (structural consistence, denoted by $\sigma_t$) exhibits an upward trend. This finding suggests that disregarding temporal characteristics may result in misleading conclusions.

This section reviews two representative information-theoretic methods, Compressed Length and Entropy Rate. \textbf{Compressed Length} adopts a structural coding perspective, treating the network as a compressible information entity. It employs efficient lossless compression algorithms to extract latent repetitive patterns and structural properties. The normalized compressed length $L^{*}$ not only quantifies the network's randomness level but also exhibits a significant empirical linear relationship with link prediction performance, providing an efficient and practical tool for assessing predictability in large-scale networks. Nevertheless, this method is still confronted with challenges relating to high computational complexity and inadequate robustness when dealing with highly sparse or extremely dense networks. Conversely, \textbf{Entropy Rate} extends the research perspective to the realm of dynamic networks. By modeling the entire temporal evolution as a joint random process and incorporating Lempel-Ziv compression principles to estimate average uncertainty per unit time, it reveals the predictive limits of temporal networks for the first time, and provides a robust theoretical foundation for modeling the predictability of temporal networks and develops a practical set of quantitative tools. The method has been demonstrated to effectively identify non-equilibrium states, sudden changes, and long-range dependencies in network evolution. An obvious limitation of the entropy rate approach is that it builds on the assumption of infinite data length, which has the potential risk of biases arising from limited observations. For both methods, the extensions to heterogeneous networks, multilayer networks, spatial networks and higher-order networks are still challenging.

\subsection{Structural Methods}
This subsection will explore two interpretable predictability models by examining network organizational principles and dynamic control mechanisms from two complementary perspectives: Structural Regularity and Structural Controllability. Both methods are underpinned by a fundamental concept: the inherent consistency of network structure and functional constraints significantly impacts the predictability of its future state.

\subsubsection{Structural Regularity}
A fundamental attribute of complex networks is the pervasive presence of local repeating subgraph patterns in their topology \cite{milo2002network,alon2007network}. It is evident that these patterns not only reflect the organizing principles of networks but also directly influence the modelability and predictive potential of their evolutionary trajectories \cite{zhang2013potential,liu2019link,he2024uncovering}. If a network consists of numerous similar substructures, the future evolution of its connections may exhibit high regularity, thereby increasing the predictability of link changes. Conversely, in cases where the structure is characterized by significant heterogeneity and an absence of recurring patterns, the future state becomes challenging to accurately predict.

Xian \textit{et al.} \cite{xian2020} proposed a method based on low-rank and sparse representation to characterize the repetition of local patterns within network structures and further quantify their overall predictability levels. The fundamental principle of this method is to linearly represent the network adjacency matrix by itself (see the detailed persepctive of linear representation in Ref. \cite{pech2019}), say 
\begin{equation}
	\bm{A}=\bm{A}\bm{Z}+\bm{E},
\end{equation}
where $\bm{Z} \in \mathbb{R}^{N \times N}$ is the representation matrix and $\bm{E}$ is the error term. Each row $\bm{A}_{i,:}$ of matrix $\bm{A}$ can be regarded as a local subgraph structure, while the $j$th column of $\bm{Z}$ describes how this subgraph is formed as a linear combination of other subgraphs. Thus, the representation matrix $\bm{Z}$ reflects the dependencies between different subgraphs in the network and their representativeness.

Real-world networks frequently demonstrate structural regularity, with local topological patterns exhibiting repetition and similarity. This regularity implies that networks can be efficiently represented by a small number of fundamental structures, that it to say, $\bm{Z}$ should be sparse and of low rank. The low rank emerges from the reduction of representational redundancy due to the presence of recurring sub-structures, while sparsity is characterized by the minimisation of non-zero parameters required. Together, Xian \textit{et al.} proposed the following optimization problem:
\begin{equation}
	\min_{\bm{Z},\bm{E}} \text{rank}(\bm{Z}) + \alpha \| \bm{Z} \|_{0} + \beta \| \bm{E} \|_{2,1} \quad  \textit{s.t.} \quad \bm{A}=\bm{A}\bm{Z}+\bm{E}.
\end{equation}
Here, $\text{rank}(\bm{Z})$ denotes the rank of matrix $\bm{Z}$, $\| \bm{Z} \|_{0}$ represents the zero-norm of matrix $\bm{Z}$ (i.e., the number of non-zero elements), and $\| \bm{E} \|_{2,1}= \sum_{j=1}^{N} \sqrt{\sum_{i=1}^{N}e_{ij}^2}$. The tunable parameters, $\alpha$ and $\beta$, are trade-off parameters.

Based on the aforementioned model, the predictability of the network can be measured during the optimization process by defining a structural regularity index, denoted by $\sigma_r$:
\begin{equation}
	\sigma_{r}=\frac{1}{\sqrt{(N-r)/N}\sqrt{\tau/(N\cdot r)}},
\end{equation}
where $r$ denotes the rank of learned representation matrix $\bm{Z}^{*}$, $\tau$ represents the number of zero elements in $\bm{Z}^{*}$, $(N-r)/N$ indicates the proportion of subgraphs in the network that can be linearly represented by other subgraphs, and $\tau/(N\cdot r)$ characterizes the density of zero elements in the simplified ladder form of matrix $\bm{Z}^{*}$ in its reduced staircase form. 

\begin{figure}[htbp]
	\centering
	\includegraphics[width=1\textwidth]{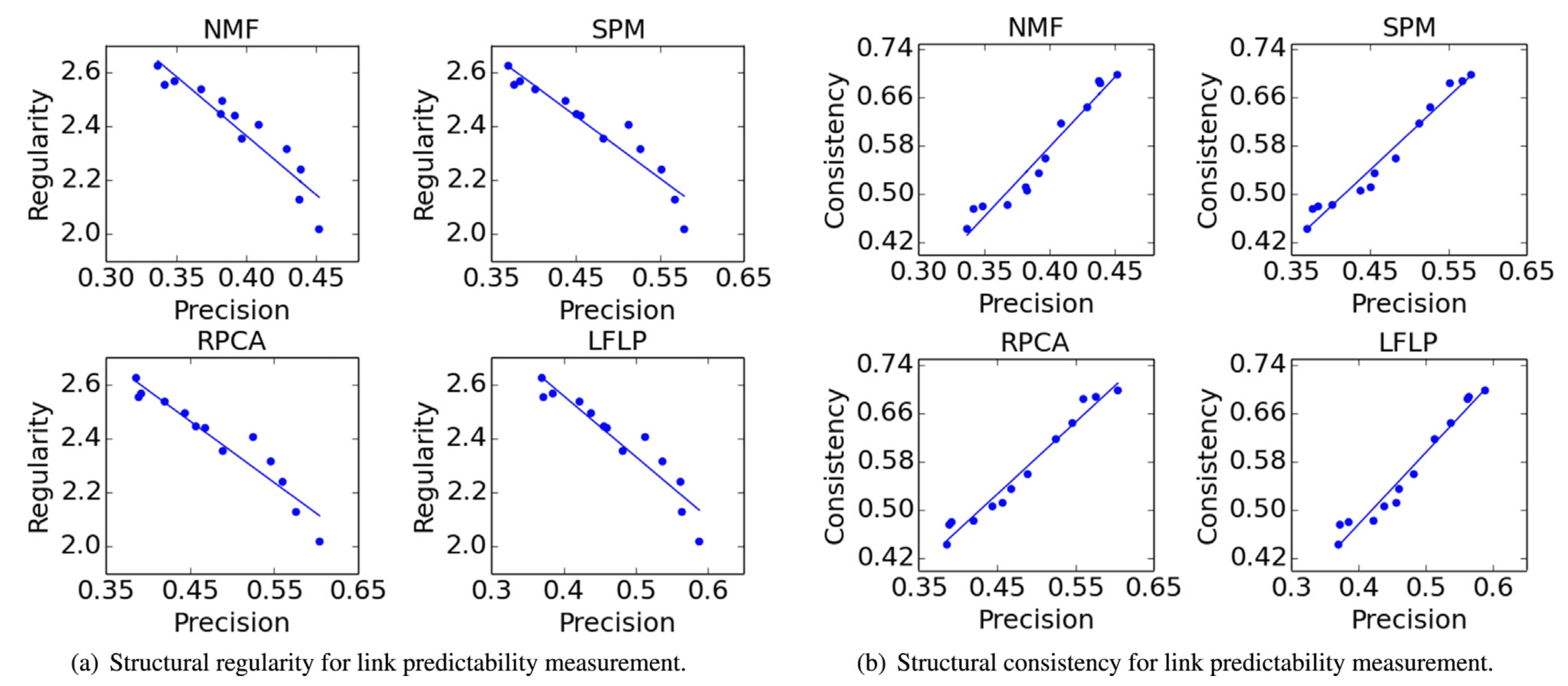} 
	\caption{Scatter plot between link predictability and prediction precision of Jazz network. For (a), the $y$-axis indicates the structural regularity $\sigma_r$ of the network, for (b), the $y$-axis indicates the network consistency $\sigma_c$ and the $x$-axis indicates the precision of link prediction methods in networks perturbed with varied percentages. The solid lines indicate the linear fittings of the results. The smaller the values of structural regularity, the more regular are the networks. \\
		\textit{Source}: The figure is reproduced from Ref. \cite{xian2020}.}
	\label{fig:sr_compare} 
\end{figure}

This metric comprehensively considers two aspects of the network: the proportion of repeated subgraphs and the sparsity of the representation matrix. These two aspects reflect the degree of regularity in the network's overall structure. The experimental results demonstrate a significant negative correlation between this metric and link prediction accuracy. In other words, the presence of more regular structures is associated with higher prediction performance, which corresponds to smaller $\sigma_r$ as illustrated in \autoref{fig:sr_compare}. In addition, the identification of the most representative subgraph structures can be achieved through the sorting of the rows of the representation matrix, denoted by $\bm{Z}^*$ based on the $L_1$ norm. These subgraphs function as "basis functions" during network evolution, constituting the fundamental building blocks of the network.

\subsubsection{Structural Controllability}
In network science, structural controllability serves as a crucial theoretical framework for understanding a system's dynamic control capabilities. It focuses on achieving complete control over the entire network state using the minimum number of externally driven nodes \cite{liu2011, yuan2013, gao2014, yan2017}. Although research in this area primarily focuses on identifying critical nodes or edges, its core objective—revealing dependencies in functional pathways—also offers new insights into predictability during network evolution. Traditionally, structural controllability and structural predictability have been regarded as distinct research directions. The former has sought to identify the key components influencing system evolution paths \cite{lin1974}. In contrast, the latter has focused on elucidating the upper bounds of connectivity possibilities between arbitrary nodes. However, Jing \textit{et al.} \cite{jing2022toward} demonstrated that these two concepts are not mutually exclusive; instead, they are intrinsically linked through a special class of edges--critical links.

In complex networks, driver nodes typically refer to the minimal set of nodes requiring external control inputs to achieve complete controllability of the network system \cite{liu2011}. In directed networks, if all unmatched nodes are designated as driver nodes and complete control over the entire network can be achieved, the system is considered structurally controllable \cite{lin1974}. Here, unmatched nodes denote those that did not become endpoints of any matched edges during the network's maximum matching process, where the objective is to select a set of edges that includes as many edges as possible, with no two edges sharing the same node. Those unmatched nodes cannot be indirectly controlled through internal network dynamics and thus require direct control via external inputs, forming a crucial component of the minimal driver set. Building on this, if the removal of a particular edge increases the number of required driver nodes, that edge is termed a \textit{critical link}; otherwise, it is classified as a trivial link \cite{poor1994}. It is imperative to acknowledge the pivotal function of critical links in preserving network controllability, whilst simultaneously underscoring their profound influence on network topology stability. Typically situated at the core of the network, these elements are responsible for the transmission of information and the provision of structural support, thereby exerting a substantial influence on the overall robustness and modelability of the structure.

To quantify the impact of key edges on network predictability, Jing \textit{et al.} \cite{jing2022toward} utilized the leave-one-out method, a technique commonly employed in link prediction. Specifically, each edge is sequentially removed from the training set; the link prediction algorithm computes its score $s_j$; and all edges are then ranked in ascending order based on these scores, forming a sequence of node pairs $\{ e_1, e_2, \dots, e_{\mathcal{L}} \}$ satisfying $s_1 \leq s_2 \leq \dots \leq s_{\mathcal{L}}$, where $1 \leq j \leq \mathcal{L}$ and $\mathcal{L} = \binom{N}{2}$. Link prediction algorithms that are contingent upon local topology (for example, CN, PA) demonstrate that the removal of any edge $e_j$ has a relatively minor impact on a small number of node pair scores. Therefore, the normalized ranking change of edge $e_j$ in the sequence is as follows:
\begin{equation}
	\delta_j=\lim_{N \rightarrow \infty}\frac{\mathcal{O}(N)}{\mathcal{L}}=0.
\end{equation}
Subsequently, for edge $e_j$, the predictability is defined as $P_j = \frac{j}{\mathcal{L}}$. Finally, by calculating the mean of the $P_j$ values across all edges, they obtained the structural predictability $\sigma_{d}$ of the entire network, as 
\begin{equation}
	\sigma_{d}=\frac{1}{M}\sum_{j=1}^{M}P_{j}.
\end{equation}

Furthermore,  Jing \textit{et al.} \cite{jing2022toward} proposed the Structural Reciprocity Index (SRI) to measure the relative predictability performance of critical links versus trivial links. It is defined as follows:
\begin{equation}
	SRI=\frac{u_1+0.5u_2}{u},
\end{equation}
where $u$ denotes the total number of comparisons (in each comparison, one critical link and one trivial link are randomly sampled), and $u_1$ and $u_2$ represent the number of comparisons where the predictability of critical links is less than and equal to that of trivial links, respectively. Indeed, SRI is the same to AUC, and a higher SRI indicates stronger interaction between structural controllability and predictability for critical links. Experimental results demonstrate that across multiple real-world and synthetic networks (as shown in \autoref{tab:scon_real}), the predictability of critical links is generally lower than that of trivial links, revealing a negative regulatory effect of structural controllability on predictability.

It has become increasingly evident that a network's modelability is contingent not solely on its statistical characteristics or information entropy levels, but is more profoundly rooted in its structural organization principles and functional control mechanisms. This section introduces two classes of predictability modeling approaches based on structural properties: structural regularity and structural controllability. These methods establish a new framework for comprehending the predictability of network evolution by addressing two dimensions: local pattern repetition and global functional path constraints. The \textbf{Structural Regularity} approach, utilising low-rank sparse representation models, elucidates the intricate relationship between the recurrence of local subgraphs within a network and its inherent predictability. This method not only provides a theoretical foundation for tasks such as link prediction, but also offers mathematical tools for identifying representative structures within networks. The \textbf{Structural Controllability} approach highlights that critical links, while determining a network's controllability, also significantly influence the predictability of its evolutionary paths. It is important to note that SRI is the first metric to quantify the coupling relationship between controllability and predictability. Despite their distinct origins, both approaches are fundamentally concerned with the same core principle, namely that predictability of a network fundamentally derives from the interplay between structural coherence and functional constraints. Collectively, these elements serve to expand the boundaries of our understanding regarding network predictability.

\begin{table}[!t]
	\centering
	\caption{\texorpdfstring{
			Basic topological statistics of eight real networks and ER networks. 
			$N$ and $M$, respectively, represent the number of nodes and links, 
			where (\textbullet) is the number of critical links. 
			$s = \frac{2M}{N(N - 1)}$ is the data sparsity. 
			$\langle k \rangle$ is the average degree. 
			$C$ is the clustering coefficient. 
			$\langle d \rangle$ represents the average distance over all node pairs in weakly connected networks, 
			where ``--'' indicates that the network is not connected. 
			$r$ is the assortativity coefficient. 
			$SRI$ is the value of the structural reciprocity index. \\
			\textit{Source}: The table is reproduced from Ref. \cite{jing2022toward}.
		}{
			Basic topological statistics of eight real networks and ER networks. 
			N and M represent the number of nodes and links. 
			s = 2M/(N(N-1)) is data sparsity. 
			<k> is average degree. 
			C is clustering coefficient. 
			<d> is average distance (``--'' if not connected). 
			r is assortativity. 
			SRI is the structural reciprocity index. 
			Source: Table reproduced from Ref. [jing2022toward].
	}}
	
	\label{tab:scon_real}
	\begin{tabular}{l *{8}{c}}
		\toprule
		\textbf{Networks} & \textbf{$N$} & \textbf{$M$(\textbullet)} & \textbf{$s$} & \textbf{$\langle k \rangle$} & \textbf{$C$} & \textbf{$\langle d \rangle$} & \textbf{$r$} & \textbf{$SRI$} \\
		\midrule
		FWEW & 69 & 916 (1) & 19.52\% & 13.28 & 0.30 & 1.89 & -0.40 & 0.81 \\
		FWMW & 97 & 1492 (1) & 16.02\% & 15.38 & 0.26 & 1.94 & -0.30 & 0.45 \\
		FWWF & 128 & 2137 (3) & 13.15\% & 16.70 & 0.17 & 1.95 & -0.23 & 0.83 \\
		Terrorist & 260 & 571 (170) & 8.48\% & 2.19 & 0.31 & -- & 0.50 & 0.74 \\
		Delicious & 300 & 957 (30) & 10.67\% & 3.19 & 0.29 & 0.53 & -0.45 & 0.72 \\
		Kohonen & 185 & 443 (36) & 13.01\% & 2.39 & 0.15 & 0.11 & -0.17 & 0.80 \\
		SciMet & 179 & 280 (28) & 8.79\% & 1.56 & 0.06 & 0.02 & -0.30 & 0.71 \\
		Guava & 620 & 1078 (51) & 2.81\% & 1.74 & 0.12 & -- & 0.14 & 0.61 \\
		\midrule
		\multirow{10}{*}{ER Networks}  
		& --  & 1002 (321) & 0.10\% & 1.00 & 0.0006 & -0.0065 & -- & 0.60 \\
		& --  & 1983 (357) & 0.20\% & 2.00 & 0.0019 & -- & -0.0032 & 0.61 \\
		& --  & 2978 (272) & 0.30\% & 3.00 & 0.0030 & 5.6898 & -0.0044  & 0.62 \\
		& --  & 4003 (111) & 0.40\% & 4.00 & 0.0040 & 4.9114 & -0.0022 & 0.69 \\
		& --  & 4997 (42.3) & 0.50\% & 5.00 & 0.0048 & 4.4103 & -0.0033 & 0.73 \\
		& --  & 5989 (19.7) & 0.60\% & 6.00 & 0.0059 & 4.0378 & 0.0004 & 0.75 \\
		& --  & 7011 (7.4) & 0.70\% & 7.00 & 0.0070 & 3.7575 & 0 & 0.76 \\
		& --  & 7958 (3.6) & 0.80\% & 8.00 & 0.0081 & 3.5613 & -0.0038 & 0.78 \\
		& --  & 8997 (1.7) & 0.90\% & 9.00 & 0.0090 & 3.3903 & -0.0001 & 0.79 \\
		& --  & 9992 (0.6) & 1.00\% & 10.00 & 0.0101 & 3.2576 & -0.0015 & 0.80 \\
		\bottomrule
	\end{tabular}
\end{table}

\subsection{Summary}
This chapter introduces multiple approaches for characterizing network predictability across three primary dimensions: spectral methods, information-theoretic methods, and structural methods. Each method possesses distinct advantages and limitations, which are summarized in \autoref{tab:methods_comparison}. In addition to the three mainstream approaches, there have been a few studies in recent years exploring network predictability from alternative perspectives. For instance, Ran \textit{et al.} \cite{ran2024maximum} proposed a novel statistical distribution-based framework to assess the maximum achievable accuracy for link prediction; García-Pérez \textit{et al.} \cite{garcia2020} constructed a network ensemble under specific constraints, where the network predictability can be analytically formulated by the precision of the optimal algorithm. Although these studies have expanded our understanding of network predictability, a systematic review of existing work reveals that the field still faces numerous unresolved challenges. 

\begin{table}[!t]
	\centering
	\caption{Comparing the three categories of methods to estimate the network predictability.}
	\label{tab:methods_comparison}
	\begin{tabularx}{\textwidth}{
			>{\raggedright\arraybackslash}p{2.5cm}
			>{\raggedright\arraybackslash}p{2.5cm}
			>{\raggedright\arraybackslash}X
			>{\raggedright\arraybackslash}X
		}
		\toprule
		\textbf{Categories} & \textbf{Methods} & \textbf{Advantages} & \textbf{Disadvantages} \\
		\midrule
		
		\multirow{3}{=}{Spectral \\Methods} 
		& Structural Consistency & Effectively evaluates overall network consistency; Sensitive to global structural changes; Applicable to networks across multiple domains. & High computational complexity; Insensitive to minor local variations. \\
		& Network Spectrum & Provides network characteristics from a global perspective; Applicable to networks across multiple domains. & Requires complete network topology information; Poor adaptability to dynamic changes.\\
		& Network Energy & Quantifies network stability and connectivity; Easy to understand and implement. & High computational cost for large-scale networks; Limited adaptability to nonlinear systems. \\
		
		\midrule
		
		\multirow{2}{=}{Information-Theoretic\\Methods}
		& Compressed Length & Suitable for large-scale networks; Provides efficient and practical predictability assessment tools. & Limited processing capability for highly sparse or extremely dense networks; Sensitive to noise. \\
		& Entropy Rate & Capable of handling time-series data from dynamic networks; Reveals non-stationarity in long-term evolution. & Higher computational complexity; demands high data quality. \\
		
		\midrule
		
		\multirow{2}{=}{Structural \\Methods}
		& Structural Regularity & Capable of identifying representative subgraphs; Provides low-rank sparse representation models. & Highly dependent on assumptions; Limited generalization ability in heterogeneous networks. \\
		& Structural Controllability & Explores the role of critical edges in network control and prediction; Offers new perspectives for understanding network evolution mechanisms. & Strongly dependent on functional paths; Requires strong assumptions. \\
		
		\bottomrule
	\end{tabularx}
\end{table}

First, there is currently no widely recognized standard to effectively compare the accuracy of different predictability metrics, as the "true value" of network predictability itself remains unclear. At present, two main approaches are relied on: the first is to conduct tests on artificial networks, observing whether the proposed metric yields low values in random networks while producing higher values in networks with stronger regularity (e.g., BA scale-free networks); the second is to apply the metric to real-world networks and verify whether its correlation with the maximum accuracy achievable by multiple link prediction algorithms is significant.

Second, it remains an open question whether there exists a new metric of network predictability that can inherit the advantages of various existing methods while avoiding their limitations. An ideal network predictability metric should possess adaptability across different network types, computational efficiency, and robustness to noise and sparsity.

Furthermore, current research still lacks diversity in terms of network types and application domains. Networks from different fields (such as social networks, biological networks, economic networks, and transportation networks) often exhibit distinctly different topological properties and evolutionary mechanisms; consequently, their performance in terms of predictability may also differ significantly. If such differences do exist, it implies that we need to develop specialized methods suitable for different domain contexts. Meanwhile, whether there are more targeted ways to characterize predictability for different types of network structures (e.g., temporal networks, directed networks, and hypergraphs) remains an open question.

The identification of the structural factors that critically influence predictability is another core challenge in this field. For a considerable period, researchers in complex networks have focused on the fundamental question of "which structures are more important". This has resulted in a multitude of research directions, including critical node identification, link prediction and evolutionary dynamics modeling. Nevertheless, even in light of these achievements, it remains challenging to address the question of which structures are truly instrumental in determining a network's predictability. Consequently, some scholars have begun exploring the underlying mechanisms linking structural elements to predictability. For instance, Chen \textit{et al.} \cite{chen2019} empirically demonstrated through empirical means how "degree-based clustering coefficients" influence the predictability of link formation. Meanwhile, Wu \textit{et al.} \cite{wu2019, ming2019} and Stavinova \textit{et al.} \cite{stavinova2022, antonov2022link} introduced novel concepts such as Structural Predictability (Link Importance) and Link Predictability Classes, respectively. These efforts may provide novel theoretical foundations and analytical tools for studying network predictability.

Network predictability not only provides a novel perspective for investigating the intrinsic structure and evolutionary mechanisms of complex networks, but also drives a paradigm shift from "passive prediction" to "active modeling." It holds broad application prospects across diverse domains, including recommendation systems, social network analysis, and bioinformatics. With the advancement of emerging technologies—such as large language models (LLMs) and causal reasoning—we have reason to anticipate that future research on complex networks will more accurately uncover the underlying patterns of network formation, thereby fostering continuous progress in the study of network predictability.

\section{Predictability of Dynamics}

In the study of complex systems, the predictability of dynamics is an important metric for assessing whether a system's future behavior can be effectively forecast from its current state. Even for systems with deterministic evolution rules, nonlinear coupling and sensitivity to initial conditions can cause long-term behavior to exhibit high complexity or even chaos, thereby severely limiting forecast lead time and accuracy. Classical approaches typically rely on dynamical or information-theoretic quantities such as the Lyapunov exponent \cite{Lyapunov1992,Wolf1985} and the Kolmogorov–Sinai entropy (KS entropy) \cite{Kolmogorov1985,sinai1959notion} to evaluate a system's response to initial perturbations and, on that basis, delineate its predictability bounds. However, in practical applications, factors such as model error, observational noise, and the system's high dimensionality further exacerbate the difficulty of prediction, making the precise characterization and effective enhancement of the predictability of dynamical systems one of the core challenges in complex-systems research.

This chapter provides a systematic discussion of the predictability of dynamical systems. Section \ref{经典回顾} reviews classical methods of predictability analysis  from three perspectives: information theory, dynamical systems theory, and statistical analysis, laying the foundation for understanding recent advances. Section \ref{最新发展} focuses on the latest developments within these theoretical frameworks. In Section \ref{AI}, we examine the application of AI methods to dynamical forecasting and the challenges they pose for predictability. Section \ref{极端事件} addresses the predictability of extreme events in high-dimensional dynamics. Finally, Section \ref{系统} takes an interdisciplinary perspective to summarize current trends and challenges in research on predictability of dynamics.

\subsection{Classical Approaches}

\label{经典回顾}

This section provides a brief review of early approaches for analyzing the predictability of dynamics, which can be roughly grouped into three categories: information-theoretic measures, indicators from dynamical systems theory, and statistical evaluation methods. Tracing the development of these classical approaches helps in understanding recent theoretical evolution and methodological innovations.

Information-theoretic measures provide a systematic framework for studying uncertainty, complexity, and information flow among variables in dynamics~\cite{kleeman2011information,boffetta2002predictability}. Shannon entropy quantifies the uncertainty of system states, while the entropy rate reflects the amount of new information generated per unit time; together they reveal the complexity of a state sequence and the difficulty of long-term forecasting \cite{shannon1948}. The Kolmogorov–Sinai (KS) entropy serves as the information-generation rate in deterministic systems: the larger the KS entropy, the faster forecast errors accumulate. Its value is approximately inversely proportional to the maximum predictability time \cite{eckmann1985ergodic}. Mutual information (MI) characterizes nonlinear statistical dependence between variables by quantifying the deviation of the joint distribution from the product of the marginals; however, due to its symmetry, it cannot reveal causal direction \cite{delsole2004predictability}. Time-delayed mutual information (TDMI) introduces a time lag to identify the system's memory structure; its decay rate can be used to gauge the effective forecast horizon and has been widely applied to time-series analysis in fields such as medicine and finance \cite{vastano1988information,fraser1986independent}. Transfer entropy (TE), based on conditional mutual information, quantifies the reduction in uncertainty about one variable's future due to another variable, identifies causal direction and information sources, and has important applications in neuroscience and in constructing financial networks \cite{schreiber2000measuring}.

Dynamical-systems theory provides a fundamental framework for revealing internal structure and evolutionary mechanisms in predictability research, enabling researchers to systematically characterize the stability of trajectory behavior, perturbation-response properties, and mechanisms of spatial propagation \cite{boffetta2002predictability,eckmann1985ergodic}. The maximal Lyapunov exponent ($\lambda_{\text{max}}$) quantifies the exponential divergence rate of trajectories with respect to initial perturbations and is a classical indicator for estimating the effective forecast horizon. However, complex systems often exhibit nonstationary behaviors such as intermittency, multiscale structure, and spatial propagation, which cannot be fully captured by the traditional Lyapunov exponent. To address this, researchers have proposed multiple variants of the Lyapunov exponent, extending the description of predictability along different dimensions \cite{boffetta2002predictability}. For example, the generalized Lyapunov exponent and Rényi entropy reveal the nonuniformity of perturbation responses and the multifractal nature of trajectory distributions, making them suitable for intermittent systems \cite{benzi1985characterisation,paladin1986intermittency}; the scale-dependent Lyapunov exponent and the finite-size Lyapunov exponent characterize growth laws under finite perturbations from the viewpoints of perturbation scale and error amplitude, respectively \cite{aurell1996growth,aurell1997predictability}; and the velocity-dependent Lyapunov exponent and the comoving Lyapunov exponent introduce a moving reference frame to quantify the direction and boundaries of spatial propagation of perturbations, and are suited to open or spatially extended systems \cite{deissler1987velocity,kaneko1986lyapunov}. Taken together, these indicators constitute a multilayered assessment framework for predictability. That said, dynamical-systems approaches primarily focus on internal structural stability, whereas actual forecasting performance remains strongly affected by the location and orientation of initial errors, as well as by observational noise and external disturbances \cite{smith1999uncertainty}.

Statistical analysis offers a structured quantitative framework for studying the predictability of dynamical systems. It enables the identification of statistical dependencies among variables, pathways of error propagation, and latent low-dimensional predictable structures even under limited data \cite{delsole2007predictability}. Canonical correlation analysis (CCA) identifies the most strongly coupled modes in a system by maximizing the correlation between linear combinations of observed and forecast variables, and is commonly used both to construct predictor sets and to validate model performance \cite{barnett1987origins}. Building on this approach, predictability component analysis (PrCA) shifts the focus from maximizing variance to maximizing predictability, transforming the variable space to extract components with the greatest predictive value; it is particularly suited for high-dimensional and highly redundant datasets \cite{schneider1999conceptual}. In linear statistical modeling, error measures based on the Mahalanobis distance incorporate the covariance structure to standardize multivariate prediction errors, thereby enhancing the robustness of error evaluation \cite{mahalanobis2018generalized}.

In summary, information theory, dynamical systems theory, and statistical analysis, which emphasize uncertainty, internal evolutionary mechanisms, and statistical structure, respectively, together provide a multidimensional foundation for early studies of  predictability of dynamics. Each framework offers distinct advantages for theoretical modeling and practical applications, and over time they have become complementary. With the advent of data-driven methods and increasing cross-disciplinary integration, these classical approaches continue to guide further developments, integrations, and innovations in the field.

\subsection{Recent Dynamical Predictability Theory}

\label{最新发展}

In recent years, the growing demand for modeling and forecasting complex systems has driven the evolution and refinement of classical theoretical methods in predictability research. Building on established frameworks, researchers have expanded both the understanding and quantification of system predictability along multiple directions, prompting updates to forecasting theory and the development of practical tools. This chapter begins with an information-theoretic perspective, systematically reviewing recent advances in the information bottleneck, entropy measures, and transfer entropy estimation, and illustrating how these approaches enhance our understanding of dynamical complexity and forecasting performance. It then examines recent theories and methods in dynamical systems, including chaotic behavior, trajectory-structure analysis, and data-driven modeling, highlighting their roles in characterizing predictability. Finally, it addresses theoretical extensions in statistical analysis for constructing predictive indices, quantifying uncertainty, and performing high-dimensional causal inference, demonstrating their broad applicability in modeling and analyzing real-world complex systems.

\subsubsection{Information-Theoretic Approaches}

Recent advances in information-theoretic approaches to the predictability of dynamics can be grouped into three related directions. First, the information bottleneck (IB) theory provides a solid theoretical foundation for compression and prediction; second, entropy measures and their estimation techniques have been steadily enriched, enhancing our ability to characterize the properties of complex systems; and third, transfer entropy estimation methods have made significant progress in noise robustness and high-dimensional modeling.

\textbf{Information Bottleneck Theory}. For the joint distribution \( P(X, Y) \), where \( X \) denotes the input variables and \( Y \) denotes the target variables, the input \( X \) often contains a large amount of redundant information unrelated to prediction (e.g., noise, background features). The information bottleneck (IB) \cite{tishby2000information} introduces an intermediate representation \( T \) so as to compress the input information while preserving as much predictive power about the target as possible. Its optimization objective is:
\begin{equation}
	\label{eq:ib_objective}
	\mathcal{L}_{\text{IB}} = I(X; T) - \beta I(T; Y),
\end{equation}
here \( I(\cdot;\cdot) \) denotes mutual information. The first term \( I(X; T) \) measures the amount of information about the input \( X \) retained in the representation \( T \), reflecting the degree of compression; the second term \( I(T; Y) \) measures the predictive power of \( T \) for the target variable \( Y \). The parameter \( \beta > 0 \) controls the trade-off between compression and prediction. As a unified information-theoretic framework for representation learning, IB provides a rigorous theoretical basis for extracting minimal sufficient representations and has had a profound impact on issues such as representation-compression mechanisms in deep learning, analyses of generalization, and modeling of dynamical systems \cite{tishby2015deep}.

When modeling real-world dynamical systems, a central challenge is how to compress past observation sequences effectively in order to predict future states. Building on the information-bottleneck concept, Creutzig \textit{et al.} \cite{creutzig2009past} proposed the Past-Future Information Bottleneck (PFIB) framework to characterize the structure of information flow from the past to the future. The modeling problem is formulated as a trade-off between compression of past inputs and prediction of future outputs. The method introduces an intermediate variable, $\hat{Y}_f$, which compresses the historical input $U_p$ while retaining as much information as possible about the future output $Y_f$. The optimization objective is expressed as
\begin{equation}
	\label{eq:pfib_objective}
	\mathcal{L}_{\mathrm{PFIB}} = I(U_p; \hat{Y}_f) - \beta I(\hat{Y}_f; Y_f),
\end{equation}
where $\beta$ controls the balance between compression and prediction. Under Gaussian-linear assumptions, the framework can be reduced analytically to a generalized eigenvalue problem and solved explicitly using singular value decomposition (SVD).

To characterize the trade-off between compressed information and predictive information in a system, PFIB introduces an "information bottleneck curve" that represents the optimal compression–prediction boundary. In Gaussian–linear systems, this curve admits an explicit analytical expression. Let the system's Hankel singular values be $\sigma_i$ and the canonical correlation coefficients be $\rho_i$; then the amounts of compressed information and predictive information are, respectively:
\begin{equation}
	\label{eq:pfib_compression}
	I(U_p ; \hat{Y}_f) = \frac{1}{2} \sum_{i=1}^{n(\beta)} \log \left[ (\beta - 1) \sigma_i^2 \right],
\end{equation}
\begin{equation}
	\label{eq:pfib_prediction}
	I(\hat{Y}_f ; Y_f) = \frac{1}{2} \sum_{i=1}^{n(\beta)} \log \left( \frac{1}{1 - \rho_i^2} \right),
\end{equation}
where $\rho_i$ and $\sigma_i$ satisfy the following transformation:
\begin{equation}
	\label{eq:rho_sigma_relation}
	\rho_i^2 = \frac{\sigma_i^2}{1 + \sigma_i^2}, \quad \text{i.e.} \quad \sigma_i^2 = \frac{\rho_i^2}{1 - \rho_i^2}.
\end{equation}
The above mutual-information sums are taken over the effective modal dimension $n(\beta)$, which increases monotonically with the compression-strength parameter $\beta$ and has the analytical form:
\begin{equation}
	\label{eq:n_beta}
	n(\beta) = \max \left\{ i \,\middle|\, \beta > 1 + \frac{1}{\sigma_i^2} \right\}.
\end{equation}
For each given $\beta$, there is an optimal intermediate-state dimension $n(\beta)$, thereby determining a unique information-bottleneck model. This structure reflects a phase-transition–like behavior as the compression constraint is gradually relaxed and representational capacity is released.

\begin{figure}[h!]
	\centering
	\includegraphics[width=0.65\textwidth]{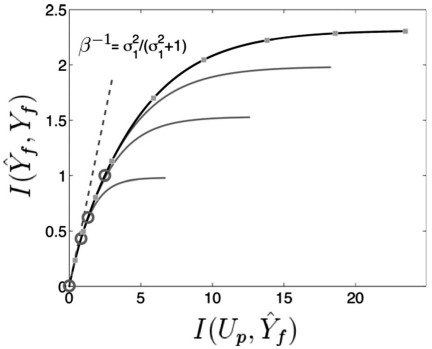}
	\caption{
		Schematic information-bottleneck curve of the PFIB applied to a spring--mass system. 
		The horizontal axis represents the historical information retained by the model, \( I(U_p; \hat{Y}_f) \), 
		while the vertical axis shows the predictive information, \( I(\hat{Y}_f; Y_f) \). 
		The black solid line indicates the theoretically optimal information-bottleneck curve, 
		the gray squares correspond to numerical model results for different compression coefficients \(\beta\), 
		and the dashed lines represent the corresponding tangents with slope \(\beta^{-1}\). 
		The gray subcurves indicate the information upper bounds for state dimensions 1, 2, and 3. 
		The figure illustrates the PFIB's modeling accuracy, the compression--prediction trade-off, 
		and dimension-driven phase-transition behavior. \\
		\textit{Source}: The figure is reproduced from Ref. \cite{creutzig2009past}.
	}
	\label{fig:pfib_info_curve}
\end{figure}

Figure~\ref{fig:pfib_info_curve} presents a schematic illustration of PFIB applied to a spring–mass system. The horizontal axis shows the historical information retained by the model \( I(U_p; \hat{Y}_f) \), and the vertical axis shows its predictive information about the future state \( I(\hat{Y}_f; Y_f) \). The black solid curve is the theoretically optimal information-bottleneck curve derived from the Hankel singular values, delineating the optimal boundary between compression and prediction. Gray squares mark the models obtained from numerical solutions at different \(\beta\), which closely track the theoretical curve, validating the effectiveness of PFIB. The dashed line is the tangent at a point on the curve; its slope is \( \beta^{-1} \), reflecting the marginal trade-off between compression and prediction. Multiple light-gray sub-curves correspond to the attainable information limits at fixed state dimensions (1, 2, 3), clearly showing the performance jumps induced by increasing model complexity and the diminishing marginal gains.

\textbf{Entropy Measures and Estimation Techniques}. As a core indicator for quantifying system complexity and uncertainty, entropy admits a wide variety of measures and estimators. Xiong \textit{et al.} \cite{xiong2017entropy} conducted a systematic assessment of different entropy-type quantities (entropy, conditional entropy, and information storage) and their estimators, the results are shown in Table~\ref{tab:entropy_estimators}. Choosing appropriate estimators together with suitable data preprocessing is crucial for accurately characterizing a system's informational features.

\begin{itemize}
	\item Entropy \(S(X)\) \cite{shannon1948} denotes the total information in the system's current state and is suitable for measuring overall uncertainty; linear estimators or  nearest-neighbor estimators are recommended, as they perform stably and are less sensitive to nonstationarity;
	\item Conditional entropy \(C(X)\) \cite{shannon1948} reflects newly generated information in the present that cannot be predicted from the past; it is suitable for capturing irregularity and unpredictability in the system but is highly sensitive to nonstationarity (e.g., trends and spikes); nearest-neighbor estimators are recommended first, followed by linear and kernel methods;
	\item Information storage \(M(X)\) \cite{lizier2012local} denotes the portion of the current state that is predictable from past states and is the most indicative of a system's predictability; it is well suited to research on dynamical forecasting; linear or nearest-neighbor estimators are recommended, though $M(X)$ is sensitive to long-range dependence and the effects of positive versus negative correlations differ.
\end{itemize}

Moreover, studies have highlighted the crucial role of data preprocessing in entropy estimation. Procedures such as detrending and standardization can effectively mitigate the effects of nonstationarity and improve estimation accuracy. Accordingly, the choice of appropriate entropy measures and their corresponding estimators, combined with proper preprocessing, is essential for uncovering the informational structure and predictability of complex systems.

\begin{table}[h!]
	\centering
	\caption{Recommended estimators and applicability of entropy measures under different conditions. \\
		\textit{Source}: The table is adapted from Ref. \cite{xiong2017entropy}.}
	\label{tab:entropy_estimators}
	\renewcommand{\arraystretch}{1.4}
	\setlength{\tabcolsep}{8pt} 
	\begin{tabular}{p{3cm} p{3.5cm} p{4cm} p{4cm}}
		\hline
		\textbf{Property} & \textbf{Entropy \(S(X)\)} & \textbf{Conditional entropy \(C(X)\)} & \textbf{Information storage \(M(X)\)} \\
		\hline
		Definition & Total information in the current state & New information in the present that cannot be predicted from the past & Information in the present that is predictable from past states \\
		\hline
		Mathematical expression & \(-\sum_x p(x) \log p(x)\) & \( S(X_t, X_{<t}) - S(X_{<t})\) & \( S(X_t) - S(X_t \mid X_{<t})\) \\
		\hline
		Recommended estimators & Linear / nearest-neighbor & Nearest-neighbor \(>\) linear \(>\) kernel methods & Linear / nearest-neighbor \\
		\hline
		Sensitivity to nonstationarity & Low & High (especially to trends and spikes) & Moderate (sensitive to trends) \\
		\hline
		Sensitivity to long-range dependence & Moderate & High & High (positive correlations enhance, negative correlations diminish) \\
		\hline
		Overall assessment & High stability; suitable for overall uncertainty, but does not reflect predictability & Captures irregularity and unpredictability, but easily perturbed & Most indicative of system predictability, but strongly affected by long-range dependence \\
		\hline
	\end{tabular}
\end{table}

In the mathematical expressions of the above entropy quantities, \(X_t\) denotes the system's state variable at the current time, \(X_{<t}\) denotes the collection of its past states, \(p(x)\) is the probability distribution function of the system state, \(S(\cdot)\) denotes entropy, \(S(X|Y)\) denotes the conditional entropy of \(X\) given \(Y\), and \(I(X;Y)\) denotes the mutual information between variables \(X\) and \(Y\). It should be noted that although this section uses the time symbol \(t\) to indicate the ordering of system states, the entropy quantities involved are, in essence, applicable to more general dynamical systems, including discrete iterative systems, continuous-time systems, and evolutionary processes on complex networks without an explicit time-series structure.

\textbf{Transfer-Entropy Estimation}. When studying the predictability of complex systems, a central issue is understanding information flow and causal structure among variables. In particular, in nonlinear dynamical systems, complex dependencies between variables can significantly affect the stability and predictability of the system's overall evolution. Transfer entropy, a causal metric within the information-theoretic framework, has been widely used in recent years to reveal asymmetric information transfer between time series \cite{schreiber2000measuring}.

Transfer entropy essentially measures the improvement in predicting one variable's future state given the past of another, thereby quantifying both the direction and the strength of information flow. Unlike correlation, transfer entropy is not about whether variables vary synchronously; rather, it asks whether the history of variable A improves the prediction of variable B's future. It thus provides an important tool for exploring causal dependencies in complex systems.

However, traditional transfer entropy estimators, such as the histogram method \cite{schreiber2000measuring}, 
kernel-density method \cite{moon1995estimation}, and \(k\)-nearest-neighbor method \cite{kraskov2004estimating}, 
often exhibit instability and substantial bias when applied to high-dimensional variables, sparse samples, or noisy environments. 
These limitations restrict their applicability in dynamical systems modeling and forecasting. To address these challenges, Diego \textit{et al.} \cite{diego2019transfer} proposed a transfer entropy estimation framework based on the Perron--Frobenius transfer operator. Instead of directly estimating high-dimensional joint probability distributions, the method leverages the system's dynamics and the induced state-transition structure to indirectly obtain the invariant measure on the attractor, from which entropy and transfer entropy are computed. Specifically, two estimators are introduced: a grid-based estimator for data-rich regimes and a triangulation-based estimator for sparse-sample regimes.

\begin{itemize}
	\item Grid estimator: The state space is partitioned into a set of grid cells \(C_i\), and the probability transition matrix is constructed by counting the frequency of trajectories moving from \(C_i\) to \(C_j\):
	\begin{equation}
		P_{ij} \approx \frac{\# \{ p_n \mid \psi(p_n) \in C_j, \, p_n \in C_i \}}{\# \{ p_m \mid p_m \in C_i \}}.
	\end{equation}
	Here, \(\#\) denotes the cardinality of a set. When the sample size is sufficiently large, this method can achieve high accuracy.
	
	\item Triangulation estimator: For sparse-sample regimes, the attractor formed by the observed points in the embedding space is partitioned into simplices \(S_a\), and a local linear mapping \(\tilde{\psi}\) is fitted within each simplex. The transition probability is then estimated based on the volume overlap with other simplices:
	\begin{equation}
		P_{ab} = \frac{m(S_b \cap \tilde{\psi}(S_a))}{m(\tilde{\psi}(S_a))},
	\end{equation}
	where \(m(\cdot)\) denotes the geometric volume. This approach remains robust even when the sample size is limited.
\end{itemize}

In this way, one can construct the system's transfer operator and thereby obtain its invariant measure on the attractor (i.e., the probability distribution of long-term behavior), which can then be used to estimate transfer entropy. The method has been validated on canonical models such as the coupled logistic map, the Rössler–Lorenz system, and Lorenz chains. Numerical results show that, compared with traditional estimators, the new approach has clear advantages in detecting causal direction, resisting noise perturbations, and handling small-sample data. In particular, for sparse time series containing only 50–100 observations, the triangulation-based estimator can still reliably identify information flow between variables, demonstrating sensitivity to weak causal structure.

In summary, the information bottleneck theory provides a systematic modeling framework for understanding the trade-off between compressed representations and predictability, and subsequent sections will further introduce AI forecasting methods that have evolved from this theory. Advances in entropy measures have improved their adaptability to different forms of system complexity, while recently proposed transfer entropy estimation techniques have extended their applicability to the analysis of real-world complex data. The ongoing evolution of these information-theoretic methods has strongly advanced tools for forecasting and causal analysis in dynamical systems, demonstrating dual value in theoretical innovation and practical adaptability.

\subsubsection{Dynamical Systems Theory}

In this section, we first review important recent theoretical perspectives on the predictability of dynamical systems, and then we summarize several forecasting and analysis algorithms for dynamical systems that have been developed from data-driven ideas.

\begin{figure}[h!]
	\centering
	\includegraphics[width=0.5\textwidth]{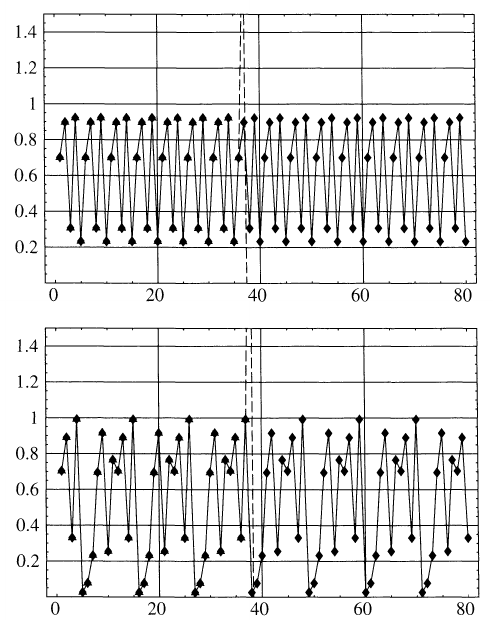} 
	\caption{
		Comparison of trajectories initialized with the same conditions using floating-point representation (triangles, dashed lines) and symbolic representation (diamonds, solid lines). The upper and lower panels correspond to different periods. The floating-point trajectories begin to diverge around \(n=36\) and eventually leave the visible range, whereas the symbolic trajectories remain stable for a longer time, illustrating improved robustness in medium- and long-term prediction. \\
		\textit{Source}: The figure is reproduced from Ref. \cite{sobottka2006periodicity}.
	}
	
	\label{representation}
\end{figure}

\textbf{Chaotic Dynamics}. The conventional view in the literature holds that the intrinsic chaotic nature of dynamical systems is the principal cause of their limited predictability. However, recent studies indicate that the way a system is represented also affects our assessment of its predictability. Sobottka \textit{et al.} \cite{sobottka2006periodicity} conducted a systematic analysis of the differences between symbolic and floating-point representations in trajectory forecasting. They found that although floating-point representations have higher numerical precision, rounding errors accumulate over time, causing trajectories to drift rapidly from the true evolution and thereby degrading medium- to long-term forecasting performance; by contrast, symbolic representations avoid numerical rounding error, maintain trajectory stability for longer, and yield more robust predictive performance (see Fig.~\ref{representation}). This finding suggests that, in analyzing the predictability of dynamical systems, one must consider not only dynamical complexity itself but also the choice of representation and its sensitivity to numerical error.

In studies of predictability in chaotic systems, accurately distinguishing different forms of dynamical behavior is of central importance. Wernecke \textit{et al.} \cite{wernecke2017test} proposed a 0–1 type test based on the analysis of pairs of initially close trajectories. The method introduces two indicators, the cross-distance scaling exponent and the finite-time cross-correlation, which together allow effective discrimination among strong chaos, partially predictable chaos (PPC), and laminar flow.

The cross-distance scaling exponent characterizes the dependence between the mean distance \(d_{12}(t) = \| x_1(t) - x_2(t) \|\) of two trajectories \(x_1(t)\) and \(x_2(t)\) after the Lyapunov time scale \(T_\lambda = 1/\lambda_{\text{max}}\) and the magnitude of the initial perturbation \(\delta\). Specifically, the initial conditions \(x_1(0)\) and \(x_2(0)\) of the trajectory pair satisfy
$\| x_1(0) - x_2(0) \| = \delta,$ where \(\delta\) is a small positive number indicating the amplitude of the initial perturbation. The two trajectories evolve independently under the same dynamical system (e.g., equation \eqref{lorenz}). The cross-distance scaling exponent \(\nu\) is characterized by
\begin{equation}
	\label{eq:scaling}
	d_{12}(t \gg T_\lambda) \propto \delta^\nu,
\end{equation}
that is, when time is much greater than the Lyapunov time, the mean distance between the two trajectories follows a power-law relation with the initial distance. If \(\nu=0\), the long-term distance is independent of the initial perturbation, which typically corresponds to strong chaos; if \(\nu \neq 0\), the degree of separation depends on the size of the initial perturbation, which usually corresponds to laminar or quasi-periodic behavior. The exponent is generally obtained by linear fitting in log–log coordinates.

Another indicator, the finite-time cross-correlation \(C_{12}(t)\), measures the statistical correlation between two trajectories at time \(t\) and is defined as
\begin{equation}
	\label{eq:corr}
	C_{12}(t) = \frac{\langle (x_1(t) - \mu)(x_2(t) - \mu) \rangle}{s^2},
\end{equation}
where \(\mu\) is the centroid of the attractor, \(s\) is the mean scale of the attracting set, and \(\langle \cdot \rangle\) denotes averaging over multiple initial conditions. Typically, in the strong-chaos regime, \(C_{12}(t)\) decays rapidly to zero after the forecast time \(T_\lambda\); in PPC, however, it can remain at a high level even when \(t \gg T_\lambda\), indicating a degree of medium- to long-term predictability.

\begin{figure}[h!]
	\centering
	\includegraphics[width=0.65\textwidth]{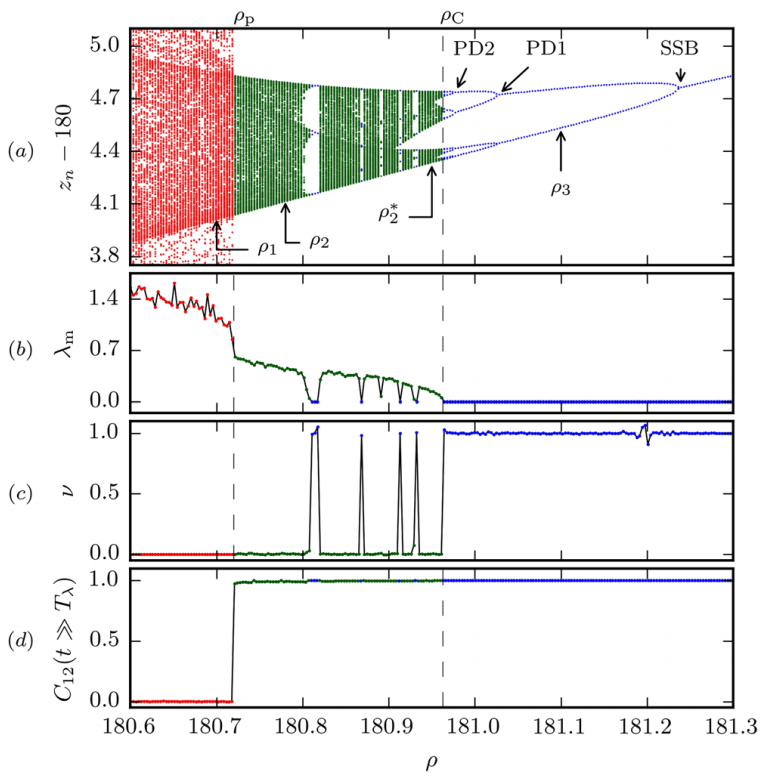} 
	\caption{
		Strong chaos in the Lorenz system (equation \eqref{lorenz}, red, $\rho < \rho_p \approx 180.72$), partially predictable chaos (green, $\rho_p < \rho < \rho_C \approx 180.96$), and laminar flow (blue, $\rho > \rho_C$). From top to bottom: (a) values of $z_n$ in the Poincaré section ($x=15$ plane, with local window shown); (b) maximal Lyapunov exponent $\lambda_m$; (c) cross-distance scaling exponent $\nu$ (see equation \eqref{eq:scaling}); (d) finite-time cross-correlation $C_{12}(t=200)$ (definition in equation \eqref{eq:corr}). PD1 and PD2 denote period-doubling bifurcation points, and SSB indicates the bifurcation corresponding to spontaneous breaking of the system symmetry $(x,y,z) \leftrightarrow (-x,-y,z)$. \\
		\textit{Source}: The figure is reproduced from Ref. \cite{wernecke2017test}.
	}
	\label{diver_chaos}
\end{figure}

The application of this method to the Lorenz system validates its effectiveness. We first consider the standard Lorenz system \(\mathbf{x} = (x, y, z)\),
\begin{equation}
	\dot{x} = \sigma(y - x), \quad \dot{y} = x(\rho - z) - y, \quad \dot{z} = xy - \beta z     
	\label{lorenz}
\end{equation}
where the standard parameters are set to \(\beta = 8/3\) and \(\sigma = 10\), with \(\rho\) serving as the bifurcation parameter.  By varying \(\rho\), as illustrated in Fig.~\ref{diver_chaos}, the system undergoes continuous transitions among laminar flow, PPC, and strong chaos, accompanied by pronounced variations in \(\nu\) and \(C_{12}(t)\). Compared with the conventional maximal Lyapunov exponent, this method demonstrates superior discriminative ability for identifying partially predictable chaos and underscores the crucial influence of attractor topology on long-term predictability. Owing to its theoretical generality and computational feasibility, the method has been applied in diverse complex-system contexts, including enzyme kinetics, financial modeling, and self-organized systems, thereby offering new perspectives and tools for analyzing the multiscale predictability of chaotic dynamics.

\textbf{Data-Driven Methods}. These methods dispense with explicit models and instead extract the system's evolution directly from observational data, making them especially suitable for systems with complex structure and unknown dynamics, and greatly expanding the applicability of predictability analysis.

In the early stage of data-driven dynamical modeling, Schmidt and Lipson \cite{schmidt2009distilling} proposed an unsupervised symbolic-regression framework based on genetic programming to automatically extract analytic conservation laws and system constraints from experimental data. Without relying on prior physical knowledge, the method jointly searches over function structures and parameters, effectively reconstructing dynamical structures such as Hamiltonians and Lagrangians. To avoid spurious physical laws caused by numerical coincidences, the authors introduced pairs of partial derivatives as a discriminative criterion, and, through validation on multiple canonical systems (e.g., the harmonic oscillator and the pendulum), demonstrated the method's potential for law discovery. In addition, they found that symbolic expression fragments extracted from simple systems can be used to initialize the search for more complex systems, thereby significantly improving modeling efficiency and revealing the emergent role of a “symbolic alphabet” in complex-system modeling.

\begin{figure}[!t]
	\centering
	\includegraphics[width=0.90\textwidth]{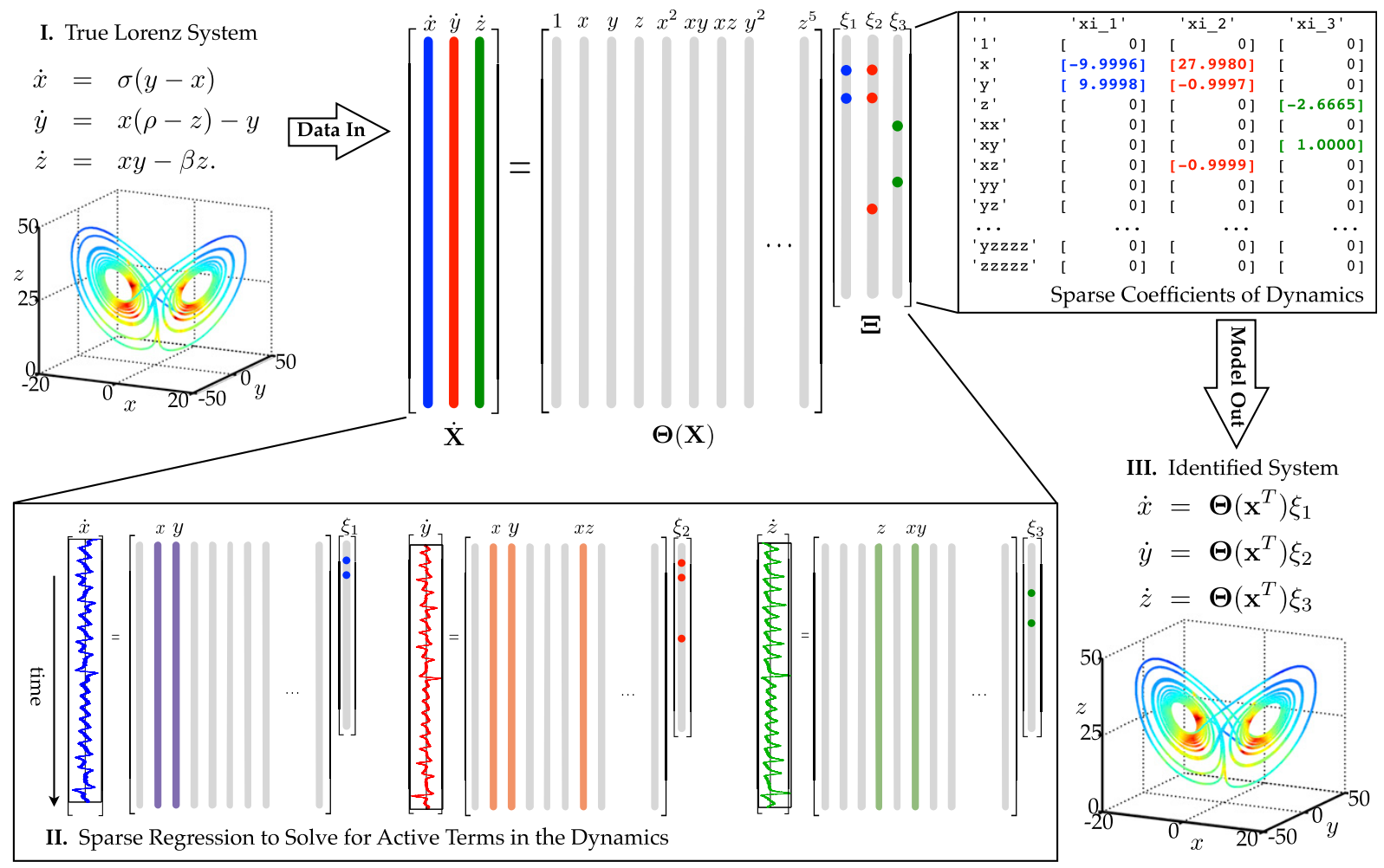} 
	\caption{
		Schematic illustration of the SINDy algorithm applied to the Lorenz system. 
		Time-series data of the system, including states \(X\) and their derivatives \(\dot{X}\), are first collected; assumptions on \(\dot{X}\) will later be relaxed. A library of nonlinear functions of the state, \(\Theta(X)\), is then constructed. Sparse regression is applied to identify the minimal set of terms in \(\Sigma\) that satisfies \(\dot{X} = \Theta(X)\Sigma\). 
		The few nonzero entries in the solution vector \(\Sigma\) indicate the relevant terms on the right-hand side of the dynamics. 
		Parameters are set as \(\sigma = 10\), \(\beta = 8/3\), \(\rho = 28\), and \((x_0, y_0, z_0)^T = (-8, 7, 27)^T\). 
		Trajectories on the Lorenz attractor are colored according to the required adaptive time step, with red indicating smaller steps. \\
		\textit{Source}: The figure is reproduced from Ref. \cite{brunton2016discovering}.
	}
	\label{sindy}
\end{figure}

In subsequent research, Brunton \textit{et al.} \cite{brunton2016discovering} proposed the Sparse Identification of Nonlinear Dynamics (SINDy) method, further advancing data-driven modeling toward a synthesis of structural sparsity and interpretability. The core hypothesis of SINDy is that the evolution of many natural systems is governed by a small number of key dynamical terms. This idea is realized by constructing sparse representations within a high-dimensional library of candidate functions. Specifically, SINDy first builds a set of candidate terms comprising various nonlinear functions, and then uses sparse regression techniques (such as LASSO or sequential thresholded least squares) to select those functions that play a decisive role in the system's dynamics, thereby yielding a concise, interpretable model with strong predictive performance. Its application to the Lorenz system is shown in Fig.~\ref{sindy}. The widespread application of SINDy across many physical and engineering systems indicates that data-driven modeling can not only reveal the structural essence of a system but also provide a new pathway for the predictability of high-dimensional complex systems.

\begin{figure}[h!]
	\centering
	\includegraphics[width=0.65\textwidth]{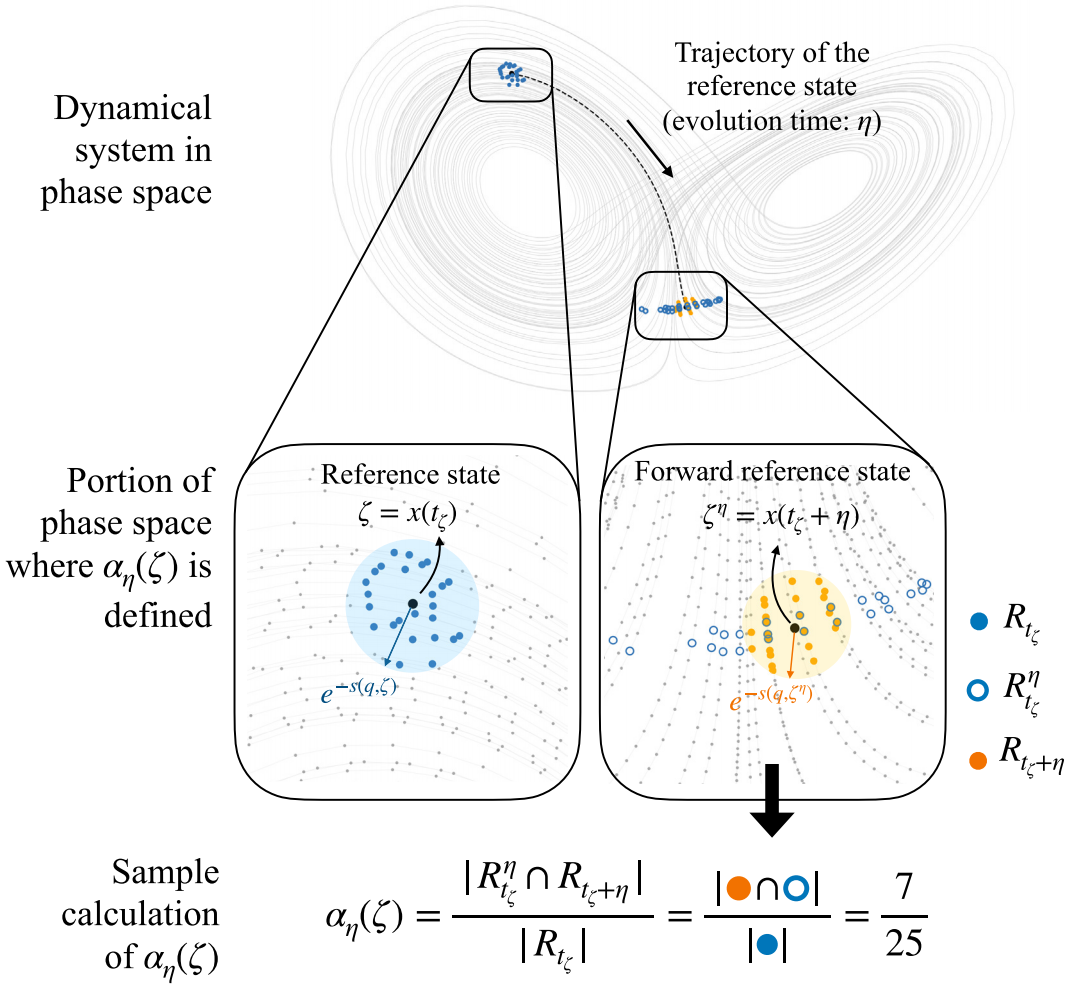} 
	\caption{Schematic illustration of the computation of \(\alpha_\eta(t)\), demonstrated in the phase space of the Lorenz-63 system. This figure presents all steps involved in computing the \textit{time-lagged recurrence} for the reference state \(\zeta\) at a forecasting horizon \(\eta\), namely \(\alpha_\eta(\zeta)\). The panels in the second row provide a zoomed-in view of the phase space region where \(\alpha_\eta(\zeta)\) is defined. Recurrences (\(R_{t_\zeta}\)), forward recurrences (\(R^{\eta}_{t_\zeta}\)), and forward-reference-state recurrences (\(R_{t_\zeta + \eta}\)) are represented by solid blue dots, empty blue dots, and orange dots, respectively. The blue circle with radius \(e^{-s(q,\zeta)}\) indicates the hypersphere used to define the neighborhood of the reference state, while the orange circle with radius \(e^{-s(q,\zeta^{\eta})}\) corresponds to the forward-reference-state. \\
		\textit{Source}: The figure is reproduced from Ref. \cite{dong2025time}.}
	\label{alpha}
\end{figure}

Beyond identifying governing equations, researchers have also sought model-free methods to assess predictability. The time-lagged recurrence (TLR) index proposed by Dong \textit{et al.} \cite{dong2025time} is a representative example. The method builds on the recurrence property of dynamical systems—namely, that a trajectory in phase space repeatedly returns to the vicinity of a region. The TLR index quantitatively characterizes local predictability at the current state by comparing the degree of overlap between the evolved set $R_{t_\zeta}^{\eta}=\{\mathbf x(t+\eta)\mid\mathbf x(t)\in R_{t_\zeta}\}$ of the current recurrence neighborhood $R_{t_\zeta}$ for the state $\zeta=\mathbf x(t_\zeta)$ after a time lag $\eta$ and the target neighborhood $R_{t_\zeta+\eta}$, with the formula:
\begin{equation}
	\alpha_{\eta}(\zeta) = \frac{|R^{\eta}_{t_{\zeta}} \cap R_{t_{\zeta} + \eta}|}{|R_{t_{\zeta}}|},
\end{equation}
where \(\alpha_{\eta}(\zeta)\) denotes the local predictability index of the reference state \(\zeta\) at lag \(\eta\); the numerator is the size of the intersection between the lagged set and the target neighborhood, and the denominator is the size of the initial recurrence neighborhood. The closer \(\alpha_{\eta}(\zeta)\) is to 1, the higher the predictability. A schematic of the computation is shown in Fig.~\ref{alpha}. Entirely grounded in the geometry of phase space, the method requires no explicit model but depends on the geometric properties of the data, making it particularly suitable for high-dimensional spatiotemporal systems. Experiments show that the TLR index effectively characterizes the predictability structure of the Lorenz-63 system and real atmospheric dynamical systems, reveals the decay of forecast skill with lead time, and holds promise for real-time prediction.

A key advantage of the TLR method lies in its natural connection to information theory. Researchers have shown that \( \alpha_{\eta}(\zeta) \) can be interpreted as the conditional probability given in Eq. \eqref{relation}, namely, the probability that, given the current state lies within the neighborhood, its future state still falls into the future neighborhood. This viewpoint links predictability directly to Shannon entropy and provides a new tool for quantifying system uncertainty.
\begin{equation}
	\label{relation}
	\alpha_\eta(\zeta)=\mathbb P \!\bigl(\mathbf x(t+\eta)\in R_{t_\zeta+\eta}\bigm|\mathbf x(t)\in R_{t_\zeta}\bigr),
\end{equation}
Studies also indicate that, compared with classical indices such as the nonlinear local Lyapunov exponent (NLLE), TLR performs better in nonlinear regimes. For instance, whereas NLLE assumes exponentially growing errors, TLR directly captures nonlinear effects through the actual evolution of recurrence points. In addition, TLR can reveal non-monotonic decay of predictability; over certain time windows the system may display temporary resurgences in predictability due to attractor structure.

\begin{figure}[h!]
	\centering
	\includegraphics[width=0.85\textwidth]{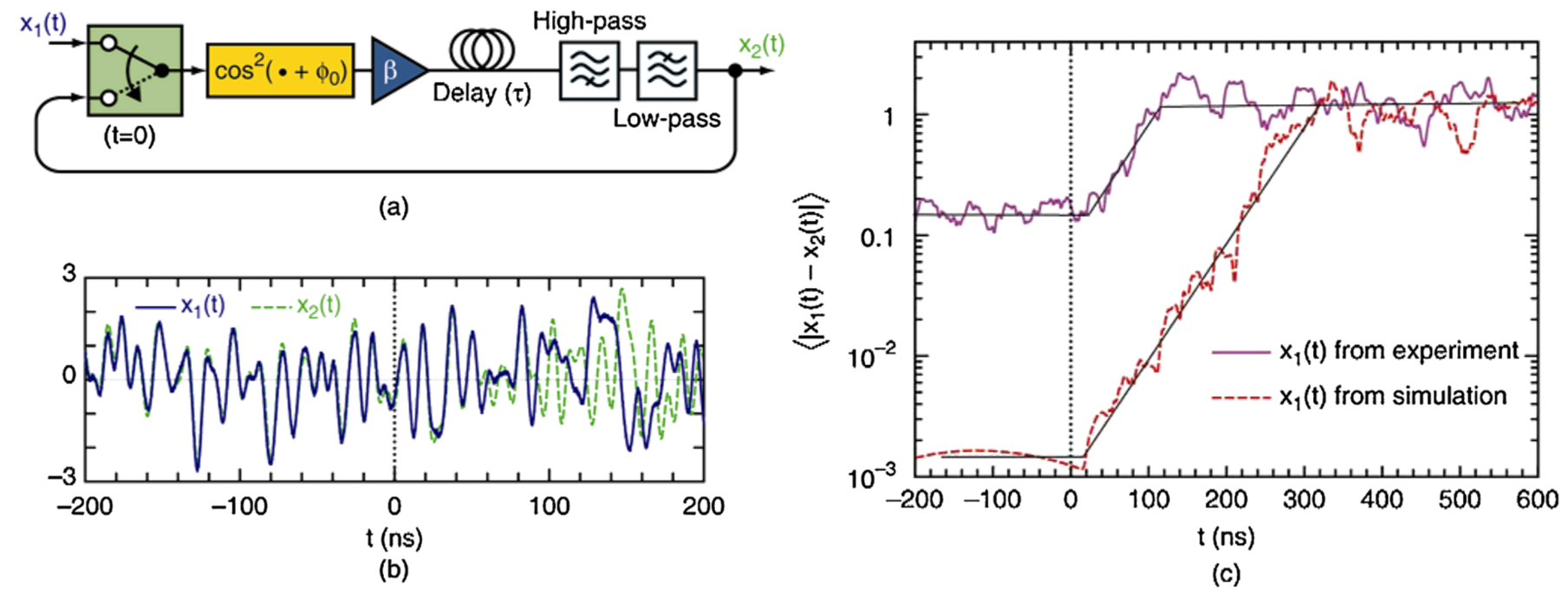} 
	\caption{Data assimilation and forecasting scheme. 
		(a) Illustration of a method in which a numerical simulation is synchronized with experimental observations. After synchronization between the two signals \(x_1(t)\) and \(x_2(t)\) is achieved, the switch is turned off at \(t=0\), allowing the numerical simulation to evolve independently. (b) Comparison of experimental and simulated time series before and after the switch is turned off, showing divergence between the two waveforms. (c) Semi-log plot of the absolute difference \(|x_1 - x_2|\), revealing exponential divergence after the switch is turned off. The absolute difference is smoothed using a 25 ns moving average to provide a reliable estimate of the slope. The dashed curve is obtained by replacing the experimental data with the numerical simulation time series, resulting in closer initial synchronization. \\
		\textit{Source}: The figure is reproduced from Ref. \cite{cohen2008using}.}
	\label{Tongbu}
\end{figure}

Further, when dealing with higher-dimensional systems with more pronounced experimental errors, Cohen \textit{et al.} \cite{cohen2008using} proposed a prediction strategy based on synchronization. Using an optoelectronic time-delay feedback loop as the experimental platform, they fed the experimental data into a numerical model so that, under suitable parameter conditions, the two remained synchronized, thereby enabling tracking of the system's future evolution. A flowchart of the method is shown in Fig.~\ref{Tongbu}. During the initial synchronization phase, by analyzing the divergence rate of the model trajectory relative to the experimental trajectory, one can estimate the local Lyapunov exponent and its reciprocal, namely, the forecast horizon at that state. The key to this strategy lies in fusing experimental data with a numerical dynamical model, thereby demonstrating the practical feasibility of synchronization phenomena for forecasting chaotic systems and offering a new approach to prediction problems in fields such as chaotic communications, sensor networks, and biomedicine.

In summary, research on the predictability of complex systems is advancing along two synergistic paths: theoretical analysis and data-driven approaches. On the one hand, perspectives such as system representation, informational complexity, and attractor structure deepen our understanding of the essence of chaos; on the other hand, data-driven methods, by directly exploiting observational data, exhibit pronounced advantages when the governing model is unknown. The two are complementary, jointly advancing both forecasting skill and mechanistic understanding.

\subsubsection{Statistical Analysis}\label{sec_Statistical_Analysis}

\begin{figure}[!t]
	\centering
	\includegraphics[width=0.65\textwidth]{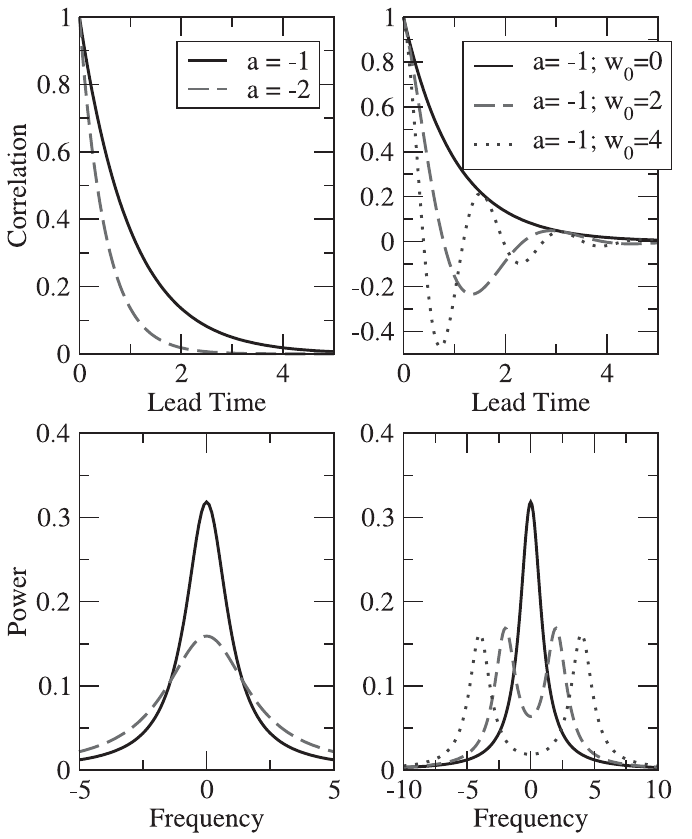} 
	\caption{
		(Top) Autocorrelation function; (bottom) power spectrum corresponding to the damped oscillatory function. \\
		\textit{Source}: The figure is reproduced from Ref. \cite{delsole2009average}.
	}
	\label{APT}
\end{figure}

First, among recent metrics for quantifying a system's overall forecasting performance, we highlight the Average Predictability Time (APT). APT was proposed by DelSole \textit{et al.} \cite{delsole2009average}; in continuous time, APT is defined as
\begin{equation}
	H = 2 \int_{0}^{\infty} H(\tau) \, d\tau,
\end{equation}
where the integrand \( H(\tau) \) is given by
$H(\tau) = \frac{1}{K} \, \mathrm{tr} \left[ \left( \Sigma_{\infty} - \Sigma_{\tau} \right) \Sigma_{\infty}^{-1} \right],$
with \( K \) denoting system dimension and \( \mathrm{tr}[\cdot] \) the matrix trace. Here \( \Sigma_{\infty} \) is the system's climate covariance matrix, reflecting its long-term statistical equilibrium, and \( \Sigma_{\tau} \) is the forecast-error covariance matrix at lead time \( \tau \). APT measures the normalized growth of forecast error relative to the system's inherent climate variance. APT is also closely related to the system's power spectrum, being proportional to the integral of the squared magnitude of the power spectral density, i.e.,
$\tau \propto \int_{-\infty}^{\infty} \left| \tilde{P}(\omega) \right|^2 d\omega.$ As shown in Fig.~\ref{APT},
this relationship indicates that systems with sharper spectral peaks have higher predictability, whereas a flatter spectrum corresponds to lower predictability. Moreover, APT can be decomposed into a set of uncorrelated components that maximize predictability time, analogous to variance decomposition in principal component analysis. This property allows APT to be used to examine how system coupling affects predictability: if a system can be represented as a collection of uncoupled stochastic models, APT attains its minimum; conversely, coupling enhances predictability and thus increases the APT value.

When assessing a system's predictability, how to measure the quality of probabilistic forecasts is itself a key issue. The logarithmic scoring rule (log score), as one of the most commonly used methods, effectively reflects a model's performance at the realized observation and enjoys strict theoretical propriety. However, when observational data contain errors or the model is biased relative to the true system, the log score tends to overreact to small differences in forecast probabilities, leading to unstable scores and thereby undermining an objective evaluation of the system's predictability.

To improve the robustness and physical interpretability of probabilistic forecast scoring, Xu \textit{et al.} \cite{xu2025probabilistic} proposed an $\epsilon$-logarithmic scoring rule to assess forecast quality within a specified tolerance. Compared with traditional pointwise scoring, this method evaluates the probability mass within an error-tolerance region, better reflecting practical situations in which observational errors and model uncertainty are ubiquitous. The score is defined as:
\begin{equation}
	L_\epsilon(\hat{p}_x, x) =
	\begin{cases}
		\log \hat{p}_x(x), & \text{if } \epsilon = 0, \\
		\log \int_{\|s - x\|_\infty \leq \epsilon} \hat{p}_x(s) \, ds, & \text{if } \epsilon > 0,
	\end{cases}
\end{equation}
where $\hat{p}_x$ is the predictive probability density function and $x$ is the realized observation. Thus, when $\epsilon=0$, the rule reduces to the traditional log score; when $\epsilon>0$, the score reflects the total probability mass of the predictive distribution within an $\epsilon$-neighborhood of the observation. Further, for a system trajectory of length $T$, one can define the average score:
\begin{equation}
	\bar{L}_\epsilon(\hat{p}_{x_1:T}, p_{x_1:T}) = \frac{1}{T} \sum_{k=1}^T \mathbb{E}_{x_{1:k-1}} \left[ L_\epsilon \left( \hat{p}_{x_k | x_{1:k-1}}, x_k \right) \right],
\end{equation}
which measures the predictor $\hat{p}$'s average forecasting performance over the entire trajectory, balancing accuracy and stability. To characterize the system's maximally achievable predictability under a given tolerance $\epsilon$, Xu \textit{et al.}\ further present the following theorem:

\begin{theorem}[Probabilistic Predictability]
	\label{thm2.1_simplified}
	Let the state-space dimension be $d_x$ and the true trajectory distribution be $p_{x_{1:T}}$. With tolerance radius $\epsilon$, the optimal average $\epsilon$-log score over all possible predictors is
	\begin{equation}
		\max_{\hat{p}} \bar{L}_{\epsilon} = d_x \log(2\epsilon) - \frac{1}{T} D_{\mathrm{KL}}\left( p_{x_{1:T}} \,\|\, \tilde{p}^\star_{x_{1:T}} \right),
	\end{equation}
	where $\tilde{p}^\star_{x_{1:T}}$ is the "blurred" predictive distribution closest to the true distribution within the error tolerance, namely, the convolutional approximation that minimizes the KL divergence:
	\begin{equation}
		\tilde{p}^\star_{x_k|x_{1:k-1}} = \arg\min_{\tilde{p} \in \mathcal{F}_\epsilon} 
		D_{\mathrm{KL}}\left(p_{x_k|x_{1:k-1}} \,\|\, \tilde{p} \right),
	\end{equation}
	\begin{equation}
		\mathcal{F}_\epsilon := \{ \hat{p} * p_{u^\epsilon} \mid \hat{p} \in \mathcal{P} \},
	\end{equation}
	where $p_{u^\epsilon}$ denotes the uniform distribution on the $\ell_\infty$-ball of radius $\epsilon$, $*$ denotes convolution, and $\ast/\ast$ denotes the corresponding deconvolution.	
\end{theorem}

From this theorem, we see that the system's optimal predictability is determined by two parts— the first term, $d_x \log(2\epsilon)$, represents the information capacity associated with the "allowed error range": the larger the tolerance, the higher the score; the second term is the KL divergence between the true distribution and the optimal blurred predictive distribution, which reflects the model's fit error relative to the true system dynamics. This theoretical result clearly delineates the trade-off between "forecast accuracy" and "error tolerance", and provides a rigorous quantitative basis for the upper limit of system predictability.

\begin{figure}[htbp]
	\centering
	\includegraphics[width=1.0\textwidth]{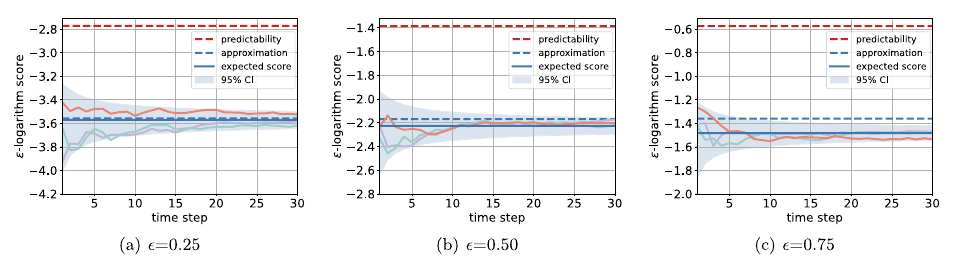} 
	\caption{
		Variation of the \(\epsilon\)-logarithm score \(\bar{L}_\epsilon\) as a function of trajectory length \(T\). 
		The blue solid line shows the average score computed over \(10^5\) trajectories, with the shaded area representing the 95\% confidence interval. 
		The red dashed line indicates the predictability limit defined in Theorem~\ref{thm2.1_simplified}, while the blue dashed line represents the theoretical approximation. 
		Colored lines depict the score evolution for three individual random trajectories. \\
		\textit{Source}: The figure is reproduced from Ref. \cite{xu2025probabilistic}.
	}
	\label{epsilon}
\end{figure}

Figure~\ref{epsilon} shows how the \(\epsilon\)-log score evolves with increasing trajectory length under different \(\epsilon\) settings. The blue solid line denotes the average score computed over \(10^5\) simulated trajectories, and the blue shading indicates its \(95\%\) confidence interval. The colored lines display the score evolution for three representative single trajectories to illustrate convergence. The results indicate that the method attains stable convergence over short horizons, with close agreement between theory and empirical results, demonstrating strong robustness and practical utility. 

\subsection{AI-Related Advances}

\label{AI}

In the preceding review of classical methods for analyzing the predictability of dynamics and their recent developments, we have seen that information theory, nonlinear dynamical systems theory, and statistical modeling provide a solid foundation for modeling and forecasting complex systems. These approaches, starting from the system's intrinsic mechanisms, characterize attractor structure, delineate stability boundaries, and quantify the degree of chaos and the level of predictability \cite{boffetta2002predictability}. However, as system dimensionality increases and data complexity grows, traditional methods face numerous challenges in handling high-dimensional, strongly nonlinear, and multiscale spatiotemporally coupled dynamical systems. For large-scale coupled systems, for example, strong sensitivity to initial conditions severely limits long-range forecasting, while numerical integration methods increasingly encounter bottlenecks due to computational cost and the accumulation of model error \cite{musielak2009high}.

In recent years, artificial intelligence (AI) methods, particularly deep learning, generative modeling, and information-driven optimization, have emerged as important tools for modeling the evolution of complex dynamical systems \cite{lecun2015deep}. Without requiring explicit differential-equation models, AI systems can learn the underlying dynamical laws from historical observations and demonstrate strong forecasting capability. For example, the Pangu-Weather model has outperformed the traditional ECMWF numerical forecasting system across multiple meteorological variables, substantially extending forecast lead times (for instance, reducing tropical-cyclone track forecast error by about 25\%) \cite{bi2023accurate}. Moreover, AI provides novel approaches for characterizing dependence structures among high-dimensional random variables. As an illustration, Belghazi \textit{et al.}\ \cite{belghazi2018mutual} proposed Mutual Information Neural Estimation (MINE), which trains a neural network to maximize a lower bound on mutual information, thereby enabling efficient estimation of mutual information between high-dimensional continuous variables and alleviating the curse-of-dimensionality limitations of classical estimators.

Overall, the rise of AI-based dynamical modeling has not only greatly surpassed traditional numerical methods in forecasting accuracy, but also provided entirely new ideas and tools for understanding the internal evolutionary mechanisms of complex systems and probing the limits of predictability \cite{carleo2019machine,yu2024learning}. This chapter is organized around three themes: first, we introduce representative AI methods from recent years and how they extend the predictability frontier of complex systems; second, we analyze the integration mechanism between the information bottleneck and AI and how it enhances dynamical forecasting; and third, we discuss the theoretical challenges that current AI models face in predictability modeling and potential directions for future development.

\subsubsection{Representative AI Methods}

The rapid development of AI methods has opened new opportunities for research on the predictability of complex systems. These approaches have not only achieved breakthroughs in forecast accuracy, but also demonstrated notable advantages in extending effective lead time, enhancing robustness to initial perturbations, and revealing latent predictable structures. This section focuses on several representative AI methods applied across different complex systems and analyzes the mechanisms by which they enhance system predictability.

\begin{figure}[htbp]
	\centering
	\includegraphics[width=0.9\textwidth]{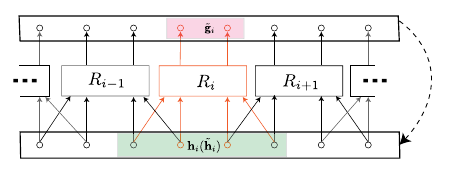} 
	\caption{
		Schematic illustration of the parallel reservoir computing architecture. Each reservoir independently processes the input from a local region and outputs the corresponding predicted state. The dashed arrows indicate the feedback connections during the prediction phase, where the reservoir output is recursively fed back as its own input. Information coupling between reservoirs is achieved through partial overlap of local inputs, thereby enhancing spatial continuity and improving overall predictive performance. \\
		\textit{Source}: The figure is reproduced from Ref. \cite{pathak2018model}.
	}
	\label{Shuiku}
\end{figure}

\textbf{Reservoir Computing}. Traditional complex dynamical systems are often high-dimensional, nonlinear, and strongly chaotic, making them difficult to capture accurately with classical modeling methods. Pathak \textit{et al.} \cite{pathak2018model} proposed a model-free forecasting framework based on reservoir computing (RC). The method partitions the system into multiple local regions and assigns an independent reservoir to each region to enable parallel prediction (see Fig.~\ref{Shuiku}); partially overlapping input windows are used to couple information across reservoirs, thereby improving global modeling of spatial continuity while effectively extending the forecast horizon. RC has been successfully applied to several canonical chaotic systems, demonstrating substantial gains in short-term predictability and parallel computational efficiency.

\begin{figure}[!t]
	\centering
	\includegraphics[width=0.8\textwidth]{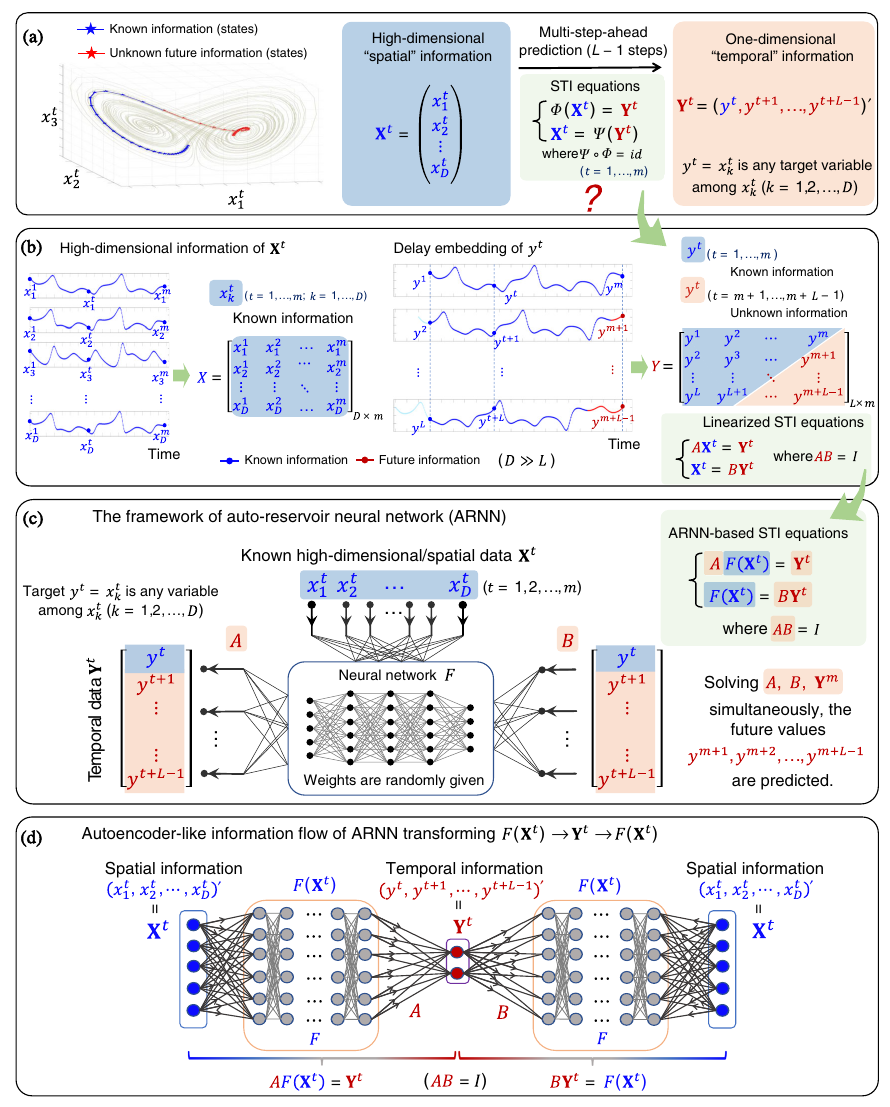} 
	\caption{
		Schematic illustration of the autoregressive reservoir neural network (ARNN).  
		(a) Given a short-term time series of a high-dimensional system, predicting the future state of any target variable is a challenging task. For the target variable to be predicted, a delay-embedding strategy is applied through the mapping $\Phi$ to construct a delay-coordinate vector $\mathbf{Y}^t$ corresponding to the observation vector $\mathbf{x}^t$. This relation defines the spatiotemporal information (STI) transformation, which possesses both original and conjugate forms (STI equations).  
		(b) The linearized STI equations also admit original and conjugate forms. Data can be represented in matrix form, where the future/unknown entries $\{y^{m+1}, y^{m+2}, \ldots, y^{m+l-1}\}$ occupy the lower-right triangular part of the matrix $\mathbf{Y}$, while the known entries $\{y^1, y^2, \ldots, y^m\}$ occupy the upper-left part.  
		(c) ARNN is a model-free approach for multi-step forecasting of the target variable $y$. Within the ARNN framework, the reservoir component consists of a random/fixed multilayer neural network $F$ with time-dependent input $\mathbf{X}^t$. The network $F$, parameterized by two weight matrices $A$ and $B$, processes the delay-embedded target vector $\mathbf{Y}^t$. This architecture is designed to solve both the original and conjugate forms of the STI equations, thereby enhancing robustness and enabling reliable prediction of the future states of $y$ even from short time series.  
		(d) In terms of information flow, ARNN exhibits an autoencoder-like structure: $F(\mathbf{X}^t) \rightarrow \mathbf{Y}^t \rightarrow \hat{\mathbf{Y}}^t \rightarrow \hat{\mathbf{X}}^t$, which differs from but parallels the conventional autoencoder flow $\mathbf{X}^t \rightarrow \mathbf{Y}^t \rightarrow \hat{\mathbf{Y}}^t \rightarrow \hat{\mathbf{X}}^t$. \\
		\textit{Source}: The figure is reproduced from Ref. \cite{chen2020autoreservoir}.
	}
	\label{arnn}
\end{figure}

\textbf{Auto-Reservoir Neural Network}.
Although classical reservoir computing (RC) can effectively handle nonlinear dynamics, its inputs typically rely on exogenous forcing, which limits the ability to capture the system's endogenous spatiotemporal relationships in depth. Building on RC, Chen \textit{et al.} \cite{chen2020autoreservoir} proposed the auto-reservoir neural network (ARNN), which constructs a spatiotemporal information (STI) mapping. This mapping feeds the delay-embedded representation of the target variable as input, thereby extracting the dominant dynamical features that drive prediction directly from the system's internal state (Fig.~\ref{arnn}). Compared with exogenous forcing, the STI mapping is grounded in endogenous intervariable relationships, making it more consistent with the nonstationary, multiscale characteristics of real-world complex systems. By encoding spatiotemporal correlations into the network, ARNN achieves stable and accurate multi-step forecasts even with limited data and strong noise. Its architecture incorporates an autoencoder-like information flow, enhancing the capture of short-term dynamics in nonstationary systems.

\textbf{Embedding and Constraints}. Beyond reservoir-based approaches, recent years have seen the emergence of a class of AI architectures that emphasize structural modeling, focusing on incorporating a system's intrinsic laws into the forecasting framework in the form of embeddings or constraints. These methods not only enhance the model's ability to stably capture long-term behavior, but also exhibit distinctive advantages in revealing predictability bounds and explaining predictive mechanisms.

\begin{figure}[htbp]
	\centering
	\includegraphics[width=0.9\textwidth]{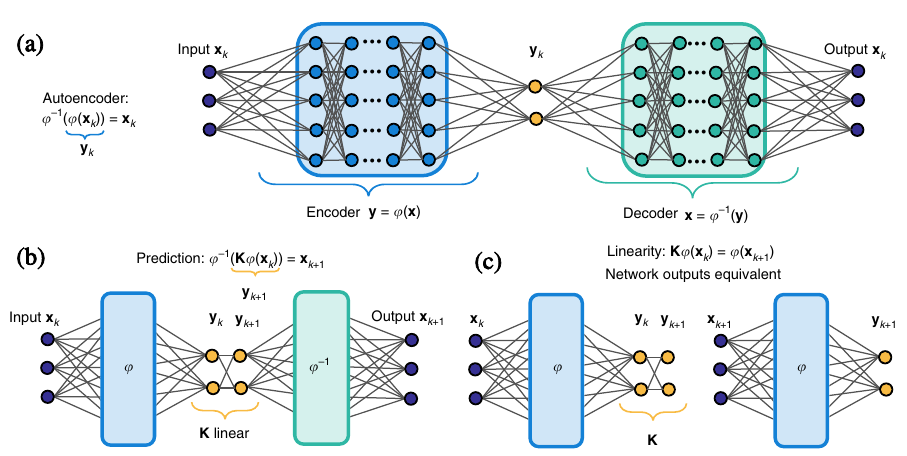} 
	\caption{
		Schematic illustration of a deep learning framework for identifying Koopman eigenfunctions \(\varphi(\mathbf{x})\).  
		\textbf{(a)} The network is based on a deep autoencoder that identifies intrinsic coordinates \(\mathbf{y} = \varphi(\mathbf{x})\) and decodes them to reconstruct the original state \(\mathbf{x} = \varphi^{-1}(\mathbf{y})\).  
		\textbf{(b), (c)} An additional loss function is introduced to learn a linear Koopman operator \(\mathbf{K}\) that advances the intrinsic variables \(\mathbf{y}\) forward in time. In practice, the operator is constrained to remain consistent with trajectory data through multiple dynamical iterations, i.e., \(\mathbf{K}^m\). In \textbf{(b)}, the loss function is evaluated in the original state space \(\mathbf{x}\), while in \textbf{(c)} it is evaluated in the intrinsic space \(\mathbf{y}\). \\
		\textit{Source}: The figure is reproduced from Ref. \cite{lusch2018deep}.
	}
	\label{koopman}
\end{figure}

Lusch \textit{et al.} \cite{lusch2018deep} proposed a forecasting model that couples a deep autoencoder with the Koopman operator. By learning bidirectional mappings between the nonlinear state variables \(\mathbf{x}\) and the embedded-space variables \(\mathbf{y} = \varphi(\mathbf{x})\), the system evolves approximately linearly in the \(\mathbf{y}\)-space as \(\mathbf{y}_{t+1} = \mathbf{K} \mathbf{y}_t\), thereby effecting a linearization of the nonlinear system. Figure~\ref{koopman} shows the core structure: the encoder–decoder framework captures the embedding relations, while the Koopman model performs the time advancement. Compared with conventional neural-network models, this approach offers greater stability for multi-step evolution modeling and is particularly suitable for dynamical systems with modal structure or periodic behavior.

Raissi \textit{et al.} \cite{raissi2019physics} proposed physics-informed neural networks (PINNs), which embed the system's governing differential equations into the neural network's loss function to guide the model toward solutions that satisfy physical constraints. Specifically, the network's predictions must simultaneously fit the observed data and minimize the residual terms of the system dynamics. This approach effectively mitigates issues such as "error drift" and "long-term instability" in conventional neural networks, and is particularly suitable when data are sparse but physical priors are strong, as in fluid mechanics, geologic simulation, and heat-conduction modeling.

\begin{figure}[!t]
	\centering
	\includegraphics[width=0.9\textwidth]{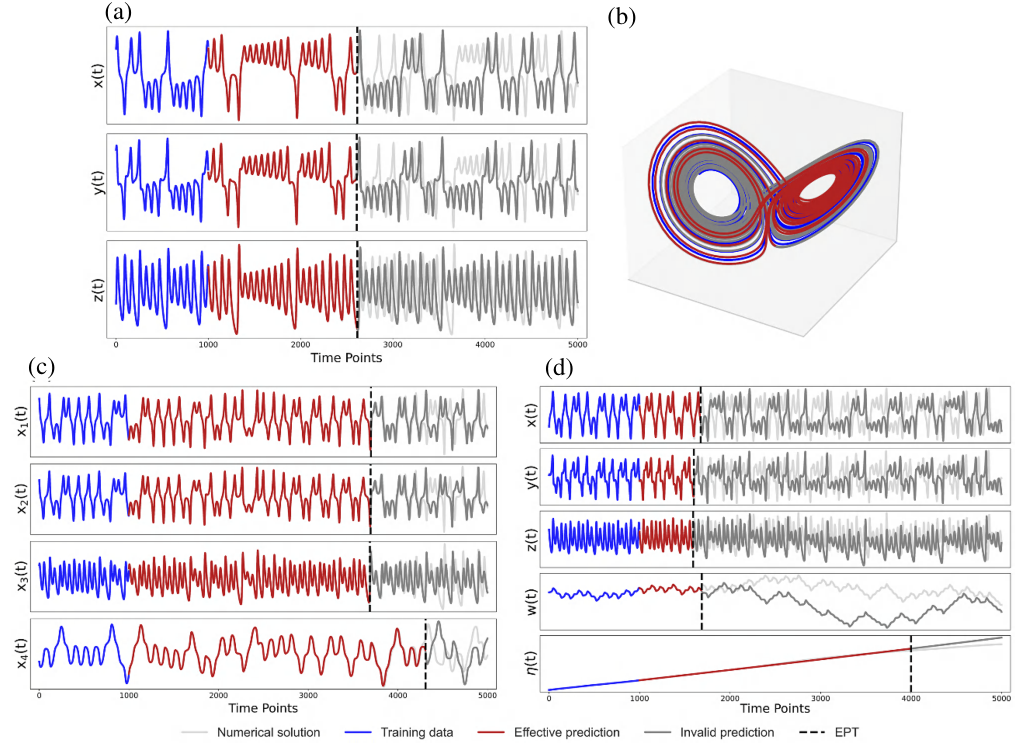} 
	\caption{
		Predictions of the DSDL model for three different chaotic dynamical systems. 
		(a) Forecast sequence for the Lorenz system. The light gray line indicates the numerical solution (true state), the blue line shows the training set, the red line represents valid predictions, and the dark gray line corresponds to invalid predictions in the test set. The vertical black dashed line indicates the effective prediction time (EPT). A training set of $10^4$ time points is used, and only the last $10^3$ points are shown in this panel. 
		(b) Predicted trajectory of the Lorenz attractor. 
		(c) Same as (a), applied to the hyperchaotic Lorenz system. 
		(d) Same as (a), applied to a conceptual ocean–atmosphere coupled Lorenz system. \\
		\textit{Source}: The figure is reproduced from Ref. \cite{wang2024interpretable}.
	}
	\label{dsdl}
\end{figure}

When dealing with chaotic or strongly nonlinear systems, the DSDL framework proposed by Wang \textit{et al.} \cite{wang2024interpretable} further strengthens interpretability of predictability structures. The method reconstructs the system's phase space via delay embedding, extracts local attractor structures through multiscale representations, and then uses interpretable mechanisms to identify key regions where forecasts fail. Figure~\ref{dsdl} illustrates its application to canonical chaotic systems, including the Lorenz system and coupled climate models. DSDL introduces the "Effective Predictable Time (EPT)" as a metric of the system's predictability limit, enabling a direct linkage between the degradation of forecast accuracy and the geometry of the attractor, and providing a new pathway for understanding and controlling the limits of prediction in complex systems.

In addressing the forecasting challenges of chaotic and strongly nonlinear systems, Wang \textit{et al.} \cite{wang2024interpretable} proposed the Dynamical System Deep Learning (DSDL) framework, which combines nonlinear dynamical systems theory with deep-learning techniques to achieve interpretable, long-range accurate prediction of chaotic dynamical systems. DSDL reconstructs the phase space based on delay-embedding theory and, together with multilayer nonlinear network construction and key-variable selection, effectively captures and reduces the dimensionality of dynamical features, thereby markedly improving interpretability and predictive performance. Figure~\ref{dsdl} shows DSDL's application to canonical chaotic systems (e.g., the Lorenz system and its variants). The method introduces the “Effective Predictable Time (EPT)”, defined as the time until the model's prediction error first exceeds a prescribed error threshold, which quantifies the time scale over which accurate prediction can be maintained in practice under chaos. EPT reflects both the system's predictability and its dynamic response to local phase-space structure, thereby revealing the spatiotemporal heterogeneity underlying forecast-skill degradation. Compared with the “Average Predictability Time (APT),” EPT places greater emphasis on model-specific predictive performance, capturing the temporal boundary of effective prediction and its local dynamical characteristics, and thus provides a more explanatory and targeted perspective for understanding the predictive limits and failure mechanisms of chaotic systems.

Overall, this class of embedding-and-constraint methods reflects a paradigm shift from purely data-driven approaches toward structural modeling. Koopman embeddings provide an interpretable linear-evolution framework; PINNs fuse data with physical models; and DSDL reveals geometric mechanisms of prediction in nonlinear systems. From different angles, they strengthen a model's ability to delineate predictability bounds and provide theoretical and methodological support for long-term controllable modeling and risk assessment in complex systems.

\textbf{Pathways to Enhancing Predictability}. To systematically organize the pathways by which the above AI methods enhance the predictability of complex systems, we construct a comparative analysis along four key dimensions—forecast accuracy, extension of effective lead time, robustness to initial perturbations, and structural interpretability of predictive mechanisms—see Table~\ref{tab:ai_predictability}. It should be noted that this article does not focus on the specific construction and training procedures of forecasting models; rather, it emphasizes how these methods extend the boundaries of predictability at both the theoretical and system levels.

\begin{table}[htbp]
	\centering
	\caption{Comparison of representative AI methods for enhancing predictability.}
	\renewcommand{\arraystretch}{1.3}
	\begin{tabular}{p{2cm}p{4cm}p{2cm}p{2cm}cc}
		\hline
		\textbf{Method} & \textbf{Primary mechanism} & 
		\textbf{Forecast accuracy} & 
		\textbf{Lead-time extension} & 
		\textbf{Robustness} & 
		\textbf{Interpretability} \\
		\hline
		Reservoir Computing 
		& Local parallel modeling with information coupling 
		& \checkmark & \checkmark & \checkmark & \\
		\hline
		ARNN 
		& STI mapping driven by endogenous structure modeling 
		& \checkmark & \checkmark & \checkmark & \checkmark \\
		\hline
		Koopman Embedding 
		& Deep encoding with linear dynamical mapping 
		& \checkmark & \checkmark & & \checkmark \\
		\hline
		PINNs 
		& Physics-constrained neural networks 
		& \checkmark & \checkmark & \checkmark & \checkmark \\
		\hline
		DSDL 
		& Attractor reconstruction with multiscale structural interpretation 
		& \checkmark & \checkmark & \checkmark & \checkmark \\
		\hline
	\end{tabular}
	\label{tab:ai_predictability}
\end{table}

In summary, AI methods are evolving from instrumental tools for improving forecast performance into analyzers for identifying mechanisms of predictability. This shift not only strengthens the quantitative and theoretical understanding of the predictability of complex systems, but also lays a methodological foundation for further integrating model interpretability with uncertainty control. It is important to emphasize that, although forecast accuracy remains an important metric in engineering applications, in predictability research what is more critical is a model's ability to control the range of predictable lead times, its structured response mechanisms to sources of error, and its capacity to identify and represent the system's intrinsic predictable structures.

\subsubsection{Integrating with Information Bottleneck}

Notably, when the information-bottleneck objective is used as a neural-network loss, it can better quantify and capture the chaotic characteristics of dynamical systems \cite{carleo2019machine}. Murphy \textit{et al.} \cite{murphy2024machine} introduced the information bottleneck (IB) principle into a measurement-optimization framework for chaotic systems, enabling efficient extraction of trajectory information without requiring knowledge of the governing dynamics. Using mutual information as the criterion, the method maximizes the dependence between the measurement sequence $U_L$ and a reference state $X_{\text{ref}}$ while minimizing the per-step information cost of the measurements; the optimization objective is:
\begin{equation}
	L = -I(U_L; X_{\text{ref}}) + \beta \sum_{i=0}^{L-1} I(U_i; X_i)
\end{equation}
Here, \( I(U_L; X_{\text{ref}}) \) denotes the mutual information between the measurement sequence \( U_L \) and the reference state \( X_{\text{ref}} \), measuring the predictive power of the measurements for the reference state—the larger the mutual information, the more information the sequence contains about the reference. The term \( \sum_{i=0}^{L-1} I(U_i; X_i) \) represents the information cost per measurement step, i.e., the amount of information transmitted at each step; minimizing it enforces efficient information transmission at every measurement. The measurement process is parameterized by a neural network, and the approach is validated on the Ikeda, Hénon, and logistic maps, where it is found that even with only $1$ bit of information per step, the method can approach the KS entropy and capture nearly all dynamical information.

Murphy \textit{et al.} \cite{murphy2022characterizing} also investigated the double-pendulum system in a related study, focusing on the gradual loss of information with increasing forecast lead time in chaotic systems. Using the information-bottleneck framework, they systematically extracted from the current state the information most predictive of the future and quantified its decay. The work combines the variational information bottleneck (VIB) with the InfoNCE loss to train neural networks to identify optimal state representations, and further employs a distributed IB method to parse the relative information contributions of different state variables, thereby enhancing interpretability of the system dynamics while maintaining predictive performance. This approach not only reveals the intrinsic link between information loss and the Lyapunov exponent, but also demonstrates the unique advantages of machine learning in characterizing the informational structure of chaotic systems.

Beyond structural understanding of chaotic systems, the information bottleneck principle has also been applied to practical time-series forecasting tasks. For example, Xu and Fekri \cite{xu2018time} proposed a probabilistic time-series forecasting method based on the recurrent information bottleneck (RIB). Its core idea is to extract, within the IB framework, a stochastic latent-state representation that is maximally relevant to the future target while remaining highly compressible. Specifically, the input data are first encoded into latent variables \(Z_t\) that satisfy the Markov chain \(Y_t \leftrightarrow X_t \leftrightarrow Z_t\); the state then transitions through a recurrent neural network (RNN) to produce hidden states \(h_t\), and the future predictive distribution is finally decoded from \(Z_t\) and \(h_t\). Mathematically, the method maximizes the mutual information \(I(Z, Y)\) between the latent variables \(Z\) and the target \(Y\) while constraining their mutual information \(I(Z, X)\) with the input \(X\) to control representation complexity, leading to the optimization problem:
\begin{equation}
	\max_{\theta} I(Z, Y; \theta) \quad \text{s.t.} \quad I(X, Z; \theta) < \epsilon
\end{equation}
By introducing a Lagrange multiplier \(\beta\), this can be converted into the following objective:
\begin{equation}
	L_{IB}(\theta) = I(Z, Y; \theta) - \beta I(Z, X; \theta)
\end{equation}
Here, \(\beta\) controls the trade-off between the amount of information about the target task and the degree of compression with respect to the input. In empirical evaluations, the method shows strong performance on both univariate (e.g., the sunspot dataset) and multivariate (e.g., California traffic) time-series forecasting tasks, surpassing several advanced methods (e.g., DBN, SAE, LSTM, and MatFact) in predictive accuracy, and enabling uncertainty quantification by constructing confidence intervals from predictive variance. This work demonstrates the substantial potential of deeply integrating information-theoretic approaches with neural architectures to improve both performance and reliability in dynamical time-series forecasting.

\subsubsection{Challenges and Theoretical Explorations}

While current research largely focuses on improving forecast accuracy, the ability of AI models to reveal the intrinsic predictability of complex systems has not been systematically examined. Recent studies increasingly address structural bottlenecks in neural networks and theoretical lower bounds on forecasting errors, particularly under limited data or high system complexity, where long-term predictive capability remains severely constrained.

Existing research indicates that the predictive capability of AI models is limited by two main sources of error: (i) input uncertainty arising from perturbations in the initial conditions, and (ii) generalization error induced by model-structure bias. This partition corresponds to Lorenz's framework for classifying predictability in classical dynamical systems. Unlike traditional settings, however, neural networks rely heavily on data-driven modeling, and these two types of error often intertwine in practice, leading to systematic degradation of predictive skill. Pan and Duraisamy \cite{pan2018long} showed that, for feedforward neural networks (FNNs) applied to low-dimensional nonlinear systems, introducing a Jacobian-regularization strategy can mitigate error amplification to some extent, with performance surpassing SINDy. Yet in higher-dimensional systems that lack low-dimensional attractor structure (e.g., buoyancy-driven mixing flows), even relatively simple non-polynomial models are extremely sensitive to the coverage quality of training data, necessitating remedial measures such as multi-trajectory training or phase-space augmentation.

To systematically characterize how initial-condition and model-parameter errors affect the forecasting capability of AI models in complex systems, Mu \textit{et al.} \cite{mu2025predictability} proposed a nonlinear-optimization-based theoretical framework. This framework constructs key indices, such as the maximum predictable lead time, an upper bound on the maximum prediction error, and the maximum admissible error tolerance, to comprehensively delineate the predictability boundary of AI models. Specifically, the maximum predictable lead time is defined as the longest time over which the model's prediction error does not exceed a prescribed threshold. When initial-state and parameter perturbations are taken into account, a constrained optimization problem is formulated to estimate its lower bound, revealing forecast stability under multiple sources of uncertainty. In parallel, optimizing an upper bound on the maximum prediction error provides a quantitative standard for assessing worst-case performance under perturbations. The framework also determines the allowable maxima of initial and parameter errors to ensure that the prediction error remains within tolerance. Finally, regarding the relative importance of error sources, the study solves for the optimal initial condition that minimizes the prediction error and measures the discrepancy between the actual initial condition and this optimum, thereby quantifying the contribution of initial-condition error to overall predictive performance. These nonlinear optimization problems not only elucidate the mechanisms of error propagation and evolution but also provide theoretical support for model evaluation, data assimilation, and structural optimization of AI models.

To evaluate the theoretical performance limits of AI models in time-series forecasting, Marzen \textit{et al.}~\cite{marzen2024complexity} proposed an assessment framework calibrated by system complexity. They use a core tool from computational mechanics, the \(\epsilon\)-machine, to generate synthetic time series with controllable causal structure and redundancy for systematic testing of forecasting models. An \(\epsilon\)-machine is a state machine that satisfies unifilarity and provides a unique minimal representation of the causal mechanism of any stationary stochastic process. Compared with conventional hidden Markov models, its state transitions remain deterministic after each symbol emission, which maximizes information retention and predictive performance. Sequences generated from \(\epsilon\)-machines explicitly capture a system’s causal depth and provide analytical lower bounds on predictive error.

Specifically, any stationary stochastic process can be described by a unique \(\epsilon\)-machine, and its minimum average prediction error is given by:
\begin{equation}
	P_e^{\text{min}} = \sum_{\sigma} \left[ 1 - \max_x p(x|\sigma) \right] p(\sigma),
	\label{eq:pe_min}
\end{equation}
where \(\sigma\) denotes the system's current causal state, \(p(x|\sigma)\) is the conditional probability of emitting symbol \(x\) in state \(\sigma\), and \(p(\sigma)\) is the stationary distribution over causal states. This formula quantifies the average prediction error that remains unavoidable even under ideal conditions. Building on this, Fano's inequality provides an upper bound on the prediction error:
\begin{equation}
	S(X_0|S_0) \le S_b(P_e) + P_e \log(|\mathcal{A}| - 1),
\end{equation}
where \(X_0\) denotes the symbol to be predicted at the current step, \(S_0\) the current causal state, \(S(\cdot|\cdot)\) the conditional entropy, \(S_b(P_e) = -P_e \log P_e - (1 - P_e) \log(1 - P_e)\) the binary-entropy function, and \(|\mathcal{A}|\) the size of the alphabet (i.e., the number of possible output symbols). This inequality gives, from an information-theoretic perspective, an upper limit on the amount of uncertainty compatible with an error rate \(P_e\).
A lower bound follows from the definition of the entropy rate:
\begin{equation}
	S(X_0|S_0) \ge h_\mu = - \sum_\sigma p(\sigma) \sum_x p(x|\sigma) \log p(x|\sigma).
\end{equation}
Taken together, these results yield a complete set of theoretical metrics that delineate the attainable performance bounds of any model when confronted with complex sequences.
\begin{figure}[htbp]
	\centering
	\includegraphics[width=0.28\textwidth]{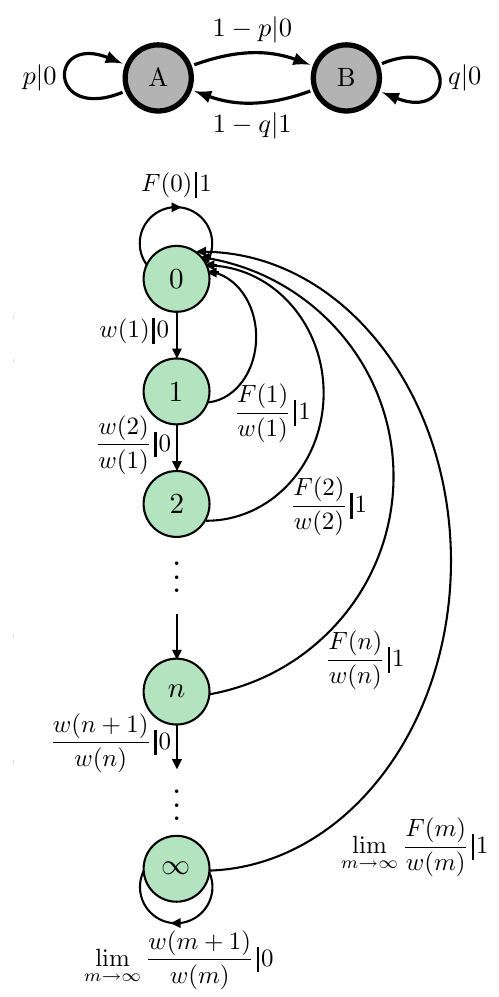}
	\caption{Non-unifilar hidden Markov model (top) and the corresponding \(\epsilon\)-machine (bottom). In the non-unifilar model, state ambiguity occurs after each symbol emission. In contrast, the \(\epsilon\)-machine ensures a one-to-one correspondence between symbol emissions and state transitions, providing optimal predictability. \\
		\textit{Source}: The figure is reproduced from Ref. \cite{marzen2024complexity}.}
	\label{fig:epsilon_machine}
\end{figure}

The core advantage of this framework is that it allows a direct comparison between the prediction error of actual neural-network models and the theoretical optimal error defined by the \(\epsilon\)-machine, thereby assessing whether a model has fully extracted and exploited the causal structure in the sequence. As shown in Fig.~\ref{fig:epsilon_machine}, conventional non-unifilar hidden Markov models exhibit state uncertainty after symbol emission. In contrast, the \(\epsilon\)-machine ensures that each state path is uniquely traceable due to structural determinism, achieving optimal prediction. Experimental results indicate that even state-of-the-art models, including next-generation reservoir computing (NG-RC) and LSTMs, produce prediction errors that remain substantially above the theoretical minimum when confronted with sequences exhibiting non-Markovianity or deep redundancy, often exceeding 50\%. Moreover, sequences with asynchrony, where the current state exerts a lagged influence on future symbols, further increase prediction errors. These findings suggest that current models have not yet fully captured the deep causal information present in complex time series.

In summary, the modeling capability of AI for the predictability of complex systems is gradually shifting from “black-box prediction” to “structural mechanism identification.” Looking ahead, research can advance along several lines: constructing a unified mapping between complexity and predictability; developing information-theoretic upper-bound theories for forecasting; incorporating error-propagation mechanisms and state-sensitivity modeling; and enhancing generalization robustness in high-dimensional, nonstationary systems. These efforts will lay the theoretical foundation for AI dynamical models that are more interpretable, stable, and task-adaptive.


\subsection{Predictability of Extreme Events}

\label{极端事件}

Despite the significant advances of AI-based dynamical models in improving forecasting capability for complex systems, their long-term performance remains constrained by several factors, including sensitivity to initial conditions, model-structure bias, and the ability to capture complex transition mechanisms in high-dimensional, nonstationary systems. Among these, the prediction of extreme events stands out, encapsulating the key challenges of predictability research. Extreme events typically manifest as low-probability, high-impact, highly abrupt dynamical transitions and are widespread in real systems such as climate extremes, ecological collapses, and financial market crashes \cite{de2013predictability,lenton2011early,mukherji2013ideas,wang2012flickering}. Their formation mechanisms often involve multiscale coupling, nonlinear feedback, and jumps between attractors, yielding evolution paths that are highly non-unique and extremely sensitive to initial perturbations. Such phenomena display statistical characteristics of non-Gaussian tails, short-lived bursts, and high intensity, thereby violating the assumptions of traditional forecasting frameworks built on stationarity. Consequently, both long-term, time-averaged stability indices (e.g., Lyapunov exponents) and purely AI methods that rely on large-sample learning may fail when confronted with extreme events. It is therefore especially important to develop specialized predictability-analysis methods for extreme transitions.

Taking the catastrophic flooding over the Yangtze River basin in 2020 as an example, the event did not arise from a canonical El Niño process but was driven by a remote air–sea coupling mechanism triggered by the record-breaking Indian Ocean Dipole (IOD) event in 2019 \cite{zhou2021historic}. The key processes included modulation of the regional thermal structure by the propagation of oceanic Rossby waves, sustained Indian Ocean warming into the subsequent summer, strengthening of the East Asian upper-level jet, and the maintenance of an anomalous anticyclone. Although this evolution involved cross-seasonal, multiscale coupling, numerical weather prediction models had already captured its emerging trend by the spring of 2020, indicating that extreme events in high-dimensional systems can be predictable under certain conditions.

To deepen understanding and improve the forecasting of such extreme events, recent research has proposed a variety of approaches that can be grouped into three main pathways. The first exploits system structure, such as multiscale separation, for reduced-order modeling and precursor detection. The second extracts key unstable modes using instantaneous stability analysis. The third identifies the optimal perturbations that most readily trigger extreme responses, along with their evolution trajectories, through variational optimization. The following sections present a systematic review and analysis of these approaches.

\subsubsection{Reduced-Order Modeling and Precursor Detection}

In multiscale coupled systems, the complexity of high-dimensional dynamics often makes extreme-event prediction difficult, whereas appropriate reduced-order modeling strategies can extract the dominant evolutionary mechanisms and significantly enhance predictability. Franzke \textit{et al.} \cite{franzke2012predictability} systematically investigated the prediction of extreme events using a stochastic climate model with a clear separation of multiple time scales. The model contains slow and fast variables. The slow variables represent components that evolve slowly in the system and dominate the dynamics at observational scales; the fast variables interact with the slow variables via cubic nonlinear coupling terms (e.g., \(X^2 Y,\, X Y^2\)) and are accompanied by stochastic perturbations to mimic rapid atmospheric dynamics and uncertainty. Because the fast variables change rapidly and are difficult to observe directly, the researchers adopted a systematic dimensional-reduction strategy: by statistically averaging over the fast variables, their influence is rendered as multiplicative noise, yielding an effective low-dimensional model involving only the slow variables. On this reduced representation, they proposed an extreme-event prediction framework based on the distance to a precursor state: by compiling the distribution of states preceding historical extremes, a precursor center \(\mathbf{x}^*\) is identified, and an Euclidean-distance metric is defined as
\begin{equation}
	D = \|\mathbf{x}^{\text{pre}} - \mathbf{x}^*\|,
\end{equation}
where \(\mathbf{x}^{\text{pre}}\) denotes the system's current state at a given time. When \(D < \delta\) (with \(\delta\) a threshold parameter), an extreme event is deemed to have been predicted.

\begin{figure}[htbp]
	\centering
	\includegraphics[width=0.75\textwidth]{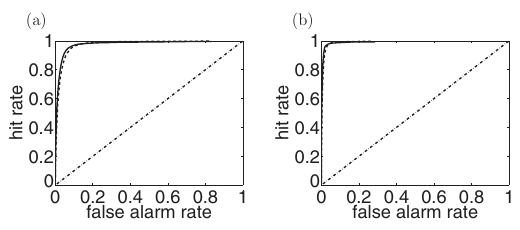}
	\caption{
		ROC characteristic curves for \(\varepsilon = 0.1\) and different prediction thresholds: (a) \(\mu = 5\), (b) \(\mu = 6\). Comparison between the full dynamics (solid line) and the reduced-order dynamics (dashed line) demonstrates the effectiveness of the reduced-order model in forecasting extreme events. \\
		\textit{Source}: The figure is reproduced from Ref. \cite{franzke2012predictability}.
	}
	\label{Jiangjie}
\end{figure}

To evaluate performance, the method employs the ROC curve to compare hit rate and false-alarm rate, without relying on the event-frequency distribution. The results show that, for both the full model and the reduced model, predictive skill is markedly better than random guessing; moreover, the more extreme the event, the easier it is to predict (Fig.~\ref{Jiangjie}). This indicates that, under multiscale coupling, appropriately designed reduced-order models can capture key precursor structures and thereby provide theoretical support for warning methods driven by real observational data. The study by Bódai \cite{bodai2015predictability} also supports this view: using the Lorenz-84 model to compare data-driven prediction (DDP) with model-driven prediction (MDP) for extreme events, they found that DDP can match MDP under ideal observational conditions, and that predictive capability actually increases with event extremity.

\subsubsection{Optimal Time-Dependent Mode}

To reveal the leading directions of instantaneous instability in nonstationary dynamical systems, Farazmand and Sapsis \cite{farazmand2016dynamical} proposed the Optimally Time-Dependent (OTD) modes. The method constructs a set of perturbation bases \(\{\mathbf{q}_i(t)\}_{i=1}^r\) that evolve in time and remain orthogonal, in order to capture the perturbation directions with the greatest growth potential at the current system state.

Specifically, consider the dynamical system
$
\frac{d\mathbf{u}}{dt} = \mathbf{F}(\mathbf{u}, t),
$
whose linearized operator along the trajectory \(\mathbf{u}(t)\) is
$
\mathbf{L}(t) = \left.\frac{\partial \mathbf{F}}{\partial \mathbf{u}}\right|_{\mathbf{u}(t)},
$
and the perturbation evolves according to the linearized equation
$
\frac{d\delta\mathbf{u}}{dt} = \mathbf{L}(t)\delta\mathbf{u}.
$
The evolution of the OTD modes \(\mathbf{q}_i(t)\) is given by
\begin{equation}
	\frac{d\mathbf{q}_i}{dt} = \mathbf{L}(t)\mathbf{q}_i - \sum_{j=1}^r \langle \mathbf{L}(t)\mathbf{q}_i, \mathbf{q}_j \rangle \mathbf{q}_j, \quad i=1,\dots,r,
\end{equation}
where \(\langle \cdot, \cdot \rangle\) denotes the inner product. This orthogonal-projection term ensures that the set \(\{\mathbf{q}_i(t)\}\) remains orthogonal during the evolution, i.e.,
\begin{equation}
	\langle \mathbf{q}_i(t), \mathbf{q}_j(t) \rangle = \delta_{ij}, \quad \forall t.
\end{equation}

\begin{figure}[htbp]
	\centering
	\includegraphics[width=0.7\textwidth]{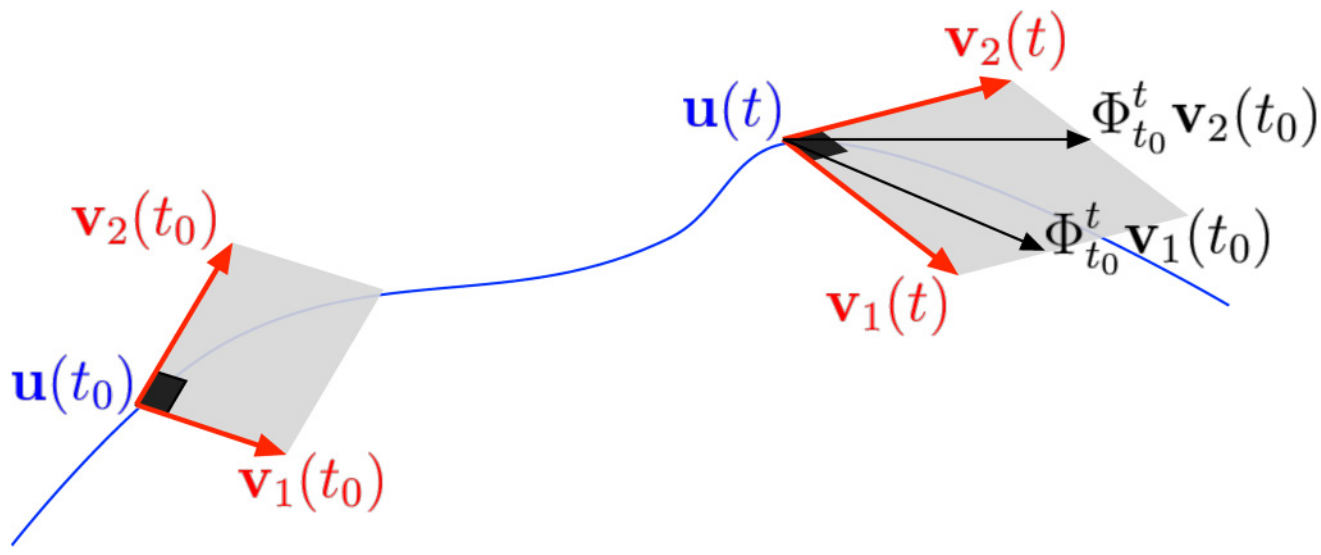}
	\caption{
		Schematic of OTD modes. The OTD mode \(\mathbf{v}_t\) remains orthogonal throughout its evolution, with right angles indicated by black squares. Although its mapping under the system's linear evolution operator \(\Phi_{t_0}^t\) does not coincide with itself, both span the same subspace, ensuring dynamic consistency of the perturbation space. \\
		\textit{Source}: The figure is reproduced from Ref. \cite{farazmand2016dynamical}.
	}
	\label{otd}
\end{figure}

Figure~\ref{otd} illustrates the core properties of the OTD modes. The OTD modes remain orthogonal during temporal evolution, and their subspace coincides with the image subspace under the system's linear evolution operator \(\Phi_{t_0}^t\), thereby enabling dynamic tracking of the system's instantaneous unstable directions. By analyzing the instantaneous growth-rate indicator of the OTD modes
\begin{equation}
	\sigma_i(t) = \frac{d}{dt} \|\mathbf{q}_i(t)\|,
\end{equation}
and in particular its maximum \(\sigma_{\max}(t) = \max_i \sigma_i(t)\), one can quantify the current strength of instability. For extreme-event forecasting, one can build a precursor subspace by combining OTD-mode features observed prior to historical extremes; by computing the similarity between the real-time modes and this precursor subspace, it is possible to determine whether the system has entered a precursor state of an extreme event and thereby issue early warnings. This approach effectively captures the instantaneous unstable structures of nonlinear high-dimensional dynamical systems, overcomes the limitations of traditional long-time–averaged stability indices, and provides a powerful tool for predictability analysis of extreme events in complex systems.

\subsubsection{Variational Optimization and Extreme-Trajectory Identification}

Farazmand and Sapsis~\cite{farazmand2017variational} proposed a variational framework for systematically probing the dynamical mechanisms underlying extreme events. The core idea of this approach is to characterize the occurrence of an extreme event as a finite-time optimization problem, in which one seeks, under physical constraints, the initial perturbation that maximizes the rapid growth of a relevant observable, such as the energy dissipation rate. Formally, the problem can be expressed as
\begin{equation}  
	\sup_{u_0 \in \mathcal{A}} \Big( I\big(\varphi(t_0+T; u_0)\big) - I\big(\varphi(t_0; u_0)\big) \Big),  
\end{equation}  
where \(\varphi(t; u_0)\) denotes the trajectory of the system starting from the initial state \(u_0\), \(I(\cdot)\) is the observable characterizing extreme events (e.g., energy input or dissipation rate), and the admissible set \(\mathcal{A}\) ensures that the perturbation remains near the system attractor and has a nonzero probability of occurrence. Solving this optimization problem allows the identification of the most explosively growing dynamical structures and their corresponding evolution pathways.

\begin{figure}[htbp]  
	\centering  
	\includegraphics[width=0.75\textwidth]{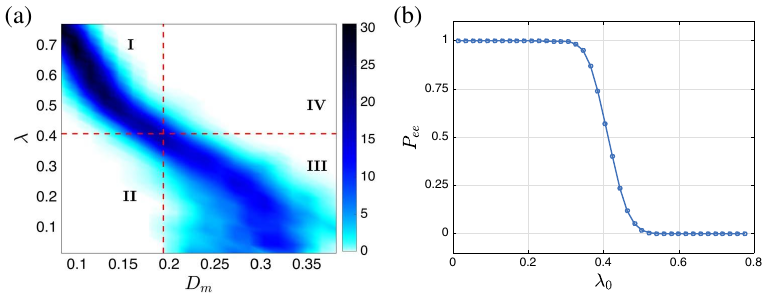}  
	\caption{  
		Prediction of extreme events.  
		(a) The vertical dashed line marks the extreme event threshold \(D_m \approx 0.194\). The horizontal dashed line marks \(\lambda = 0.4\). The quadrants correspond to: I, correct rejections; II, false positives; III, hits; and IV, false negatives.  
		(b) The probability of extreme events \(P_{ee}\) corresponding to the extreme event threshold \(D_e = 0.194\). \\
		\textit{Source}: The figure is reproduced from Ref. \cite{farazmand2017variational}.  
	}  
	\label{variation}  
\end{figure}

Applied to the two-dimensional Kolmogorov flow, the framework revealed that extreme dissipation events are triggered by a specific triad interaction \((0,k_f), (1,0), (1,k_f)\). In particular, the mode \((1,0)\) transfers energy to the dominant mode \((0,k_f)\) immediately prior to the extreme event, inducing a simultaneous surge in energy input and dissipation. Based on this mechanism, the mode amplitude \(l(t) = |a(1,0,t)|\) was proposed as a precursor indicator of extreme events. Statistical analysis demonstrates a pronounced nonlinear relationship between this indicator and the conditional probability of future extreme dissipation events: when \(l(t) < 0.3\), the system is almost certain to experience an extreme dissipation event shortly thereafter, whereas when \(l(t) > 0.5\), the probability of a future extreme event is close to zero. These results not only reveal the dynamical origin of extreme dissipation but also demonstrate the potential of the variational optimization framework to provide interpretable predictive indicators in complex turbulent systems (see Fig.~\ref{variation}).

Overall, this approach establishes a closed-loop workflow from dynamical mechanism identification and optimal perturbation computation to precursor indicator design and statistical validation, uncovering the predictability foundations of extreme events in high-dimensional turbulent systems. Its core principle, based on dynamical excitation modes, combines physical interpretability with computational feasibility, and possesses significant potential for generalization to other fluctuating and dissipative dynamical systems.

\subsubsection{Summary and Outlook}

The three approaches above provide distinct advantages for improving the predictability of extreme events in high-dimensional systems. Reduced-order modeling exploits a system's multiscale structure to extract slow variables and construct precursor indices. By retaining the main dynamical features while compressing dimensionality, it allows extreme events to display clearer precursor patterns in a low-dimensional space. This approach is well suited to systems with clear time-scale separation, incurs low computational cost, and can be easily combined with observational data for prediction, though it relies on stable statistical properties. Instantaneous-mode analysis identifies the currently most unstable perturbation directions in real time and captures local instabilities preceding critical transitions, making it particularly effective for high-dimensional systems with rapidly varying backgrounds and strong nonstationarity. This method provides strong responsiveness but is computationally more demanding and requires access to linearization information. Variational optimization directly targets the system's internal mechanisms to determine initial perturbations that most readily trigger extreme responses. The resulting predictive indicators often have clear physical meaning and high specificity, making this method suitable for systems with well-understood structure and known models, although its implementation is more involved and depends strongly on model accuracy. Overall, these methods excel in different respects, reflecting the inherently multiscale and multi-mechanism nature of extreme-event predictability. Future research could explore hybrid strategies, such as using instantaneous modes to aid dimensional reduction or to design optimized initial states, thereby enhancing the stability and broad applicability of predictions.

\subsection{Cross-Disciplinary Perspectives}

\label{系统}

Having established the theoretical frameworks and analytical methods for assessing predictability, we turn to recent progress in understanding the predictability of dynamics from a multidisciplinary perspective, encompassing climate systems, ecosystems, epidemiology, and neural dynamics. Across these diverse domains, despite differences in temporal and spatial scales or data modalities, researchers systematically leverage the governing laws of system evolution to characterize regularity, stochasticity, and the intrinsic limits of predictability. The overarching goal is to elucidate mechanisms that constrain forecastability, thereby providing both rigorous theoretical foundations and actionable insights for modeling and anticipating the behavior of complex systems.

In climate systems, Xu \textit{et al.} \cite{xu2020potential} showed that the warming trend may weaken the dynamical predictability of key climate variables. Based on analyses of CMIP5 simulations, the global mean potential precipitation predictability (PPP) exhibits a declining trend under high-emissions scenarios, especially pronounced in the tropics; this result is closely tied to changes in sea-surface-temperature (SST) dynamics and a weakening of its predictability. On geological time scales, Westerhold \textit{et al.} \cite{westerhold2020astronomically} constructed a high-resolution climate sequence spanning 66 million years, revealing four typical response states of the Cenozoic climate system to astronomical forcing and emphasizing the key regulatory role of polar ice-sheet changes in long-term climate dynamical predictability. At the theoretical level, DelSole and Tippett \cite{delsole2018predictability} extended predictability metrics from a linear framework to nonstationary systems, introduced the concept of “total climate predictability,” and systematically characterized the relationship between forced predictability and initial-condition predictability, thereby providing a unified analytical tool for describing climate dynamics under strong nonlinearity and nonstationary coupling.

The dynamical predictability of ecosystems likewise exhibits high complexity. Norden \textit{et al.} \cite{norden2015successional} found that tropical-forest successional trajectories diverge markedly even under similar initial conditions, indicating that the predictability of plant-community dynamics is constrained by the joint effects of an intrinsically high-dimensional state space and environmental perturbations. Approaching from a structural perspective, Luo \textit{et al.} \cite{luo2015predictability} argued that the compartmental structure of ecological processes and the flow pathways among carbon pools constitute structural constraints on an ecosystem's predictive capacity, emphasizing the need to identify highly predictable key variables to improve modeling effectiveness. Rego \textit{et al.} \cite{rego2018chaos} demonstrated a "chaos suppression" phenomenon in nonlinear eco–evolutionary systems: the introduction of external forcing can stabilize internal chaotic fluctuations and thereby enhance the predictability of long-term evolutionary trajectories. These results suggest that, in ecosystems, dynamical predictability is determined not only by intrinsic nonlinearity but also by the profound influence of exogenous disturbances and coupling mechanisms.

In a broader array of complex systems, research on dynamical predictability continues to expand its boundaries. For example, Blonder \textit{et al.} \cite{blonder2017predictability} showed in community ecological dynamics that system responses to climatic forcing often involve long lags and memory effects, posing a major challenge for predictive models. Scarpino and Petri \cite{scarpino2019predictability} used permutation-entropy analysis to quantify dynamical entropy barriers in the spread of infectious diseases and emphasized the profound constraints that network heterogeneity and model structure impose on the upper limits of prediction. In neuroscience, Lau \textit{et al.} \cite{lau2022brain} evaluated multiscale complexity metrics of EEG signals as predictors of individual neural states, revealing a potential association between dynamical features and clinical status, while also noting that the field still faces weak theoretical underpinnings and a lack of unified measurement standards.

In summary, research on the predictability of dynamics in complex settings has revealed several shared mechanisms: nonlinear evolutionary laws, structural constraints, exogenous forcing, and observational uncertainty jointly determine the upper bound of forecasting skill. At the same time, studies show that dynamical characteristics across different systems (e.g., stability, chaos, lagged responses) display diversity and specificity in their predictability, prompting a methodological shift from uniform metrics toward system-tailored approaches. Looking ahead, integrating dynamical systems theory, information-theoretic measures, and machine-learning techniques is expected to further expand both the research frontier and the practical capabilities of predictability in complex systems.

\section{Applications}

\subsection{Human Mobility}
Among the various domains of human behavioral predictability, human mobility stands out as one of the most active and extensively studied areas. It has yielded numerous foundational theoretical advances and spurred diverse interdisciplinary applications. Song \textit{et al.} \cite{song2010limits} were the first to establish a framework for deriving the maximum predictability of human movement by combining entropy measures with the Fano inequality, using large-scale mobile communication data. Their findings revealed that, despite substantial individual differences in mobility patterns, the theoretical upper bound of predictability remains remarkably high (approximately $93\%$), indicating that human mobility is far from random and instead governed by highly structured regularities. This finding established a theoretical benchmark for subsequent research. However, Lin \textit{et al.} \cite{lin2012predictability}, analyzing high-resolution positioning data, demonstrated that the upper bound is not a fixed constant but varies significantly with spatial and temporal sampling granularity: at finer resolutions, predictive accuracy declines markedly. This highlights the sensitivity of maximum predictability to the choice of spatial and temporal scale. Furthermore, Cuttone \textit{et al.} \cite{cuttone2018understanding} introduced an exploration-return framework to explain the behavioral mechanisms underlying these predictability limits. They observed that individuals consistently exhibit a tendency to explore new locations--with approximately $20-25\%$ of trips directed to previously unvisited places--while simultaneously developing stable visitation patterns centered on a small set of familiar locations. This dynamic balance between novelty exploration and habitual return fundamentally shapes the upper bound of long-term mobility predictability, suggesting that gains in predictive performance are inherently constrained by underlying behavioral mechanisms.

Building on these works, subsequent studies have further explored how individual behavioral traits and social relationships jointly influence the predictability of mobility. Teixeira \textit{et al.} \cite{teixeira2019deciphering} refined the entropy-based framework of Song \textit{et al.} \cite{song2010limits} by introducing two intuitive metrics--stationarity and regularity--to interpret inter-individual differences in predictability. The former quantifies the short-term persistence of remaining in the same location, whereas the latter reflects the degree to which long-term activities are concentrated in a few habitual places. Analyses based on GPS and call detail record (CDR) data showed that these two indicators can explain between $76\%$ and $94\%$ of the observed entropy variation, thereby elucidating the behavioral roots of predictability differences through the dual lenses of temporal stability and spatial regularity. Extending this perspective to social relationships, Chen \textit{et al.} \cite{chen2022contrasting} compared how friends' trajectories and non-friends' co-location patterns enhance individual mobility predictability. Their results demonstrated that friends' trajectories provide supplementary information beyond an individual's own behavioral history, boosting model accuracy to levels approaching the theoretical upper bound of predictability, whereas non-friends yield only marginal improvements. Further analyses revealed that these gains primarily depend on the overlap ratio of visited locations, indicating that mobility predictability is not merely an individual attribute but is deeply embedded in social relational structures.

In studies of human mobility predictability, external contextual factor and data environments have been shown to profoundly shape the limits of prediction, beyond the influence of individual behavioral patterns. Human movement rarely occurs in isolation-it is inherently embedded in specific temporal, spatial, and situational contexts. Zhang \textit{et al.} \cite{zhang2022beyond} proposed a context-transition predictability framework that integrates contextual information such as time segments, location categories, and activity types into entropy-based modeling. Applying their method to check-in data from New York and Tokyo, they quantified the transition probabilities between different contextual states, thereby significantly reducing the entropy of raw mobility sequences. For example, under the composite context of “morning-residential area-commuting,” individual mobility patterns exhibit greater stability and compressibility. Empirical results demonstrate that incorporating contextual constraints markedly increases the maximum predictability of individuals and maintains strong robustness in cross-city prediction tasks. This finding suggests that the structural regularities of external environments can substantially compensate for the limitations of single-sequence trajectories, offering a promising pathway for modeling mobility predictability across regions and scenarios. Meanwhile, Teixeira \textit{et al.} \cite{teixeira2018predictability} highlighted the challenges posed by heterogeneous external conditions. By comparing check-in data from Instagram and Foursquare, they found substantial differences in estimated predictability: Instagram tends to capture leisure, tourism, and irregular activities, while Foursquare reflects more routine and everyday behaviors. Simply merging these heterogeneous datasets does not improve predictability; on the contrary, it may degrade performance due to contextual inconsistency. This suggests that, rather than indiscriminately increasing data volume, it is more crucial to ensure contextual alignment across data sources; otherwise, inconsistent behavioral signals may compromise the stability and reliability of predictive models.

\begin{figure*}[!t]
	\centering
	\setlength{\abovecaptionskip}{0pt}
	\setlength{\belowcaptionskip}{0pt}
	\includegraphics[width=0.9\textwidth]{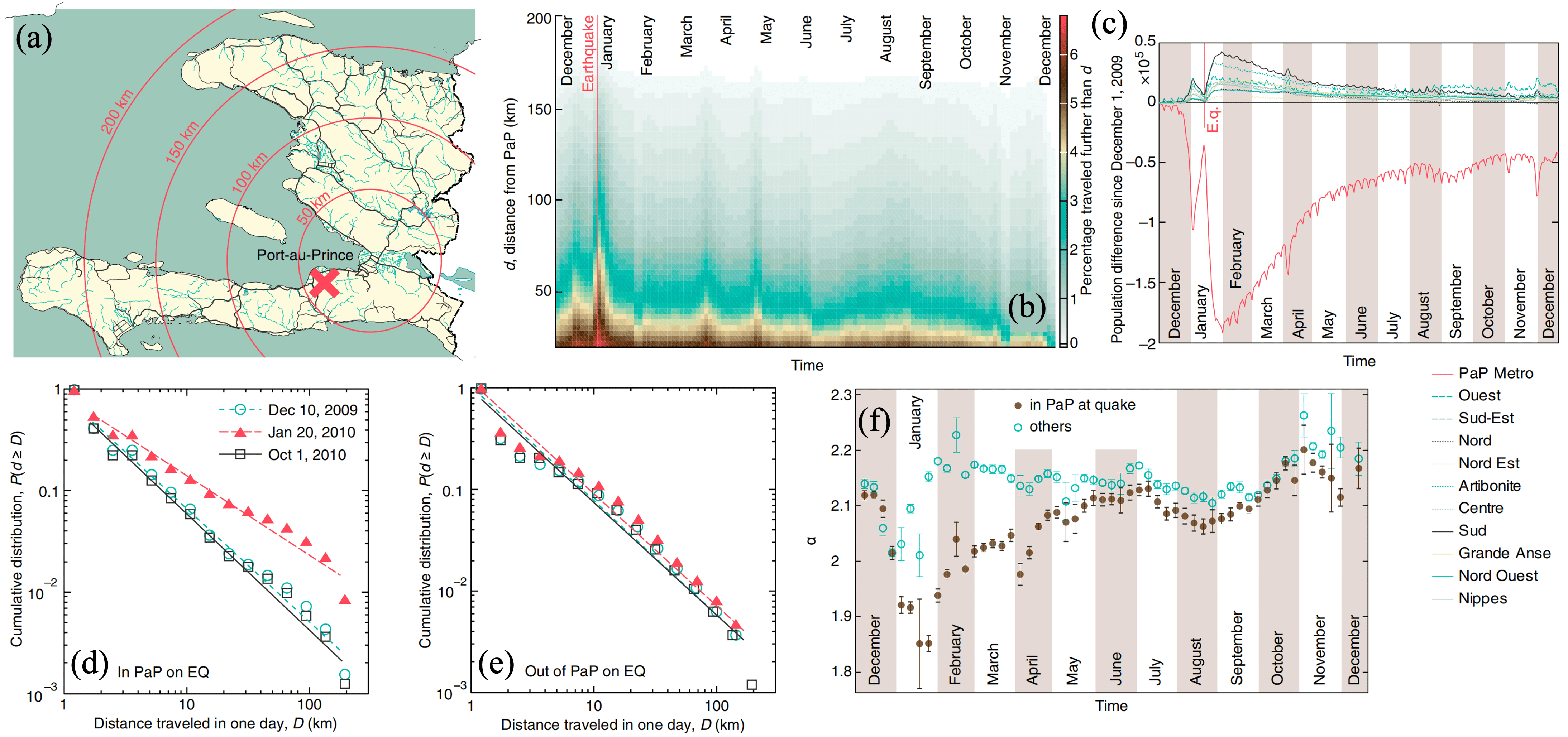} 
	\caption{Population mobility patterns following the 2010 Haiti earthquake.
		(a) Administrative map of Haiti and the location of Port-au-Prince (PaP); red circles indicate zones of different radii.
		(b) Distribution of daily travel distances, where color represents relative proportions; the vertical axis denotes distance from PaP, the horizontal axis denotes time, and the red line marks the moment of the earthquake.
		(c) Population change by province relative to the December 2009 baseline, showing post-disaster outward migration and gradual recovery.
		(d) Distribution of single-day travel distances for residents within PaP on the day of the earthquake.
		(e) Distribution of single-day travel distances for residents outside PaP on the same day.
		(f) Temporal evolution of the power-law exponent $\alpha$, reflecting how mobility patterns changed during the recovery process. \\
		\textit{Source}: The figure is reproduced from Ref. \cite{lu2012predictability}.}
	\label{fig:haiti_population_movement}
	\vspace{0pt}
\end{figure*}

Traditionally, natural disasters have been perceived as sudden events that disrupt social norms and introduce significant unpredictability into human behavior. Leveraging the unprecedented opportunity presented by the 2010 Haiti earthquake, Lu et al. \cite{lu2012predictability} conducted a systematic investigation into the predictability of human mobility during extreme events. Their analysis revealed that the disaster did not eradicate trajectory regularities; instead, it underscored the profound constraints imposed by social structures on behavioral patterns. In collaboration with Digicel--the largest mobile network operator in Haiti--they collected call detail records (CDRs) from approximately 1.9~million users, spanning 42~days before and 341~days after the earthquake, covering nationwide daily mobility trajectories (see Fig.~\ref{fig:haiti_population_movement}(a)). 
This large-scale, high-spatiotemporal-resolution longitudinal dataset enabled an in-depth analysis of post-disaster population movement patterns and their predictability boundaries. At the aggregate level, the study found that 19~days after the earthquake, the population of Port-au-Prince had decreased by~23\%, while the average daily travel distance increased significantly, exhibiting a short-term “fat-tailed” distribution with a higher proportion of long-distance trips (see Fig.~\ref{fig:haiti_population_movement}(b--d)). Nevertheless, this anomalous distribution gradually returned to a steady state within 2--3~months (Fig.~\ref{fig:haiti_population_movement}(f)), demonstrating strong resilience of population mobility systems. Even under extreme shocks, population movement was not entirely random but instead gradually reverted to the stable mobility patterns observed prior to the disaster.

\begin{figure*}[!t]
	\centering
	\setlength{\abovecaptionskip}{0pt}
	\setlength{\belowcaptionskip}{0pt}
	\includegraphics[width=0.8\textwidth]{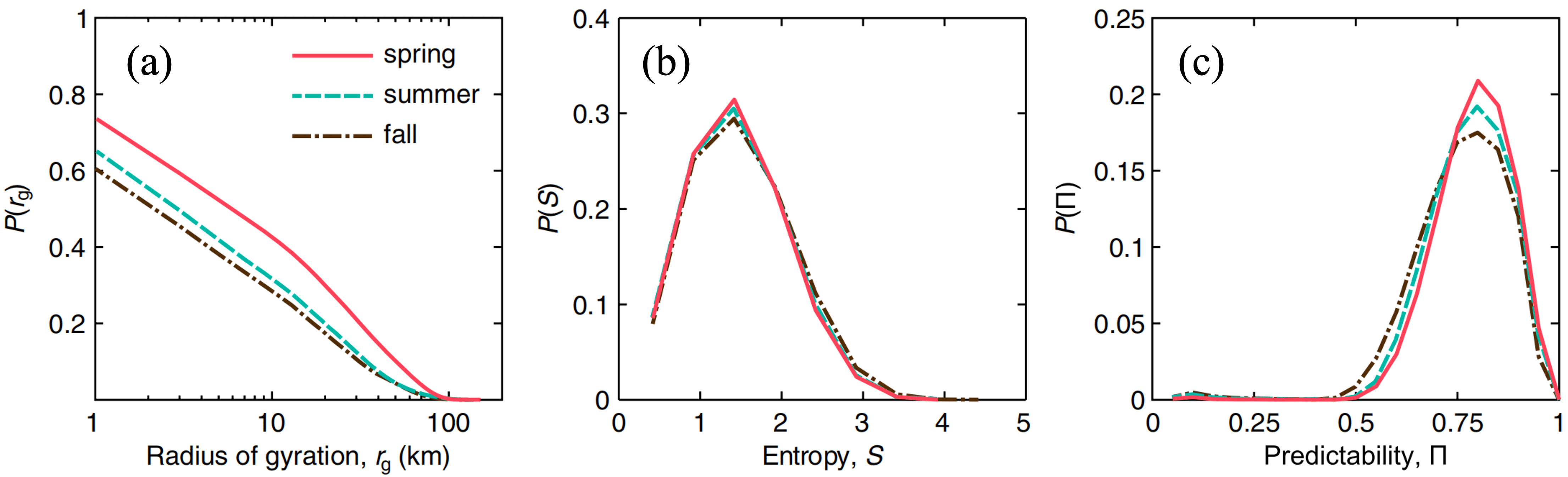} 
	\caption{Trajectory analysis and predictability assessment based on mobile communication data.
		(a) Cumulative distribution of users' radius of gyration $r_g$;
		(b) Distribution of entropy $S$;
		(c) Distribution of maximum predictability $\Pi$. \\
		\textit{Source}: The figure is reproduced from Ref. \cite{lu2012predictability}.}
	\label{fig:predictability_entropy_analysis}
	\vspace{0pt}
\end{figure*}

More importantly, the study evaluated individual-level predictability in the disaster context using indicators such as radius of gyration, entropy, and maximum predictability (see Fig.~\ref{fig:predictability_entropy_analysis}(a--c)). The results showed that individuals' maximum predictability remained at approximately~0.8 after the earthquake, even slightly higher than the pre-disaster baseline. This implies that, despite an expanded mobility range and increased heterogeneity, people's core activities remained concentrated in a limited set of locations--most of which overlapped with their pre-disaster residences or the locations of relatives and friends. In other words, historical mobility trajectories and social structures served as stable supports for maintaining predictability under disaster conditions. Further analyses of evacuation and return behaviors revealed that outward migration reached its peak three days after the earthquake. Notably, the destinations of this migration strongly overlapped with those visited during holidays such as Christmas and New~Year, a finding that highlights the decisive role of social ties in shaping evacuation behavior. Additionally, both the timing of return and duration of stay followed power-law distributions. This observation indicates that even in chaotic environments, migration behavior adhered to deep statistical regularities. Overall, this study makes an important contribution by demonstrating that disaster shocks do not fully randomize human mobility. On the contrary, individual trajectories retain strong regularity and predictability. Beyond extending the applicability of research on human mobility predictability, this work also provides empirical evidence to guide public health interventions, disaster response efforts, and resource allocation strategies. Specifically, rescue agencies can utilize communication data to rapidly identify post-disaster population flows and their scales, thereby enhancing the scientific accuracy and timeliness of emergency management practices.

\begin{figure*}[!t]
	\centering
	\setlength{\abovecaptionskip}{0pt}
	\setlength{\belowcaptionskip}{0pt}
	\includegraphics[width=0.98\textwidth]{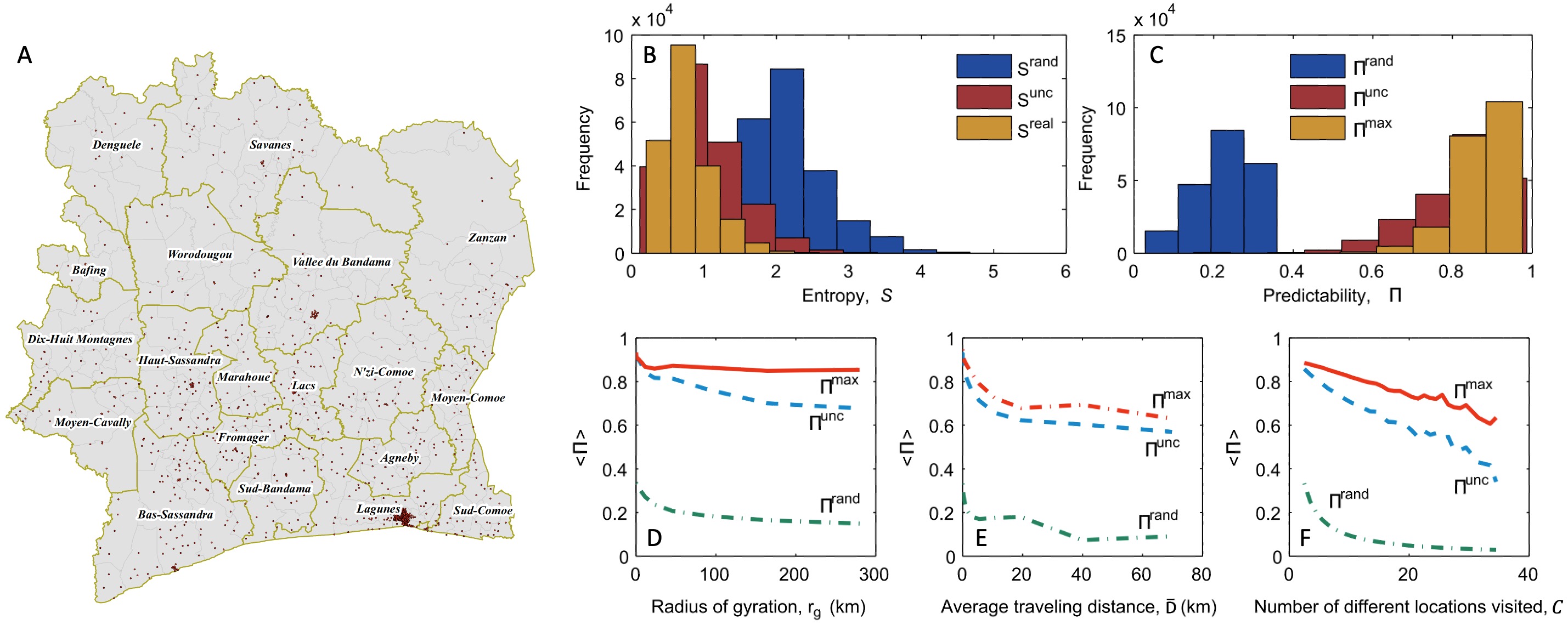} 
	\caption{Analysis of human mobility predictability in Côte d'Ivoire.
		(a) Administrative divisions and distribution of mobile communication towers, covering 237 sub-administrative regions;
		(b) Frequency distributions of $S^{\mathrm{rand}}$, $S^{\mathrm{unc}}$, and $S^{\mathrm{real}}$;
		(c) Frequency distributions of $\Pi^{\mathrm{rand}}$, $\Pi^{\mathrm{unc}}$, and $\Pi^{\max}$;
		(d) Relationship between radius of gyration $r_g$ and predictability $\Pi$;
		(e) Relationship between average travel distance $\bar{D}$ and $\Pi$;
		(f) Relationship between the number of visited locations $C$ and $\Pi$. \\
		\textit{Source}: The figure is reproduced from Ref. \cite{lu2013approachingTest}.}
	\label{fig:ci_entropy_predictability}
	\vspace{0pt}
\end{figure*}

Having established the existence of a high theoretical upper bound for human mobility predictability, a central question emerges: Can real predictive models practically approach this upper bound? 
To address this issue, Lu~\textit{et al.}~\cite{lu2013approachingTest} utilized mobile communication data from 500,000 users in Côte d'Ivoire, covering 237 sub-administrative regions nationwide (see Fig.~\ref{fig:ci_entropy_predictability}(a)), to systematically investigate the achievability of the theoretical limit under large-scale empirical conditions. They further used large-scale communication records. By integrating spatial information content, visit frequency distributions, and temporal dependencies, they derived and empirically validated the distribution of maximum predictability $\Pi^{\max}$ (Fig.~\ref{fig:ci_entropy_predictability}(b--c)). The results revealed that for most individuals, $\Pi^{\max}$ remained stable within the range of~0.85--0.9, with over~80\% of users exhibiting predictability above~0.8—significantly higher than that of random baselines. Further analyses uncovered the influence of spatial mobility features on predictability: an increased radius of gyration $r_g$ (up to several hundred kilometers) had almost no effect on $\Pi^{\max}$ (Fig.~\ref{fig:ci_entropy_predictability}(d)); a larger average travel distance $\bar{D}$ caused a slight decline (Fig.~\ref{fig:ci_entropy_predictability}(e)); and an increased number of visited locations $C$ led to an approximately linear reduction in predictability (Fig.~\ref{fig:ci_entropy_predictability}(f)). This underscores that exploratory behavior is a key factor weakening predictability.

When comparing theoretical limits with actual predictive models, Lu~\textit{et al.}~introduced Markov Chain (MC) models of different orders for systematic evaluation~\cite{lu2013approachingTest}. The findings indicated that a first-order Markov Chain, MC(1), can already capture the main dependencies in behavioral sequences, achieving prediction accuracy close to~93\% of the theoretical upper bound. 
Higher-order models (MC(2), MC(3), and beyond), although theoretically incorporating richer temporal information, yielded only marginal improvements in practice due to rapid state-space explosion and data sparsity--occasionally even leading to performance degradation. This suggests that approaching $\Pi^{\max}$ depends not on model order, but on effectively capturing repetitive and regular mobility patterns. In most cases, MC(1) is already sufficient to approximate the upper bound. Furthermore, the study examined the impact of temporal resolution on model performance. At the hourly prediction scale, although entropy and uncertainty were higher, the relative proximity of MC(1) to $\Pi^{\max}$ remained stable. At the daily scale, the model's performance improved slightly, but the differences were limited, demonstrating the robustness of the Markov Chain approximation framework across temporal scales. Further analysis combining spatial behavior features revealed that even when $\Pi^{\max}$ remained high under conditions of a large $C$ or long-distance travel, the actual predictive accuracy of MC models declined as exploratory behavior increased. In other words, although the models effectively captured repetitive and routine mobility, insufficient ability to predict exploratory locations significantly widened the gap between theoretical and empirical predictability.

\subsection{Digital Activities}
With the growing digitalization of daily life, individuals leave increasingly rich and fine-grained behavioral traces in online environments. Sinatra and Szell~\cite{sinatra2014entropy} analyzed eight months of comprehensive data from the large-scale multiplayer online game Pardus, which encompasses tens of thousands of users' mobility, social interaction, trading, and conflict behaviors. By employing entropy measures and the Fano inequality, they assessed the predictability of different types of actions. The results indicated that players' spatial mobility exhibited an upper bound of predictability close to 90\%, which is comparable to that observed in real-world human trajectories. In contrast, social, economic, and conflict interactions generally exhibited upper bounds of around 80\%, with noticeable variation across behavior categories. This study demonstrated that even in virtual societies, individuals' activities follow stable and regular patterns.

After confirming the overall high predictive potential of digital activities, Stavinova~\textit{et al.}~\cite{stavinova2021predictability} shifted the focus to individual heterogeneity. They proposed the concept of predictability classes: they used a long short-term memory (LSTM) model to predict customers' future banking transactions and group users based on their prediction error levels. Empirical results revealed that certain customers exhibit highly predictable transactional behavior with stable model performance, whereas others behave almost randomly, making them difficult to model. A key contribution of this approach is its ability to assess a user's predictability level prior to modeling, which in turn enables banks and e-commerce platforms to allocate modeling resources more strategically and avoid excessive computational and marketing costs for low-predictability customers.

Subsequently, Bezbochina~\textit{et al.}~\cite{bezbochina2022dynamic} further introduced the concept of predictability portraits, emphasizing that predictability is a dynamic attribute that evolves over time. 
Based on an incremental learning framework, they continuously tracked customers' predictability across different temporal windows and conducted an empirical analysis under the external shock of the COVID-19 pandemic. The findings showed that overall predictive performance declined during the pandemic. However, highly predictable groups recovered rapidly after short-term fluctuations, whereas low-predictability groups remained unstable over extended periods. 
These results suggest that predictability is not a fixed property; instead, it fluctuates with changes in external environments--providing valuable insights for financial risk identification and customer segmentation.

\begin{figure*}[!t]
	\centering
	\setlength{\abovecaptionskip}{0pt}
	\setlength{\belowcaptionskip}{0pt}
	\includegraphics[width=0.98\textwidth]{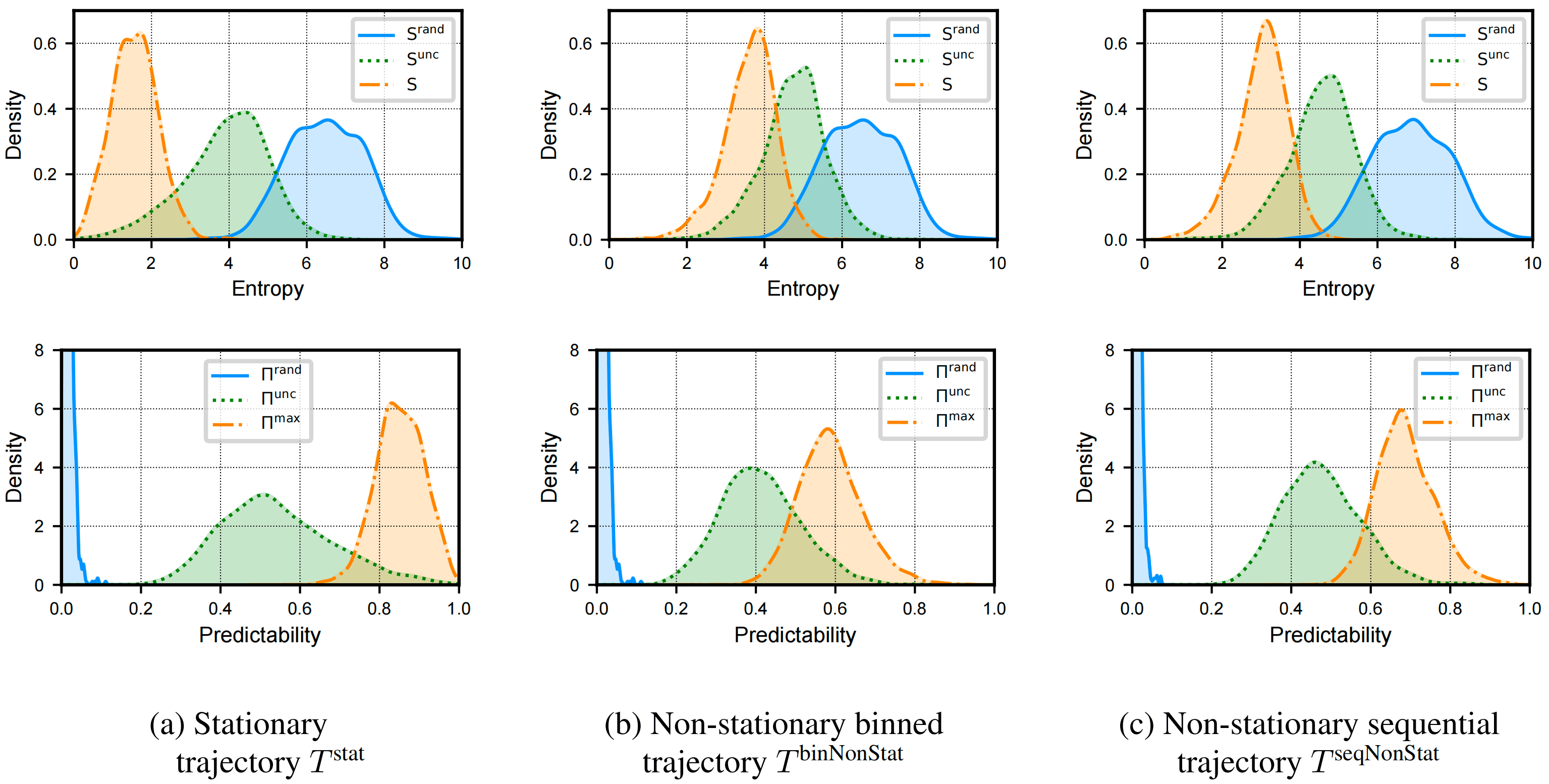} 
	\caption{Entropy and predictability distributions across different trajectory types.
		(a) Stationary trajectory ($T^{\mathrm{stat}}$);
		(b) Non-stationary binned trajectory ($T^{\mathrm{binNonStat}}$);
		(c) Non-stationary sequential trajectory ($T^{\mathrm{seqNonStat}}$).
		The top row shows the distributions of three entropy measures: random entropy $S^{\mathrm{rand}}$ (blue), temporally uncorrelated entropy $S^{\mathrm{unc}}$ (green), and real entropy $S$ (orange).
		The bottom row presents the distributions of three predictability measures: random predictability $\Pi^{\mathrm{rand}}$ (blue), temporally uncorrelated predictability $\Pi^{\mathrm{unc}}$ (green), and maximum predictability $\Pi^{\max}$ (orange).
		Results indicate that different trajectory modeling approaches yield substantial variations in the estimation of entropy and predictability. \\
		\textit{Source}: The figure is reproduced from Ref. \cite{kulshrestha2021web}.}
	\label{fig:trajectory_entropy_predictability}
	\vspace{0pt}
\end{figure*}

Kulshrestha~\textit{et al.}~\cite{kulshrestha2021web} extended the research framework of human mobility predictability to the online domain, raising a core question: Does web browsing behavior exhibit regularity and predictability similar to those of human mobility trajectories? To address this question, the authors proposed a method for constructing “digital trajectories,” which systematically disentangles temporal and sequential information to identify the true sources of predictability in online browsing behavior. Methodologically, three types of trajectories were designed (see Fig.~\ref{fig:trajectory_entropy_predictability}):  
(a) Stationary trajectory $T^{stat}$: The time axis is divided into fixed-length windows, and within each window, only the website on which the user spent the longest time is recorded--thus preserving both visit order and dwell rhythm; (b) Binned non-stationary trajectory $T^{binNonStat}$: This trajectory is also based on time-window partitioning, but only records whether a website was visited within a window, excluding dwell time information; (c) Sequential non-stationary trajectory $T^{seqNonStat}$: This trajectory is constructed by linking all visit events in their actual temporal order, with no time binning--thus retaining only the sequence of events. This decomposition allows researchers to precisely distinguish the contributions of “sequential inertia” and “temporal rhythm” to the predictability of web behavior. 
Subsequently, the authors employed multiple entropy measures to quantify uncertainty and applied the Fano inequality to convert these measures into an upper bound of predictability, $\Pi^{\max}$.

Empirical results showed that under the stationary trajectory (Fig.~\ref{fig:trajectory_entropy_predictability}(a)), $\Pi^{\max}$ reached 0.8-0.9--indicating that the combination of temporal order and dwell time significantly enhances predictability. Under the sequential non-stationary trajectory (Fig.~\ref{fig:trajectory_entropy_predictability}(c)), $\Pi^{\max}$ remained around 0.6-0.8, suggesting that even in the absence of dwell duration, sequential dependencies remain a strong source of predictability. In contrast, under the binned non-stationary trajectory (Fig.~\ref{fig:trajectory_entropy_predictability}(b)), predictability decreased significantly to 0.5-0.7--confirming the importance of rhythm information. In further analyses, $S^{\mathrm{unc}}$ denotes temporal-uncorrelated entropy, which measures user uncertainty when only visit frequencies are considered and temporal order is ignored. $S$ represents the temporal-correlated entropy, which is evaluated using the Lempel-Ziv algorithm~\cite{kontoyiannis2002nonparametric}. The difference between them, $S^{\mathrm{unc}} - S$, quantifies the reduction in uncertainty contributed by temporal information. Results revealed a strong positive correlation between this difference and $\Pi^{\max}$, indicating that the more repetitive and regular a user's web access pattern, the higher the predictability of that pattern. In robustness tests, the authors found that the upper bound of predictability is highly sensitive to data resolution: overly fine-grained temporal binning resulted in lower $\Pi^{\max}$ due to data sparsity, where aggregating URLs at the domain or category level significantly improved predictability. Furthermore, an analysis of 2,148 German users showed an overall average predictability of approximately 85\%, with substantial inter-individual variation. For instance, predictability levels were generally higher among female and younger users than among male and middle-aged users--driven by behavioral mechanisms such as more concentrated dwell times and lower interest diversity.

After establishing the measurement framework and completing individual-level clustering, researchers further explored the underlying causes of variability in the predictability of digital activity and the potential translation of this variability into value in applied contexts. 
Schnauber-Stockmann~\textit{et al.}~\cite{schnauber2023routines} approached this issue from a psychological perspective, aiming to unpack the generative mechanisms underlying individual differences in digital behavior predictability. They proposed a “routine activation-execution” model and validated it using large-scale web-tracking data. The study found that when users' daily activities are initiated at fixed time points (routine activation), the predictability of these activities increases significantly; in contrast, the specific execution patterns following activation (routine execution)--such as click paths or browsing order--have a relatively limited impact on predictability. This suggests that the predictability of digital behavior primarily reflects the regularity of when actions occur rather than how they are performed. Such a distinction provides a new explanatory perspective for understanding individual differences in behavioral predictability.

Meanwhile, Yeo~\textit{et al.}~\cite{yeo2018conversion} investigated the issue from the perspective of e-commerce conversion prediction and proposed a dual-layer market–individual framework. The core idea is to first estimate the overall conversion rate at the market level to capture macro-level trends, and then introduce a “customer predictability” metric at the individual level. This metric integrates users' interaction density and behavioral stability to assess whether personalized modeling is worthwhile. The results showed that highly predictable customers exhibit significantly higher conversion rates when exposed to personalized recommendations, whereas low-predictability customers perform better under aggregate market models--thus avoiding resource inefficiency. This framework explicitly demonstrates that predictability is not merely a passive statistical property; instead, it can also serve as a pre-modeling constraint, guiding resource allocation and improving marketing efficiency.

\subsection{Communications}
Large-scale empirical studies have shown that communication behavior and network usage patterns are not random; instead, they exhibit stable regularities. Cao~\textit{et al.}~\cite{cao2017human} analyzed three months of campus WLAN logs and found that the upper bound of predictability for users' access behavior across multiple temporal scales (minutes, hours, and days) generally remained between 70\%–90\%--reflecting pronounced temporal rhythms and spatial regularities. At the mechanistic level, Jensen~\textit{et al.}~\cite{jensen2010predictability} revealed, based on mobile phone call and contact data, that the selection of communication partners is simultaneously constrained by social relationships and geographic proximity. This coupling effect significantly reduces conditional entropy, thereby rendering communication sequences more predictable. From an engineering perspective, Song~\textit{et al.}~\cite{song2006predictability} evaluated temporal and spatial prediction methods using WLAN mobility data. Through bandwidth reservation simulations, they demonstrated that even limited prediction accuracy can significantly reduce VoIP call drop rates--thus validating the practical value of predictability in Quality of Service (QoS) assurance. Furthermore, Guo~\textit{et al.}~\cite{guo2021can} proposed a unified framework that directly compares real entropy measures with model-based predictive performance. Using large-scale communication and base-station load data, they found that the theoretical upper bound of predictability ranges from 0.6 to 0.9, whereas existing mainstream models still fall short of this limit. The study also emphasized the critical role of temporal granularity: coarser quantization intervals improve prediction accuracy but at the cost of information loss, thereby forming a trade-off between “complexity” and “accuracy”.

After confirming the high predictability of communication behavior, Zhang~\textit{et al.}~\cite{zhang2014analysis} further investigated the mechanisms underlying these regularities. Using three large-scale datasets (email, SMS, and phone call records), they constructed partner selection sequences and analyzed them systematically. They proposed that communication predictability primarily stems from two complementary mechanisms: intrinsic patterns and bursty interactions. The intrinsic patterns refer to stable partner transition relationships; for example, “after interacting with partner A, a user is more likely to interact next with partner B.” This sequential dependency among communication partners provides a structural constraint that enhances predictability. 
Bursty interactions, by contrast, refer to multiple consecutive communications with the same partner within a short time window (e.g., sending several text messages in succession). Such behavior markedly compresses uncertainty and thus facilitates prediction. The results revealed distinct dependencies on these two mechanisms across different communication modalities. SMS exhibited the highest predictability, which was almost entirely driven by bursty interactions; when bursts were removed, predictability declined sharply. Email communication, by contrast, relied primarily on intrinsic partner-transition patterns, with a weaker influence from bursts. Phone calls exhibited a combination of both mechanisms, although neither played as dominant a role as they did in SMS or email communications. Meanwhile, the number of communication partners had a direct impact on predictability: fewer partners led to lower sequence entropy and significantly higher predictability. Further bootstrap-based randomization tests confirmed the validity of these mechanisms. The results showed that real communication sequences consistently exhibited higher predictability than randomized baselines, with the largest gaps observed in SMS data. This finding indicates that bursty interactions are not statistical artifacts but rather key behavioral drivers underpinning the predictability of communication dynamics.

At the macro level, research on predictability has progressively expanded to various types of infrastructure networks, extending from individual communication sequences to the load and state sequences generated during system operations. In the context of network operations, Zhou~\textit{et al.}~\cite{zhou2012predictability} quantified cell-level traffic predictability using large-scale cellular base station records. The results showed that prior information such as time, spatial location, and service type can all significantly enhance prediction accuracy, and approximately the most recent 15~hours of historical data is sufficient to capture the main regularities. 
Within the same cell, voice and SMS traffic were found to be strongly correlated; furthermore, incorporating neighboring-cell traffic yielded additional predictive gains, with effectiveness varying by service type and time of day. In the context of network security, Chen~\textit{et al.}~\cite{chen2015spatiotemporal} analyzed attack traffic collected by honeypots. They constructed spatiotemporal transition sequences and computed their entropy. The results revealed significant differences in regularity between deterministic and random attacks, with the maximum predictability of attack frequency sequences exceeding 90\%. 
These findings provide methodological support for early warning and adaptive defense mechanisms against cyberattacks. In the context of sensor networks, Zhou~\textit{et al.}~\cite{zhou2017measuring} proposed the Flexible Multiscale Entropy (FMSE) measure, which addresses the instability of traditional multiscale entropy methods when analyzing short time series. When applied to multidimensional sensor data, FMSE demonstrated higher stability and discriminative power, allowing for a more accurate characterization of system-state complexity and predictability. This advancement provides a valuable new tool for infrastructure monitoring and anomaly detection.

\begin{figure*}[!t]
	\centering
	\setlength{\abovecaptionskip}{0pt}
	\setlength{\belowcaptionskip}{0pt}
	\begin{subfigure}[t]{0.48\textwidth}
		\centering
		\includegraphics[width=\textwidth]{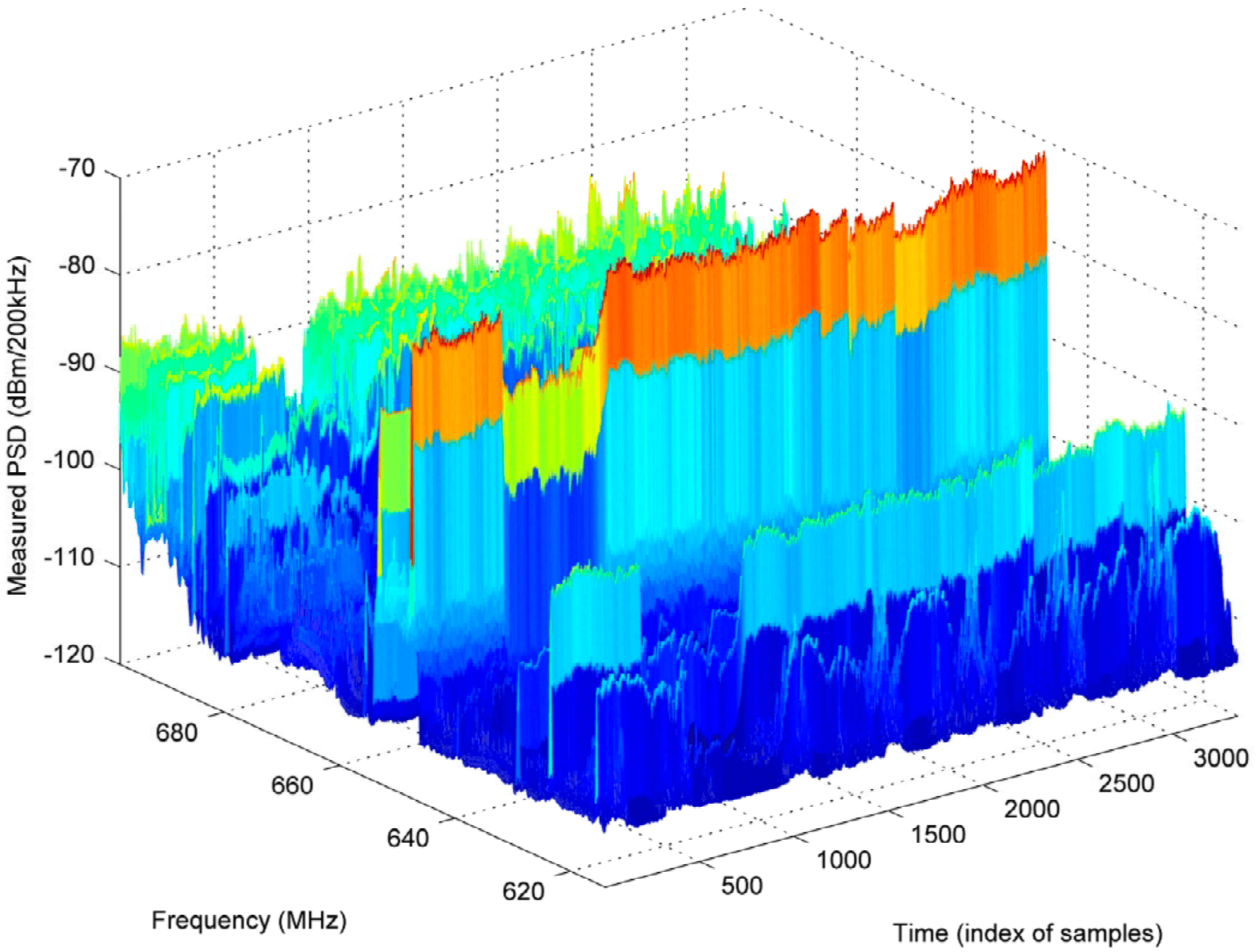}
		\label{fig:left}
	\end{subfigure}
	\hspace{0.02\textwidth} 
	\begin{subfigure}[t]{0.455\textwidth}
		\centering
		\includegraphics[width=\textwidth]{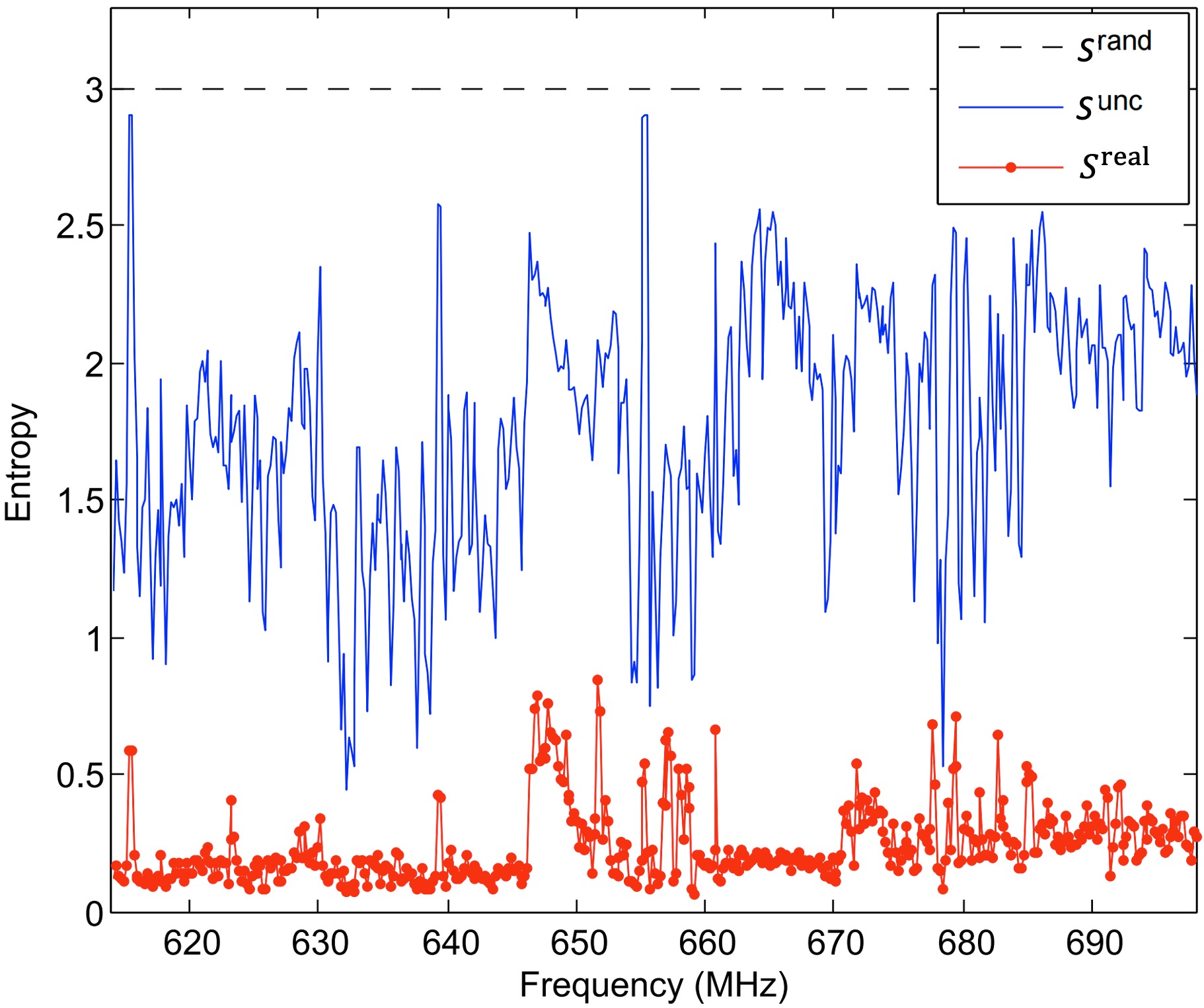}
		\label{fig:right}
	\end{subfigure}
	\caption{Evolution and entropy analysis of received signal strength (RSS) in television frequency bands.
		(Left) Three-dimensional evolution trajectories of measured RSS over one week (614–698 MHz). Each 200 kHz sub-band contains 3,360 samples, and approximately 480 samples are plotted per day, illustrating the dynamic variations in frequency–time–power spectral density.
		(Right) Entropy analysis of RSS dynamics, including random entropy $S^{\mathrm{rand}}$, temporally uncorrelated entropy $S^{\mathrm{unc}}$, and real entropy $S^{\mathrm{real}}$, where $S^{\mathrm{rand}} = \log_2(Q=8) = 3$ bits. \\
		\textit{Source}: The figure is reproduced from Ref. \cite{ding2015limits}.}
	\label{fig:rss_entropy_tvband}
\end{figure*}

Furthermore, researchers have turned their attention to the dynamic predictability of the radio frequency spectrum. Ding~\textit{et al.}~\cite{ding2015limits}, inspired by the entropy–Fano framework proposed by Song~\textit{et al.}~\cite{song2010limits}, developed an analytical framework to assess the predictability of the Radio Spectrum State (RSS). Using one week of measured data collected at RWTH Aachen University in Germany, the study covered multiple representative frequency bands, including TV, ISM, cellular (GSM uplink/downlink), and the 2.3\,GHz band. With a frequency resolution of 200\,kHz, Ding~\textit{et al.} constructed high-dimensional time–frequency power spectral density (PSD) representations to capture the dynamic characteristics of different frequency bands (see Fig.~\ref{fig:rss_entropy_tvband}, left). These three-dimensional evolutionary trajectories revealed the temporal–spectral–power dynamics of the RSS, forming the foundation for subsequent information-theoretic analysis.

Ding~\textit{et al.} employed the Fano inequality to estimate the theoretical upper limit of spectral evolution predictability and quantified the regularity of RSS dynamics by calculating the random entropy $S^{\mathrm{rand}}$, the temporally uncorrelated entropy $S^{\mathrm{unc}}$, and the real entropy $S^{\mathrm{real}}$ (see Fig.~\ref{fig:rss_entropy_tvband}, right). The results showed that RSS evolution is far from pure noise; instead, it exhibits clear regularities: across multiple frequency bands, the predictability upper bound $\Pi^{\max}$ consistently exceeded 90\%, which is substantially higher than the Gaussian noise baseline of approximately 80\%. 
Among these bands, the TV band showed the highest stability (minimum value $>0.88$), the ISM band ranked moderate, and the cellular downlink--due to its regular scheduling mechanisms--exhibited higher predictability. The cellular uplink was more random but still exhibited discernible patterns in certain sub-bands, whereas the 2.3\,GHz band showed the weakest regularity, with some sub-bands approaching noise levels. Further analysis revealed that $\Pi^{\max}$ is sensitive to both sampling intervals and state quantization schemes: overly fine quantization introduces noise and reduces predictability, whereas moderate temporal aggregation and state compression can effectively improve the upper bound.

\subsection{Recommender Systems}
Recommender systems are deeply integrated into daily life, capturing large volumes of user interaction data. 
These digital traces not only reflect individual preferences but also mirror broader patterns of economic activity. 
Therefore, investigating the predictability of these traces helps clarify users' habitual behaviors \revadd{and offers a micro-level perspective for observing regularities in economic activities}{5}. 
The entropy-Fano framework provides a rigorous theoretical basis for addressing this problem and has recently been extended to various recommendation tasks. Järvinen~\textit{et al.}~\cite{jarv2019predictability} introduced this framework into the sequential recommendation setting and analyzed the predictability limits of session-based next-item recommendation. Their study demonstrated that even amid complex and diverse interaction data, recommendation performance exhibits a distinct theoretical upper bound. Xu~\textit{et al.}~\cite{xu2023quantifying} pointed out that recommendation systems differ fundamentally from human mobility behaviors: 
while mobility trajectories typically involve a limited candidate set (on the order of hundreds), the number of item candidates in recommender systems can reach hundreds of millions. Additionally, users continuously explore new items, leading to a dynamically expanding candidate set. 
If the total number of items is directly adopted as the candidate size $C$, the resulting theoretical upper bound will be significantly overestimated. To address this issue, Xu~\textit{et al.}~\cite{xu2023quantifying} proposed that although purchase behavior is no longer constrained by geographical topology, it remains subject to logical constraints: specifically, users' selections are concentrated within a small subset of items relevant to their interests. 
Based on this insight, they further argued that when the accuracy of Top-$\ell$ recommendation saturates as $\ell$ increases, the $\ell$ value near the inflection point can be considered a reasonable estimate of the candidate set size. This effectively avoids significant biases in upper-bound estimation.

Xu~\textit{et al.}~\cite{xu2024limits} further extended the above framework to Top-$\ell$ recommendation by explicitly incorporating the Top-$\ell$ candidate probability distribution into the Fano scaling formulation. This extension not only offers a more compact theoretical representation but also avoids the systematic overestimation inherent in the traditional entropy-Fano approach. Under the traditional method, even highly stochastic behavioral sequences were often assigned unrealistically high upper bounds (i.e., above 0.5). 
In contrast, the new Top-$\ell$ extension produces more reasonable and empirically consistent limit estimations.

Furthermore, Xu~\textit{et al.}~\cite{xu2025upper} investigated the predictability of rating prediction tasks. This task presents two major challenges: (1) Historical behaviors are represented as (item, rating) pairs, whereas the prediction target focuses solely on ratings. This creates an asymmetry that violates the assumptions of the traditional framework; (2) Ratings are numerical rather than categorical, making Shannon entropy unsuitable for quantifying their uncertainty. 
To address these issues, Xu~\textit{et al.}~\cite{xu2025upper} proposed a new framework based on conditional entropy. This framework transforms historical behaviors into the entropy of “rating distributions given items” and integrates sample entropy and weighted entropy to capture both numerical characteristics and concentration tendencies of ratings.
Experimental results revealed substantial differences in the predictability upper bounds across datasets. In some datasets, such as Book-Crossing and Epinions, even the best-performing algorithms still exhibited a substantial gap from the theoretical upper bound. 
This indicates that predictability analysis can help identify datasets in which there remains considerable potential for performance improvement. Collectively, these studies outline a clear evolutionary trajectory: from validating the predictive limits of sequential recommendation, to introducing logical constraints and extending the framework to Top-$\ell$ recommendation, and finally to addressing numerical rating prediction. Throughout this process, the entropy-Fano framework has been continuously refined and expanded, progressively encompassing the major task types in recommender systems and forming a comprehensive analytical system for predicting upper bounds of predictability.

\begin{figure*}[!t]
	\centering
	\setlength{\abovecaptionskip}{0pt}
	\setlength{\belowcaptionskip}{0pt}
	\includegraphics[width=0.7\textwidth]{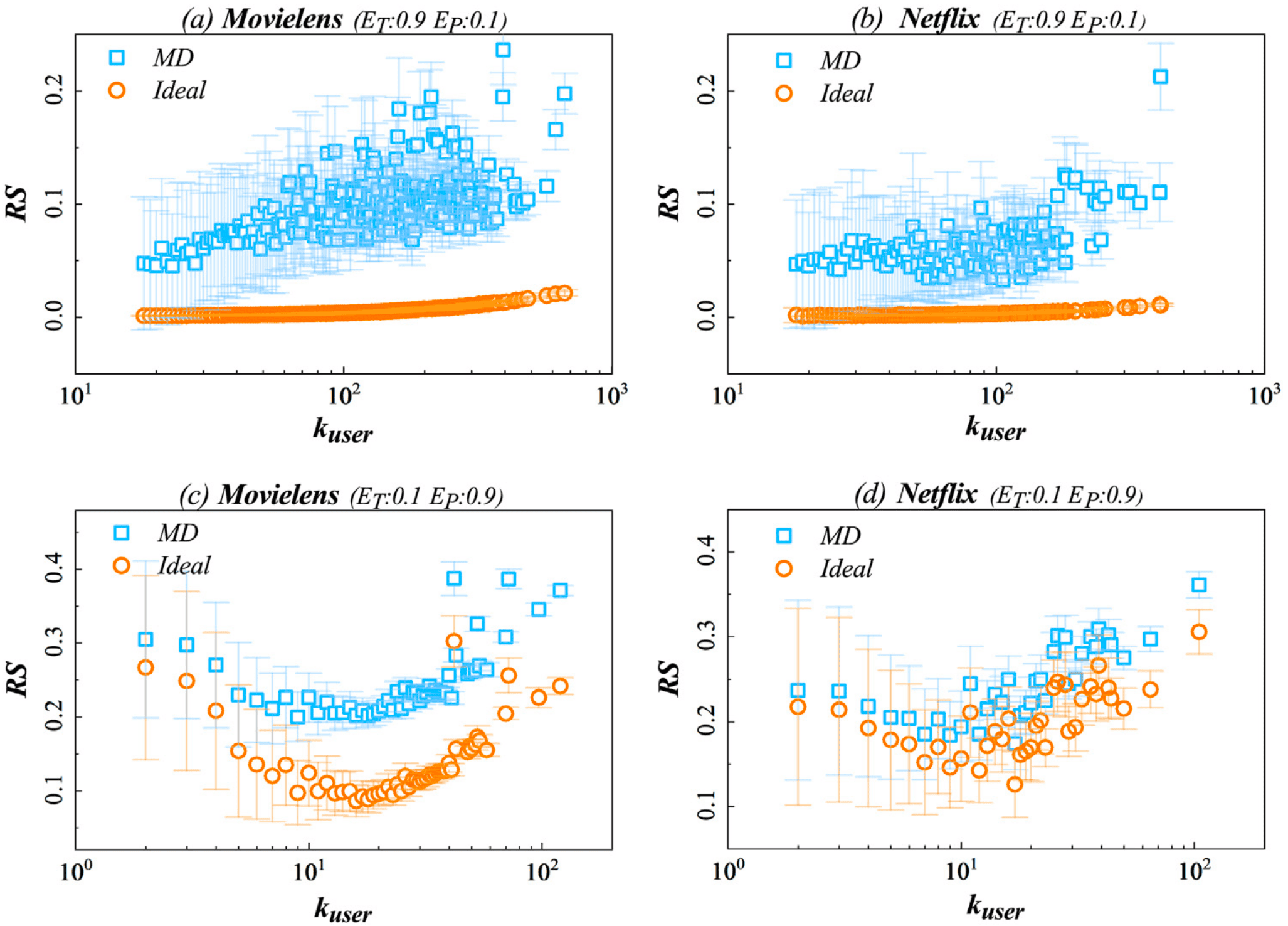} 
	\caption{Predictability analysis of diffusion-based recommender algorithms.
		The figure compares the performance of the ideal method (Ideal) and the mass diffusion algorithm (MD) in terms of ranking score (RS) under different user degrees ($k_{user}$).
		(a)(b) Results on the Movielens and Netflix datasets under dense-split conditions (training-to-test ratio $E_T : E_P = 0.9 : 0.1$).
		(c)(d) Results under sparse-split conditions ($E_T : E_P = 0.1 : 0.9$).
		The results show that in dense networks, MD exhibits a clear performance gap from the Ideal benchmark, leaving room for improvement; whereas in sparse networks, MD's performance approaches the theoretical upper bound. \\
		\textit{Source}: The figure is reproduced from Ref. \cite{zhang2019predictability}.}
	\label{fig:recommendation_predictability}
	\vspace{0pt}
\end{figure*}

Unlike studies based on the entropy-Fano framework, Zhang~\textit{et al.}~\cite{zhang2019predictability} quantitatively characterized the predictability of recommendation performance from the perspective of algorithmic mechanisms. Zhang~\textit{et al.} proposed the Ideal framework, which abstracts the process of diffusion-based recommendation algorithms (DBA) as resource propagation on a user–item bipartite network, and defines the theoretical performance upper bound under idealized assumptions. In the experiments, the effectiveness of recommendations was evaluated using the Ranking Score (RS) metric--originally proposed by Zhou~\textit{et al.}~\cite{zhou2007bipartite,zhou2008effect,zhou2009accurate}--which measures the average ranking position of target items in the recommendation list. The RS metric ranges from 0 to 1, where smaller values indicate better ranking performance. The comparative algorithm used was the Mass Diffusion (MD)~\cite{zhou2007bipartite,zhou2010solving}, representing one of the canonical paradigms of diffusion-based recommendation algorithms (DBAs). Figure~\ref{fig:recommendation_predictability} presents the RS results of Ideal framework and the MD method on the Movielens and Netflix datasets under different user degrees $k_{\mathrm{user}}$. Under dense-split conditions ($E_T:E_P=0.9:0.1$, Fig.~\ref{fig:recommendation_predictability}(a\&b)), the RS of the Ideal approaches zero, whereas the MD method exhibits significantly higher RS values. This indicates that diffusion-based algorithms still have room for improvement when sufficient observations are available. Under sparse-split conditions ($E_T:E_P=0.1:0.9$, Fig.~\ref{fig:recommendation_predictability}(c\&d)), the two curves converge, suggesting that MD performance approaches the theoretical upper limit. Overall, the performance of diffusion-based algorithms varies with data density: under dense partitions, a noticeable gap remains relative to the theoretical bound, whereas under sparse partitions, the results of Ideal and MD become much closer.

\subsection{Stock Price}
In financial market--a prototypical complex system--researchers have extensively explored the predictability of stock prices and volatility. 
On the one hand, some scholars, adhering to the Efficient Market Hypothesis, argue that price fluctuations are approximately random and thus inherently unpredictable. On the other hand, a growing body of empirical evidence suggests that despite the high volatility and nonlinear nature of financial markets, market behavior exhibits several robust statistical regularities. These include the power-law distribution of returns, volatility clustering (i.e., the tendency for large fluctuations to occur in succession), and long-range correlations in return sequences~\cite{mantegna1999introduction,chakraborti2011econophysics,hidalgo2021economic}. Fiedor~\cite{fiedor2014frequency} systematically compared high-frequency and daily returns using entropy rate estimation and the BDS test, finding that high-frequency data significantly deviate from the independent and identically distributed (i.i.d.) hypothesis, with short-period price series displaying stronger predictability. 
This result highlights the critical role of sampling frequency in shaping predictive potential, showing that at the second- or minute-level granularity, price sequences still retain short-term dependencies. 
Furthermore, Fiedor and Hołda~\cite{fiedor2016effects} examined bankruptcy events in the Warsaw Stock Exchange and found that extreme shocks to firms alter the price-formation process. This causes market predictability to exhibit dynamic fluctuations and forward-looking responses. These findings suggest that predictability itself is dynamic, often undergoing pronounced variations during major market events.

To address the limitations of single-frequency analysis, Zhao~\textit{et al.}~\cite{zhao2019quantifying} proposed the Multiscale Entropy Difference (MSED) method to systematically quantify predictive potential across price series ranging from tick-level to daily frequencies. 
Their results showed that short-term price movements behave more like random processes, whereas long-term sequences exhibit stronger regularities. At first glance, this appears contradictory to Fiedor's findings~\cite{fiedor2014frequency}; however, the discrepancy stems from differing perspectives: the former emphasizes local dependency structures observable at high-frequency granularity, whereas the latter focuses on global regularities across time scales. Together, these studies reveal a key insight: market predictability varies with both sampling resolution and temporal scale. Beyond stock prices, the predictability of volatility has also garnered significant attention. 
D'Amico~\textit{et al.}~\cite{d2019stock} integrated GARCH models with information entropy to develop a theoretical upper-bound framework for volatility prediction. Their findings showed that during periods of heightened volatility, the gap between model performance and theoretical limits widens substantially--indicating that predictive difficulty increases during unstable market phases. This provides theoretical support for understanding the predictive boundaries of financial crises and market turbulence.

In the latest empirical study from the Chinese market, Chen~\textit{et al.}~\cite{chen2024predictability} analyzed tick-level data (with a 3-second resolution) for 3,834 stocks from the Shanghai and Shenzhen exchanges, and used Lempel-Ziv compression entropy combined with the Fano inequality to derive the theoretical upper bound of price sequence predictability. The results showed that approximately 74\% of the stocks exhibited real entropy values below~2, which corresponds to theoretical predictive accuracies above~70\%. This indicates that at high-frequency granularity, most stock price sequences exhibit pronounced short-term memory and structured volatility characteristics. To evaluate the gap between theoretical limits and model performance, the authors employed two predictive approaches: a second-order Markov Chain and a Diffusion Kernel (DK) model. Price changes were discretized into intervals of varying granularity: at fine granularity (with a price step of $T=0.01$~CNY), the state space became denser and noisier, whereas at coarser granularity (with a price step of $T=0.05$~CNY), state transitions were smoother and volatility was partially averaged. 
Results revealed that both models achieved average accuracies of around~0.75 under fine granularity and up to~0.89 under coarse granularity. Although predictive accuracy improved significantly with coarser discretization, it remained consistently below the theoretical upper bounds derived from the entropy-Fano framework in all cases--with a stable gap of about~8-12~percentage points. Further analysis revealed that this gap widened substantially for high-priced, low-liquidity, and high-volatility stocks, while narrowing for low-priced, low-volatility stocks. Additionally, both average price and volatility were found to be significantly negatively correlated with predictive accuracy.

In the field of economic policy, Bekiros~\textit{et al.}~\cite{bekiros2016incorporating} incorporated Economic Policy Uncertainty (EPU) into a stock market prediction framework to examine its role under different market regimes. Using a nonlinear quantile regression approach, the study systematically compared predictive performance during high-volatility and stable market periods. The results demonstrated a strong regime-dependent effect of EPU: during turbulent market phases, policy uncertainty amplifies fluctuations in risk premiums and significantly reduces market predictability, whereas during stable periods, the predictive influence of EPU weakens or even becomes negligible. This finding suggests that the predictive boundaries of financial markets are shaped not only by model complexity but also by institutional transparency and policy stability. Thus, enhancing transparency and consistency in economic policy can strengthen market predictability at the macro level, providing institutional support for financial risk management and macroeconomic regulation.

\subsection{Economic Growth}

The theory of \textit{Economic Complexity} was first introduced by Hidalgo and Hausmann to explain the structural origins of long-term developmental disparities among nations. In their seminal work, Hidalgo~\textit{et al.}~\cite{hidalgo2007product} proposed the concept of the “Product Space,” modeling the global trade system as a bipartite network of countries and products. They revealed the path-dependent nature of industrial structure: countries are more likely to diversify into products similar to those they already possess the capabilities for, while struggling to leap to structurally distant, high-complexity industries. Subsequently, Hidalgo and Hausmann~\cite{hidalgo2009building} introduced the “Method of Reflections”, which recursively maps and computes the complexity levels of countries and products, establishing a quantitative framework for measuring the sophistication of production systems. Their findings demonstrated a strong correlation between economic complexity, national income, and future growth potential. Overall, these studies revealed that economic growth is constrained by structural path dependence and capability accumulation, providing a theoretical foundation for understanding the predictable patterns arising from the path dependence of economic development.

At the macro level of complex economic systems, Cristelli~\textit{et al.}~\cite{cristelli2013measuring,cristelli2015heterogeneous} proposed a representative dynamical framework to identify the boundaries of predictability in national economic growth trajectories. Unlike traditional approaches that rely on unified regression equations to explain economic growth, this framework emphasizes that economic development is not driven by a single mechanism but results from the joint influence of structural heterogeneity and cross-country historical trajectories. To this end, the authors mapped countries onto a two-dimensional plane of fitness (a measure of national productive capability) and per capita GDP, reconstructing the underlying evolutionary dynamics to establish the Selective Predictability Scheme (SPS).

Methodologically, the first step of the study defines a country–product bipartite matrix $M_{cp}$, which serves as the foundation for calculating indicators such as fitness ($F_c$) and product complexity ($Q_p$). An element $M_{cp}=1$ indicates that country $c$ has a comparative advantage in product $p$, and $M_{cp}=0$ otherwise. 
Comparative advantage is typically determined using the Revealed Comparative Advantage (RCA) index proposed by Balassa, defined as:
\begin{equation}
	\mathrm{RCA}_{cp} = 
	\frac{\displaystyle \frac{E_{cp}}{\sum_{p'} E_{cp'}}}
	{\displaystyle \frac{\sum_{c'} E_{c'p}}{\sum_{c',p'} E_{c'p'}}},
	\label{eq:rca}
\end{equation}
where $E_{cp}$ represents the export volume of product $p$ from country $c$. The numerator denotes the share of product $p$ in country $c$'s export structure, while the denominator denotes its overall share in global exports. If $\mathrm{RCA}_{cp}>1$, the country's export share in that product exceeds the world average, implying a comparative advantage; otherwise, it does not. The resulting bipartite matrix $M_{cp}$ captures the connections between countries and the products in which they hold competitive strength. Based on this matrix, Tacchella~\textit{et al.}~\cite{cristelli2015heterogeneous} proposed the following nonlinear iterative equations to compute country fitness ($F_c$) and product complexity ($Q_p$):
\begin{equation}
	F_c^{(n)} = \sum_p M_{cp} Q_p^{(n-1)}, \quad 
	Q_p^{(n)} = \frac{1}{\sum_c M_{cp} \frac{1}{F_c^{(n-1)}}}.
	\label{eq:fitness}
\end{equation}
In Eq.~\ref{eq:fitness}, $F_c$ denotes the fitness of country $c$, reflecting its overall industrial competitiveness and accumulated capabilities, while $Q_p$ denotes the complexity of product $p$, measuring the scarcity and sophistication of its required capabilities. 
The two quantities are mutually coupled through the country–product network $M_{cp}$, capturing the co-evolutionary relationship between countries' capabilities and product structures.

\begin{figure*}[!t]
	\centering
	\setlength{\abovecaptionskip}{0pt}
	\setlength{\belowcaptionskip}{0pt}
	\includegraphics[width=0.98\textwidth]{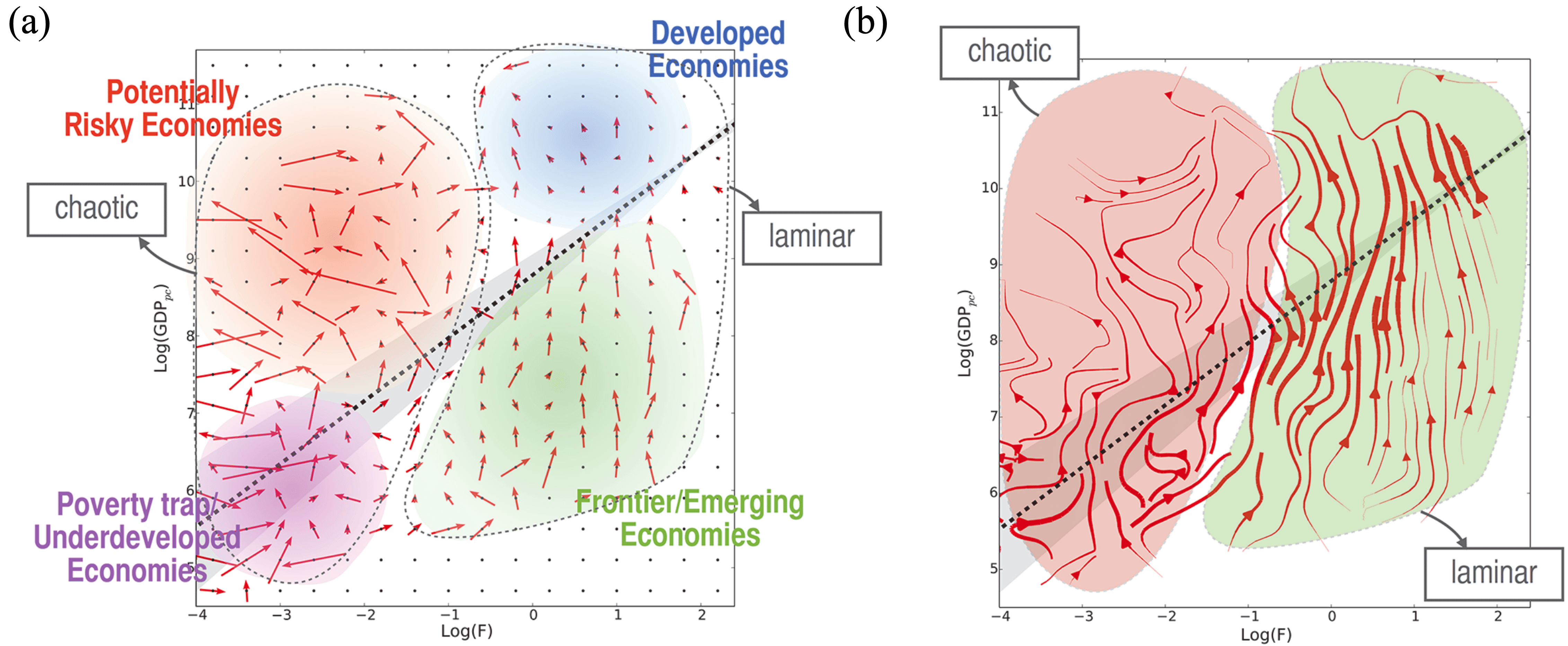} 
	\caption{Four evolutionary regimes of heterogeneous dynamics in economic complexity.
		(a) Coarse-grained dynamical analysis on the fitness–income plane, revealing two typical regions: a laminar regime, where national evolution is primarily driven by fitness and follows a highly predictable trajectory; and a chaotic regime, characterized by low predictability and complex, irregular dynamics. Within the laminar regime, distinct evolutionary paths can be observed for developed and emerging economies.
		(b) The interpolated dynamical flow field further illustrates the contrast in predictability between laminar and chaotic regimes, highlighting the heterogeneous and region-dependent nature of economic evolution. \\
		\textit{Source}: The figure is reproduced from Ref. \cite{cristelli2015heterogeneous}.}
	\label{fig:economic_complexity_regimes}
	\vspace{0pt}
\end{figure*}

$F_c^{(n)}$ indicates that a country's fitness depends on the set of products it is capable of producing and the complexity of those products. If $M_{cp}=1$, it means that country $c$ has the capability to produce product $p$; the more high-complexity products ($Q_p$) a country exports, the higher its $F_c$. This reflects the concept of capability accumulation—national economic potential arises from a diverse and advanced portfolio of industries. The update rule for $Q_p^{(n)}$ reveals the logic behind product complexity: a product's complexity is inversely related to the fitness of the countries that can produce it. 
If a product is produced by many low-fitness countries, the denominator term $\tfrac{1}{F_c}$ increases, causing $Q_p$ to decline significantly. Conversely, if only high-fitness countries can produce a product, its $Q_p$ remains high. This inverse relationship--"determined by the weakest producer"--embodies a nonlinear penalization mechanism that prevents ubiquitous products from being overestimated in complexity. Through iterative computation, the system gradually converges to a stable hierarchical structure: high-fitness countries and high-complexity products form the network core, while low-fitness countries and low-complexity products occupy the periphery. The resulting $F_c$ can thus be interpreted as the depth of a country's productive capability within the global economic system, whereas $Q_p$ characterizes the relative complexity of products in the international division of labor. Combined with per capita GDP $Y_t$, each country's state at time $t$ can be represented as a point $(F_t, Y_t)$, forming the fitness–income plane that captures the evolutionary dynamics of the global economic system.

The second step involves reconstructing the empirical dynamical vector field of national development. By tracking each country's trajectory in the $(F, Y)$ space over multiple years, the authors calculated the average direction and velocity of movement within local neighborhoods. 
The resulting vector field provides an empirical representation of a macrodynamical system, where the horizontal axis corresponds to the trend of fitness change and the vertical axis corresponds to the change in per capita income (see Fig.~\ref{fig:economic_complexity_regimes}(a)). Unlike traditional regression-based approaches, this method does not rely on a predefined functional form but instead extracts the dynamical laws of national development directly from the data. Building upon this, the authors proposed the Selective Predictability Scheme (SPS). The core idea draws inspiration from Lorenz's method of analogues: predicting a country's future trajectory by identifying historically similar cases. However, this method is only effective in low-dimensional and trajectory-concentrated regions (the laminar regime). In these regions, fitness and income exhibit a stable dynamical relationship, allowing for accurate forecasting of future directions, 
whereas in highly dispersed regions (the chaotic regime), historical analogues fail to provide meaningful guidance, rendering prediction inherently unreliable (Fig.~\ref{fig:economic_complexity_regimes}(b)).

Experimental results revealed that national development trajectories on the fitness–income plane exhibit four representative regimes (Fig.~\ref{fig:economic_complexity_regimes}(a)). In the laminar regime, trajectories are highly concentrated and exhibit stable trends, indicating strong predictability. This region includes both emerging economies (green area) that achieve sustained growth through industrial capability accumulation and developed countries (blue area) characterized by high income and high predictability. In contrast, trajectories in the chaotic regime are highly uncertain: one group consists of “poverty-trap” countries (purple area) that remain stagnated at low fitness and low income; another group comprises “resource-dependent” countries (red area) whose growth relies on resource exports and external shocks—although short-term surges may occur, their long-term trajectories fluctuate significantly, making prediction nearly impossible.

This categorization directly illustrates that economic growth predictability is not universal but contingent upon a country's position within the fitness–income space. In other words, predictability is structural and condition-dependent. For countries located in laminar regions, the method of historical analogues is effective and predictive potential is high; for those in chaotic regions, the method of historical analogies fails and prediction becomes infeasible. Overall, the study by Cristelli~\textit{et al.}~\cite{cristelli2015heterogeneous} represents the first systematic effort to place national economic growth trajectories within a complex-systems framework. By reconstructing the dynamics of the fitness–income plane, they proposed the Selective Predictability Scheme (SPS), revealing that the predictability of economic growth is not universal but depends on a country's structural position: in the laminar regime, developmental paths are stable and predictable, whereas in the chaotic regime, trajectories are dispersed and nearly unpredictable. This work not only transcends the limitations of traditional regression models but also provides new theoretical support for understanding the heterogeneous evolution of macroeconomic systems.

Tacchella~\textit{et al.}~\cite{tacchella2018dynamical} further developed the Selective Predictability Scheme (SPS) into an operational macroeconomic forecasting tool. They noted that traditional GDP forecasting models typically rely on numerous socioeconomic variables and empirical parameters, whereas within the framework of economic complexity, the primary dynamical features of macroeconomic growth can be effectively captured using only two key indicators: a country's economic fitness and its per capita GDP. To this end, they proposed the bootstrap-based SPSb model and introduced an improved Velocity-SPS version that incorporates historical growth trends to enhance prediction stability and accuracy. The experiments utilized long-term trade and GDP data from 169 countries and compared its performance across multiple five-year forecasting windows with the official projections of the International Monetary Fund (IMF). Results demonstrated that the mean absolute error of Velocity-SPS was approximately 25\% lower than that of the IMF forecasts, and the model exhibited greater robustness for highly complex economies. Moreover, the study found that the error distributions of SPS and IMF models were nearly independent, indicating that the two approaches capture complementary aspects of economic dynamics—suggesting that their combination could further improve overall forecasting accuracy.

Overall, the work~\cite{tacchella2018dynamical} not only validated the feasibility of GDP forecasting based on complexity metrics at the global scale but also revealed a consistent relationship between a country's predictability of growth and its level of economic complexity: 
high-complexity economies exhibit smoother growth trajectories and higher predictive accuracy, whereas low-complexity economies display greater uncertainty in their evolutionary paths. This study highlights the potential of complex-systems approaches in macroeconomic forecasting, offering a new theoretical and methodological paradigm beyond traditional regression-based models of economic growth.

\subsection{Spreading}

The works of Colizza~\textit{et al.}~\cite{colizza2006role,colizza2006modeling} are among the earliest and most influential explorations into the predictability of epidemic spreading, which introduced a macro-level modeling and quantitative framework for analyzing cross-regional disease transmission. Using the International Air Transport Association (IATA) global flight database and population data from thousands of cities, Colizza~\textit{et al.} constructed a multi-city coupled stochastic epidemic model that integrates local SIR dynamics \cite{HARKO2014184} and intercity travel fluxes to simulate the global spread of infectious diseases. Unlike traditional models that assume homogeneous random mixing among individuals, this framework explicitly incorporates the topology and traffic heterogeneity of the air transportation network, thereby enabling a more realistic reconstruction of the predictability of cross-border epidemic spread.  

Colizza~\textit{et al.} designed several quantitative indicators to measure propagation uncertainty:
\begin{itemize}
	\item Shannon entropy, which quantifies the dispersion of epidemic spread across multiple cities. 
	High entropy indicates a highly dispersed spread pattern and low predictability, whereas low entropy implies that infections propagate along a few dominant routes, corresponding to higher predictability.  
	\item Backbone overlap, which measures the consistency of key transmission pathways across multiple stochastic simulations. 
	If the same backbone routes consistently emerge, this indicates that the epidemic follows a stable core spreading structure.  
	\item Similarity overlap function, which evaluates the spatial overlap of infection distributions across multiple realizations, thereby assessing the overall predictability of epidemic outcomes.  
\end{itemize}

These indicators collectively reveal several key regularities.
First, the heterogeneity of the air transportation network constitutes the core source of epidemic predictability. In homogeneous random networks, epidemic spread is almost entirely stochastic and thus lacks long-term predictive value; however, in real-world aviation networks, even when traffic flows are artificially homogenized, epidemics still tend to propagate along a limited number of stable backbone routes. This finding indicates that topological structure--particularly the uneven degree distribution--determines the skeleton of global transmission.
Second, despite the inherently stochastic nature of individual transmission events, the epidemic exhibits a stable, backbone-like spreading pattern at the macroscopic level: key transmission pathways recur consistently across independent simulations, thereby substantially reducing overall uncertainty. Third, the initial outbreak location plays a decisive role in shaping predictability. When an epidemic originates from a major hub airport, the spread rapidly branches and diverges, leading to relatively low predictability; by contrast, outbreaks starting in secondary cities tend to propagate through only a few fixed routes, resulting in markedly higher predictability. 

These findings carry significant implications. On the theoretical side, they address a fundamental question: why highly stochastic individual transmissions can still exhibit regular patterns at the macroscopic level, demonstrating that the structural constraints of complex networks strongly compress the uncertainty of epidemic spread. On the practical side, the results suggest a more efficient containment strategy. Instead of imposing blanket travel restrictions, it is more effective to identify and intervene in the critical channels along the backbone transmission paths, thereby reducing the uncertainty of epidemic propagation while minimizing the associated social and economic costs.

\begin{figure*}[!t]
	\centering
	\setlength{\abovecaptionskip}{0pt}
	\setlength{\belowcaptionskip}{0pt}
	\includegraphics[width=0.8\textwidth]{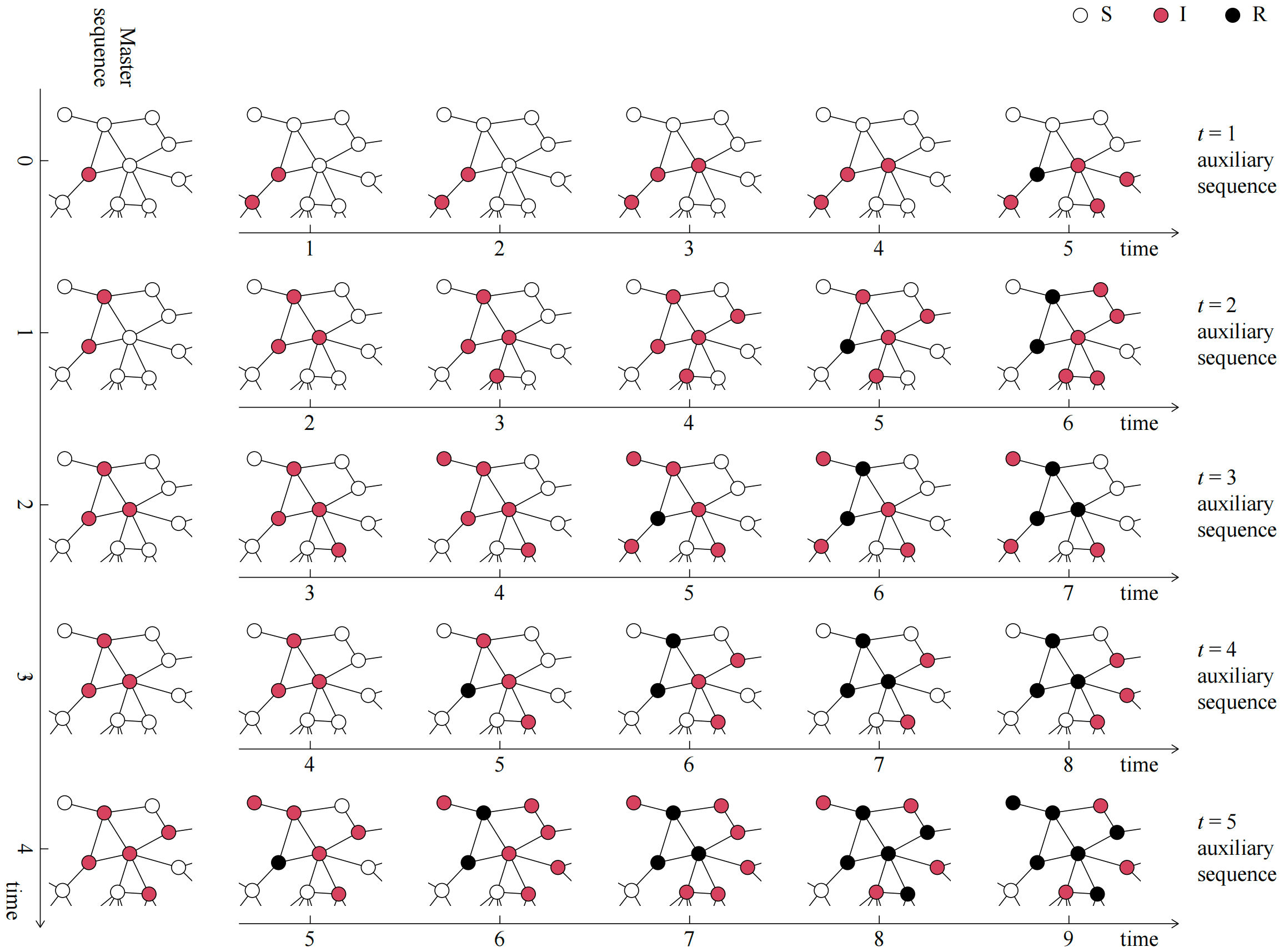} 
	\caption{Schematic illustration of the methodology for studying predictability decay based on the SIR model.
		A master sequence of epidemic propagation is first simulated, and the system state is recorded at each time step. Each recorded state is then used as the initial condition to generate 1,000 auxiliary sequences. By comparing the standard deviations of the final outbreak sizes and termination times across these auxiliary simulations, the level of predictability at that specific time point can be quantified.
		White circles represent susceptible individuals (S), red circles represent infected individuals (I), and black circles represent recovered individuals (R). \\
		\textit{Source}: The figure is reproduced from Ref. \cite{holme2015time}.}
	\label{fig:sir_predictability_decay}
	\vspace{0pt}
\end{figure*}

Holme and Takaguchi~\cite{holme2015time} introduced a temporal-evolution perspective and proposed a framework that quantifies predictability through auxiliary trajectory divergence, thereby revealing the dynamic evolution of epidemic predictability across different transmission stages. Using the classical SIR model, they simulated epidemic spreading on various network topologies, including large-world, small-world, random regular, and scale-free networks with different power-low exponents, where the large-world network corresponds to the Watts–Strogatz model with rewiring probability $p=0$. The central question addressed was: to what extent can the final epidemic size and duration be predicted at different points in the spreading process? To answer this, the authors designed a main–auxiliary sequence experimental framework (see Fig.~\ref{fig:sir_predictability_decay}). 
Specifically, they first executed a complete SIR simulation (the main sequence) and, at each time step, captured the system’s current state as an initial condition, from which thousands of auxiliary trajectories were generated in parallel. By comparing the resulting distributions of the final epidemic size ($\Omega$) and duration ($\tau$) across these auxiliary trajectories, they quantified the predictive capacity of each time step. If all auxiliary trajectories converged to similar outcomes, the current state would exert a strong constraint on future evolution, implying high predictability. Conversely, large divergences among outcomes indicated high uncertainty and low predictability. In other words, the standard deviation of auxiliary trajectory outcomes (e.g., $\sigma_{\Omega|s,t}$) serves as a quantitative measure of uncertainty, 
transforming future trajectory divergence into a direct metric of epidemic predictability.

\begin{figure*}[!t]
	\centering
	\setlength{\abovecaptionskip}{0pt}
	\setlength{\belowcaptionskip}{0pt}
	\includegraphics[width=0.98\textwidth]{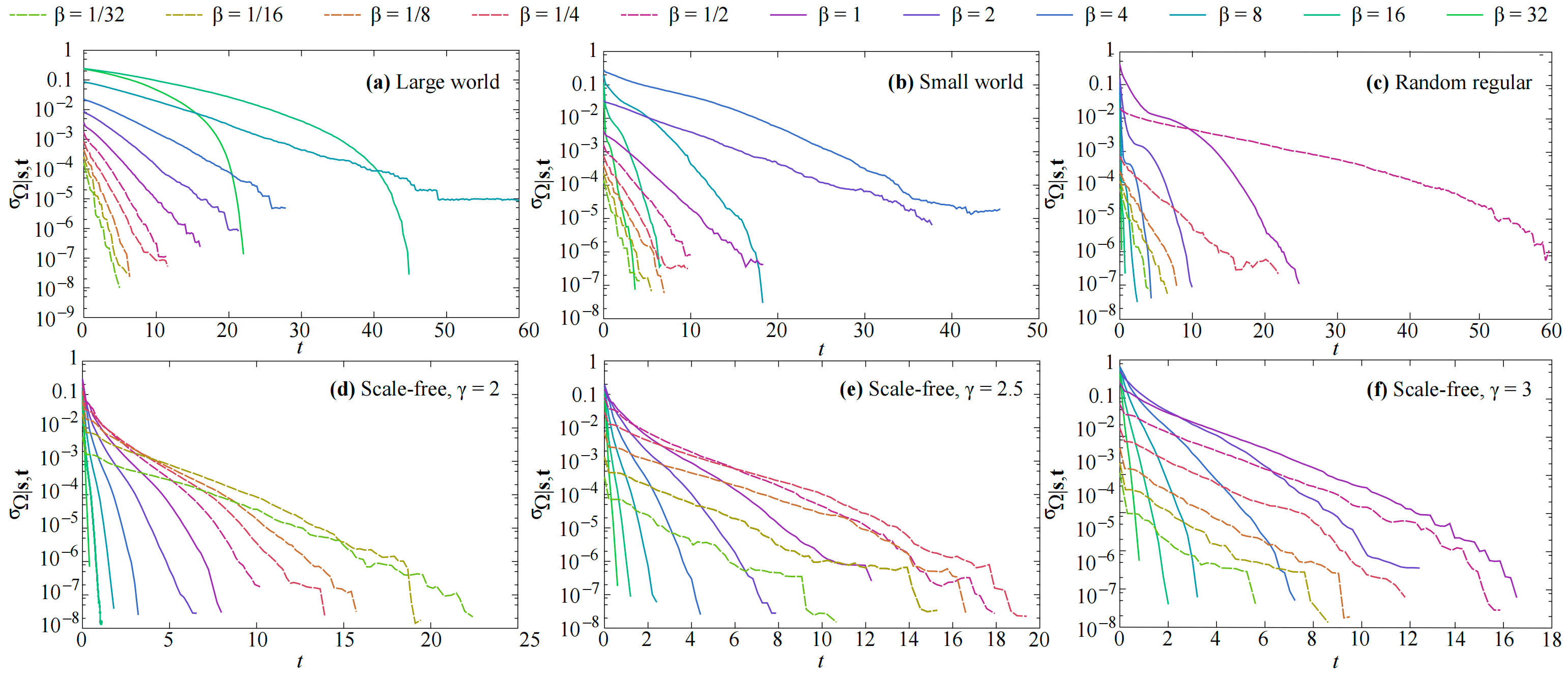} 
	\caption{Temporal evolution of the standard deviation of epidemic sizes under different network structures, illustrating the decay of predictability.
		(a) Large-world network; (b) small-world network; (c) random regular network; (d) scale-free network ($\gamma = 2$); (e) scale-free network ($\gamma = 2.5$); (f) scale-free network ($\gamma = 3$).
		The vertical axis represents the standard deviation of the final epidemic size at a given time $t$, denoted as $\sigma_{\Omega|s,t}$, which reflects the level of uncertainty in epidemic outcomes. The horizontal axis represents time $t$. As the epidemic spreads, an increase in $\sigma_{\Omega|s,t}$ indicates a decrease in predictability, while after the epidemic ends, the standard deviation returns to zero and is no longer shown. Different infection rates ($\beta$) significantly affect the rate of predictability decay. \\
		\textit{Source}: The figure is reproduced from Ref. \cite{holme2015time}.}
	\label{fig:predictability_decay_networks}
	\vspace{0pt}
\end{figure*}

Holme and Takaguchi~\cite{holme2015time} analyzed the predictability boundaries of epidemic spreading from a temporal-evolution perspective.
Their findings revealed a general law: even under a “full-information scenario,” where the exact state of every individual is known, epidemic predictability decays rapidly over time. Predictive accuracy is relatively high in the early stages, but once large-scale transmission begins, uncertainty accumulates quickly, making forecasts substantially more difficult. Figure~\ref{fig:predictability_decay_networks} illustrates the temporal evolution of the standard deviation of the final epidemic size, $\sigma_{\Omega|s,t}$, across different network topologies, where a larger value indicates greater difficulty in predicting future outcomes. As time progresses, $\sigma_{\Omega|s,t}$ gradually decreases, indicating a steady improvement in predictability, and eventually converges to zero when the epidemic subsides, as all simulation outcomes become consistent.

A further comparison across network structures reveals that the rate of predictability decay is strongly influenced by topological features.
In large-world networks, predictability decays most slowly, whereas in Watts–Strogatz networks ($p=0.01$) and random regular networks, the decay occurs more rapidly. In scale-free networks, the effect of the power-law exponent $\gamma$ on predictability dynamics is particularly pronounced: a smaller $\gamma$ (indicating stronger heterogeneity) corresponds to slower predictability decay. Moreover, the infection rate $\beta$ modulates the shape of the decay curve—under moderate $\beta$, uncertainty persists the longest, yielding a wider predictability window; at low or high $\beta$, predictability declines rapidly due to early extinction in the former and rapid saturation in the latter.
Overall, these results demonstrate that network topology and transmission rate jointly shape the temporal evolution of predictability: systems with higher connectivity and shorter path lengths converge to deterministic behavior more quickly, whereas systems with strong structural heterogeneity retain higher uncertainty for longer periods. Holme and Takaguchi’s work was the first to conceptualize epidemic predictability as a dynamically evolving process, systematically revealing the interplay among network topology, transmission rate, and information availability \cite{holme2015time}. 
This perspective not only explains the paradox of “why real-world epidemics are often unpredictable while models appear deterministic,” but also provides practical implications: epidemic forecasting must account for phase-dependent predictability windows and can be improved by increasing surveillance data or using suitable aggregation indicators to slow down the decay of predictability, thereby building a more robust predictive framework.

Unlike previous studies that indirectly inferred predictability from entropy through the Fano inequality, Scarpino \textit{et al.}~\cite{scarpino2019predictability} proposed a direct quantification approach based on permutation entropy ($S^{\mathrm{perm}}$), defining predictability as $\Pi = 1 - S^{\mathrm{perm}}$. This measure captures short-term regularities in time series with high sensitivity, providing an intuitive representation of the interplay between structural order and stochastic variability in epidemic evolution. Using long-term surveillance data covering nine infectious diseases across multiple U.S. states, the study revealed a universal temporal pattern of predictability: epidemics typically exhibit high predictability in their early phases ($\Pi \approx 1$), which rapidly decays and stabilizes as the outbreak progresses. Further normalization analyses uncovered a cross-disease regularity—when time is rescaled by the basic reproduction number ($R_0$), predictability curves for different diseases converge to a similar form. At the same time, external interventions such as vaccination were shown to significantly alter the entropy structure of the time series, thereby reducing predictability. The key contribution of this work lies in uncovering a universal law of predictability decay over time across multiple diseases and linking it closely to $R_0$, thus providing a unified framework for cross-disease comparison of epidemic predictability.

Salganik \textit{et al.}~\cite{salganik2006experimental} conducted a large-scale online experiment that empirically revealed, for the first time, the mechanisms of inequality and unpredictability in cultural markets. They created an artificial music market involving 14,341 participants, where users selected downloads from a set of 48 unfamiliar songs under two conditions: independent choice and social influence.
The experiment was organized into multiple parallel worlds, allowing social dynamics to evolve independently under identical initial conditions. The results showed that social influence significantly amplified the inequality of success distribution: popular songs became disproportionately successful, while less favored ones were largely ignored. Moreover, the final rankings diverged dramatically across parallel worlds, indicating a high degree of systemic unpredictability.
Further analysis revealed that while song quality exerted some influence, medium-quality songsexhibited the most variability—they could become either major hits or complete failures. This finding demonstrated that the success of cultural products depends not only on intrinsic quality but also on the stochastic accumulation of social interactions.
This study provided the first experimental validation of how social influence generates both inequality of success and outcome unpredictability, laying the empirical and theoretical groundwork for subsequent research on cultural market predictability (e.g., Shulman \textit{et al.}~\cite{shulman2016predictability}).

Shulman \textit{et al.}~\cite{shulman2016predictability} considered two fundamental questions about predictability: (i) To what extent can the rise of cultural trends actually be forecast? (ii) Does the apparent success of existing models reflect a genuine understanding of the mechanisms driving popularity, or merely a dependence on superficial correlates and early signals? To answer these questions, Shulman \textit{et al.} conducted a large-scale empirical assessment of the limits of predictability across four major online ecosystems—Last.fm, Flickr, Goodreads, and Twitter—each representing distinct forms of social and cultural interaction. They organized predictive variables into four conceptual domains: temporal dynamics (e.g., the rate at which content reaches its first few adopters), early adopter characteristics (e.g., user influence and connectivity), structural properties (e.g., the density of links among adopters), and similarity measures (capturing the resemblance between new and prior content). Through a series of balanced classification experiments, the authors compared the contribution of each domain to the prediction of eventual popularity, offering a systematic account of the factors that shape the predictability of cultural diffusion within large-scale digital environments.

The results revealed that early diffusion speed almost entirely determined the predictability. Using only one variable—the time required to reach the first five adopters—models achieved over 70\% prediction accuracy across all four platforms (and up to 83\% on Twitter), nearly matching the performance of full-feature models. By contrast, non-temporal features contributed little and lacked cross-platform consistency—for example, network density had opposite effects across different platforms. These findings indicate that the predictability of cultural popularity depends primarily on the momentum of early diffusion rather than the intrinsic quality of the content itself. Overall, this study makes a significant contribution by clarifying the boundaries of predictability in digital cultural systems: short-term prediction based on early diffusion speed is highly effective, but predictability quickly diminishes once temporal signals are removed. This finding highlights the distinction between “being able to predict” and “understanding the mechanism,” emphasizing that future research should move beyond statistical correlations to develop causal and mechanistic explanatory frameworks that can more precisely identify which factors fundamentally determine predictability in cultural diffusion.

\subsection{Earth System}

\begin{figure*}[!t]
	\centering
	\includegraphics[width=0.7\textwidth]{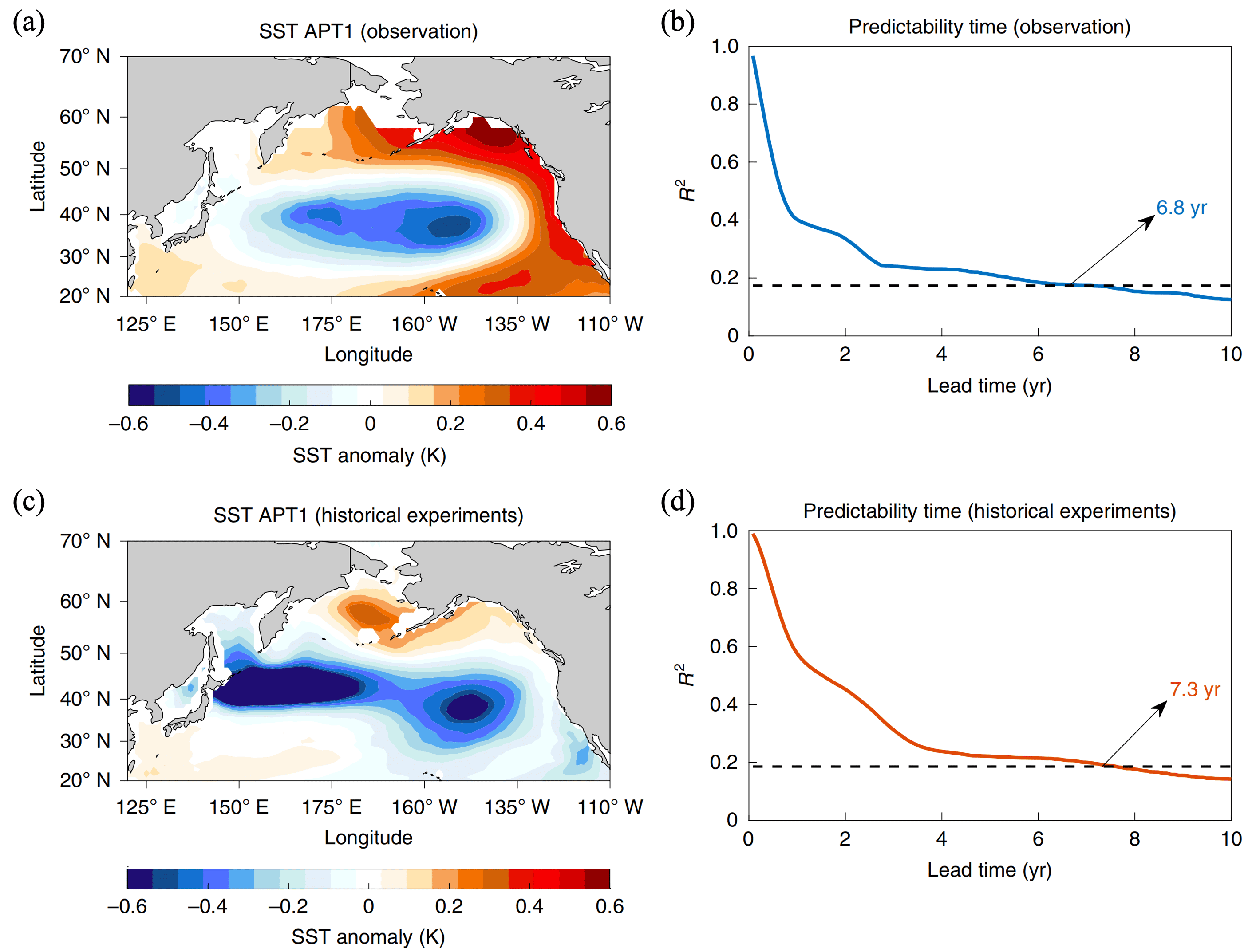} 
	\caption{APT-based evaluation of the predictability of the Pacific Decadal Oscillation (PDO). 
		(a,b) Spatial pattern (a) and predictability time (b) of the first APT mode derived from observational data, corresponding to the PDO. 
		Monthly sea surface temperature (SST) anomalies from seven reanalysis products during 1871--2010 were treated as observations, detrended using quadratic regression, concatenated, and subjected to APT analysis. 
		The quality of prediction is measured by the coefficient of determination ($R^2_r$), which quantifies the correlation between predicted and actual values. 
		Because prediction accuracy decreases with increasing lead time, the predictability time is defined as the lead time at which $R^2_r$ drops to the 5\% significance level. 
		For the reanalysis data, the predictability time is 6.8~years, as shown in (b). 
		(c,d) Same as (a) and (b), but for five CMIP5 historical experiment models covering 1860--2005, with a predictability time of 7.3~years. 
		The dashed lines in (b) and (d) mark the 5\% significance threshold. \\
		\textit{Source}: The figure is reproduced from Ref. \cite{li2020pacific}.}
	\label{PDO}
\end{figure*}

In recent years, the average predictability time (APT) method has been widely applied in studies of climate system predictability. In this subsection, APT is defined as the forecast lead time at which model predictive skill (e.g., $R^2$) decreases to the 5\% significance level \cite{delsole2013decadal,srivastava2017decadal}. This definition is related but different from that introduced in Section \ref{sec_Statistical_Analysis} \cite{delsole2009average}. Notice that, although these two definitions are mathematically different, they are both used to quantify the temporal scale of predictability and proposed by the same research group \cite{delsole2009average,delsole2013decadal,srivastava2017decadal}. Unlike the traditional Empirical Orthogonal Function (EOF) analysis, which seeks to maximize explained variance, the APT method is a feature extraction tool optimized to identify spatiotemporal modes that maximize predictability. In the work of Li~\textit{et al.}~\cite{li2020pacific}, APT was used to systematically evaluate changes in the predictability of the Pacific Decadal Oscillation (PDO) under increasing greenhouse gas concentrations, as illustrated in Fig.~\ref{PDO}. Specifically, the authors selected seven reanalysis datasets covering the period 1871–2010 and detrended the sea surface temperature (SST) fields to remove long-term warming trends caused by external forcing (e.g., greenhouse gas emissions and solar radiation variations), thereby isolating de-forced SST anomalies and identifying the dominant PDO mode from observations.
Furthermore, five CMIP5 models \cite{AnOverviewofCMIP5andtheExperimentDesign} (GISS-E2-H, GISS-E2-R, HadGEM2-ES, IPSL-CM5A-LR, and MPI-ESM-LR) were used for consistent comparison under both the historical experiments (1860–2005) and three Representative Concentration Pathway (RCP) scenarios (RCP~2.6, RCP~4.5, and RCP~8.5). The results showed that the APT in the historical experiments was approximately 7.3~years, whereas it exhibited a pronounced decreasing trend under future scenarios—6.7~years (RCP~2.6), 5.3~years (RCP~4.5), and 2.3~years (RCP~8.5)—corresponding to average reductions of 22.1\%, 38.4\%, and 73.3\% relative to the control experiment.

The study further elucidated the physical mechanisms behind the decline in predictability. As greenhouse gas concentrations increase, ocean stratification strengthens significantly, meaning that the density gradient between surface and deep waters intensifies. This change accelerates the propagation speed of planetary-scale Rossby waves in tropical and mid-latitude regions. Rossby waves play a key dynamical role in modulating the periodicity and amplitude of the PDO; thus, faster propagation compresses the PDO cycle, shortens its growth phase, and reduces oscillation amplitude, collectively diminishing overall predictability. These findings have important implications for climate prediction and risk management. As a key linkage between oceanic internal variability and atmospheric circulation, the PDO strongly influences global and regional climate anomalies—including the Asian monsoon and temperature and precipitation patterns along the North American west coast, among others. A marked reduction in PDO predictability implies a significantly weakened ability to anticipate decadal-scale climate variability based on PDO phase information. This not only imposes higher accuracy requirements on climate prediction models but also heightens challenges in ecosystem management, agricultural planning, fisheries assessment, and early warning of extreme climate events. Moreover, given that the PDO is closely coupled with global surface temperature variability, a reduction in its predictability may further exacerbate uncertainty in understanding the temporal rhythm of global warming. Hence, this study not only provides theoretical support for understanding future climate system predictability but also underscores the critical role of oceanic internal processes in shaping climate responses.

\begin{figure*}[!t]
	\centering
	\includegraphics[width=0.6\textwidth]{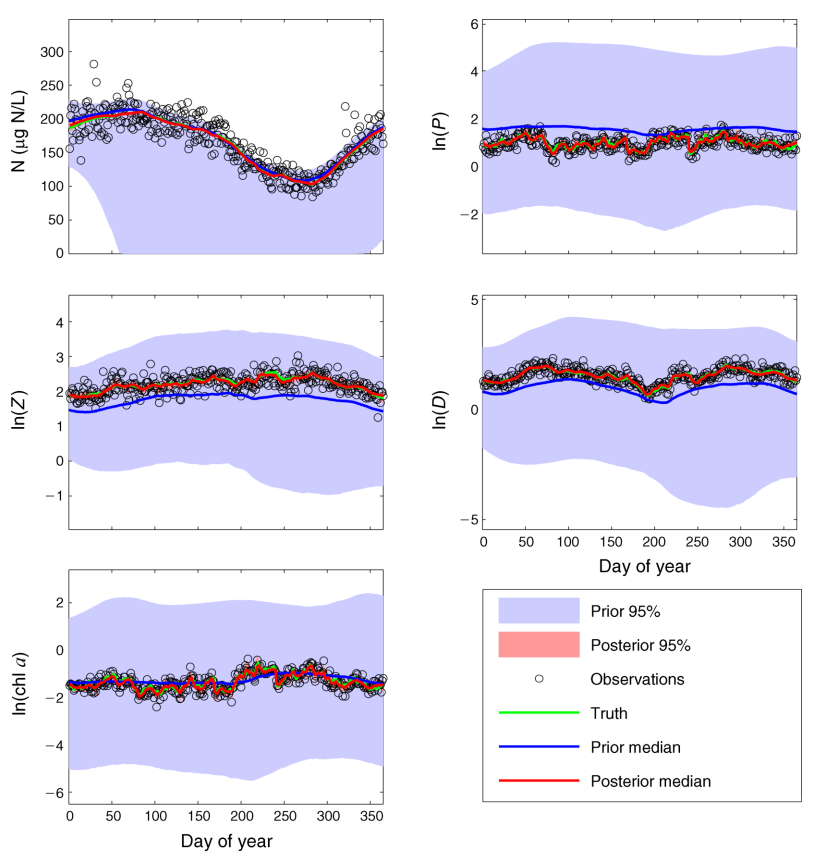} 
	\caption{
		Twin experiment: a time series of the prior and posterior distributional properties of the state variables for phytoplankton ($P$), zooplankton ($Z$), nutrients ($N$), and detritus ($D$) in the common currency of nitrogen (measured as µg N/L) and the instantaneous concentration of chlorophyll \textit{a} pigment (chl \textit{a}; µg chl \textit{a}/L). 
		Note that the posterior credible intervals are so close to the posterior medians that they are almost indistinguishable. \\
		\textit{Source}: The figure is reproduced from Ref. \cite{parslow2013bayesian}.
	}
	\label{bayes}
\end{figure*}

In more application-oriented contexts, the problem of spatiotemporal forecasting (STF) in Earth systems reveals the complex and multifaceted nature of predictability across different temporal and spatial scales. 
Xu~\textit{et al.}~\cite{xu2021spatiotemporal} provided a systematic review of this topic, emphasizing that the sources of predictability in the Earth system arise from multiple layers of influencing factors, including sensitivity to initial conditions, internal variability (e.g., ENSO \cite{science_1132588} and MJO \cite{zhang2005madden}), external forcing (e.g., greenhouse gas emissions, volcanic eruptions, and solar radiation variability), as well as model structural and parametric uncertainties. Across temporal scales, these factors exhibit a hierarchical organization. At short timescales (e.g., days to weeks in weather forecasting), uncertainty in initial values dominates; system evolution is highly sensitive to initial states, reflecting the classical characteristics of chaotic systems. This short-term predictability is referred to as "initial-value predictability." At longer timescales (e.g., seasonal to decadal climate prediction), external forcing mechanisms such as greenhouse gas accumulation or solar activity cycles gradually exert stronger influence, defining so-called "forced predictability." This distinction provides a theoretical foundation for understanding the lower bounds of predictability in contemporary climate research.

Across spatial scales, the predictability of Earth systems demonstrates dual features of spatial correlation and heterogeneity. 
On one hand, according to Tobler's First Law of Geography~\cite{miller2004tobler}, geographical proximity generally leads to spatial correlation in predictability, that is, neighboring regions tend to exhibit similar predictability patterns. For instance, ENSO-induced precipitation anomalies in the tropical Pacific often synchronize climatic responses along the western coast of South America and throughout Southeast Asia. On the other hand, such spatial correlations are frequently disrupted by topographic variations, surface heterogeneity, land–sea distribution, and regional feedback mechanisms, giving rise to pronounced spatial heterogeneity. This phenomenon is particularly evident between tropical and mid-to-high-latitude regions: tropical precipitation tends to exhibit higher predictability due to its direct association with ENSO, whereas mid-latitude climates—dominated by localized processes—are substantially harder to predict.

\begin{figure*}[!t]
	\centering
	\includegraphics[width=0.75\textwidth]{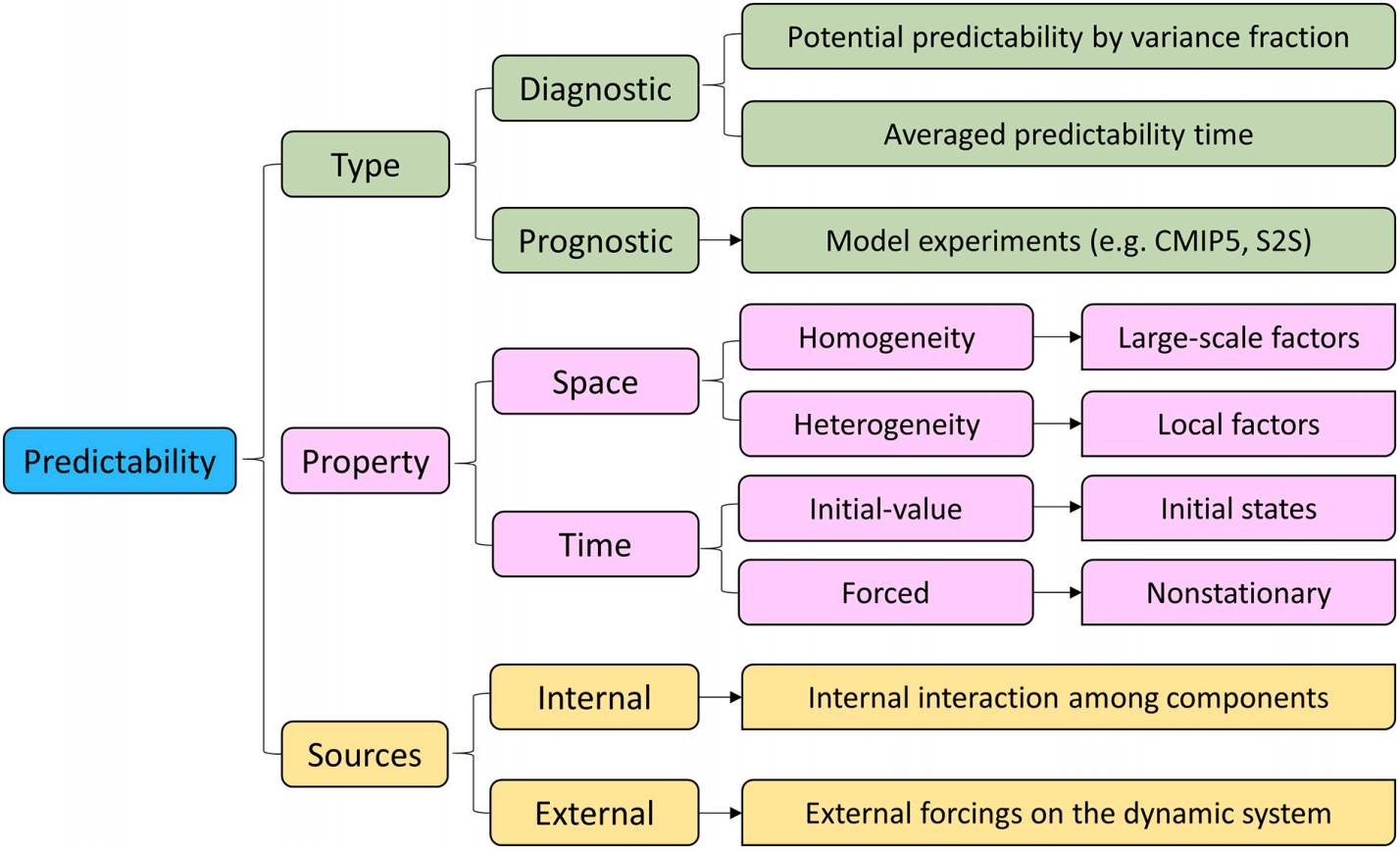}
	\caption{Types, properties, and sources of predictability in the Earth system. \\
		\textit{Source}: The figure is reproduced from Ref. \cite{xu2021spatiotemporal}.}
	\label{predictability}
\end{figure*}

Xu~\textit{et al.}~\cite{xu2021spatiotemporal} highlighted that predictability itself should be treated as a measurable property rather than a static assumption. In diagnostic analyses, quantities such as APT are often used to measure the inherent predictability embedded in historical observational data, thereby assessing potential predictability. In prognostic analyses, ensemble forecasting and multi-model ensembles are employed to evaluate the actual performance of models under varying initial conditions and physical assumptions. Xu~\textit{et al.}~\cite{xu2021spatiotemporal} further integrated these metrics with the sources of predictability, proposing a “Type–Source–Representation” framework to construct a comprehensive spatiotemporal predictability map (as illustrated in Fig.~\ref{predictability}). This framework not only clarifies how different sources of predictability manifest within models but also provides theoretical support for the design of multi-scale, application-oriented prediction systems.

\subsection{Politics}
\begin{figure*}[!t]
	\centering
	\setlength{\abovecaptionskip}{0pt}
	\setlength{\belowcaptionskip}{0pt}
	\includegraphics[width=0.9\textwidth]{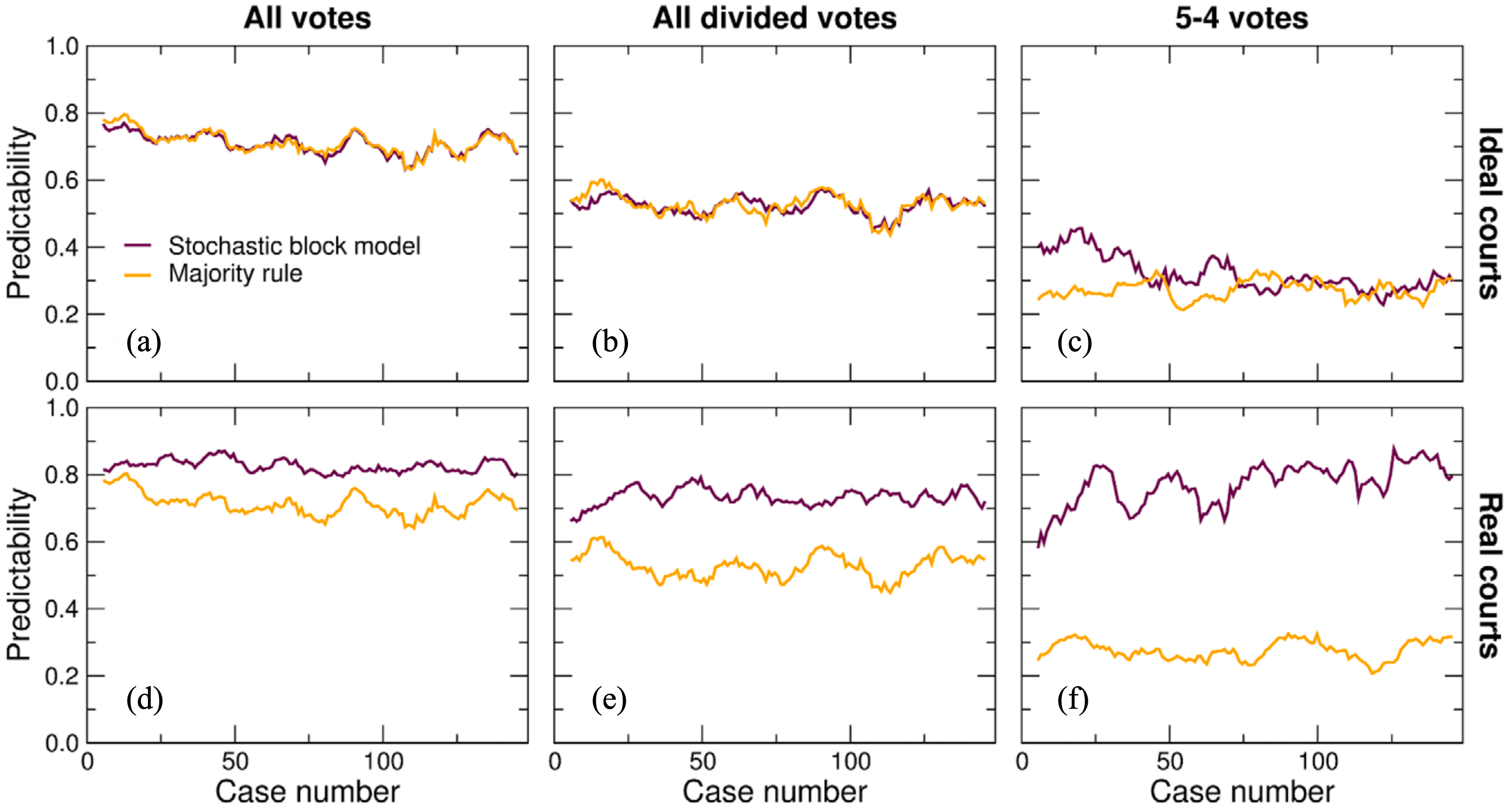} 
	\caption{Analysis of the average predictability of U.S. Supreme Court decisions.
		(a–c) Idealized court scenarios; (d–f) real court scenarios. The horizontal axis represents case indices, and the vertical axis represents predictability. The purple curve corresponds to the SBM, and the orange curve represents the Majority Rule.
		(a,d) Include all court decisions; (b,e) exclude unanimous and 8–1 decisions, retaining only divided rulings; (c,f) consider only 5–4 split decisions. \\
		\textit{Source}: The figure is reproduced from Ref. \cite{guimera2011justice}.}
	\label{fig:supreme_court_predictability}
	\vspace{0pt}
\end{figure*}

In political systems, the central question of predictability research lies in clarifying the extent to which institutional processes and behavioral mechanisms exhibit regularity, and under what conditions they face inevitable limits to prediction. Guimerà and Sales-Pardo~\cite{guimera2011justice} provided a representative case of predictability in institutionalized political behavior. Focusing on the U.S. Supreme Court, they posed a fundamental question: Can future judicial decisions be predicted solely from historical voting patterns among justices, without any knowledge of case content? This perspective moves beyond traditional debates centered on legal doctrine or ideology and instead applies complex network methods to uncover the latent structure underlying judicial behavior. Methodologically, Guimerà and Sales-Pardo modeled the Court’s voting record as a bipartite network, where justices and cases form two distinct node sets and votes define the links between them. Using a Stochastic Block Model (SBM) \cite{guimera2009}, they identified long-term, stable coalitional patterns among justices and used these inferred structures to predict future votes. Two contrasting scenarios were examined: in an “ideal court,” where all justices vote independently, predictive performance should not exceed that of a simple majority rule—predicting each justice’s vote as the majority decision of the other eight members (see Fig.\ref{fig:supreme_court_predictability}(a\&c)). In contrast, for the real Court, if persistent dependencies or alignment patterns exist among justices, the SBM’s predictive accuracy should significantly surpass this baseline, as illustrated in Fig.\ref{fig:supreme_court_predictability}(d\&f).

The results showed that the complex model achieved substantially higher predictive accuracy than the simple majority rule. This performance gap, termed the predictability gap, reflects latent regularities in judicial voting behavior. The difference was particularly pronounced in divided decisions--for instance, after excluding unanimous rulings and highly consistent 8–1 decisions (Fig.~\ref{fig:supreme_court_predictability}(b\&e)), and especially in the most contentious 5–4 split cases (Fig.~\ref{fig:supreme_court_predictability}(c\&f)), where the SBM’s predictive accuracy far exceeded that of the majority rule. These findings indicate the presence of stable voting coalitions among justices, suggesting that their decisions are not entirely independent.
Longitudinal analysis further revealed a declining trend in overall predictability from the Warren Court (1950s) to the Rehnquist Court (2000s). Moreover, systematic variations emerged under different political administrations: predictability levels were generally higher during Republican presidencies than during Democratic ones. At the individual level, justices exhibited marked heterogeneity—some maintained highly stable voting patterns, while others showed greater variability across cases. The key contribution of this work lies in translating the regularities of judicial decision-making into quantifiable metrics through complex network analysis, thereby introducing the notion of relative predictability. The results demonstrate that even within a highly institutionalized judicial system, decisions are not entirely autonomous but are constrained by enduring group structures and institutional dynamics. This study not only provides a novel quantitative lens for examining judicial independence and politicization but also illustrates the existence of high predictability under strongly institutionalized settings within broader political systems.

\begin{figure*}[!t]
	\centering
	\includegraphics[width=0.5\linewidth]{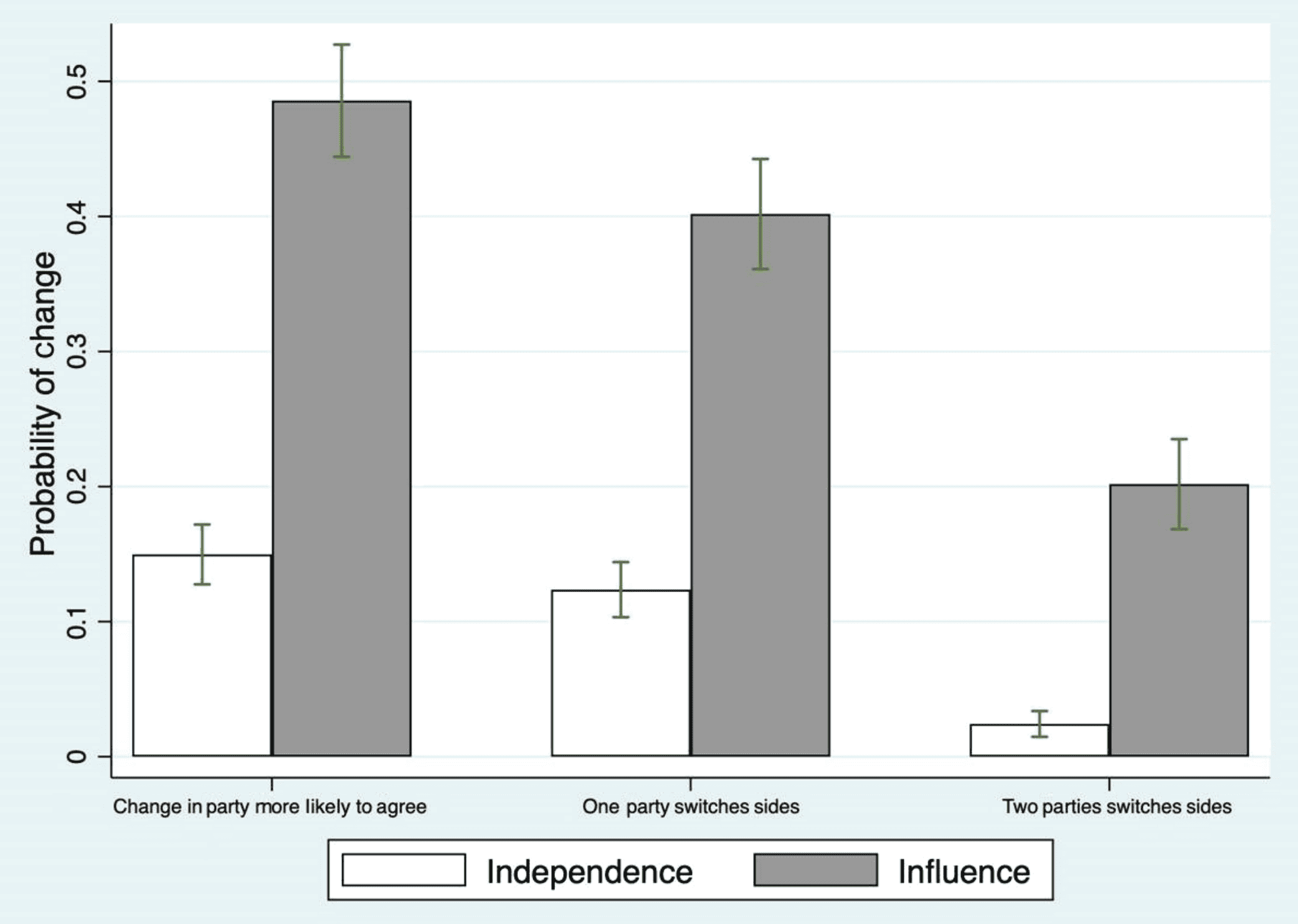}
	\caption{The effect of social influence on the unpredictability of partisan positions.
		The figure compares three types of partisan position shifts under two conditions: independence (white bars) and social influence (gray bars)—namely, (i) increased within-party alignment, (ii) one-party defection, and (iii) simultaneous defection of both parties.
		Results show that under the influence condition, the probabilities of these drastic shifts increase significantly, with the likelihood of changes in overall party alignment approaching 0.5.
		This indicates that while social influence accelerates the formation of polarization, it also makes the outcomes more dependent on early random perturbations, thereby reducing the overall predictability of the system. \\
		\textit{Source}: The figure is reproduced from Ref. \cite{macy2019opinion}.}
	\label{ig:social_influence_unpredictability}
\end{figure*}

The study by Macy \textit{et al.}~\cite{macy2019opinion} provides crucial evidence for understanding the boundaries of predictability in political polarization. Contrary to traditional views that attribute polarization primarily to deep-seated ideological divisions, the authors propose that polarized outcomes may instead arise from the path dependence of opinion cascades—where small, random differences in early opinions are progressively amplified via social influence, leading to high level of instability and inherent unpredictability. Macy \textit{et al.}~\cite{macy2019opinion} introduced an innovative multiple-worlds paradigm. In this large-scale behavioral experiment, 4,581 participants were assigned to several parallel “worlds” and asked to respond to 20 emerging political and technological issues (e.g., robot lawyers, climate engineering, and genetic privacy). Under the independent condition, participants answered without social input, whereas under the influence condition, they were shown the distribution of positions previously expressed by members of their own political party, thus allowing social influence to shape their responses (see Fig.~\ref{ig:social_influence_unpredictability}, compare white and gray bars). This experimental framework, inspired by cascade experiments in cultural market research, represents the first large-scale application of controlled parallel-world designs to the study of political behavior.

The results revealed three key phenomena. First, social influence significantly amplified partisan divisions, markedly increasing the probability of within-party consensus, with the corresponding bar heights approaching 0.5 (Fig.~\ref{ig:social_influence_unpredictability}, left panel). Second, under the influence condition, the probabilities of single-party defection and simultaneous cross-party defection both increased substantially (Fig.~\ref{ig:social_influence_unpredictability}, middle and right panels), indicating that polarization outcomes were highly unstable across parallel worlds—the same issue could evolve in entirely opposite directions under identical initial conditions. Third, early opinions proved decisive: the positions of the first ten participants exhibited much greater predictive power for the eventual outcome than inherent partisan preferences, underscoring the system’s extreme sensitivity to initial perturbations. Taken together, these findings demonstrate that while social influence can rapidly generate strong polarization, it simultaneously renders the system exceedingly sensitive to early random fluctuations, thereby reducing overall predictability. Methodologically, this study introduced a novel experimental paradigm; mechanistically, it uncovered the pivotal role of opinion cascades in the formation of polarization; and theoretically, it challenged the intuitive assumption that strong polarization implies high predictability. Instead, it suggests that polarized outcomes may stem more from historical contingency than from fixed ideological divides. This insight deepens our understanding of the dynamics of political polarization and offers an important caution for policymakers and social governance: polarization is not necessarily controllable or predictable but may arise from the amplification of early chance events.

Chadefaux~\cite{chadefaux2017conflict} conducted a comprehensive evaluation of conflict prediction, outlining the advantages and limitations of four major methodological pathways: expert judgment, structural econometric modeling, event data and text analytics, and game-theoretic or agent-based approaches. Expert forecasts rely on intuition and contextual knowledge but are prone to subjective bias and poor reproducibility. Structural models based on macro-level variables can identify long-term risk factors yet struggle to capture short-term triggers of conflict onset. Event-based and text-analytic methods offer higher temporal sensitivity but often suffer from data collection bias and semantic ambiguity. Game-theoretic and agent-based models shed light on underlying mechanisms of strategic interaction, though the presence of mixed strategies and bounded rationality frequently undermines predictive stability. From a more fundamental standpoint, Chadefaux argued that the challenge of conflict prediction arises not merely from methodological limitations but from the intrinsic strategic and informational nature of conflict itself. When actors possess private information and can manipulate or conceal signals, systematic discrepancies emerge between observable indicators and underlying realities. Moreover, once predictive assessments become known to adversaries, they may adapt their behavior strategically, thereby invalidating the forecast through reflexive feedback. Accordingly, Chadefaux stressed the importance of distinguishing model-level errors from phenomenon-level randomness and called for future research to delineate the conditions under which conflict dynamics are genuinely predictable versus those in which uncertainty is structurally unavoidable.

\subsection{Education}
\begin{figure*}[!t]
	\centering
	\setlength{\abovecaptionskip}{0pt}
	\setlength{\belowcaptionskip}{0pt}
	\begin{subfigure}[t]{0.36\textwidth}
		\centering
		\includegraphics[width=\textwidth]{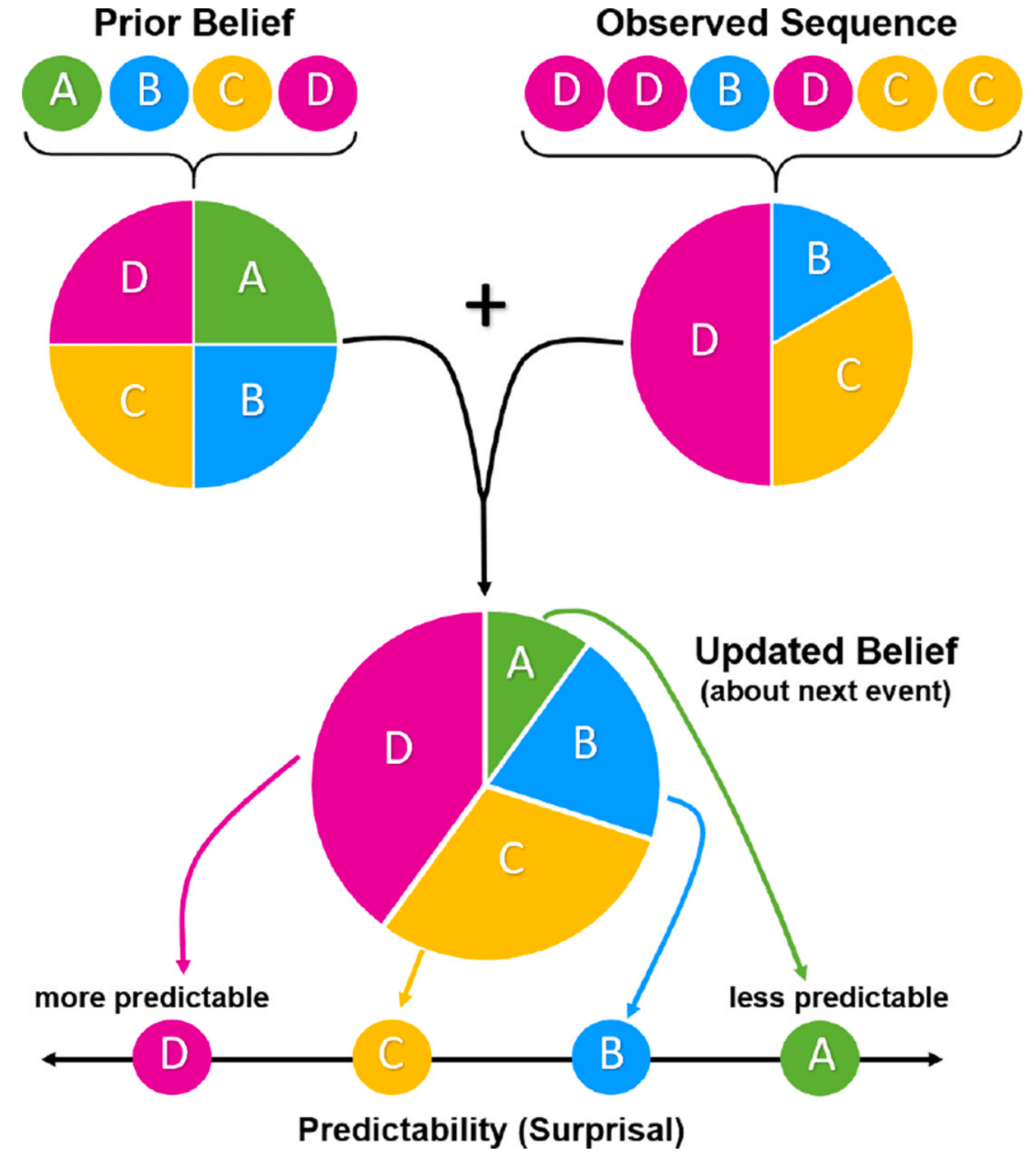}
		\label{fig:left}
	\end{subfigure}
	\hspace{0.02\textwidth} 
	\begin{subfigure}[t]{0.33\textwidth}
		\centering
		\includegraphics[width=\textwidth]{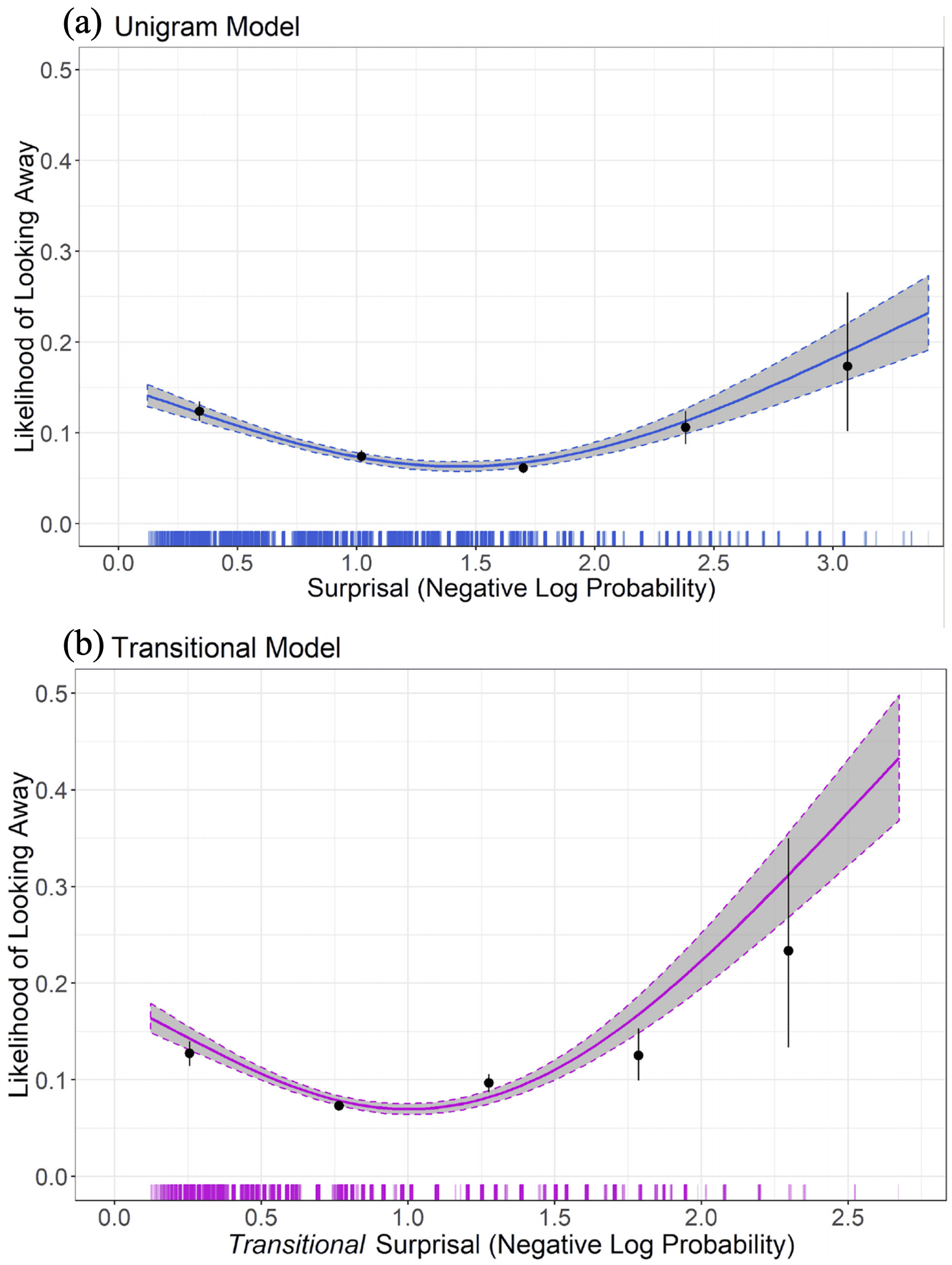}
		\label{fig:right}
	\end{subfigure}
	\caption{Effects of the ideal observer model and predictability on attentional allocation.
		(Left) Schematic of the ideal observer model: prior beliefs are combined with observed event sequences to update expectations about the next event, thereby inferring the predictability (or surprisal) of different outcomes.
		(Right) Experimental results showing how the likelihood of gaze deviation varies with predictability.
		(a) Relationship between surprisal (negative log-probability) and gaze deviation under the unigram model;
		(b) Relationship between transitional surprisal and gaze deviation under the transitional model.
		Results indicate that high predictability (low surprisal) corresponds to reduced attentional deviation, whereas low predictability (high surprisal) increases the probability of attentional shifts. \\
		\textit{Source}: The figure is reproduced from Ref. \cite{cubit2021visual}.}
	\label{fig:ideal_learner_predictability}
\end{figure*}

Predictability in education involves both the internal mechanisms of individual learning and the regularities of group interaction networks, together revealing the multilayered boundaries of predictability within educational systems. At the individual level, studies have shown that learners allocate attention to information in systematic rather than random ways. Learners are typically not drawn to extremely simple or completely unpredictable stimuli but instead focus on inputs of moderate predictability—a phenomenon known as the Goldilocks effect. Cubit \textit{et al.}~\cite{cubit2021visual} extended the examination of this effect beyond infancy to preschool-aged children, aiming to test whether this attentional preference represents a robust mechanism across developmental stages. In their study, 72 children aged 3–6 years participated in a naturalistic viewing task with their gaze patterns were continuously recorded via eye-tracking technology. During the experiment, children were shown sequences of novel objects, and each event’s surprisal—the degree of unpredictability given the preceding sequence—was quantified using an ideal learner model (see Fig.~\ref{fig:ideal_learner_predictability}, left). Events that were fully predictable had low surprisal values, whereas entirely unexpected events exhibited high surprisal. Cubit \textit{et al.} then examined the probability of gaze aversion across different surprisal levels to characterize how children’s attention varies as a function of predictability.

The experimental results, shown in Fig.~\ref{fig:ideal_learner_predictability} (right), reveal a characteristic U-shaped relationship between predictability and attention. Children were more likely to avert their gaze when stimuli were either highly predictable or completely unpredictable, whereas their attention was most sustained under conditions of moderate predictability. This pattern was consistent across both the unigram model based on single-event probabilities (Fig.~\ref{fig:ideal_learner_predictability}(a)) and the transitional model based on transition probabilities (Fig.~\ref{fig:ideal_learner_predictability}(b)), as well as across children of different ages and cognitive ability levels. In other words, regardless of age, executive function, attentional control, or processing speed, children exhibited similar attentional dynamics in response to predictability. This finding carries important theoretical and pedagogical implications. On the theoretical side, it confirms the robustness of the Goldilocks effect in early childhood, suggesting that a preference for moderately predictable stimuli may reflect a low-level, automatic information-processing mechanism rather than a reliance on higher-order cognition. On the educational side, it supports the “optimal information complexity” hypothesis, which posits that learners are naturally drawn to information that is neither overly simple nor entirely random, thereby maximizing informational value while avoiding cognitive overload. Together, these insights deepen our understanding of children’s learning and attention mechanisms and suggest that educational materials of moderate complexity and structured regularity are most effective in sustaining engagement and promoting learning.

At the group level, Ciobanu \textit{et al.} \cite{ciobanu2014interaction} analyzed 64 days of multimodal mobility data collected from students and faculty at the University Politehnica of Bucharest—including Bluetooth scans, Wi-Fi access points, and communication logs—to investigate the interaction dynamics of academic opportunistic networks. Their findings revealed that interpersonal encounters are not random noise but are shaped by strong temporal and spatial regularities. Temporally, by computing the entropy of “whether a contact occurs in the next time interval,” the authors found that most nodes had entropy values below 0.35, indicating a high degree of regularity in interaction timing—that is, past contact patterns were strongly predictive of when the next interaction would occur. In contrast, predicting who the next interaction partner would be proved more difficult, with entropy values reaching as high as 4.25, reflecting the low predictability of partner selection. Further analysis showed that the distribution of encounter frequencies closely followed a Poisson process, with pronounced periodicity when aggregated by hour of the day or day of the week. Spatially, individuals tended to appear within a limited set of access points, and their location entropy was substantially lower than that of random baselines, reflecting stable spatiotemporal routines. Collectively, these findings demonstrate that even within dynamic academic settings, interaction behavior exhibits quantifiable regularities and definable boundaries—providing empirical support for modeling resource allocation and behavioral patterns in both opportunistic networks and educational environments.

In educational and developmental contexts, the predictability of policy and family environments constitutes a critical determinant of learning and health trajectories. Glynn \textit{et al.} \cite{glynn2024leveraging} introduced predictability into public health and child development research, proposing that environmental unpredictability serves as a major risk factor for neurodevelopmental and long-term health outcomes. Drawing on interdisciplinary empirical studies and animal experiments, they examined indicators such as household income volatility, residential instability, and high-entropy structure of parental behavioral patterns. The results revealed that elevated levels of environmental unpredictability are strongly associated with cognitive deficits, abnormal stress responses, and developmental delays in children, with these effects remaining remarkably consistent across species. Glynn \textit{et al.} argued that enhancing predictability should be a central objective of public policy. For instance, “fair workweek” legislation, which stabilizes parental work schedules, can increase the predictability of family life, while “unpredictability screening” in pediatric care can help identify at-risk children at early stages. These examples highlight that policies function not only as an external system of regulation but also as a direct source of predictability that shapes individual developmental trajectories.

\subsection{Academic Career}

\begin{figure*}[!t]
	\centering
	\setlength{\abovecaptionskip}{0pt}
	\setlength{\belowcaptionskip}{0pt}
	\includegraphics[width=0.98\textwidth]{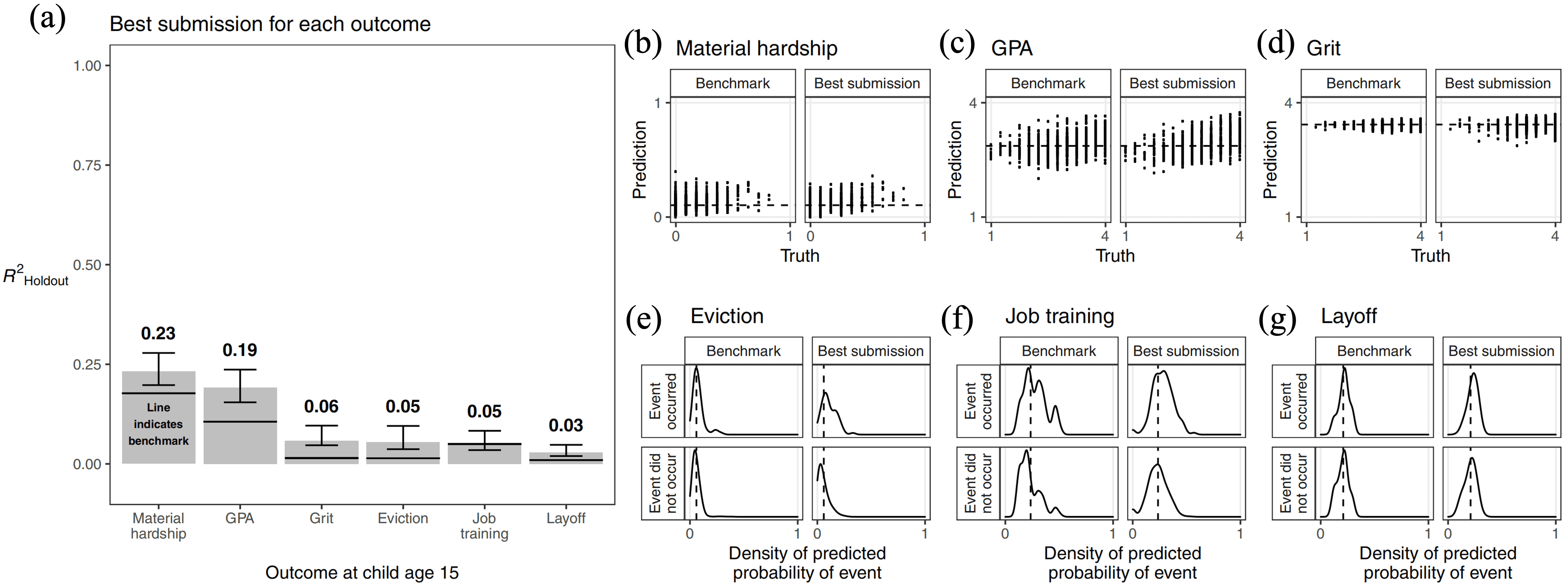} 
	\caption{Comparison between the best submitted models and baseline models across different prediction tasks.
		(a) Performance on six outcome variables measured by $R^2_{\mathrm{Holdout}}$. Gray bars represent the baseline model, black horizontal lines indicate baseline performance, and colored bars denote the best submissions, with error bars showing 95\% confidence intervals. Results show significant improvements of the best models over the baseline for Material Hardship and GPA, while gains are limited for Grit, Eviction, Job Training, and Layoff.
		(b–d) Comparison between predicted and true values for Material Hardship, GPA, and Grit respectively; ideal predictions lie along the diagonal line.
		(e–g) Predicted probability distributions for Eviction, Job Training, and Layoff; the top row corresponds to event occurrence and the bottom row to non-occurrence, with dashed lines indicating the mean values in the training data. \\
		\textit{Source}: The figure is reproduced from Ref. \cite{salganik2020measuring}.}
	\label{fig:prediction_challenge_results}
	\vspace{0pt}
\end{figure*}

Research on predictability has also extended to the long-term trajectories of academic impact and professional development. Salganik \textit{et al.} \cite{salganik2020measuring} introduced a unique large-scale collaborative experiment—the Fragile Families Challenge—to quantify the predictability of life-course outcomes in the social sciences. The challenge was built upon the Fragile Families and Child Wellbeing Study, a longitudinal panel survey that tracked over 4,000 U.S. families for 15 years (1999–2014). The study conducted six follow-up waves on the same cohort, systematically documenting the social, economic, and psychological characteristics of children from birth through adolescence. Based on these data, 160 participating research teams were asked to predict six outcome variables at age 15 using background features collected from birth to age 9. These outcomes included children’s own academic and psychological measures (e.g., GPA and grit) as well as the socioeconomic conditions of their households and primary caregivers (e.g., eviction, material hardship, unemployment, and job training participation), thereby capturing the long-term coupling between individual development and family environment. The main results are summarized in Fig.~\ref{fig:prediction_challenge_results}. First, in terms of overall explanatory power, the best-performing submissions achieved only modest predictive accuracy: the top \(R^2_{\mathrm{Holdout}}\) values were approximately 0.23 for material hardship and 0.19 for GPA, while the remaining four outcomes did not exceed 0.06 (Fig.~\ref{fig:prediction_challenge_results}(a)). Even compared with minimalist baselines, the performance gains of complex models were marginal. Second, for continuous prediction tasks (Figs.~\ref{fig:prediction_challenge_results}(b-d)), scatter plots of true versus predicted values showed strong mean-reversion patterns, indicating that models failed to meaningfully distinguish individual-level variation. Third, in binary classification tasks (Figs.~\ref{fig:prediction_challenge_results}(e-g)), the predicted probability distributions for event and non-event samples were highly overlapping. Only the best submissions achieved slight separation, reflecting marginal improvements in calibration but limited overall discriminative power.

Taken together, these results reveal the structural constraints underlying life-course prediction. First, the long temporal lag between input features and outcomes leaves many stochastic shocks unobserved, substantially lowering the theoretical ceiling of predictability. Second, the outcome variables themselves are sparse and noisy, making the base rates of event-type outcomes difficult to model accurately. Third, the striking similarity in error structures across hundreds of teams suggests the existence of systematically “hard-to-predict” individuals—an intrinsic limitation that cannot be mitigated by simply changing algorithms. Hence, the primary bottleneck lies not in the choice of modeling techniques but in the data and the generative mechanisms that produce it. Overall, this work provides one of the first empirical quantifications of the predictive boundaries in social science. Even under ideal conditions, the best-performing models achieved an \(R^2_{\mathrm{Holdout}}\) of around 0.2 for several outcomes—far below common expectations for machine learning systems.
This finding serves as a cautionary note for both researchers and policymakers: rather than pursuing ever more complex algorithms, attention should shift toward identifying which populations and contexts are inherently more predictable and toward improving data quality to overcome fundamental information bottlenecks.

\begin{figure*}[!t]
	\centering
	\setlength{\abovecaptionskip}{0pt}
	\setlength{\belowcaptionskip}{0pt}
	\includegraphics[width=0.6\textwidth]{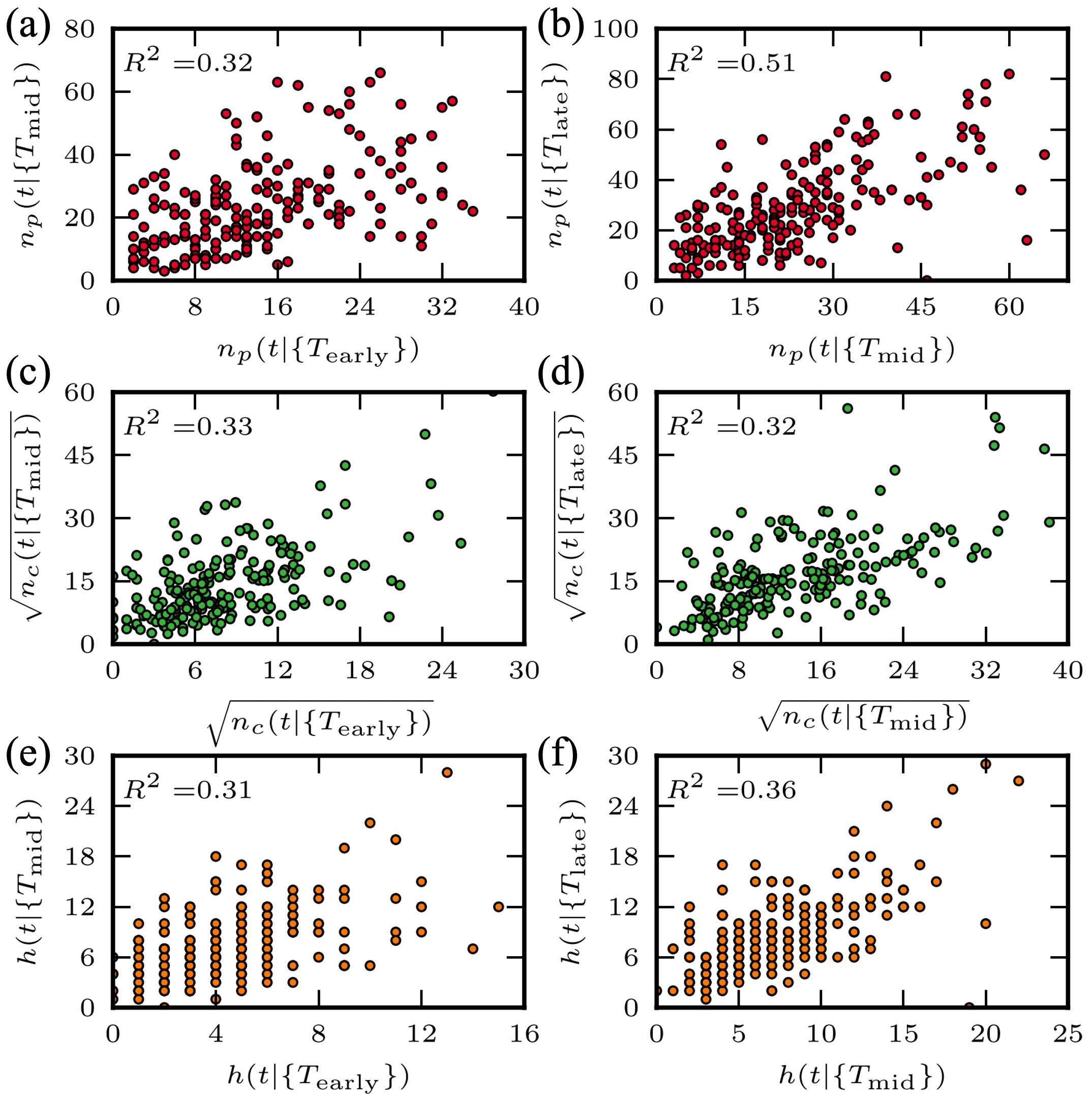} 
	\caption{Correlations of scientific performance across different career stages of physicists.
		(a,b) Correlations in publication counts between early–mid and mid–late career stages;
		(c,d) correlations in the square root of total citation counts between early–mid and mid–late stages;
		(e,f) correlations in the non-cumulative $h$-index between early–mid and mid–late stages.
		Each scatter point represents an individual author's corresponding academic metric, and the coefficient of determination ($R^2$) is reported in each panel. \\
		\textit{Source}: The figure is reproduced from Ref. \cite{penner2013predictability}.}
	\label{fig:physics_career_predictability}
	\vspace{0pt}
\end{figure*}

Academic evaluation and recruitment often center on the question of whether a researcher’s future scientific impact can be predicted. Consequently, many approaches rely on cumulative indicators--such as the \textit{h}-index--and use regression models to extrapolate their future trajectories. Penner \textit{et al.} \cite{penner2013predictability} compiled longitudinal career data for 762 scientists across physics, biology, and mathematics, encompassing both established scholars and early-career researchers. They first replicated the Acuna model \cite{acuna2012predicting}, which predicts a researcher’s future \textit{h}-index based on current bibliometric variables (e.g., present \textit{h}-index, total publications, journal count, number of highly cited papers, etc.), using an elastic-net regression framework. At first glance, the model appeared to perform impressively, yielding coefficients of determination ($R^2$) typically above 0.75. However, closer inspection revealed that this apparent “high predictability” primarily arises from the cumulative and autocorrelated nature of the \textit{h}-index itself, rather than the model’s actual ability to capture future scientific potential. To demonstrate this, the authors designed two types of null models: (1) a random increment model, in which \textit{h}-index growth was fully randomized; and (2) a paper-shuffling model, in which each researcher’s publication record was randomly permuted. Even under these conditions—where genuine scholarly structure was completely eliminated—the regression models still achieved comparably high $R^2$ values. This clearly indicated that predictions based on cumulative indicators exhibit an inherent “illusory predictability.”

A further stratified analysis revealed that this problem is particularly pronounced among early-career researchers. When the data were segmented by career stage, prediction errors were found to be substantially higher for junior scientists than for senior ones. This pattern is clearly illustrated in Fig.~\ref{fig:physics_career_predictability}: regardless of the metric used—publication count (Figs.~\ref{fig:physics_career_predictability}(a\&b), square root of total citations (Figs.~\ref{fig:physics_career_predictability}(c\&d)), or non-cumulative \textit{h}-index (Figs.~\ref{fig:physics_career_predictability}(e\&f))—the correlations between the early–mid and mid–late career stages remain consistently weak. In particular, early-career scholars exhibit highly unstable research trajectories, rendering their scientific performance effectively unpredictable. Building on these findings, the authors proposed a more meaningful modeling target: instead of forecasting cumulative \textit{h}-index values, one should focus on non-cumulative incremental measures such as $\Delta h$. Yet, when the dependent variable was replaced by $\Delta h$, the model’s predictive power dropped sharply, further highlighting the intrinsic uncertainty and cross-disciplinary heterogeneity of academic career development. Overall, the core contribution of this study lies in rigorously demonstrating—through both empirical analysis and null-model comparison—that regression-based forecasts of future scientific impact derived from the \textit{h}-index suffer from fundamental methodological flaws, particularly for early-career researchers, for whom predictive reliability is nearly nonexistent. This finding serves as a direct caution for academic hiring and evaluation practices, while also serving as a canonical case for delineating the limits of predictability in academic careers.

\subsection{Music Appreciation}
\begin{figure*}[!t]
	\centering
	\setlength{\abovecaptionskip}{0pt}
	\setlength{\belowcaptionskip}{0pt}
	\includegraphics[width=0.8\textwidth]{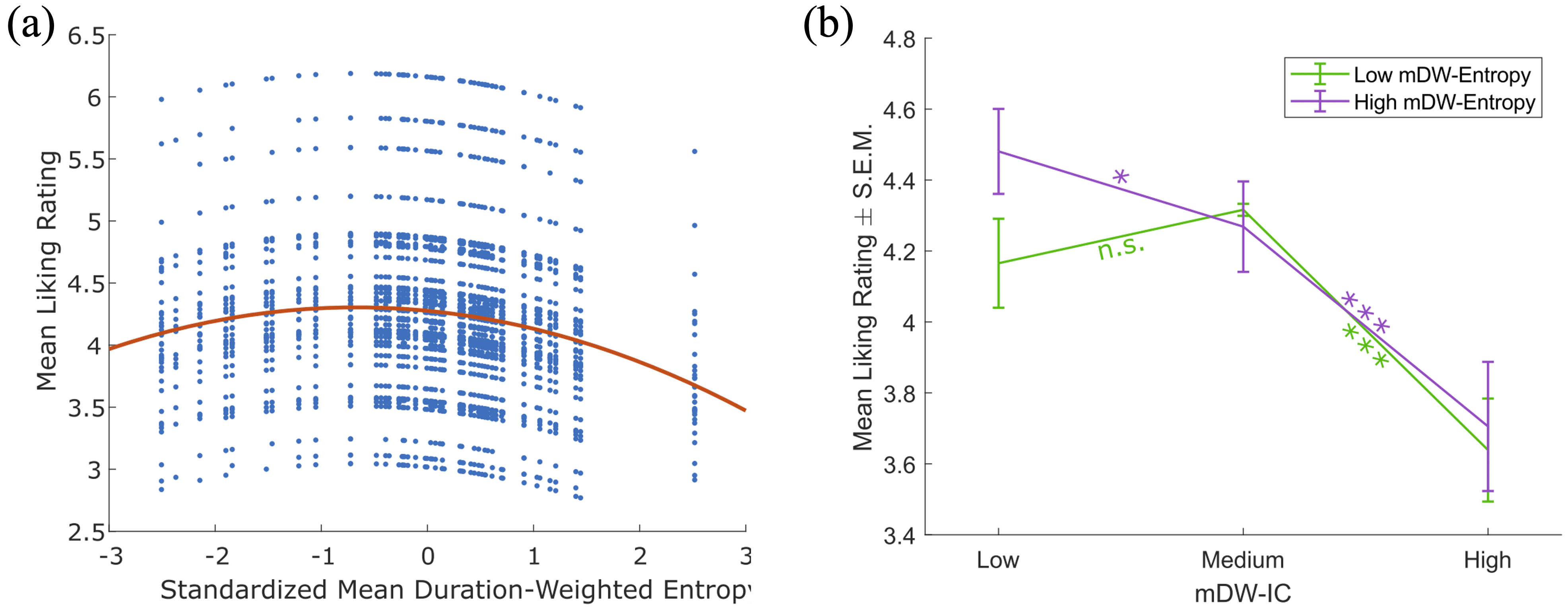} 
	\caption{Behavioral effects of unpredictability and uncertainty on subjective liking.
		(a) Relationship between standardized mDW-Ent and liking ratings, showing a significant quadratic effect;
		(b) interaction effect of different levels of mDW-IC on liking ratings under high and low mDW-Ent conditions.
		Results indicate that under high uncertainty, stimuli with low unpredictability elicit the highest liking, whereas under low uncertainty, liking becomes more sensitive to variations in mDW-IC.
		Significance levels: n.s. = not significant, * $p<0.05$, ** $p<0.01$, *** $p<0.001$. \\
		\textit{Source}: The figure is reproduced from Ref. \cite{gold2019predictability}.}
	\label{fig:music_behavioral_effects}
	\vspace{0pt}
\end{figure*}

In the domains of culture and art, predictability research addresses both the regularities underlying musical experience and the evolutionary mechanisms of cultural transmission. Gold \textit{et al.}~\cite{gold2019predictability} systematically investigated the roles of predictability and uncertainty in shaping aesthetic liking in music.  
The core question was: why do people find certain melodies more “pleasant” than others? To answer this, the authors employed the Information Dynamics of Music model (IDyOM) to precisely quantify the information content (IC) and entropy of musical events, representing unpredictability and uncertainty, respectively. The experimental results revealed a classic inverted U-shaped relationship: listeners generally preferred melodies within a moderate complexity range. Music that was too predictable was perceived as monotonous, while completely random melodies sounded chaotic. As shown in Fig.~\ref{fig:music_behavioral_effects}(a), liking ratings exhibited a significant quadratic dependence on mean duration-weighted entropy (mDW-Ent), peaking at intermediate complexity levels and forming a characteristic nonlinear curve. Further mechanistic analysis, illustrated in Fig.~\ref{fig:music_behavioral_effects}(b), revealed that under high-uncertainty conditions, low unpredictability (i.e., more stable structures) elicited the highest liking ratings. Conversely, under low-uncertainty conditions, liking ratings were more sensitive to changes in mean duration-weighted IC (mDW-IC). In other words, human musical enjoyment does not arise purely from unexpected surprises but rather from a dynamic balance between stability and novelty. The study also identified familiarity effects and individual differences. In a second experiment, repeated exposure to the same musical fragments reduced overall liking but did not alter the fundamental inverted U-shaped relationship, indicating that the preference for moderate complexity is a robust mechanism rather than a mere novelty-driven effect. Moreover, musically trained participants exhibited greater sensitivity to predictive structures, showing sharper preference curves, suggesting that musical expertise influences how individuals engage predictive mechanisms during aesthetic experience.  

\begin{figure*}[!t]
	\centering
	\setlength{\abovecaptionskip}{0pt}
	\setlength{\belowcaptionskip}{0pt}
	\includegraphics[width=0.4\textwidth]{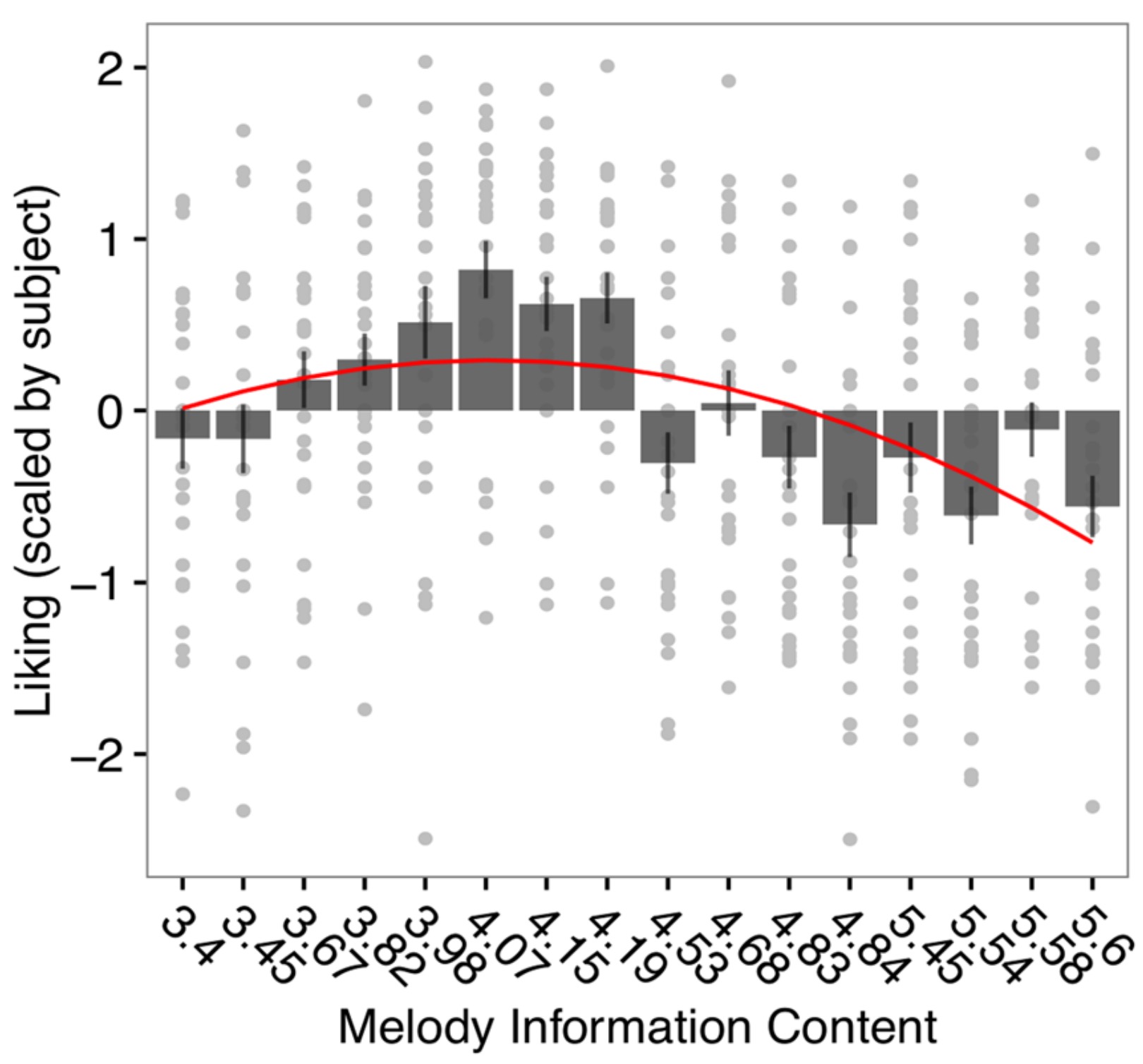}
	\caption{Relationship between melodic predictability and subjective liking ratings.
		The horizontal axis represents the IC of melodies, and the vertical axis shows the standardized liking ratings.
		Results indicate an inverted U-shaped relationship, where liking increases with melodic IC up to a moderate level and then declines as IC continues to rise. \\
		\textit{Source}: The figure is reproduced from Ref. \cite{bianco2019music}.}
	\label{fig:bianco_liking_predictability}
	\vspace{0pt}
\end{figure*}

Bianco \textit{et al.}~\cite{bianco2019music} conducted a two-stage experiment to systematically examine the roles of musical predictability and subjective liking in shaping aesthetic experience and learning performance. The experimental materials consisted of newly composed melodic sequences, in which the IC of contextual notes was strictly manipulated to be either high or low, while the final target note was held constant. This design ensured that any observed effects were primarily driven by differences in the predictability of the preceding musical context. At the subjective level, the results revealed a classic inverted U-shaped relationship between melodic liking and predictability (see Fig.~\ref{fig:bianco_liking_predictability}). Listeners consistently preferred melodies of moderate complexity—highly predictable melodies were perceived as monotonous, whereas completely random ones lacked structure and appeal. This pattern suggests that the optimal point for auditory liking lies not at the extremes of certainty or uncertainty but rather in the balanced middle ground between the two. The study thereby empirically validated the “optimal complexity hypothesis,” which posits that aesthetic enjoyment is strongest when familiarity and novelty are in dynamic balance.

At the physiological level, the authors further verified this subjective pattern using pupillary dilation responses. They found that predictable melodies generally elicited greater pupil dilation than unpredictable ones, and this difference was more pronounced among participants who reported higher subjective liking, indicating a synergistic relationship between predictability and liking in modulating physiological arousal. At the learning level, the experiment also demonstrated that predictable melodic contexts significantly improved performance accuracy and rhythmic consistency, particularly during early learning stages. Under unpredictable conditions, however, higher subjective liking partially compensated for reduced learning efficiency. Overall, this study revealed a multilayered coupling between musical predictability, aesthetic liking, and learning performance. The inverted U-shaped association between subjective liking and predictability provides key evidence for understanding the mechanisms underlying aesthetic enjoyment and offers experimental validation for the psychological foundations of predictive processing theory in musical experience.

\section{Discussions}

In this review, we have provided a systematic overview of major advances in the study of the predictability of complex systems, with particular emphasis on three core directions: time series, complex networks, and dynamical systems. We have compared different methods in terms of their theoretical foundations, scope of applicability, and forecasting performance, highlighting both their strengths and limitations. Although recent years have witnessed significant progress in the development of indicator systems, the expansion of methodological frameworks, and interdisciplinary applications (e.g., human mobility, financial markets, disease transmission, and climate systems), several fundamental issues concerning the predictability of complex systems remain unresolved. In particular, under conditions of nonstationarity, multiscale dynamics, and high-dimensional strong coupling, key challenges remain. These include how to reliably quantify predictability within a unified framework, how to delineate the boundaries of predictability, and how to develop robust strategies for its enhancement. Together, these issues continue to constitute the central challenges in the field.

In the context of time-series predictability, each of the symbolic and numerical approaches exhibits distinctive strengths and limitations. Symbolic time-series methods discretize system states into a finite alphabet and carry out forecasting by leveraging historical regularities and contextual information, while theoretical upper bounds on predictability can be estimated using tools such as Fano’s inequality or Bayesian error rates. These methods are advantageous for characterizing discrete regularities but tend to depend heavily on historical distributions, making them less effective at handling nonstationarity or multiscale variability. By contrast, numerical time-series approaches can capture fine-scale dynamics under continuous observations and employ tools such as permutation entropy, the $\kappa$ index, and hold-out strategies to quantify the balance between regularity and randomness. However, these methods are highly sensitive to data length, embedding parameters, and noise levels, and they lack a unified framework for robustness assessment. Overall, both symbolic and numerical approaches encounter performance bottlenecks when dealing with nonstationary, multiscale, and noisy environments. This indicates an urgent need for future research to develop general-purpose indicators that integrate symbolic and numerical information, reconcile continuity with discreteness, and provide theoretical guarantees of robustness, thereby offering more reliable support for time-series forecasting in complex systems.

In the study of the predictability of complex networks, different methods exhibit complementary strengths and weaknesses. Spectral approaches leverage the eigenvalues and eigenvectors of adjacency or Laplacian matrices to reveal global structural regularities, thereby quantifying the predictability of potential links. They are effective in capturing overall structural coherence but are less sensitive to local dynamic evolution. Information-theoretic methods, by contrast, quantify network uncertainty and redundancy from probabilistic and coding perspectives, though their stability is limited in sparse networks, extremely dense regimes, or noisy conditions. Structure-based methods such as low-rank sparse representations, structural consistency analysis, or critical-link analysis can highlight the contributions of subgraphs, modules, or control pathways to overall predictability, but they often lack robust generalization in heterogeneous, multilayer, or dynamic networks. Overall, existing approaches face significant limitations in addressing heterogeneity, dynamic evolution, and noisy environments--this underscores the need for future research to develop a unified predictability framework that integrates global and local perspectives, static and dynamic features, as well as stability and generalization. Such a framework would in turn enhance our understanding of the intrinsic regularities underlying network evolution.

For high-dimensional and strongly nonlinear dynamical systems, the problem of predictability is particularly formidable. Traditional indicators such as the maximal Lyapunov exponent, Kolmogorov-Sinai (KS) entropy, and mutual information can effectively characterize orbit sensitivity and variable dependence in low-dimensional systems, but their performance deteriorates markedly under conditions of high dimensionality, multiscale interactions, and nonlinear coupling. Perturbation growth often exhibits spatiotemporal nonuniformity, frequent local extreme events, and complex error-propagation pathways, making it difficult for any single global indicator to provide a comprehensive assessment of system predictability. Against this backdrop, artificial-intelligence methods have substantially expanded predictive capability. Reservoir computing and autoregressive reservoir neural networks achieve high-dimensional forecasting by learning local spatiotemporal features; models combining deep autoencoders with Koopman operators approximate nonlinear evolution with near-linear dynamics in embedded spaces; and physics-informed neural networks as well as deep-learning approaches for dynamical systems improve predictive consistency and effectiveness by incorporating physical constraints or local attractor structures. However, these methods remain constrained by insufficient interpretability, uncertain generalization, and sensitivity to data availability and prior assumptions, making it difficult to ensure robust long-term forecasting. In the realm of extreme-event prediction, reduced-order modeling, transient-instability analysis, and variational optimization provide new tools for identifying critical perturbation directions. For example, optimal time-dependent (OTD) modes and variational principles have been employed to capture rapidly growing disturbances and the strongest finite-time responses, respectively. Yet, these methods are still limited by the curse of dimensionality, data scarcity, and nonlinear complexity. Overall, the predictability of high-dimensional and nonlinear systems is constrained by the scalability of classical indicators, the interpretability and generalization of AI-based approaches, and insufficiencies in data and modeling for extreme-event forecasting. Future research will need to establish a unified framework that integrates the strengths of classical dynamics, statistical analysis, and AI methods. Such a framework should be able to capture transient local features while preserving global stability, and it should ensure interpretability and robustness under high-dimensional, nonstationary, and multiscale conditions.

It is worth noting that cross-domain applications place higher and more demanding requirements on the predictability of complex systems. Different systems exhibit pronounced differences in metrics, time scales, and predictability limits: in human mobility, predictability is constrained by sampling granularity, behavioral patterns, and social structures; in business and financial systems, sampling frequency and volatility determine the range and scale dependence of forecasting potential; in disease transmission and climate systems, strong spatiotemporal coupling causes predictability to decay rapidly across phases or regimes; while in domains such as education, culture, and arts, forecasting relies more on cognitive mechanisms and collective evolutionary processes. These cases demonstrate both the cross-domain applicability of predictive frameworks and the diversity of boundary conditions and mechanisms. Different approaches thus provide complementary perspectives: information-theoretic measures emphasize trade-offs between input compression and predictive accuracy; complex-network and controllability analyses supply structural priors for dynamical modeling; and the integration of multiscale time-series models with network–dynamics models holds promise for enhancing predictive performance through cross-domain information sharing. Future progress will require breakthroughs in unified metrics, standardized benchmarks, and deeper integration of mechanisms with data, so as to achieve forecasting that is both robust and interpretable.

A core direction for future research is the establishment of a “complexity–predictability mapping framework”. Complexity should not be understood merely as a single-dimensional indicator, such as entropy, the Lyapunov exponent, or dimension. Instead, it should be viewed as a multidimensional construct encompassing nonlinear coupling, randomness, volatility, emergence, fragility, adaptability, and resilience—attributes that may jointly and nonlinearly determine the boundaries of system predictability. Accordingly, it is essential to develop theoretical frameworks that integrate multidimensional complexity and establish explicit links to predictive limits, thereby uncovering common regularities across systems and guiding cross-system analyses. At the same time, the deep integration of theoretical methods and artificial intelligence remains in its early exploratory stage. Current attempts primarily focus on feeding theoretical features into AI or adding constraints to models, whereas more promising directions include distilling dynamical and statistical mechanisms into higher-order features for AI to learn, embedding conservation laws, nonlinear couplings, or causal relationships explicitly into network architectures, and building ensemble frameworks that combine theory-driven and data-driven approaches so as to improve both interpretability and robustness of predictive models. What is more interesting is that a new category of complex systems will certainly emerge in the future, in which all individuals are AI agents with generative capabilities. These agents will also possess the abilities of perception, learning, and decision-making, and form complex relationships and interactions with one another. Through the interactions of these agents, as well as the emergent rules and even institutions, we can reflect on history and look ahead to the future. We are curious about two questions: first, whether some new unpredictable phenomena will arise in such new-type complex systems; second, which—between such new-type complex systems and real human societies—exhibits higher predictability.

Against this backdrop, several pressing challenges come to the forefront. These include the operational quantification of multidimensional complexity, the extraction of key dynamical information under high-dimensional and nonstationary conditions, understanding how error-propagation pathways and extreme events constrain predictive limits, the construction of collaborative mechanisms between theory and AI, cross-domain validation and standardization, and high-sensitivity forecasting of extreme and rare phenomena. Progress along these directions is expected to advance research on the predictability of complex systems from an empirically driven endeavor toward a more systematic “science of prediction.” Such a shift would not only elucidate the regularities of system behavior but also provide theoretical and methodological support for the management, intervention, and optimization of cross-domain complex systems.

\appendix
\section*{Appendix A. Notation Summary}
\addcontentsline{toc}{section}{Appendix A. Notation Summary}

This appendix summarizes the main mathematical notations used throughout the paper. 
The definitions are grouped by thematic sections to facilitate cross-referencing and consistency across the text.

\begin{longtable}{p{0.25\linewidth} p{0.7\linewidth}}
	\caption{Notation Summary}\label{tab:notation-all}\\
	\hline
	\textbf{Symbol} & \textbf{Meaning} \\
	\hline
	\endfirsthead
	
	\multicolumn{2}{l}{\footnotesize\itshape (continued) Notation Summary} \\
	\hline
	\textbf{Notation} & \textbf{Meaning} \\
	\hline
	\endhead
	
	\hline
	\multicolumn{2}{r}{\footnotesize Table continues on the next page.} \\
	\endfoot
	
	\hline
	\endlastfoot
	
	\chapterdivider{Chapter 2: Predictability of Time Series}
	$\mathcal{Z} = \{Z_1, Z_2, \dots, Z_L\}$ & Time series (of length $L$) \\
	$Z_t$ & State at time step $t$ \\
	$h_{t-1} = \{Z_{t-1}, \dots, Z_1\}$ & Historical sequence (up to $t-1$) \\
	$\pi(h_{t-1})$ & One-step predictability upper bound: probability of the most likely state given history \\
	$\Pi(h_{t-1})$ & Average predictability under history $h_{t-1}$ \\
	$\Pi$ & Theoretical upper limit of overall time-series predictability (long-term average accuracy) \\
	$\Pi^{\max}$ & Maximum predictability (solved from $S = S_F(\Pi^{\max})$) \\
	$P(z \mid h_{t-1})$ & True conditional distribution of next state $Z_t = z$ \\
	$P_\alpha(z \mid h_{t-1})$ & Predictive distribution estimated by algorithm $\alpha$ \\
	$z_{\mathrm{MS}}$ & Most likely state \\
	$r_g$ & Radius of gyration \\
	$S(\cdot)$ & Entropy \\
	$S^{\text{rand}}$ & Random entropy (assuming all states are equally likely) \\
	$S^{\text{unc}}$ & Temporal-uncorrelated entropy \\
	$S^{\text{real}}$ & Real or empirical entropy (estimated from data) \\
	$S^{\text{est}}$ & Estimated entropy (e.g., Lempel–Ziv estimate) \\
	$S_F(p)$ & Fano function: $-p\log_2 p - (1-p)\log_2(1-p) + (1-p)\log_2(C-1)$ \\
	$S^{\text{est}}(A|B)$ & Cross-entropy rate: information content of sequence $A$ relative to $B$ \\
	$S^{\mathrm{perm}}$ & Permutation entropy \\
	$C$ & Size of candidate set \\
	$C_s$ & Original candidate set size (number of visited locations) \\
	$C_r$ & Refined candidate set size (number of directly reachable locations) \\
	$\ell$ & Number of returned items in Top-$\ell$ recommendation \\
	$r_i$ & Probability ratio of the $i$-th candidate relative to the maximum-probability item \\
	$st(\mathcal{T})$ & Stability index: proportion of unchanged consecutive states \\
	$reg(\mathcal{T})$ & Regularity index: degree of concentration in visited locations \\
	$\mathcal{R}_B$ & Bayes error rate \\

	\chapterdivider{Chapter 3: Predictability of Complex Networks}
	$G(V, E)$ & Undirected and unweighted network \\
	$N = |V|$ & Number of nodes in the network \\
	$M = |E|$ & Number of edges in the network \\
	$V = \{v_1, v_2, \dots, v_N\}$ & Node set of network $G$ \\
	$E = \{e_1, e_2, \dots, e_M\}$ & Edge set of network $G$ \\
	$U$ & Set of all possible edges in the network \\
	$U - E$ & Set of potential (unobserved) edges \\
	$E^T$ & Training set of edges used for link prediction \\
	$E^P$ & Testing set of edges used for link prediction \\
	$E^R$ & Remaining edges after perturbation $\Delta E$ \\
	$E^L$ & Top $L$ ranked edges forming a testing subset \\
	$\bm{D} = \{r_1, r_2, \dots, r_M\}$ & Ranking list of edge scores \\
	$C(N)$ & Set of all coordinate positions up to $N$ \\
	$\bm{A}$ & Adjacency matrix of the network \\
	$\tilde{\bm{A}}$ & Perturbed adjacency matrix \\
	$\bm{Z}$ & Representation matrix used in Structural Regularity computation \\
	$\bm{B}$ & Two-dimensional extended matrix \\
	$\bm{\widetilde{B}}$ & Filtered matrix \\
	$\bm{F}$ & Length of compressed binary sequence \\
	$E(G)$ & Energy of network $G$ \\
	$\widehat{E}(G)$ & Normalized network energy \\
	$R$ & Maximum compression length \\
	$R^2$ & Coefficient of determination \\
	$\lambda_i$ & $i$-th eigenvalue of adjacency matrix $\bm{A}$ \\
	$\lambda_1$ & Largest eigenvalue of $\bm{A}$ \\
	$\bm{x}_i$ & $i$-th eigenvector of $\bm{A}$ \\
	$\tau$ & Number of zero elements in $\bm{Z}$ \\
	$r$ & Rank of matrix $\bm{Z}$ \\
	$b_{\theta}$ & Threshold parameter in filtered matrix $\bm{\widetilde{B}}$ \\
	$\langle k \rangle$ & Average degree of the network \\
	$\langle d \rangle$ & Average shortest path length between nodes \\
	$a_i$ & Activation rate of edge $e_i$ \\
	$\delta_j$ & Normalized ranking variation of edge $e_j$ \\
	$p$ & Probability of link formation between two nodes \\
	$p^{H}$ & Perturbation ratio when sampling $\Delta E$ \\
	$\alpha, \beta$ & Fitting parameters in structural regularity ($\sigma_r$) computation \\
	$\Lambda^{v}_v$ & Maximum window size required for a new pattern to appear at position $v$ \\
	$u$ & Total number of samples in structural reciprocity index (SRI) \\
	$\sigma_c$ & Structural consistency index \\
	$\sigma_s$ & Network spectrum index \\
	$\sigma_e$ & Network energy index \\
	$\sigma_l$ & Compressed length index \\
	$\sigma_{l^*}$ & Normalized compressed length index \\
	$\sigma_t$ & Entropy rate index \\
	$\sigma_r$ & Structural regularity index \\
	$\sigma_d$ & Structural controllability index \\
	
	\chapterdivider{Chapter 4: Predictability of Dynamics}
	$P(X,Y)$ & Joint distribution of input variable $X$ and target variable $Y$ \\
	$\mathcal{L}_{\text{IB}}$ & Objective function of Information Bottleneck theory \\
	$I(\cdot;\cdot)$ & Mutual information \\
	$U_p$ & Historical input sequence \\
	$\hat{Y}_f$ & Intermediate representation sequence \\
	$Y_f$ & Future output sequence \\
	\(X_t\) & System state at time $t$ \\
	\(X_{<t}\) & Set of past states \\
	$H$ & Average Predictability Time (APT) \\
	$H(\tau)$ & $H(\tau) = \frac{1}{K} \, \mathrm{tr} \!\left[ \left( \Sigma_{\infty} - \Sigma_{\tau} \right) \Sigma_{\infty}^{-1} \right]$ \\
	$L_\epsilon(\hat{p}_x, x)$ & $\epsilon$-logarithmic scoring function \\
	$D_{KL}(\cdot||\cdot)$ & Kullback–Leibler divergence \\
	$P_e^{\min}$ & Minimum mean prediction error \\
	$S_b(\cdot)$ & Binary entropy function\\
	$P_{ee}$ & Probability of an extreme event \\
	$R_{t_{\zeta}}$ & Recurrence region \\
	$\rho_i$  & Canonical correlation coefficient \\
	$n(\beta)$ & Effective modal dimension \\
	$\Phi(\cdot)$ & Information storage (context-dependent) \\
	\(p(x)\) & Probability density function of system states \\
	$\#$ & Cardinality of a set \\
	$m(\cdot)$ & Geometric volume \\
	$x_1(t),x_2(t)$ & Trajectory functions \\
	$\lambda_m$ & Maximum Lyapunov exponent \\
	$T_{\lambda}$ & Lyapunov time, $T_{\lambda}=1/\lambda_{m}$ \\
	$v$ & Cross-distance scaling exponent \\
	$\mu$ & Centroid of the attractor \\ 
	$\langle\cdot\rangle$ & Statistical average \\
	$\rho$ & Coupling parameter in the Lorenz system \\
	$\delta$ & A small positive number, also used as a threshold \\
	$\sigma_i(t)$ & Instantaneous growth rate of OTD modes \\
	$\eta$ & Time delay \\
	\(\alpha_{\eta}(\zeta)\) & Local predictability index of reference state \(\zeta\) at delay \(\eta\) \\
	\( \mathrm{tr}[\cdot] \) & Matrix trace operator \\
	$d_x$ & Dimension of the system’s state space \\
	$p_{x_{1:T}}$ & True trajectory distribution \\
	$\tilde{p}^\star_{x_{1:T}}$ & Fuzzy predictive distribution closest to the true distribution within error bounds \\
	$\mathbf{x}^*$ & Precursor center of an extreme event \\
	$\mathbf{x}^{\text{pre}} $ & Current state of the system \\
	\(\{\mathbf{q}_i(t)\}_{i=1}^r\) & Orthogonal basis of perturbations \\
\end{longtable}

\bibliographystyle{elsarticle-num}

\bibliography{Reference}

\end{document}